%% file: Paper3.tex
\documentclass[fleqn,usenatbib]{mnras}

\usepackage[T1]{fontenc}
\usepackage{ae,aecompl}

\usepackage{graphicx}	
\usepackage{amsmath}	
\usepackage{pifont}
\usepackage{booktabs}
\usepackage{lscape}
\usepackage{longtable}
\usepackage{tabularx}
\usepackage{multirow}
\usepackage{layouts}
\usepackage{nicefrac}
\usepackage[normalem]{ulem}
\newcommand{\cmark}{\ding{51}} 
\newcommand{\xmark}{\ding{55}} 
\input{Commands.tex}

\setcitestyle{citesep={,}}
\newcolumntype{Y}{>{\centering\arraybackslash}X}
\usepackage[english]{babel}
\addto\extrasenglish{%
}
\addto\extrasenglish{%
}
\addto\extrasenglish{%
}
\title[Ionised jets from MYSOs]{A Galactic survey of radio jets from massive protostars}

\author[S. J. D. Purser, S. L. Lumsden, M. G. Hoare and S. Kurtz]{S. J. D. Purser$^{1,2}$\thanks{E-mail:
purser@cp.dias.ie}, S. L. Lumsden$^{2}$, M. G. Hoare$^{2}$ and S. Kurtz$^{3}$\\
$^{1}$School of Cosmic Physics, Dublin Institute for Advanced Studies, 31 Fitzwilliam Place, Dublin 2, Ireland\\
$^{2}$School of Physics and Astronomy, University of Leeds, Leeds LS2 9JT, UK\\
$^{3}$Instituto de Radioastronom\'{i}a y Astrof\'{i}sica, Universidad Nacional Aut\'{o}noma de M\'{e}xico, 58089 Morelia,  Michoac\'{a}n, M\'{e}xico}

\date{Accepted XXX. Received YYY; in original form ZZZ}

\pubyear{2018}

\usepackage{newtxtext}
\usepackage[varvw]{newtxmath}
\begin{document}
\label{firstpage}
\pagerange{\pageref{firstpage}--\pageref{lastpage}}
\maketitle

\begin{abstract}
In conjunction with a previous southern-hemisphere work, we present the largest radio survey of jets from massive protostars to date with high-resolution, ($\sim 0\farcs04$) Jansky Very Large Array (VLA) observations towards two subsamples of massive star-forming regions of different evolutionary statuses: 48 infrared-bright, massive, young, stellar objects (MYSOs) and 8 infrared dark clouds (IRDCs) containing 16 luminous ($\Lbol>10^3\Lsol$) cores. For $94\%$ of the MYSO sample we detect thermal radio ($\alpha \geq -0.1$ whereby $S_\nu \propto \nu^\alpha$) sources coincident with the protostar, of which $84\%$ (13 jets and 25 candidates) are jet-like. Radio luminosity is found to scale with $\Lbol$ similarly to the low-mass case supporting a common mechanism for jet production across all masses.  Associated radio lobes tracing shocks are seen towards $52\%$ of jet-like objects and are preferentially detected towards jets of higher radio and bolometric luminosities, resulting from our sensitivity limitations. We find jet mass loss rate scales with bolometric luminosity as $\jml \propto \Lbol^{0.9\pm0.2}$, thereby discarding radiative, line-driving mechanisms as the dominant jet-launching process. Calculated momenta show that the majority of jets are mechanically capable of driving the massive, molecular outflow phenomena since $p_{\rm jet}>p_{\rm outflow}$. Finally, from their physical extent we show that the radio emission can not originate from small, optically-thick \textsc{Hii} regions. Towards the IRDC cores, we observe increasing incidence rates/radio fluxes with age using the proxy of increasing luminosity-to-mass $\left( \nicefrac{L}{M} \right)$ and decreasing infrared flux ratios $\left(\nicefrac{S_{70\um}}{S_{24\um}}\right)$. Cores with $\nicefrac{L}{M}<40\,\Lsol\Msol^{-1}$ are not detected above ($5.8\GHz$) radio luminosities of $\sim1\mJy\kpc^2$.
\end{abstract}


\section{Introduction}
\label{sec:intro}
Massive star ($> 8\Msol$) formation is a topic whose understanding is limited, and complicated, by many observational issues. Galactic sites of massive forming stars are distant ($>700\pc$) and rarely isolated, the formation process lasts only $\sim10^5\yr$ \citep{Davies2011} in a completely enshrouded (by optically thick dust and gas) state and is rare in comparison to low mass star formation \citep{KroupaIMF}. Consequently, multiple formation mechanisms are consistent with current observations, including the turbulent core model \citep{McKeeTan2003}, competitive accretion model \citep{Bonnell2001a}, or varying contributions of each.

Furthering our understanding of this topic necessitates the use of large surveys, utilising observations at both the highest angular resolutions possible and at wavelengths which are optically thin to the enshrouding material. Hence, statistically-sized surveys using radio interferometers are one of the best possible means to refine our understanding of star formation at these mass regimes.

A ubiquitous feature of low \citep{Anglada1995,Furuya2003,AMI2012} and high \citep{Purser2016} mass forming stars are collimated, high-velocity ($\gtrsim300\kmps$), ionised jets of material ejected along the rotation axes of accretion discs which surround forming stars. Many mechanisms may launch these jets from radiation \citep[via line driving][]{Proga1998} to magneto-centrifugal forces. For the latter, opinion is divided whether X-wind \citep{Shu1994} or disc-wind \citep{BlandfordPayne1982} models are likely to be more dominant. One of the main challenges with high mass star formation is that, while low mass stars are almost completely convective, high mass stars are more radiative. This challenges the production of protostellar magnetic fields and therefore the launch of jets by such fields, which in themselves are a necessary component of the X-wind model. However a study by \citet{Hosokawa2010} showed that a massive young stellar object (MYSO) bloats during its formation and consequently evolves through both convective and radiative phases, thereby suggesting the production of stellar magnetic fields is possible. Thus, the main focus of jet studies has always been in discerning between these possible mechanisms.

Our uncertainties have been exacerbated by the factors discussed above, but also due to the small number of large scale surveys for jets conducted, until very recently. \citet{Guzman2012} surveyed a sample of 7, southern, MYSOs selected on the basis of their thermal radio spectral indices ($\alpha>-0.1$ whereby $S_{\nu} \propto \nu^\alpha$), modest radio luminosities in comparison to their large ($>2\times10^4\Lsol$) bolometric luminosities, and reddened infrared spectra. This led to the identification of 2 ionised jets, 3 hypercompact \textsc{Hii} (\textsc{HCHii}) and 2 ultracompact \textsc{Hii} (\textsc{UCHii}) regions. More recently the POETS survey \citep{Moscadelli2016,Sanna2018} approached sample selection differently by selecting distance-limited ($<9\kpc$), radio-weak ($<50\mJy$) YSOs exhibiting strong water maser emission (a possible jet-shock tracer). They found a similar incidence rates for jets in their sample ($45\%$) and correlation between the jets' radio luminosities and their parental MYSOs' bolometric luminosities,  as \citet{Purser2016}. \citet{Rosero2019} also reported similar jet incidence rates ($>30\%$) as well as the same scaling relation for jet and YSO luminosity as those works listed above.

A southern-hemisphere, radio survey \citep[][hereafter P16]{Purser2016}\defcitealias{Purser2016}{P16} conducted observations at 4 frequencies from $5$ to $22\GHz$ towards a sample of 49 objects, of which 28 were identified as either ionised jets or candidates (the rest being \textsc{Hii} regions, radiatively driven disc-winds (e.g.\ S140 IRS 1) or of an unknown nature). Their sample drew directly from the Red MSX Source (RMS) survey \citep{Lumsden2013} and was comprised of a smaller, distance-limited ($<7\kpc$) subsample (34 objects) spaced evenly over a wide range of bolometric luminosities, and with a greater constraint on radio-to-infrared flux ratios than \citet{Guzman2012}. Interestingly, that work showed that approximately half of the identified jets were associated with non-thermal emission, a ratio also observed by \citet{Moscadelli2016} and a recent, northern-hemisphere survey of non-thermal emission towards MYSOs \citep{Obonyo2019}. Whilst the central, thermal radio-jet is always coincident with the MYSO's infrared position, the non-thermal emission was observed as radio lobes aligned along the jet's axis, spatially distinct from the thermal radio/IR source. This emission was determined to be synchrotron emission, showing the presence of magnetic fields, in agreement with magnetohydrodynamic launch/collimation mechanisms of jets. It is thought that these magnetic fields are shock-enhanced and originally stem in either/both the disc or ambient material \citep{Frank2000,GardinerFrank2000}. Due to the typical spectral indices found towards these lobes ($\bar{\alpha}=-0.55$) this emission was determined to originate in a shock-accelerated \citep[via the 1\textsuperscript{st} order Fermi mechanism, or diffusive shock acceleration,][]{Bell1978} population of relativistic electrons. These shocks are likely interactions of the jet with the ambient medium, or internal working surfaces within the jet resulting from variability. For the star formation paradigm as a whole they play an important role in local feedback through their production of low-energy cosmic rays \citep[e.g. the study of DG Tau A by][for the low-mass case]{Ainsworth2014} and contributions to turbulent support of the parental cloud.

As a complement to \citetalias{Purser2016}, this work performs a similar, RMS survey-derived, survey towards a northern sample of $48$ MYSOs utilising the VLA in its most extended configuration and completing a Galactic radio survey of jets. The main goal is to establish a sample of identified, northern, ionised jets to augment the southern sample, as well as provide a set of Q-band, matching-beam observations for a future C-band e-MERLIN legacy survey\footnote{\url{http://www.e-merlin.ac.uk/legacy/projects/feedbackstars.html}}. Further to this, we investigate the emergence of collimated outflow phenomena towards even earlier stages of massive star formation in the cores of IRDCs, of which we have chosen 8 fields from previous millimetre surveys. Full details of the sample, its constituting two subsamples and their selection procedure are discussed in \autoref{sec:sampleselection}. In \autoref{sec:obs} we describe the VLA observations conducted towards our sample and the performance of the VLA when observing ionised jets. These results are subsequently presented in \autoref{sec:results}. In \autoref{sec:analysis} we investigate the properties of the radio emission and compare them to molecular outflows and accretion processes. Finally, we summarise the preceding sections and establish the conclusions stemming from them in \autoref{sec4:conclusions}.

\section{The Sample}
\label{sec:sampleselection}
Membership criteria for the observational sample of this work have been tailored to incorporate a wide range of evolutionary states, from the earliest phase of massive star formation represented by the IRDC stage, to the mid-infrared bright, pre-UC\textsc{Hii} phase. Increasing luminosities arise from growth of the central object as it evolves, while changes in infrared colours are brought about by the evolution of a YSO's spectral energy distribution (SED). Selection criteria for the sample are therefore based on bolometric luminosity and infrared colour, assumed to be indicators of mass and evolutionary status, respectively.

\subsection{The IRDC sample}
\label{sec:irdcsample}
Evolutionarily-speaking, the first subsample is based upon the work by \citet{Rathborne2010}, who surveyed a number of IRDCs at mm wavelengths, and subsequently derived many of their filial cores' physical properties. They employed the same classification system as \citet{Chambers2009}, whereby each core was categorised based upon observed, infrared evolutionary indicators. Specifically these indicators include the presence of excessive $4.5$, $8$ or $24\um$ emission \citep[the former giving rise to the widely observed `extended green objects', or EGOs, in the GLIMPSE survey; see][]{Cyganowski2011}). Core classifications include (in order of increasing evolutionary status) quiescent (Q), intermediate (I), active (A) or red (R). Details for each can be found in Table 4 of \citet{Chambers2009}.

A requirement of the IRDCs selected for observation was that they contain cores of a variety of classifications and therefore numerous stages can be seen simultaneously within the primary beam of our C-band observations. Quantitatively, the selection criteria included IRDCs harbouring cores with luminosities $>10^3\Lsol$, infrared flux ratios, $\nicefrac{S_\mathrm{70\um}}{S_\mathrm{24\um}}$, $\gtrsim50$ and distances $<7\kpc$. This led to a sample size of 8 IRDCs, incorporating 45 cores within the field of view (see \autoref{sec:obsinfo}), of which 16 have bolometric luminosities greater than the cut-off (see \autoref{tab:irdctargets}). All cores of the sample with derived values for $\nicefrac{S_\mathrm{70\um}}{S_\mathrm{24\um}}$ are shown in blue in \autoref{fig:firlumratios}.

\input{Tables/IRDCTargets}
\input{Tables/MYSOTargets}

\subsection{The MYSO sample}
\label{sec:mysosample}
\begin{figure}
\centering
\includegraphics[width=3.32in]{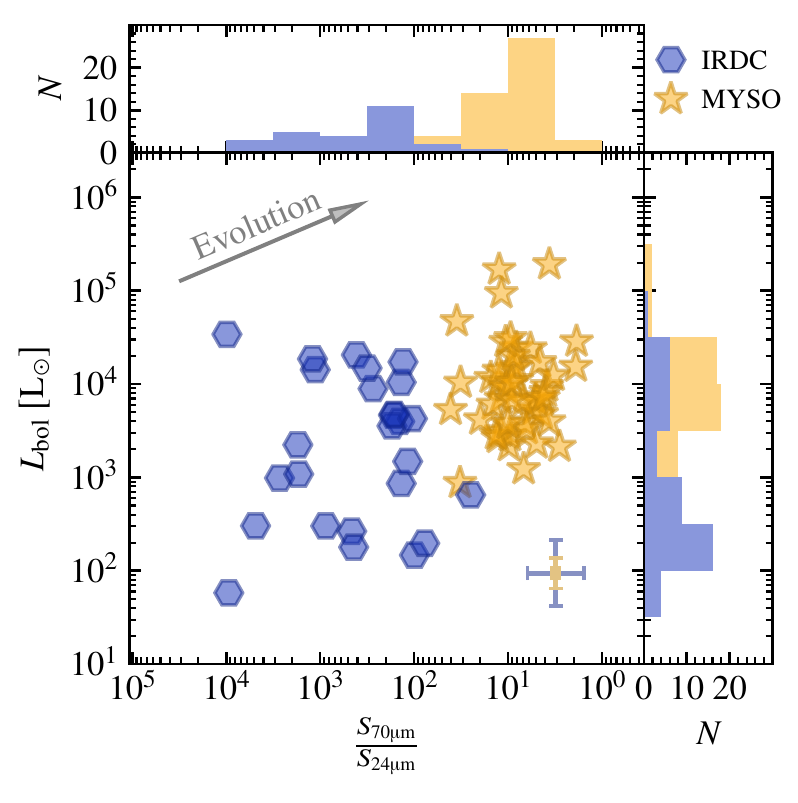}
\caption[Mid-infrared flux ratios against bolometric luminosity for the IRDC cores/MYSOs of the sample]{Mid-infrared, $S_\mathrm{70\um}$ to $S_\mathrm{24\um}$ flux ratios against bolometric luminosity for IRDC cores and MYSOs. The arrow represents the general evolutionary trend for an IRDC core through to an MYSO. Histograms of both mid-infrared ratios and bolometric luminosity are shown.}
\label{fig:firlumratios}
\end{figure}

For the second subsample, the RMS survey database\footnote{\url{http://rms.leeds.ac.uk/cgi-bin/public/RMS_DATABASE.cgi}} was used to draw mid-infrared bright targets, representing a more-evolved phase of massive star formation than the IRDCs of \autoref{sec:irdcsample}. For both their luminosities and distances, the same criteria as for the IRDCs \citep[$>10^3\Lsol$ and $<7\kpc$,][]{Mottram2011a} are imposed. Other requirements include a $21\um$ to $8\um$ flux ratio $>2$ \citep{Lumsden2013} and no previous radio detection \citep[or flux of $<1\mJy$,][]{Urquhart2009VLA} in order to ensure the sources are not in the UC\textsc{Hii} phase. These extra criteria ensure selection of different evolutionary stages compared to those comprising the sample of \autoref{sec:irdcsample}. Comparing their $S_\mathrm{70\um}/S_\mathrm{24\um}$ ratios to the IRDCs, the mid-infrared bright MYSOs have values of $\lesssim100$ (shown in yellow in \autoref{fig:firlumratios}) meaning that between the MYSO and IRDC subsamples, a continuous range in evolutionary status is represented (highlighting the previous point). Near-infrared ancillary data \citep{Cooper2013} also show that many are still accreting due to the presence of CO bandhead emission \citep[e.g.][]{Ilee2013} and likely driving ionised winds from their weak Br$\gamma$ emission (relative to that expected from \textsc{Hii} regions). In total, this provides a sample size of 48 MYSOs which are listed, along with their basic properties, in \autoref{tab:mysotargets}.

\subsection{Distances, core masses, bolometric luminosities and infrared ratios}
\label{sec:distslumsirratios}
To ensure proper weighting of data-points during the employment of fitting algorithms, rather than use a typical error of $34\%$ for $\Lbol$ \citep{Mottram2011b}, we instead recalculated $\delta\Lbol$ for each target. To do this, standard errors were required for both bolometric flux and distance.

Bolometric fluxes and their errors were calculated by integrating the area under the infrared SED of each core of the IRDC subsample, for which a greybody function was used to fit to available infrared fluxes \citep[see \textcolor{blue}{\S\,$3.2$} of][]{Rathborne2010}. In cases where the SED could not be reliably fit, bolometric fluxes were derived from the given luminosities/distances of \citet{Rathborne2010}.

Core masses were calculated using \autoref{eq:coremass} \citep{Hildebrand1983}, with the same set of assumptions as used by \citet{Rathborne2006} (a dust opacity at $1.2\,\mathrm{mm}$, $\kappa_{\rm 1.2 \, mm}=0.1\,\mathrm{cm^{2}\,g^{-1}}$ and dust-to-gas-mass ratio, $R=100$). Temperatures from our SED fits were used or, if not available, those derived by \citet{Rathborne2010}. Our own distance estimates were also used (see below).

\begin{equation}
	M_{\rm core}=\frac{R \, F_{1.2\,\rm mm}D^2}{\kappa_\nu B_\nu\left(T_{\rm dust}\right)} \label{eq:coremass}
\end{equation}

\noindent where $R$ is the dust-to-gas-mass ratio, $F_\mathrm{1.2 \, mm}$ is the core's integrated flux at $1.2\,\mathrm{mm}$, $D$ is distance, $\kappa_{1.2\, \rm mm}$ is the dust opacity at $1.2\,\mathrm{mm}$ and $B_\nu\left(T_{\rm dust}\right)$ is the blackbody function.

Infrared flux ratios ($\nicefrac{S_{70\um}}{S_{24\um}}$) were derived from fitted SEDs. However, in some cases the measured $24\um$ fluxes were not fit well by the greybody function and the ratio was instead calculated from the measured $24\um$ flux and the $70\um$ flux calculated from the greybody SED fit (which ignored flux measurements at $24\um$) instead. For the MYSO sample, bolometric fluxes and their associated errors were taken from \citet{Mottram2011a}, while $\nicefrac{S_{70\um}}{S_{24\um}}$ ratios were calculated using $S_{70\um}$ and $S_{24\um}$ values interpolated between IRAS, IGA, AKARI and/or WISE fluxes.

Distance errors were slightly more complex to quantify. While for distances determined via parallax measurements these errors are easily calculated, for those distances determined via kinematic analyses (i.e. the whole IRDC sample and most of the MYSO sample) another approach had to be used. A work by \citet{Wenger2018} used Monte Carlo techniques to determine kinematic distances, and their errors, to high-mass star formation regions in the Galaxy (their `Method C'), whose determined values were within a median difference of $14\%$ of those determined via parallax. To use this Monte Carlo technique, values for $v_\mathrm{LSR}$ were taken from \citet{Rathborne2006} and the RMS survey \citep{Urquhart2008}. As a consistency check, distances calculated from this approach were compared to those derived using the alternate model of \citet{Reid2019}. The two distances strongly agreed with eachother, with a mean/standard deviation of ${d_{\rm W18}}/{d_{\rm R19}} = 1.05  \pm 0.25$ where $d_{\rm W18}$ and $d_{\rm R19}$ are the distances computed via \citet{Wenger2018} and \citet{Reid2019}, respectively.

It is important to note that although \citet{Mottram2011a} used previous kinematic distance estimates (with a general error of $\delta D=\pm1\kpc$) to determine extinction, and therefore $F_\mathrm{bol}$, the values derived for $F_\mathrm{bol}$ generally agreed for near/far kinematic distances (their Figure 2). Therefore using their values for $F_\mathrm{bol}$ should not significantly affect the estimates for $\Lbol$ and $\delta\Lbol$ presented in this work. 

With these calculations of both bolometric fluxes and distances, as well as their errors, the calculation of bolometric luminosity and its error is a trivial exercise. Results for these are shown in \autoref{tab:irdctargets} and \autoref{tab:mysotargets} for the IRDC and MYSO sample respectively. It should be noted that for 26 of the 48 MYSOs we used more accurate distances from the literature (i.e.\ maser parallax measurements). 

\section{Observations}
\label{sec:obs}

\subsection{Observational information}
\label{sec:obsinfo}
All observations were taken with the Very Large Array (VLA) in its A-configuration, between 13\textsuperscript{th} October 2012 and 27\textsuperscript{th} December 2012 for $5.8\GHz$ observations (project code 12B-140; PI - M. G. Hoare) and between 16\textsuperscript{th} March 2014 and 27\textsuperscript{th} July 2015 for $44\GHz$ observations  (project codes 14A-141 and 15A-238; PIs - S. L. Lumsden and S. J. D. Purser, respectively). The WIDAR correlator was set up in full continuum mode, with bandwidths of $2\GHz$ ($16$ spectral windows of $64\times2\MHz$ channels with 8-bit samplers) and $8\GHz$ ($64$ spectral windows of $64\times2\MHz$ channels with 3-bit samplers) centred on frequencies of $5.8$ and $44\GHz$ respectively, which we refer to as C and Q-bands. In the A-configuration, the VLA has a minimum and maximum baseline length of $680\,\mathrm{m}$ and $36400\,\mathrm{m}$, corresponding to C and Q-band largest, recoverable, angular scales of $8\farcs9$ and $1\farcs2$ respectively. Synthesised beam widths of $\sim0\farcs3$ and $\sim0\farcs04$ are typical, while primary beam sizes at FWHM are $\sim430\arcsec$ and $\sim60\arcsec$.

In total, 56 target fields were observed at C-band and 49 fields at Q-band comprising observations of the 8 IRDCs and 48 MYSOs at both C and Q-bands, of which 38 were observed only at the lower frequency. Cores used as pointing centres for their IRDC complexes are denoted with an asterisk next to their name in \autoref{tab:irdctargets}. However, G033.69$-$00.01, which is a relatively extended complex, was observed at C-band with a pointing centre maximising the fraction of the IRDC in high-response areas of the primary beam ($\alpha_{\rm J2000}=18\rahr52\ramin54\fs0$ and $\delta_{\rm J2000}=00\degr39\arcmin39\arcsec$). At Q-band, the smaller primary beam necessitated a mosaic of 3 pointings with one pointing centre mirroring that at C-band, and the other two using coordinates of $\alpha_{\rm J2000}=18\rahr52\ramin49\fs9$, $\delta_{\rm J2000}=00\degr37\arcmin57\arcsec$ and $\alpha_{\rm J2000}18\rahr52\ramin58\fs1$, $\delta_{\rm J2000}=00\degr41\arcmin20\arcsec$. For the MYSO sample, the pointing centres are the coordinates of the MYSOs themselves as given in \autoref{tab:mysotargets}.

Depending on the LST of the observation, different flux and bandpass calibrators were used to bootstrap the flux density scale and calibrate the frequency-dependent gains. For calibration of the time-varying gains, a phase calibrator was observed every $12\min$ at C-band, for between $30-60\s$ (dependent on calibrator flux). At Q-band, due to increased atmospheric instability, a calibrator cycle time of $2\min$, including slew times and phase-calibrator scan lengths ($30-40\s$), was adopted. On-source times for the science targets were $24-26$ and $15-17\min$ achieving a theoretical rms noise of $6$ and $27\uJy\,{\rm beam}^{-1}$ at C and Q-bands respectively. Target fields in the MYSO sample labelled between G$118$ and G$148$ only received half of the Q-band observing time required to achieve this sensitivity, resulting in a theoretical rms noise of $38\uJy\,{\rm beam}^{-1}$. Synthesised beam sizes and rms noise levels achieved in the final images are shown in each plot (and caption) in \textcolor{blue}{Figures} \ref{cplot:G018.82}$-$\ref{cplot:G196.4542} of \textcolor{blue}{Appendix} \ref{sec:appendixfigures}.

For the flagging, editing, calibration and subsequent deconvolution/imaging, the \textsc{casa} software package \citep{CASARef} was used in conjunction with the \textsc{casa} pipeline (version $4.7.2$). Manual flagging was performed first before running the pipeline, after which output calibration tables were manually inspected. In cases where erroneous calibration solutions were found, flagging was repeated and the pipeline was rerun until the time-varying gains, bandpass solutions and bootstrapped flux densities were of a high quality. At C-band, spectral cube imaging of the $2\MHz$ ($\Delta v \sim90\kmps$) channels corresponding to the $6.7\GHz$ $\mathrm{CH}_3\mathrm{OH}$ maser line was conducted with maser positions recorded (green `$\,\times\,$' in any contour plots) and the relevant channel(s) subsequently flagged prior to continuum imaging. Absolute position is accurate to within $10\%$ of the beam width, or $\sim30\mas$ at C-band and $\sim4\mas$ at Q. For all images primary-beam correction was applied.

Resulting imaging parameter, such as restoring beam dimensions, and image noise levels are summarised in \autoref{tab:BeamsNoises} of \autoref{sec:appendixtables}.

\subsection{Performance of the array}
\label{sec:arrayperformance}
An important diagnostic of collimated jets are spectral changes in flux and deconvolved major axis length. At different frequencies, \citet{Reynolds1986} showed that the relative contributions of optically-thick and thin emission vary with a corresponding positional shift of the surface at which optical depth is unity (i.e. $\tau_\nu \sim 1$). These changes are quantified with the spectral indices $\alpha$  and $\gamma$ (whereby $\theta_{\rm maj} \left( \nu \right) \propto \nu^\gamma$ and $\theta_{\rm maj}$ is the observed jet length).

Unfortunately, when observing sources which may display extended emission, the degree to which different spatial scales are filtered out by interferometric observations is not well quantified for specific cases. With observational setups like those discussed above, where the same antenna configuration is used for radically different observing frequencies, this effect can be exacerbated. Since accurate spectral analysis is essential for radio jets we must therefore check if multi-frequency, A-configuration VLA-observing would significantly alter our measured values for $\alpha$ or $\gamma$.

To check the array performance of the VLA for our observations, the physical modelling, radiative-transfer and synthetic observation code, \texttt{RaJePy}\footnote{\url{https://github.com/SimonP2207/RaJePy}} (Purser, in prep) was used. A bi-conical jet profile (i.e. $\alpha=+0.6$ and $\gamma=-0.7$) with a mass-loss rate of $\jml = 5\times10^{-6}\, \Msol \yr^{-1}$, opening angle $\theta_{\rm op}=30\degr$ and distance of $D=3\kpc$ was employed for the physical model which is representative of jets in general \citepalias{Purser2016}. This profile was modelled up to an extent of $0\farcs5$ for both jet and counter-jet (total length of $1\arcsec$) using a cell size of $2\au$ ($\sim0.7\mas$), ensuring good sampling of the flux distribution by the beam ($1\arcsec$ is $\sim4$ C-band beams across). Synthetic observations were performed at each major frequency band of the VLA from L ($1-2\GHz$) to Q-bands with A-configuration antenna positions and typical continuum bandwidths ($1-8\GHz$).

\begin{figure}
	\centering
	\includegraphics[width=\columnwidth]{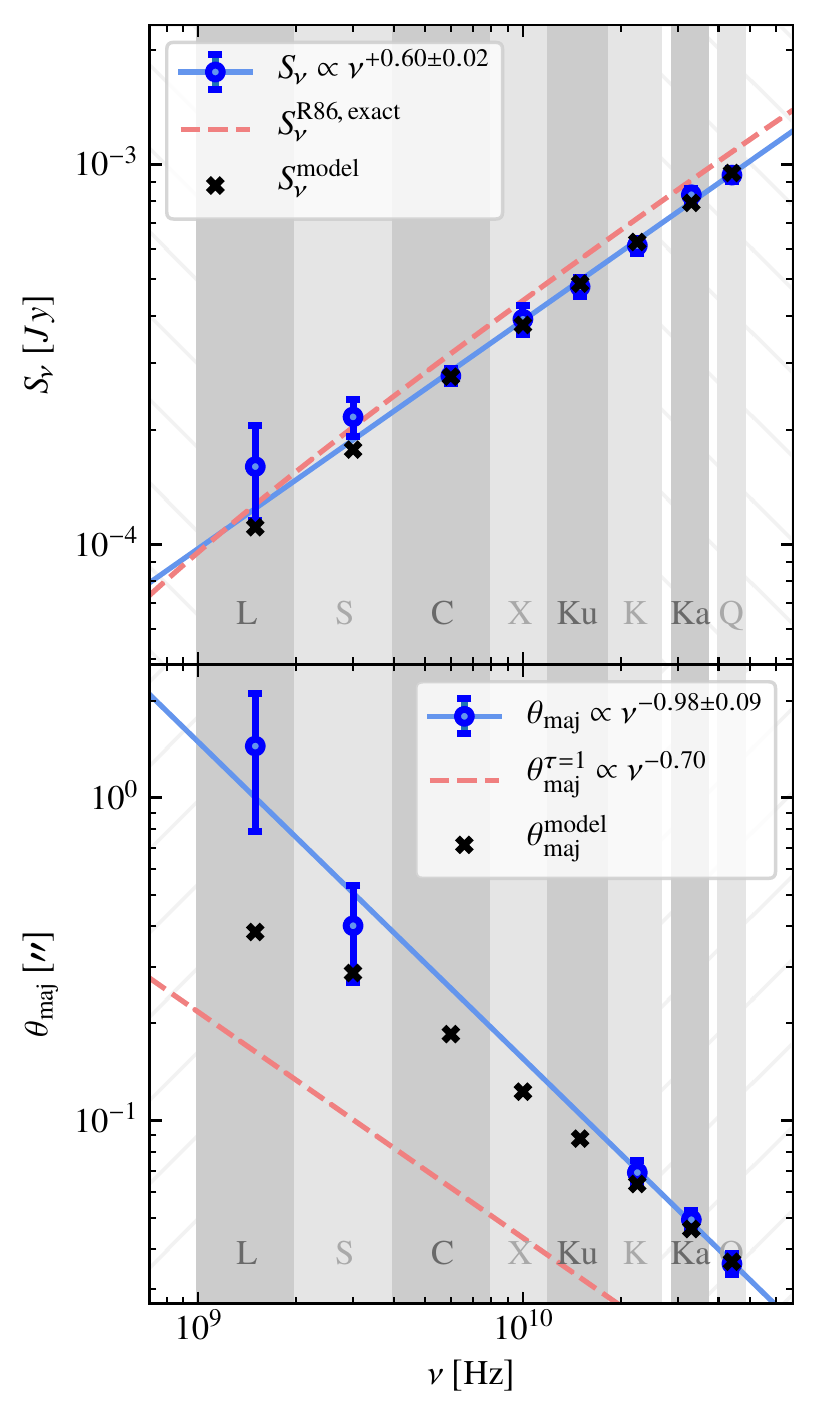}
	\caption[VLA A-configuration performance in the recovery of the flux-distribution for an ionised jet]{Performance of the A-configuration VLA towards a jet model with $\jml = 5\times10^{-6}\, \Msol \yr^{-1}$, $\theta_{\rm op}=30\degr$ and $D=3\kpc$. Fluxes and deconvolved sizes derived from fitting a 2-D Gaussian in the image plane are shown in the top and bottom panels, respectively. Blue markers/errorbars show results for the synthetic observations, the LSQ fit to which is shown as a solid, blue line. Black, '$\times$' markers show the fit results towards the sky-model convolved with the restoring beams of the observations. Dashed, red curves show the flux predicted by Equation 8 of \citet{Reynolds1986} (top panel) and the separation of the jet/counter-jet's $\tau=1$ surfaces as given by their Equation 12. Observed bandwidths of each observing band are shown as shaded areas.}
	\label{fig:vlaperformance}
\end{figure}

In \autoref{fig:vlaperformance}, the results of a single-component (Gaussian) fit to the recovered flux distributions of the synthetic observations are shown (blue points/errorbars). This fit failed at C, X and Ku-bands due to lower signal-to-noise, mixed with increasingly compact emission. For effective comparison, the same fit was performed on the sky model convolved with the clean beam of the observations (crosses). Recovered fluxes range from $98\pm6\%-144\pm40\%$ of the sky model's showing them to be well-recovered by the VLA owing to the increasingly compact (with frequency) flux distribution and comprehensive $uv$-coverage of the array. A least-squares (LSQ) fit to the fluxes yields $\alpha=0.60\pm0.02$, in good agreement with that expected from the analytical model of \citet{Reynolds1986}. From our synthetic observations the derived value of $\gamma=-1.0\pm0.1$ diverges from $\gamma = -0.7$ expected for the $\tau = 1$ surface from \citet{Reynolds1986}. We believe this results from approximating a jet's flux distribution as a Gaussian, yet overall behaviour of $\theta_{\rm maj}$ is still a useful indicator that the extent of optically thick emission is contracting towards the jet-base with frequency.

For the purposes of this work, we find that the VLA in its A-configuration performs well towards science targets of this type. With insignificant levels of flux being lost due to spatial-filtering, spectral analysis of ionised jets from C/Q-band data such as that presented here is a useful diagnostic of physical conditions within the jets, and requires no baseline-matching between C and Q-bands.

\section{Results}
\label{sec:results}
Every field was imaged out to $10\%$ of the primary beam's peak response at each observed frequency ($\sim6\arcmin$ and $\sim1\arcmin$ from the pointing centre at C and Q-bands respectively) using a typical robustness of 0.5 and cell sizes of $0\farcs07$ and $0\farcs01$ respectively. \textcolor{blue}{Appendix} \ref{sec:appendixfigures} contains the resulting clean maps of radio flux in \textcolor{blue}{Figures} \ref{cplot:G018.82}$-$\ref{cplot:G033.69} of \textcolor{blue}{Appendix} \ref{app:irdcfigures} for the IRDC sample and \textcolor{blue}{Figures} \ref{cplot:G033.6437}$-$\ref{cplot:G196.4542} of \textcolor{blue}{Appendix} \ref{app:mysofigures} for the MYSO sample. Links to each plot are available in \autoref{tab:BeamsNoises} for ease.


For the measurement of fluxes and physical sizes, the same methods discussed in \textcolor{blue}{\S$3.3$} of \citetalias{Purser2016} were adopted (i.e. fitting of the emission with a Gaussian profile in the image plane). A full list of sources detected (i.e. $>5\upsigma$ where $\upsigma$ is the rms noise in each image) in the field are recorded, along with their derived fluxes and physical sizes, in \textcolor{blue}{Tables} \ref{tab:CBandPosFlux}$-$\ref{tab:DeconvSizes}. As a note, errors in flux used the local rms noise in their calculation, thereby accounting for the increased, effective noise towards the edge of the primary beam. Calculated spectral indices for both flux ($\alpha$) and deconvolved, major axis-length ($\gamma$), between $5.8\GHz$ and $44\GHz$, are recorded in \autoref{tab:AlphaGamma}. At C-band, for the MYSO sample, only sources within $1\arcmin$ from the pointing centre are recorded (i.e. within the field imaged at Q-band, for spectral comparison). For reader ease, links to the discussion notes for each, individual object of both subsamples can be found in the  last column of \autoref{tab:BeamsNoises}. All clean images, data products and tables are also available online\footnote{\url{https://github.com/SimonP2207/RadioJetsFromYSOs.git}}.

Classification of the compact radio sources follows the same algorithm discussed in \textcolor{blue}{\S$\,4.1$} of \citetalias{Purser2016}. In light of the results of \autoref{sec:arrayperformance} values for $\gamma$ are not restrictive \citep[i.e. in that they must be related to $\alpha$ as per][]{Reynolds1986} and only negative values are required for jets. Resulting classifications are summarised in \autoref{tab:classnumbers}, with a detailed breakdown in \autoref{tab:AlphaGamma}. Detailed discussion of the classifications and results for each member of both samples are contained in \textcolor{blue}{Appendix} \ref{sec:appendixnotes} for the interested reader.

As a further note, we expect pollution of the sample by extragalactic sources to be minimal. An analysis of the radio catalogue of \citet{Bonaldi2019} shows that, within the C-band primary beam, we expect only 6 AGNs above a flux limit of $24\uJy$ ($\sim4\sigma$) and below a size-limit of $1\arcsec$ (and therefore of similar appearance to our targets of interest). At Q-band, due to the small primary beam and higher sensitivity limit, this number is negligible.


\input{Tables/ClassificationNumbers}

\section{Analysis}
\label{sec:analysis}
\subsection{IRDCs and their radio evolution}
\label{sec:irdcanalysis}

\begin{figure*}
\centering
\includegraphics[width=6.97522in]{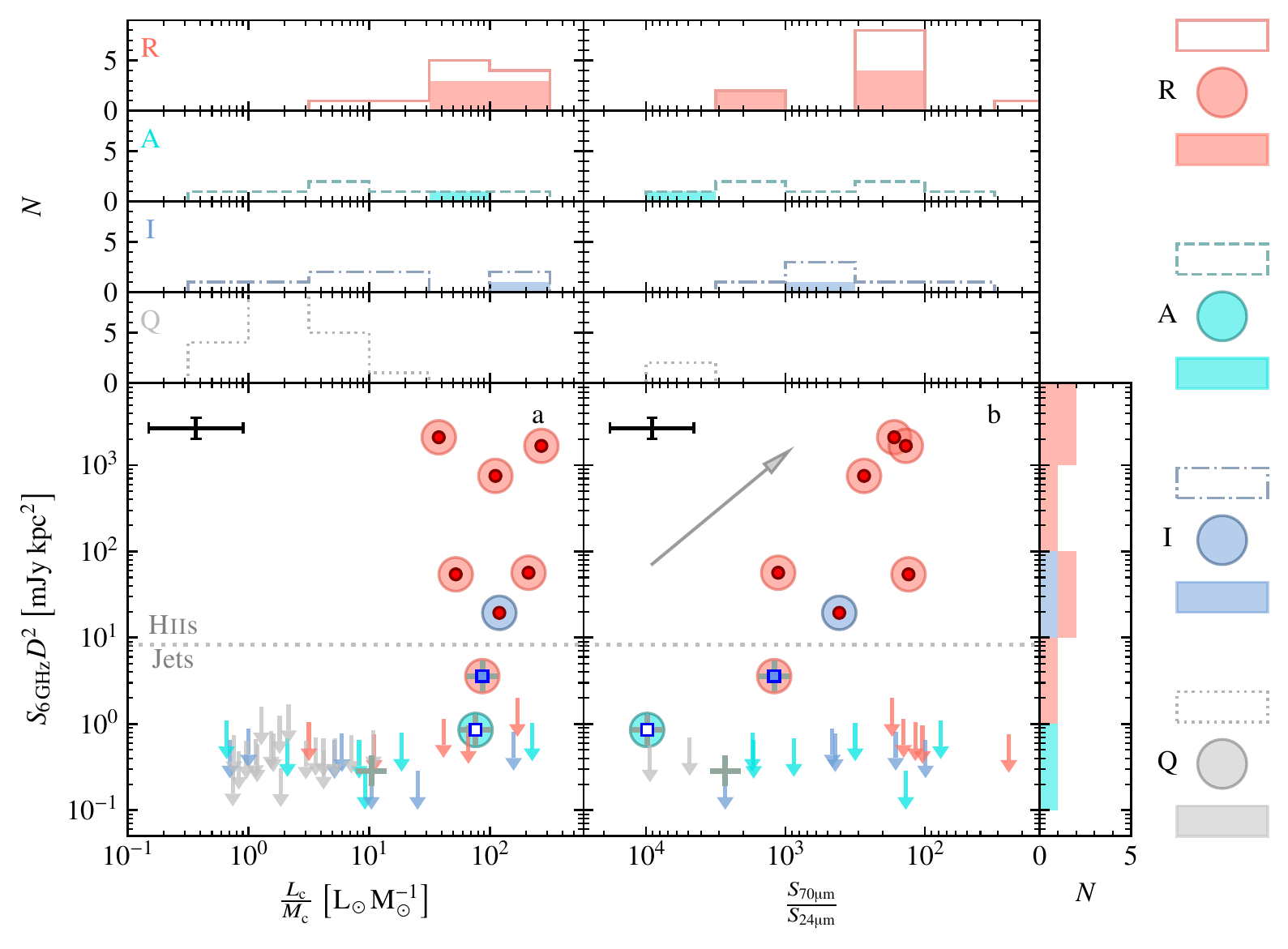}
\caption[Radio luminosity against parental core bolometric luminosity-to-mass ratio for the IRDC sample]{Plots of the $6\GHz$ distance luminosities against parental core luminosity-to-mass ratios \citep[panel a; data from][]{Rathborne2010} and $70\um$ to $24\um$ flux ratios (panel b), for the IRDC sample. Cores harbouring radio-detections are shown as circles, the colour-scheme for which is based upon their classification (`R', `A', `I' or `Q') according to \citet{Rathborne2010}. Non-detections (upper limits) follow the same colour scheme based upon core classification. Smaller markers illustrate our radio source classification following the scheme of \autoref{fig:distlumvsbollum} while the thick, `$+$' marker denotes the IRDC cores coincident with detected CH$_3$OH masers. Histograms (colours as discussed) are shown, with solid bars showing cores with detected radio sources, while hollow bars include detections \textit{and} radio non-detections. The dotted line delineates detected radio source classification from this work as \textsc{Hii} regions, or jets/jet candidates. Representative (median) errors are shown in the top left of each plot.}
\label{fig:DLvsLMRatio}
\end{figure*}

For their $1.2\mm$-detected cores, \citet{Rathborne2010} employed the classification scheme of \citet{Chambers2009}, as summarised in \autoref{sec:irdcsample}. Based upon their infrared properties this scheme establishes a measure of the cores' evolutionary states, from completely inactive (`quiescent') to harbouring active sites of star formation (`red'). Exactly when a YSO jet's radio emission `switches on' during this evolution is an open question. Here we use both the core luminosity-to-mass and the $70\um$ to $24\um$ flux ratios as quantitative, evolutionary indicators. Both can be considered proxies for a core's transition from a `cold' (or quiescent) to a `hot' (intermediate/active/red) molecular core and therefore age. In light of this and to investigate the onset of radio emission \autoref{fig:DLvsLMRatio} therefore plots detected radio flux against these indicators, showing that radio detections are only found towards some of the `I', `A' or `R' cores whose luminosity-to-mass ratios are $\gtrsim40$. No weak ($\gtrsim 30 \uJy\,{\rm beam}^{-1}$) radio emission is detected towards any quiescent (`Q') core (see \autoref{tab:classnumbers} for a summary). 

An intermediate-class core (MM2 of G024.60+00.08) displays maser emission yet no corresponding radio-continuum source is detected. Weak radio-continuum emission typical of ionised jets is seen towards the `R' and `A' cores G024.33+00.11 MM1 (the most massive of the sample) and G028.37+00.07 MM1 (which harbours two jet candidates), the most and third-most luminous cores within the sample. Those two cores harbouring jets also display maser emission, and their cores have higher luminosity-to-mass ratios than the maser-only source. Strong radio emission from \textsc{Hii} regions is observed towards six (five `R' class and one `I' class) cores possessing high $\nicefrac{L_{\rm C}}{M_{\rm C}}$ ratios. Those cores containing \textsc{Hii} regions also have higher values for $\nicefrac{S_{70\um}}{S_{24\um}}$ and radio flux (as expected), than those with jets (see panel b of \autoref{fig:DLvsLMRatio}).

Evolutionarily, these findings make sequential sense. First, collapse-induced heating liberates volatile species into the gas phase via desorption from icy mantles \citep{Viti2004}, providing the conditions for maser emission from the desorbed CH$_3$OH. As the core collapses further, accretion and ejection phenomena in the form of discs and jets (the `Class I' phase of low-mass star formation) produce weak radio emission, after which the newly formed massive protostar's ultraviolet Lyman flux increases to the point whereby an \textsc{Hii} region is formed.

Due to the small number of radio sources detected towards the IRDC sample, only this brief analysis could be conducted. However those radio-detected IRDC cores harbouring jets and non-detections will help guide future surveys in terms of sensitivity requirements, especially in the SKA era.

\subsection{Radio luminosity against bolometric luminosity}
\label{sec:radiobollum}
One of the key results of \citetalias{Purser2016} was the segregation of MYSOs harbouring ionised jets from those powering \textsc{Hii} regions, in radio/bolometric luminosity parameter space. While the jets were found to occupy a region that adhered to the low-mass power-law for jets found by \citet{Anglada1995}, \textsc{Hii} regions were roughly as radio-luminous as their ultraviolet Lyman fluxes \citep[inferred from the models of][]{Davies2011} would predict. However there were some that were significantly lower than their predicted radio flux. These under-luminous \textsc{Hii} regions were still classified as such since classification was based on not just radio flux, but also upon morphology and infrared properties. This approach caught those \textsc{Hii} regions that were either resolved out by the interferometer and/or compact/optically-thick at the observed frequencies and is adopted in this work.

\begin{figure*}
\centering
\includegraphics[width=\textwidth]{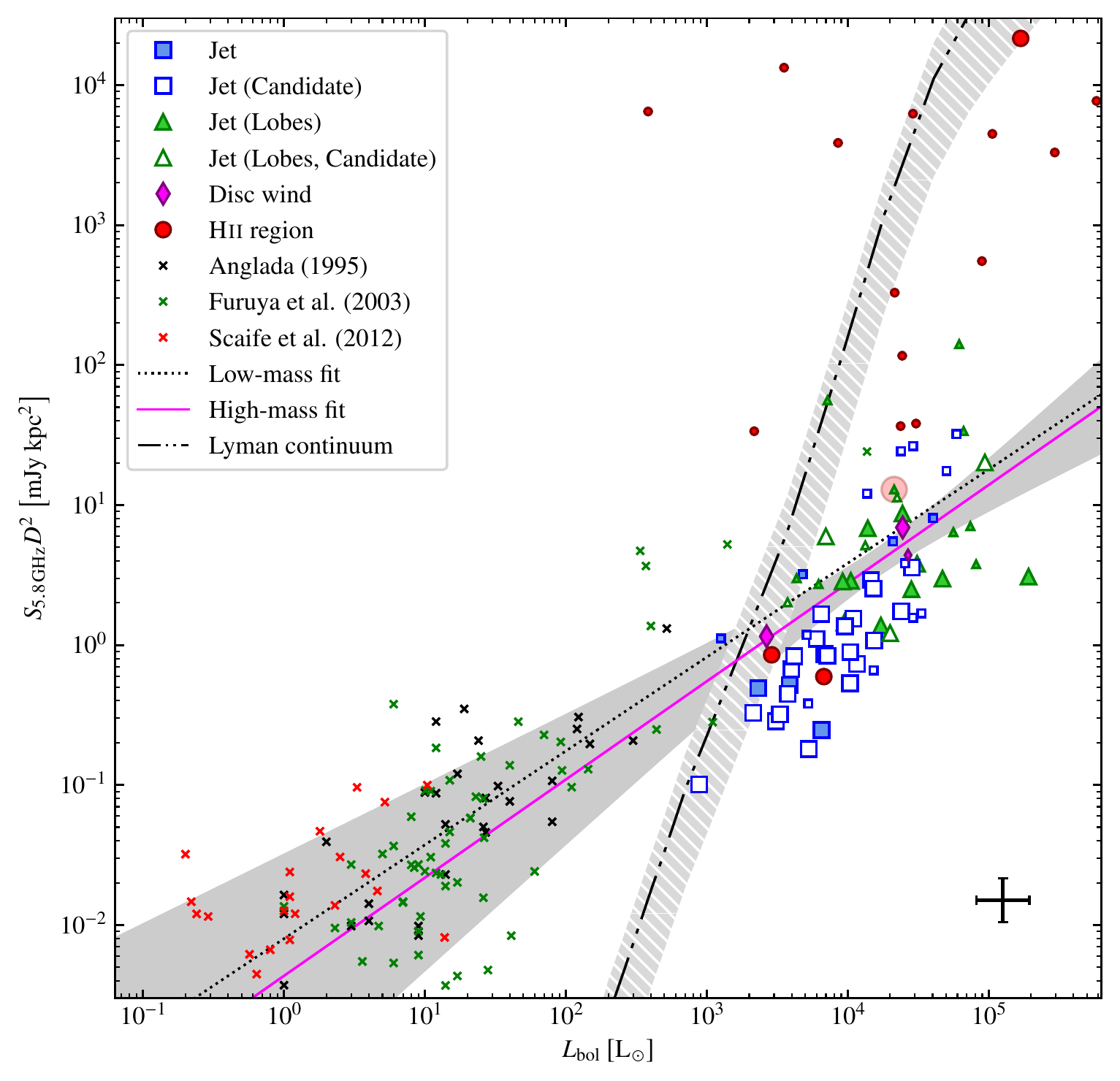}
\caption[Radio luminosity at $5.8\GHz$ against bolometric luminosity]{A plot of the radio luminosities at $5.8\GHz$ against bolometric luminosity for those objects detected from the sample discussed in \autoref{sec:mysosample}. Different sub-classifications are marked as blue squares for jets, hollow blue squares for jet candidates, green triangles for jets with lobes, hollow green triangles for candidate jets with lobes, red circles for \textsc{Hii} regions and magenta diamonds for disc winds (see legend). Those objects from \citetalias{Purser2016} are shown using the same scheme as discussed but half the marker size of the objects from this work. A contemporaneous jet/\textsc{Hii} region is highlighted by a red, transparent circle \citep[G345.4938+01.4677,][]{Guzman2016}. Note that for jets with lobes, only the central, thermal jet is plotted. Small `$\times$' markers are low-mass radio jets, normalized to $5.8\GHz$ \citep[black, green and red for][respectively]{Anglada1995,Furuya2003,AMI2011a}, assuming $\alpha=0.6$ where the information is not available. The dotted line represents the power law we derived for the low-mass case (\autoref{eq:dislumlm}), while the magenta line is that fitted to jets (\textit{not} candidates) in this sample (\autoref{eq:dislumhm}) with the $1\sigma$ confidence interval shaded in grey. Radio luminosity of an optically-thin, \textsc{Hii} region resulting from the ionising, ultraviolet, Lyman continuum of a ZAMS star of that $\Lbol$ (calculated using the stellar models of \citet{Davies2011} for $\Lbol > 10^3 \Lsol$, and \citet{Thompson1984} otherwise) is shown as a dot-dashed line. Lighter, grey, hatched shading represents the area where the radio-luminosity is between $20-180\%$ of that expected from those models. Representative (median) errors are shown in the bottom right of the plot.} 
\label{fig:distlumvsbollum}
\end{figure*}

To compare our results to those of \citetalias{Purser2016}, the C-band distance luminosities for the sample of \citetalias{Purser2016} were calculated by using their derived values for spectral index. In cases identified as \textsc{Hii} regions where the loss of flux with increasing resolution becomes an issue (i.e. $\alpha_\mathrm{measured}<-0.1$), an optically thin spectral index is assumed and the flux at $5.8\GHz$ is extrapolated from that measured at $5.5\GHz$. In \autoref{fig:distlumvsbollum}, the calculated distance luminosities at $5.8\GHz$ from \citetalias{Purser2016} (smaller markers) are plotted against bolometric luminosity, as well as for all radio detections towards the MYSO sample of \autoref{sec:mysosample} (larger markers). Fitting the jets (not candidates) from this work and those from \citetalias{Purser2016} (whose $D$, $\Lbol$ and $\delta \Lbol$ were also recalculated using the methods of \autoref{sec:distslumsirratios}) with a power law gives the relation shown as the magenta line in \autoref{fig:distlumvsbollum} and explicitly stated in \autoref{eq:dislumhm}. For this process the BCES algorithm of \citet{Akritas1996a} was used since it takes into account errors in both $\Lbol$ (independent variable) and $S_{\rm 5.8\,GHz}D^2$ (dependent variable). For comparison, we also fit the low-mass sample with the same algorithm, the results of which are plotted as a dotted line in \autoref{fig:distlumvsbollum}, and given in \autoref{eq:dislumlm}. Both derived relations for low and high-mass YSOs agree within errors. This shows that jet radio luminosity scales with bolometric luminosity in the same way across 6 orders of magnitude, from $10^{-1}$ to $10^6\Lsol$. As in \citetalias{Purser2016}, this suggests that those jets associated with high-mass MYSOs may be produced via `scaled-up' processes of their lower-mass counterparts. 



\begin{align}
\log_{10}\left[ \frac{S_{\rm 5.8\GHz}D^2}{\mJy\,\kpc^2} \right] &= (0.70\pm0.24)\cdot\log_{10}\left[ \frac{\Lbol}{\Lsol} \right] - (2.36\pm1.02) \label{eq:dislumhm}\\
\log_{10}\left[ \frac{S_{\rm 5.8\GHz}D^2}{\mJy\,\kpc^2} \right] &= (0.67\pm0.06)\cdot\log_{10}\left[ \frac{\Lbol}{\Lsol} \right] - (2.10\pm0.07) \label{eq:dislumlm}
\end{align}

\begin{figure}
	\centering
	\includegraphics[width=\columnwidth]{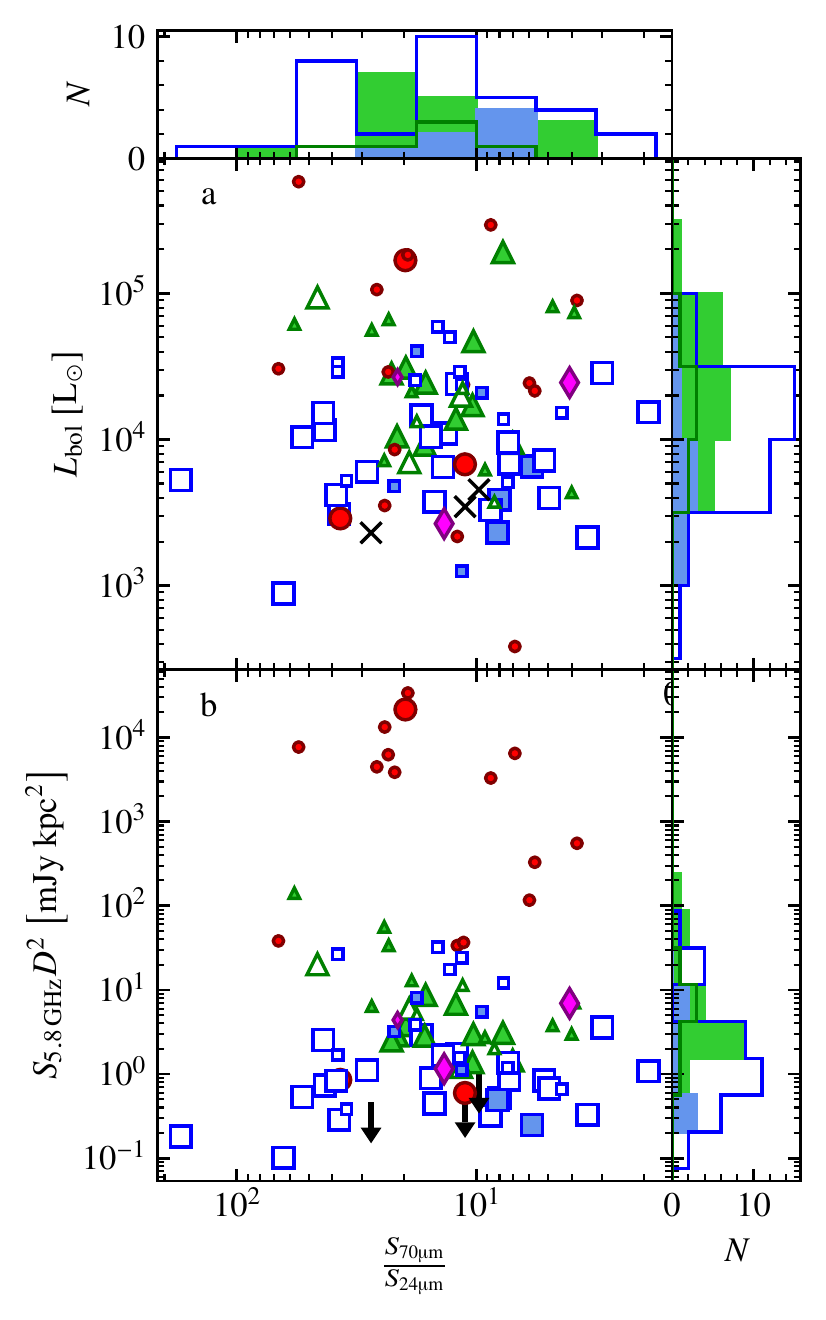}
	\caption[Subplots of bolometric and radio luminosity against mid-infrared $70\um$ to $24\um$ flux ratio]{A plot of bolometric luminosity (panel a) and distance luminosity at $5.8\GHz$ (panel b), against the $70\um$ to $24\um$ flux ratios. Black `$\times$' symbols mark non-detections in panel a, which are consequently marked as upper limits in panel b. All other symbols have the same meaning as in \autoref{fig:distlumvsbollum}. Histograms are plotted for confirmed jets (solid bars) and candidate jets (line-only bars) using the same colour-scheme employed for the markers.}
	\label{fig:DLvsIRratio}
\end{figure}

\begin{figure*}
	\centering
	\includegraphics[width=\textwidth]{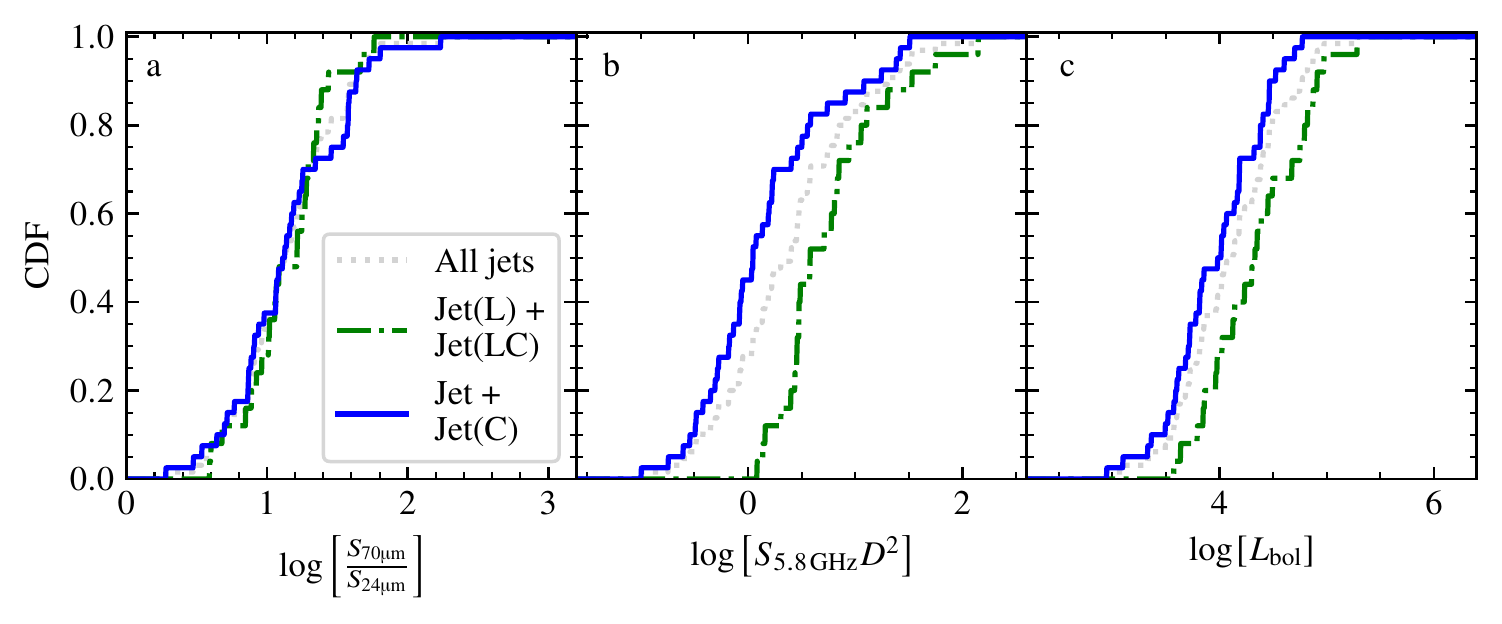}
	\caption[Cumulative distribution functions (CDFs) of the mid-infared ratios, radio luminosities and bolometric luminosities of northern hemisphere MYSOs with jets]{Calculated cumulative distribution functions for the $70\um$ to $24\um$ flux ratios (panel a), $5.8\GHz$ radio luminosities (panel b) and bolometric luminosities of MYSOs harbouring ionised jets. CDFs for jets with lobes including candidate jets with lobes (dash-dotted green line), jets without lobes including candidates (solid blue line) and all jets and candidates (dotted grey line) are plotted.}
	\label{fig:CDFsFIRratiosRadiolums}
\end{figure*}

\subsection{Evolution and relationship of jets and shock-ionised lobes}
\label{sec:lobes}
Theoretical works have highlighted how accretion is a variable process. For example, \citet{Meyer2019} show how accretion rates increase in the first $\sim 10^4 \yr$ of an MYSO's lifetime and how accretion rate variability becomes larger in amplitude towards the end of the MYSO stage. Due to the intrinsic connection between accretion and ejection, increased variability in the jet over time is expected. This may manifest itself at radio wavelengths as a time-varying radio flux for the thermal, radio jet, or an increase in the presence/change in the shock-ionised lobes along the jet's axis due to evolving mass-loss characteristics (such as precession or varying ejection velocities). While the former requires multi-epoch observations to investigate, the latter can be examined by analysis of the shock-ionised lobes and their correlation to evolutionary indicators. With this in mind, we therefore investigate the jet sample and possible evolutionary trends connected to the presence of shock-ionised lobes below.

In \autoref{fig:DLvsIRratio} the bolometric (panel a) and $5.8\GHz$ distance luminosities (panel b) are plotted against the $70\um/24\um$ flux ratios. Superficially, those jets associated with lobes appear to occupy a narrower range of infrared flux ratios than those without lobes, with a mean/standard deviation for $\log_{10}(\nicefrac{S_{70\um}}{S_{24\um}})$ of $1.14\pm0.29$ and $1.15\pm0.40$, respectively. For the bolometric luminosity, the mean/standard deviation for $\log_{10}(\Lbol)$ is $4.34\pm0.45$ and $3.96\pm0.43$, while for the distance luminosity, $\log_{10}(S_{5.8\GHz}D^2)$ is $0.75\pm0.50$ and $0.15\pm0.60$ for the jets associated with lobes and those without, respectively. These statistics are reflected in the corresponding cumulative distribution functions (CDFs) plotted in \autoref{fig:CDFsFIRratiosRadiolums} and indicate that jets associated with lobes are found in a narrower range of infrared ratios, and at higher radio and bolometric luminosities.

\begin{figure*}
	\centering
	\includegraphics[width=\textwidth]{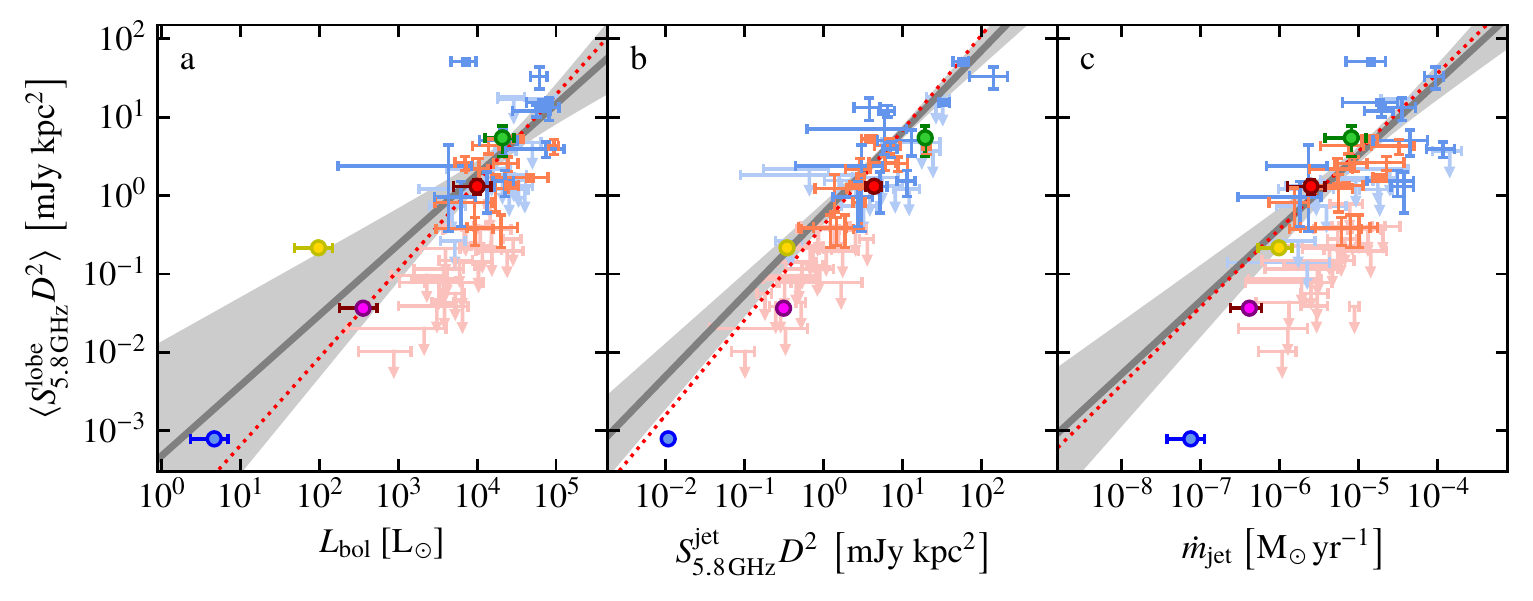}
	\caption{Log-mean $5.8\GHz$ lobe luminosities against MYSO bolometric luminosity (panel a), central jet $5.8\GHz$ luminosities (panel b) and central jet mass loss rates (panel c). Faded points represent $3\sigma$ upper limits on lobe luminosities for jets not associated with any lobes. Grey lines show the lines of best fit whose derived parameters are given in \autoref{tab:loberesults}, with the $1\sigma$ confidence interval shaded light grey. Jet candidates, while plotted, are not included in this analysis. The red dotted line is the Akritas-Theil-Sen line \citep{Akritas1995} which includes censored (i.e.\ upper-limits) data. Coloured circles represent YSOs from the literature for comparison which are, DG Tau A \citep[blue marker;][]{Purser2018b}, the Serpens triple radio source \citep[yellow marker;][]{RodriguezKamenetzky2016}, G035.02+0.35N \citep[red marker;][]{Sanna2019}, HOPS 370 \citep[purple marker;][]{Osorio2017} and HH 80-81 \citep[green marker;][]{Marti1993}.}
	\label{fig:lobefluxes}
\end{figure*}

\begin{table*}
\centering
\caption{Fitted power-law parameters (of the form $\log_{10}y=m \log_{10}x+c$), calculated partial correlation coefficients whilst controlling for distance ($\tau_{\rm kendall}^{xy.D}$) and their associated $p$-values. All fitted (not including upper limits) power-laws are represented in the relevant panels of \autoref{fig:lobefluxes} as grey lines. For those fits not including upper-limits the BCES algorithm is used, otherwise the Akritas-Theil-Sen estimator \citep{Akritas1996a} is used.}
\begin{tabularx}{\textwidth}{YYYYYYYY}
\toprule
$x$ & $y$ & $N$ & Include Upper limits? & $m$ & $c$ & $\tau^{xy.D}_{\rm kendall}$ & $p$-value \\ 
\midrule
\multirow{2}{*}{$L_\mathrm{bol}$}                      & \multirow{2}{*}{$\langle S_\mathrm{5.8\,GHz}^\mathrm{lobe}D^2\rangle$} & \multirow{2}{*}{18} & \xmark &$+0.90\pm0.48$ & $-3.33\pm2.15$ & $0.360$ & $ 0.044$ \\
 & & &\cmark & $+1.13$ & $-4.38$ & $0.513$ & $<0.001$ \\ \cline{4-8}
\multirow{2}{*}{$S_\mathrm{5.8\,GHz}^\mathrm{jet}D^2$} & \multirow{2}{*}{$\langle S_\mathrm{5.8\,GHz}^\mathrm{lobe}D^2\rangle$} & \multirow{2}{*}{19} & \xmark & $+1.04\pm0.18$ & $-0.23\pm0.15$ & $0.615$ & $<0.001$ \\ 
 & & &\cmark & $+1.24$ & $-0.40$ & $0.658$ & $<0.001$ \\ \cline{4-8}
\multirow{2}{*}{$\dot{m}_\mathrm{jet}$}                & \multirow{2}{*}{$\langle S_\mathrm{5.8\,GHz}^\mathrm{lobe}D^2\rangle$} & \multirow{2}{*}{19} & \xmark & $+0.93\pm0.30$ & $ +5.18\pm1.51$ & $0.485$ & $ 0.005$ \\ 
 & & &\cmark & $+1.02$ & $+5.67$ & $0.558$ & $<0.001$ \\
\bottomrule
\end{tabularx}
\label{tab:loberesults}
\end{table*}

To more thoroughly examine this conclusion two-sample, Kolmogrov-Smirnov (K-S) tests were performed to see if the jets with, and those without, lobes (for both our MYSO subsample and that of \citetalias{Purser2016}) were drawn from the same distributions in infrared flux ratios, radio luminosities and bolometric luminosities. K-S tests quantify this based upon a test statistic (values of $0\rightarrow10$) and an associated $p$-value, defined as the probability of falsely rejecting the null hypothesis that both populations are drawn from the same distribution. For the mid-infrared ratios a K-S test statistic of $0.195$ and $p$-value of $0.537$ were calculated. For the radio luminosities these were calculated to be $0.580$ and $<0.001$, while for bolometric luminosity values of $0.330$ and $0.055$ were calculated respectively. Although this shows that the two samples could be drawn from the same mid-infrared ratio distribution and bolometric luminosity, the opposite is true for the radio luminosities. 

To understand why shocked lobes are more prevalent towards brighter MYSOs, we compare lobe properties to those of the powering MYSO's thermal jet. Therefore, in \autoref{fig:lobefluxes}, the log-mean (i.e.\ averaged over the lobes associated with each individual jet) $5.8\GHz$ distance-luminosities for lobes associated with each jet from this work and those of \citetalias{Purser2016} are plotted against bolometric luminosity (panel a), central jet luminosity (panel b) and jet mass loss rate (panel c; see \autoref{sec:jmls} for details on the calculation of $\jml$). Upper limits ($3\sigma$) on lobe fluxes/luminosities for those jets without any associated lobes are also plotted. For the sample of \citetalias{Purser2016}, if no spectral indices were available (i.e.\ detection at only one frequency) extrapolation of lobe luminosities to $5.8\GHz$ used an average value for spectral index of $\bar{\alpha}=-0.55\pm0.31$ (as per the findings of \citetalias{Purser2016}). For all compared parameters, power-laws were fitted and, since all compared variables use distance in their respective calculations, partial correlation coefficients\footnote{Partial correlation tests measure the degree of correlation between two variables whilst removing the influence of another variable affecting both dependent and independent variable} (controlling for distance) were calculated. Results for these are tabulated in \autoref{tab:loberesults} with fitted power-laws (using BCES as in \autoref{sec:radiobollum}) also shown in \autoref{fig:lobefluxes}. From those results, we show a statistically-significant correlation of lobe luminosity with both jet luminosity and jet mass loss rate. No correlation between lobe and bolometric luminosity was found. For comparison, we also plot known examples of jets with lobes (HH 80-81, G35.2N, Serpens triple radio source, DG Tau A and HOPS 370) from the literature, across the YSO mass spectrum, as coloured circles in all panels of \autoref{fig:lobefluxes}. As can be seen, these objects adhere well to the derived power laws, regardless of their values for $\Lbol$.

Interestingly, from panels b and c of \autoref{fig:lobefluxes}, the lobe-flux-density, $3\sigma$ upper-limits for jets without lobes appear to be lower than would be expected from their jets' fluxes or mass loss rates. This may simply result from less luminous lobes falling below our sensitivity limits. Alternatively, jets with lower mass-loss rates may be less likely to produce lobes, implying a more intrinsic difference between them and higher-luminosity YSOs. In order to distinguish between these possibilities, we fit the lobe luminosities including both detections, as well as upper-limits of lobe luminosity for those jets without lobes. Since we are fitting singly-censored data (i.e.\ upper limits), the BCES algorithm is inadequate and instead the Akritas-Theil-Sen estimator \citep{Akritas1995} is used, which is insensitive to outliers. Where this discrepancy is a detection issue, fitted parameters of these two algorithms should not change. As shown in \autoref{tab:loberesults}, this is what is observed. When repeated for average lobe fluxes (i.e.\ \textit{not} luminosity and therefore distance \textit{independent}) against MYSO infrared flux and jet flux, the same result is seen. Thus, we establish that for the lower-luminosity radio jets, the non-detection of shock-ionised lobes is due to the sensitivity limit of our observations. Of course, this does not preclude the possibility that the lower-luminosity jets are less likely to produce lobes, a possibility which more sensitive radio observations are required to elicit.

\subsection{Spectral indices and dust contribution}
\label{sec:alpha+dust}
\subsubsection{Thermal, protostellar radio emission}
At cm-wavelengths emission from an MYSO is generally dominated by free-free emission from ionised gas. However, thermal emission from dust grains begins to dominate the spectral energy distribution in the mm-regime. At Q-band therefore, thermal dust emission may contribute to the measured flux. Fortunately, power-law contributions can be validly assumed for both the ionised and dust components since turnover frequencies of ionised jets are generally higher than $44\GHz$ (from general consensus of observations) and mm-wavelengths fall under the Rayleigh-Jeans approximation ($h \nu \ll k T_{\rm dust}$). While the ionised jet's flux follows a power-law, the dust's flux is related to frequency by \autoref{eq:dustflux} (where $\beta$ is the dust opacity index) with the total flux (ionised gas and dust) given in \autoref{eq:doublepowerlaw}. In the ISM, the average dust opacity index, $\beta$, is $\beta=1.8\pm0.2$, while in protoplanetary discs, where grain agglomeration leads to increased dust grain sizes, this value can fall to $\beta\approx1$ \citep{Draine2006} or even less if observing an optically-thick, hot accretion-disc. Typically, for MYSOs the value for $\beta$ falls in the range $1 \leq \beta \leq 2$ \citep[e.g.\ $\beta=1$, $\beta=1.3$ and $\beta=1.5$ for][respectively]{Zhang2007,GalvanMadrid2010,Chen2016}.

\begin{align}
S_\nu^\mathrm{dust} &\propto \nu^{\beta+2}\label{eq:dustflux}\\
S_\nu &=  c_1  \nu^\alpha + c_2  \nu^{\beta+2} \label{eq:doublepowerlaw}
\end{align}

\noindent where $c_1=S_0^\mathrm{jet} \nu_0^{-\alpha}$, $c_2=S_0^\mathrm{dust} \nu_0^{-\beta - 2 }$ and $S_0^\mathrm{jet}$ and $S_0^\mathrm{dust}$ are the flux contributions at some reference frequency, $\nu_0$, from the ionised and dust components respectively.

For establishing dust contributions to the Q-band fluxes we employ four methods which are listed, and subsequently discussed, below:
\begin{enumerate}
	\item[(1)] Interpolation of $S_{\rm 44 \GHz}$ using matching-resolution mm/sub-mm observations
	\item[(2)] Extrapolation of $S_{\rm 44\GHz}$ using fluxes from L to Ka-bands
	\item[(3)] Analysis of position angle differences from C to Q-band
	\item[(4)] Comparison of spectral index distribution across the sample with that of `dust-free' surveys
\end{enumerate}

\textit{Method 1}: Matching-resolution, sub-mm/mm fluxes are available in only two cases, G160.1452+03.1559 and G173.4815+02.4459 (within the field of view for G173.4839+02.4317). For each of those two MYSOs, it is therefore possible to constrain the SED of the central, thermal source and directly derive the power-laws governing dust and ionised emission. Using the method of least squares in conjunction with \autoref{eq:doublepowerlaw} we deduce that dust contributes $22\pm9\%$ of the Q-band flux for G160.1452+03.1559 and $1\pm4\%$ for G173.4815+02.4459. Further details are available in \autoref{sec:G160.1452} and \autoref{sec:G173.4839} of {\color{blue}Appendix} \ref{sec:appendixnotes}, respectively. 

\begin{figure*}
	\centering
	\includegraphics[width=4.75in]{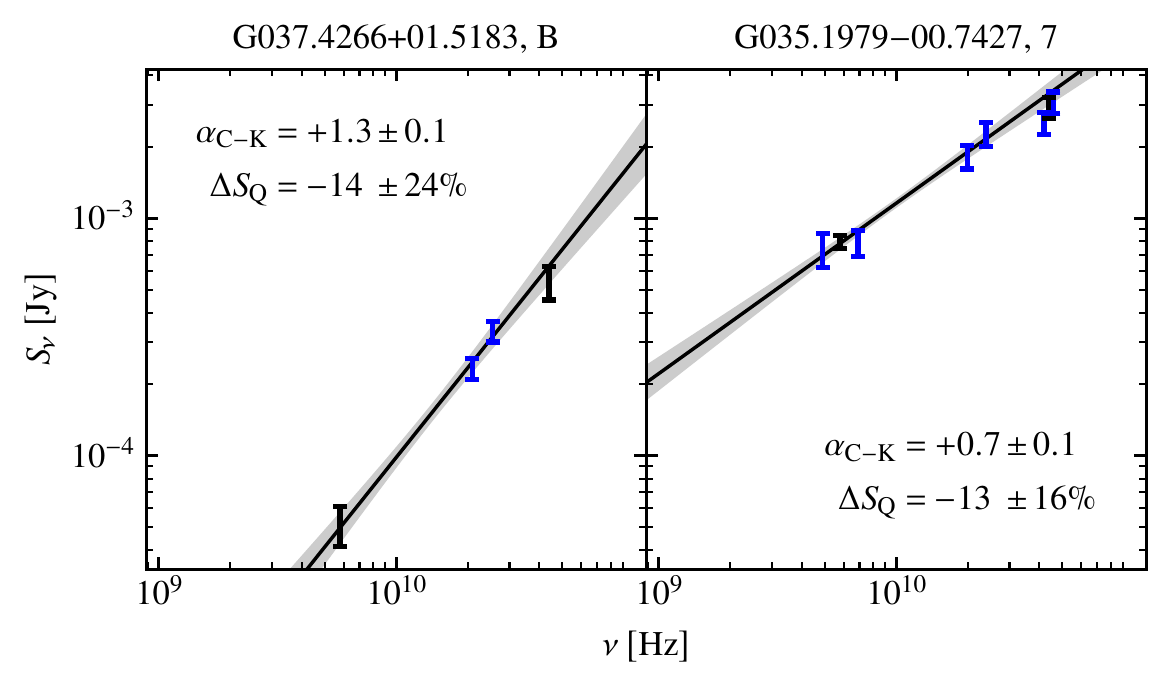}\\
	\includegraphics[width=6.75in]{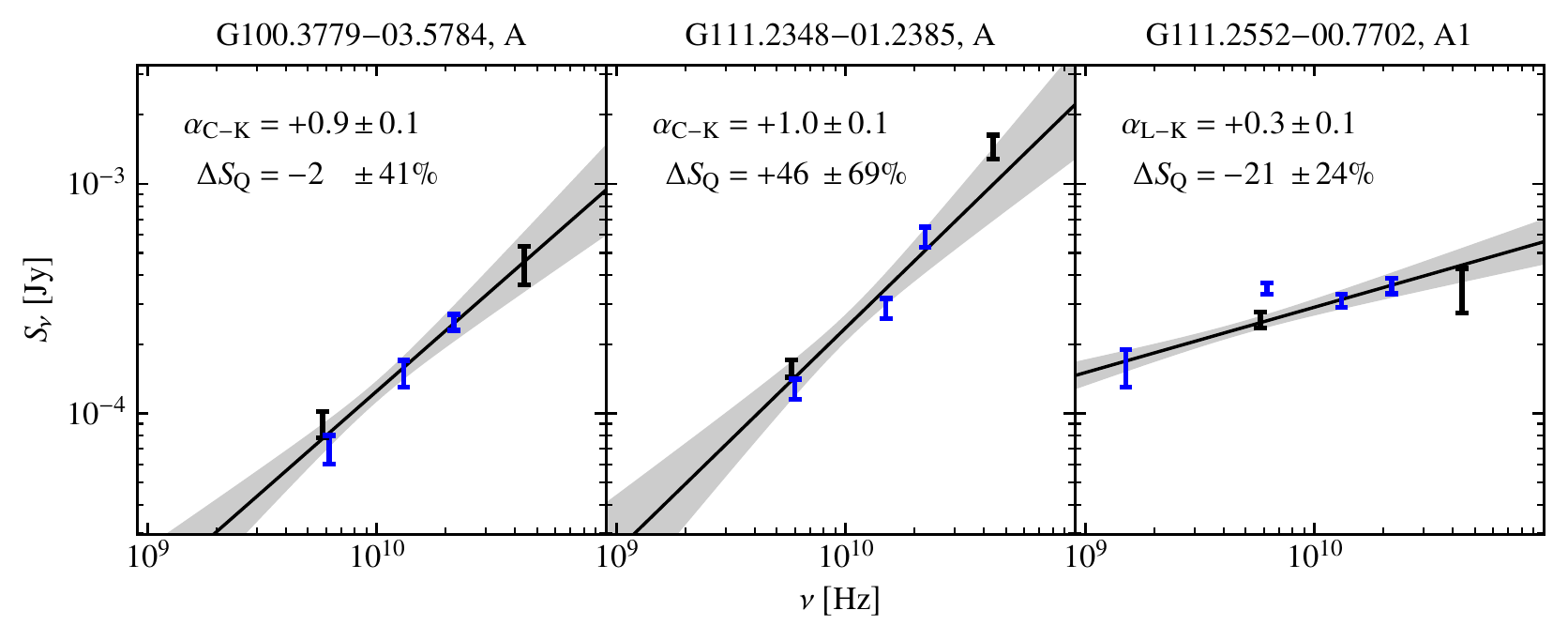}
	\caption[Radio SEDs for 5 MYSOs in the sample]{Radio SEDs for the thermal radio emission of 5 MYSOs in our sample, including fluxes taken from the literature (blue error-bars). Spectral indices derived from fluxes between L and K-bands are indicated, as well as the excess Q-band flux, $\Delta S_{\rm Q}$. Grey shading indicates the $1\sigma$ confidence intervals on the power-laws derived (solid, black lines). \textit{References}: Top row, left panel - \citet{Rosero2016}. Top row, right panel - \citet{Rosero2019} for component 7 of G035.1979--00.7427 Bottom row - \citet{Sanna2018} and \citet{Obonyo2019} (L-band flux of G111.2552--00.7702). In the case of component A1 of G111.2552--00.7702, the C-band flux recorded by \citet{Sanna2018} is disregarded in the power-law fit due to source confusion.}
	\label{fig:radioSEDs}
\end{figure*}

\textit{Method 2}: In \autoref{fig:radioSEDs}, five MYSOs' radio SEDs are plotted which include cm-fluxes from four other surveys. Using the method of least squares a power-law was fitted to those fluxes recorded at frequencies ranging from $1.5-25.5\GHz$ (where dust emission is assumed to be negligible), with derived values for $\alpha$ indicated on each panel. From those power-laws the predicted, dust-free Q-band flux was calculated (errors are the prediction interval) along with its difference with the observed flux, $\Delta S_{\rm Q}$ (also indicated in each panel). Values in the range $-21\pm24\% \leq \Delta S_{\rm Q} \leq +46 \pm 69\%$ show that, within errors, no excess flux from dust is therefore observed in these examples.

\begin{figure}
	\centering
	\includegraphics[width=\columnwidth]{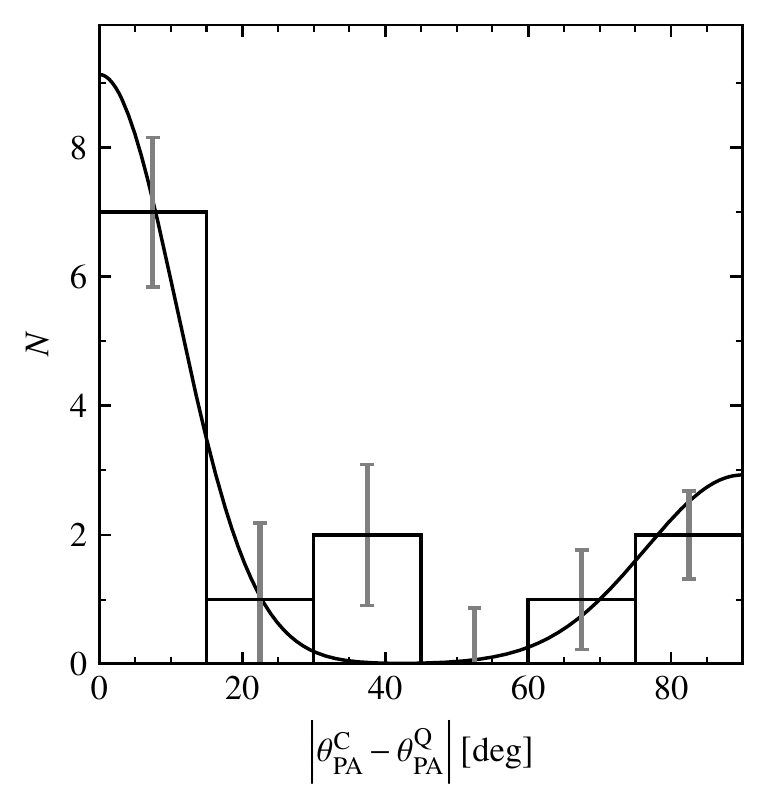}
	\caption[Histogram of differences in major axis position angle from C to Q-band across the sample]{Histogram of differences in major axis position angle from C to Q-band across the sample. Binning is defined as $s + w(n-1)\leq \alpha < s + wn$, where $s$ is the left edge value for the left-most bin ($| \theta_{\rm PA}^{\rm C} - \theta_{\rm PA}^{\rm Q} |=0\degr$), $n$ is the bin number (starting at $1$ with the left-hand bin) and $w$ is the bin width ($15\degr$). The black line shows the fit of a double-Gaussian distribution to the histogram.}
	\label{fig:jetdpa}
\end{figure}

\textit{Method 3}: Across our sample, 13 MYSOs with jet-like emission were measured to have finite, deconvolved sizes at both C and Q-bands. In those examples the difference in major-axis position angles, $| \theta_{\rm PA}^{\rm C} - \theta_{\rm PA}^{\rm Q} |$, could therefore be calculated. It is expected that $| \theta_{\rm PA}^{\rm C} - \theta_{\rm PA}^{\rm Q} | = 0\degr$ if jet-emission dominates and $| \theta_{\rm PA}^{\rm C} - \theta_{\rm PA}^{\rm Q} | = 90\degr$ when dust-contributions dominate. Between these two values, both the jet and dust contribute significant emission. In \autoref{fig:jetdpa}, we therefore plot the histogram for $| \theta_{\rm PA}^{\rm C} - \theta_{\rm PA}^{\rm Q} |$ showing that jet-emission dominates in most cases. To quantify the average fraction of the Q-band flux from dust emission, we have fitted this histogram with two Gaussian distributions whose means are fixed at $0\degr$ and $90\degr$ (solid, black line in \autoref{fig:jetdpa}). Using the ratio of the areas under each distribution as a proxy for relative flux contributions, we estimate that dust contributes an average of $29\pm23\%$ to the Q-band fluxes across our sample.

\begin{figure*}
\centering
\includegraphics[width=2\columnwidth]{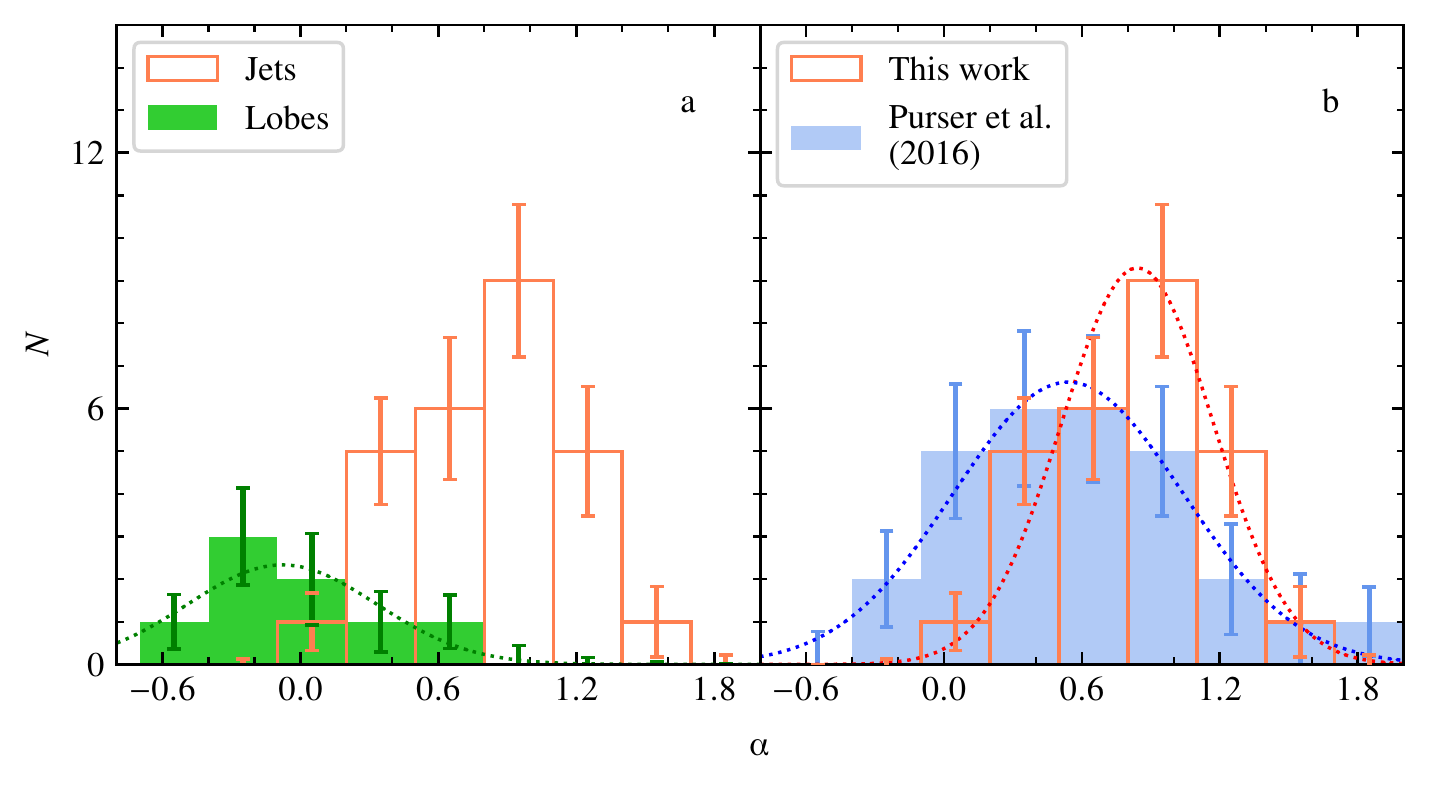}
\caption[Spectral index histograms for jet-like sources]{Histograms of spectral indices for: a.) All lobes (green) and jet-like sources (red) for which a spectral index between C and Q-bands was derived. b.) All jet-like sources with a derived spectral index from this work (red) and those from \citetalias{Purser2016} (blue). Normal distributions have been fitted to the jets from this work (red dotted line) and from \citetalias{Purser2016} (blue dotted line). Binning is defined as $s + w(n-1)\leq \alpha < s + wn$, where $s$ is the left edge value for the left-most bin ($\alpha=-0.7$), $n$ is the bin number (starting at $1$ with the left-hand bin) and $w$ is the bin width ($0.2$ for panel a and $0.3$ for panel b).}
\label{fig:alphahist}
\end{figure*}

\textit{Method 4}: Histograms of the spectral indices found towards the ionised jets (\autoref{fig:alphahist}) were compared to those of the lower frequency ($\nu_{\rm obs}<23\GHz$) survey of \citetalias{Purser2016} (their Figure 7), where sources have a minimal dust contribution. The resulting histograms are plotted in \autoref{fig:alphahist}. Comparison between spectral indices recorded in \citetalias{Purser2016} and those here shows that those jets observed in this work tend to have higher values for $\alpha$ than those of \citetalias{Purser2016}. Fitting a normal distribution to each sample yields a mean spectral index of $\bar{\alpha}_\mathrm{P16}=0.54$ and $\bar{\alpha}_\mathrm{VLA}=0.84$ (this work) for the jets from \citetalias{Purser2016} andt his work, respectively. Due to the use of the same classification scheme, there should be no intrinsic differences between southern and northern hemisphere jets and this difference is therefore attributable to dust contributions at Q-band which increase the derived values for $\bar{\alpha}$. Using the derived values for $\bar{\alpha}$, dust therefore contributes an average of $46\pm10\%$ of the total emission at Q-band, across our sample. This value is similar to the results of method 3 and to values obtained by \citet{SanchezMonge2008}. In that work 2 of a sample of 4 MYSOs showed dust contributions to the Q-band flux of $38\%$ and $44\%$ for IRAS 04579+4703 (our G160.1452+03.1559; see \autoref{sec:G160.1452}) and IRAS 22198+6336, respectively.

Considering the results of each method outlined above, dust contributions do not dominate at Q-band. However, measurable changes in the average spectral index and changes in the deconvolved position angles from C to Q-bands are observed showing dust still contributes. Considering method 4 had the largest sample sizes of any of the methods, we therefore conclude that within our sample dust contributes an average of $44\pm10\%$ of the Q-band flux. Individual contributions can not be discerned at this point however, and therefore recorded spectral indices are not adjusted.

As a note, we also investigated the relationship between $| \theta_{\rm PA}^{\rm C} - \theta_{\rm PA}^{\rm Q} |$ and $\alpha$, as we would expect higher values of $\alpha$ in cases of increased dust contributions, however no correlation was observed.

\subsubsection{Shock-ionised lobes}
As for the shock-ionised lobes' spectral index distribution, the population with detections at both C and Q-bands, and therefore with calculated values for $\alpha$ between these frequencies, is relatively small (4 lobes). An L-band survey by \citet{Obonyo2019} observed some of our sample's sources and, in conjunction with our results, derived spectral indices for some of the associated lobes. These have been included along with ours and plotted as a histogram in panel (a) of \autoref{fig:alphahist} (green bars). Subsequently we calculate that the associated lobes \citep[4 from this work and 4 from][]{Obonyo2019} belong to a normal distribution with a mean value of $\bar{\alpha}=-0.08$ and standard deviation of $\sigma_\alpha=0.41$ (green, dotted line in panel a of \autoref{fig:alphahist}), much flatter than the average spectral index found by \citetalias{Purser2016} ($\bar{\alpha}=-0.55$). We believe that this results from several factors (1) only lobes with shallower spectral indices would be detected in this work, (2) shock-ionised lobes have $\alpha\sim-0.5$ \citepalias{Purser2016}, lowering their signal-to-noise at Q-band (3) Shock-ionised lobes are extended on the arcsecond-scale and therefore flux-loss at Q-band is an issue. Due to these observational selection effects, as well as small sample size, no further analysis was performed.

\subsection{Mass loss and accretion}
\label{sec:jmls}

Jets' mass loss rates are an important parameter which can act as a discriminator between different jet-launching mechanisms. When compared with accretion rates \citep[using the so called, `magnetic lever arm' parameter, $\lambda=\nicefrac{\dot{m}_{\rm acc}}{\jml}$;][]{Frank2014PPVI} we can discriminate between different magneto-centrifugal launching mechanisms. Whilst `X-winds' typically are expected to have $\nicefrac{1}{\lambda}\sim0.3$ \citep{Shu1997}, values of $\nicefrac{1}{\lambda}\sim0.1$ are anticipated for the disk-winds described by \citet{Pelletier1992}. Radiative launching mechanisms have also been suggested and we can determine their significance by comparing $\jml$ with $\Lbol$ since for radiatively-launched/line-driven jets $\jml \propto \Lbol^{\nicefrac{3}{2}}$ \citep{Proga1998}.

\begin{figure}
	\centering
	\includegraphics[width=\columnwidth]{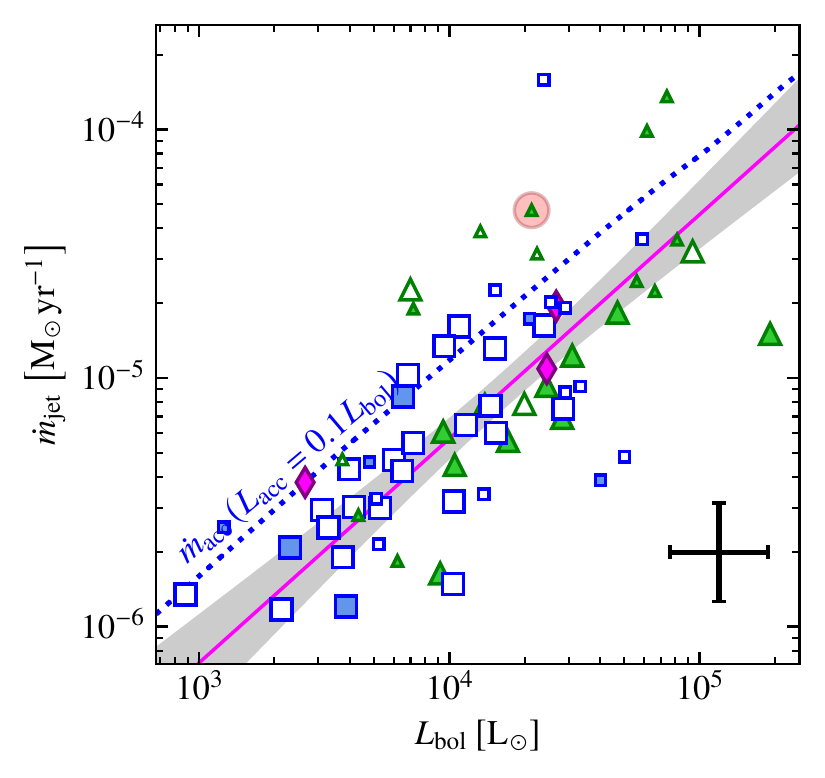}
	\caption[Jet mass loss rate against bolometric luminosity for jet-like sources]{A plot of the jet mass loss rate against bolometric luminosity for jet-like sources detected. Symbols have the same meaning as in \autoref{fig:distlumvsbollum} and the magenta line is the fit (\autoref{eq:jmlequationvlaatca}) to both the data presented here and that of \citetalias{Purser2016} with the corresponding $1\sigma$ confidence interval shaded in light grey. The dotted line represents the accretion rate \citep[using \autoref{eq:accrate} and assuming $L_{\rm acc}=0.1\Lbol$,][]{Cooper2013}.}
	\label{fig:jmlvslbol}
\end{figure}

For this work one of two techniques was used to compute $\jml$ based upon the observed values for $\alpha$ (details of which are can be found in \textcolor{blue}{Appendix} \ref{sec:appendixJMLs}). Calculated values for $\jml$ possess a range from $(1.2\pm0.7)\times10^{-6}$ to $(1.8\pm0.4)\times10^{-5}\Msol\yr^{-1}$,  a mean (of log values) of  $(8.7\pm6.1)\times10^{-6} \Msol\yr^{-1}$ and a median of $\left(6.1\pm2.5\right)\times 10^{-6}\Msol\yr^{-1}$. A full table of calculated values of $\jml$ for each individual case is available in \autoref{tab:JMLsOpangs} of \textcolor{blue}{Appendix} \ref{sec:appendixtables}. In \autoref{fig:jmlvslbol} we have plotted $\jml$ against $\Lbol$ for all jets from our sample and those from \citetalias{Purser2016}. A power-law was fitted to the sample of this work and that of \citetalias{Purser2016}, not including candidates, using the same methods discussed in \autoref{sec:radiobollum}. Consequently, \autoref{eq:jmlequationvlaatca} shows the derived relation for the combined sample which is plotted as a solid, magenta line in \autoref{fig:jmlvslbol}. For this relation we derive a partial correlation coefficient, whilst controlling for distance, of $\tau_{\rm kendall}=0.554$ with an associated $p$-value of $\sim1\times10^{-4}$ showing a significant correlation between $\jml$ and $\Lbol$.

\begin{align}
\log_{10} \left[ \frac{\jml}{\Msol\yr^{-1}} \right] &= (0.90\pm0.19)\cdot\log_{10}\left[ \frac{\Lbol}{\Lsol} \right] - (8.85\pm0.78)\label{eq:jmlequationvlaatca}
\end{align}

As discussed above, jets are proposed to be launched magnetocentrifugally or radiative line-driving. For the latter, it is expected that $\jml\propto \Lbol^{\nicefrac{3}{2}}$ \citep{Proga1998} yet we find that $\jml \propto \Lbol^{0.90\pm0.19}$. We take this as evidence negating line-driving as the dominant launching mechanism of jets from MYSOs.

To discern between competing magnetocentrifugal mechanisms, establishing the ratio of accretion to ejection rates is of paramount importance. Using \autoref{eq:accrate}, accretion rates were calculated from the accretion luminosities (whereby $\nicefrac{L_\mathrm{acc}}{L_\mathrm{bol}}=0.1$) following the work of \citet{Cooper2013} who assumed the empirical relationship between Br$\gamma$ and accretion luminosity (see their Figure 8). For that calculation, the results of \citet{Davies2011} were used to compute $R_\star$ and $M_\mathrm{\star}$ assuming the ZAMS configuration can approximate the MYSOs' protostellar structure.

\begin{equation}
\dot{m}_\mathrm{acc}=\frac{R_\star L_\mathrm{{acc}}}{\mathrm{G} M_\star}
\label{eq:accrate}
\end{equation}

\noindent where $\dot{m}_{\rm acc}$ is the accretion rate, $R_\star$ is the radius of the MYSO, $L_{\rm acc}$ is the accretion luminosity and $M_\star$ is the mass of the MYSO.

In \autoref{fig:jmlvslbol} the accretion rate is shown (blue, dotted line) and compared to our correlation for the jets' mass loss rates suggesting a reasonably constant ratio of $\nicefrac{\jml}{\dot{m}_\mathrm{acc}}$ across the high-mass regime. Finding the ratios of calculated values for $\jml$ and $\dot{m}_{\rm acc}$ gives an average value for $\nicefrac{\jml}{\dot{m}_\mathrm{acc}}$ of $0.19$ with a standard deviation of $0.19$, higher than those found towards low mass cases \citep[$\sim0.01-0.1$;][]{Hartigan1994}. However, due to the large approximations involved in the calculations of $\jml$ and $\dot{m}_{\rm acc}$, determining the dominant model of jet launching in MYSOs can not be achieved from the results here. To constrain the models further, a more accurate follow-up survey to constrain the accretion rates of each object, as well as jet velocities and ionisation fractions, is required to definitively measure this ratio.

\subsection{Jets and molecular outflows}
\label{sec4:jetmoms}
How molecular outflows are driven is as yet unknown, with a possibility being entrainment by jets. \citet{Guzman2012} argued that the typical momenta of jets, when compared with that of the large scale outflows, was too small to drive them. On the other hand, \citet{Sanna2016} showed that the ratio between the momentum of a $4\times10^4\Lsol$ MYSO's jet and that of its associated molecular outflow, over the dynamical timescale of the outflow, was of order unity. This indicated that the jet was mechanically able to fully drive the outflow. Since this was a single object study, its application to MYSOs and their molecular outflows in general necessitates a larger sample.

A distance-limited survey of 89 MYSOs by \citet{Maud2015CO} showed 59 to be associated with massive, molecular outflows and derived relationships for outflow force and momentum with $\Lbol$ (their Table 6). Consequently they also established an average dynamical timescale ($\bar{t}_{\rm dyn}$) for the molecular outflows of $8.4\times10^4\yr$, roughly the same as known timescales for massive star formation \citep[$\sim10^5\yr$,][]{McKeeTan2003,Mottram2011b}. Combining the work presented here with that of \citetalias{Purser2016}, an opportunity to compare the momenta of the molecular outflows \citep[from][]{Maud2015CO} and jets is available.

\begin{figure}
\centering
\includegraphics[width=\columnwidth]{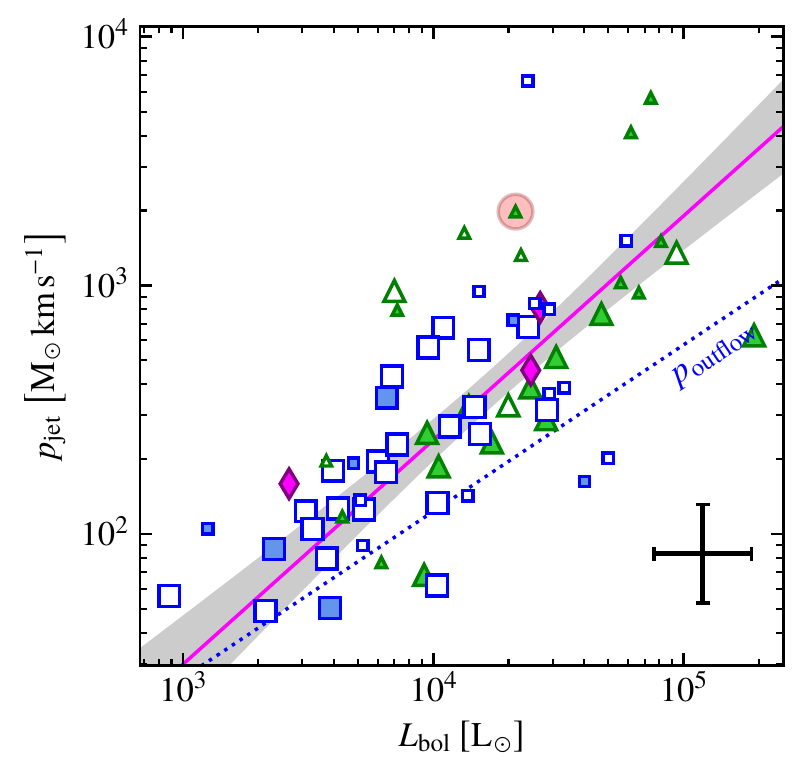}
\caption[Total jet momentum against bolometric luminosity]{A plot of the total jet momentum against bolometric luminosity. Symbols have the same meaning as in \autoref{fig:distlumvsbollum}, while the magenta line is the fit to both the data presented here and that of \citetalias{Purser2016} with the $1\sigma$ confidence interval shaded in light grey. The dotted line is the relationship between $p_\mathrm{outflow}$ and $\Lbol$ from Table 6 of \citet{Maud2015CO}.}
\label{fig4:jetmomentum}
\end{figure}

To calculate the momenta of the jets, their force (i.e.\ momentum rate), $F_{\rm jet}$, which is the product of jet velocity and mass loss rate, is integrated over time. Assuming $v_{\rm jet}=500\kmps$ and a jet lifetime of $8.4\times10^4\yr$ (i.e.\ $\bar{t}_{\rm dyn}$ of the molecular outflows), we calculate the total momenta of the jets, $p_{\rm jet}$. In \autoref{fig4:jetmomentum} $p_{\rm jet}$ is plotted against $\Lbol$ with the relationship for the molecular outflows from \citet{Maud2015CO} also shown (blue dotted line). Taking the ratio of these two power-laws, as with the single object study of \citet{Sanna2016}, $\nicefrac{p_{\rm jet}}{p_{\rm outflow}}\gtrsim 1$ ($0.9-2.1$ in the range $10^3 \leq L_{\rm bol}\leq 10^5$), supporting the idea that jets are the driving forces behind the molecular outflows. To further investigate this result, proper motion studies should be utilised to calculate jet velocities, and therefore total momentum, more accurately.

\subsection{Measured sizes and implications}
\label{sec:trappedhiis}
Early expansion of \textsc{Hii} regions may be governed by the interplay between pressure outwards and gravitational forces inwards \citep{Keto2002}. Towards more-evolved \textsc{Hii} regions, the gas pressure far exceeds the gravitational forces (i.e.\ $\tfrac{GM^2}{rnkT}\ll 1$) at their Str\"{o}mgren radii where ionisation is halted due to equality of Lyman fluxes and recombination rates. However, an MYSO with limited UV photon flux has a much smaller Str\"{o}mgren radius, whereby $\tfrac{GM^2}{rnkT}\gg 1$, and therefore the \textsc{Hii} region can become trapped only to expand when its radius exceeds that of the `gravitational radius', $r_{\rm g}$ (\autoref{eq:gravradius}).

\begin{equation}
r_{\rm g}=\frac{GM_\star}{{2 v_\mathrm{s}}^2}
\label{eq:gravradius}
\end{equation}

\noindent where $M_\star$ is the MYSO mass and $v_\mathrm{s}$ is the sound speed ($v_\mathrm{s}=\sqrt{\tfrac{kT}{m_{\rm H}}}$ where $T\sim10^4\K$ and $m_{\rm H}$ is the mass of hydrogen).

Pertinent to this work, these lines of thought lead to the question of whether observed thermal, free-free, radio emission originates in a trapped \textsc{Hii} region, or an ionised jet. To differentiate between the two possibilities, the gravitational radius of the MYSO and physical extent of the ionised gas must be compared, with similarity between the two quantities favouring a trapped \textsc{Hii} region. It is possible to infer MYSO mass from bolometric luminosity using the models of \citet{Davies2011}, and therefore calculate gravitational radius using \autoref{eq:gravradius}. A measure of the plasma's physical extent can be deduced from the radio images, assuming the radio component can be reasonably described by a 2-D Gaussian function.

\begin{figure}
\centering
\includegraphics[width=\columnwidth]{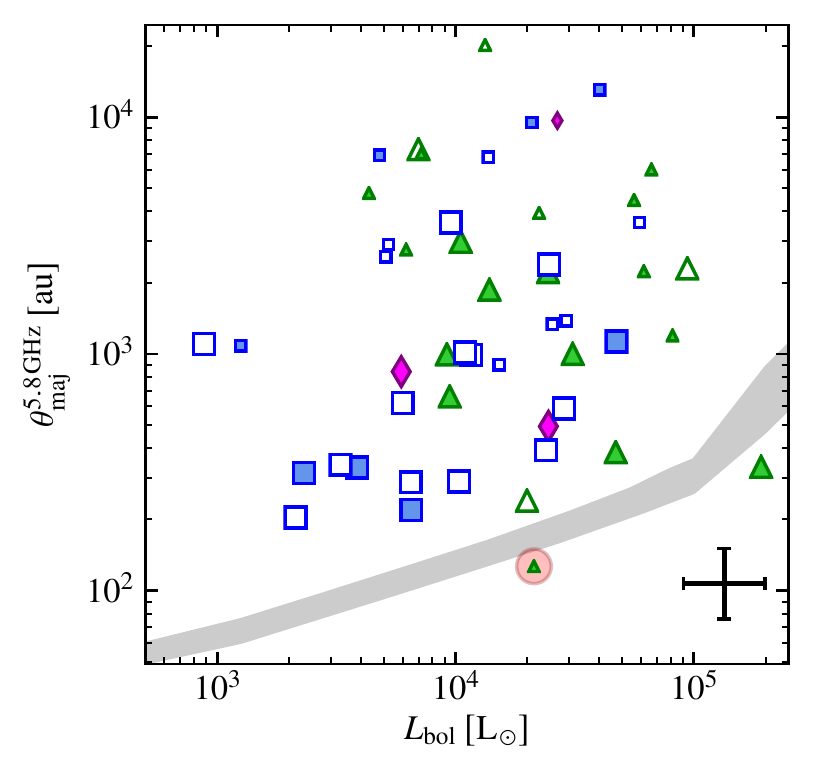}
\caption{Deconvolved major axis length at $5.8\GHz$, as derived using \textsc{imfit}, against bolometric luminosity for all sources displaying jet-like characteristics. Symbols are the same as in \autoref{fig:distlumvsbollum}, while the grey area represents twice the gravitational radius (\autoref{eq:gravradius}) for the relevant bolometric luminosity, with a $34\%$ error, \citep{Davies2011}.}
\label{fig:gravradii}
\end{figure}

\begin{figure}
\centering
\includegraphics[width=\columnwidth]{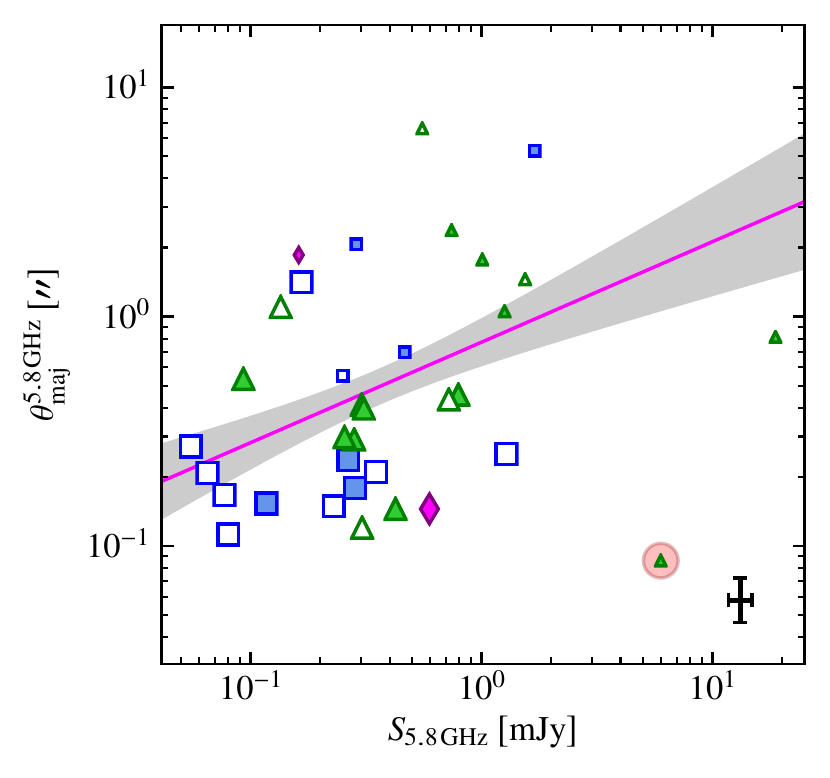}
\caption{Deconvolved major axis length at $5.8\GHz$ (in arcseconds), as derived using \textsc{imfit}, against radio flux at $5.8\GHz$ for all sources displaying jet-like characteristics. Our fit (\autoref{eq:thetamajdistlum}) to both the jet-like sources of this work and those of \citetalias{Purser2016} is shown as a magenta line with the fit's $1\sigma$ confidence interval shaded light grey. All markers used are the same as in \autoref{fig:distlumvsbollum}.}
\label{fig:thetamajdistlum}
\end{figure}

In \autoref{fig:gravradii}, the deduced major axes are plotted against bolometric luminosity for all sources with jet-like characteristics. Also plotted are the expected gravitational radii calculated using \autoref{eq:gravradius} in conjunction with the models of \citet{Davies2011}. In cases where measurement of $\theta_{\rm maj}$ was only possible at one frequency, a standard value of $\gamma=-0.7$ was used to extrapolate to $5.8\GHz$. No relation between jet major axis length and bolometric luminosity is found with a partial $\tau_{\rm kendall}$ (whilst controlling for distance) of 0.026 and corresponding $p$-value of 0.796. It is apparent that the major axes far exceed the gravitational radii. From \citet{Keto2002}, the ionisation front of an expanding \textsc{Hii} region moves with an approximate velocity of $\sim5\kmps$, corresponding to mean and median dynamical times for the ionisation front of $\sim640\yr$ and $1000\yr$ respectively. Considering the whole process of massive star formation is thought to last $\sim10^5\yr$ \citep{Davies2011}, only a small percentage of the jet-candidates observed could potentially be recently untrapped \textsc{Hii} regions, while the vast majority must have an elongated, ionised jet component to explain the large extent over which ionised material is found. As a further note, it is interesting that the contemporaneous \textsc{Hii}/jet from \citetalias{Purser2016}, G345.4938+01.4677, has a major axis length which coincides with that expected of a trapped \textsc{Hii} region model (highlighted marker in \autoref{fig:gravradii}).

Another question, that an analysis of jet morphology may answer, is that of jet collimation and on what scales it typically occurs. From Equations 12 and 16 of \citet{Reynolds1986} it can be shown that

\begin{align}
\left[\frac{\theta_{\rm maj}}{\au}\right] &\propto \left( S_\nu D^2 \right)^\frac{1}{1+\epsilon+q_{\rm T}}\nonumber\\
\Rightarrow \left[\frac{\theta_{\rm maj}}{\prime\prime}\right] &\propto  S_\nu ^\frac{1}{1+\epsilon+q_{\rm T}} D^{\frac{2}{1+\epsilon+q_{\rm T}}-1}\label{eq:thetamajfluxreynolds}
\end{align}

\noindent where $\epsilon$ is the power-law coefficient for variation of jet-width along its length, and $q_{\rm T}$ is a similar coefficient for temperature. In sensible physical models of jets, these parameters are constrained to be $0 \leq \epsilon \leq 1$ and $q_{\rm T} \leq 0$.

For the standard, non-recombining, conical jet model whereby the jet material adheres to ballistic trajectories (i.e.\ no longer influenced by magnetic fields), $\epsilon = 1$ and $q_{\rm T} = 0$. If the conical model were the dominant model to generally describe the jets in our sample, it is expected that $\theta_{\rm maj} \propto S_\nu ^{\nicefrac{1}{2}}$ following from \autoref{eq:thetamajfluxreynolds}. Whereas for jets still under collimation, we would expect $\epsilon < 1$ and therefore a steeper relation since $\left(1+\epsilon+q_{\rm T}\right)^{-1}>\nicefrac{1}{2}$ (assuming significant cooling does not occur on these scales).

\begin{align}
\log_{10}\left[ \frac{\theta_{\rm maj}}{\prime\prime} \right] &= (0.44 \pm 0.16)\cdot\log_{10}\left[ \frac{S_{\rm 5.8\GHz}}{\mJy} \right] - (0.11 \pm 0.12)\label{eq:thetamajdistlum}
\end{align}

In \autoref{fig:thetamajdistlum} therefore, we examine major axis length and its relationship with radio flux at $5.8\GHz$ (to avoid emission from dust at Q-band). A power law is fitted to ascertain the dominant jet model, the results of which are given in \autoref{eq:thetamajdistlum}. Since we find a power-law coefficient of $0.44\pm0.16$ and following from the above discussion, this may be evidence showing that the standard, conical model for ionised jets is the dominant one on the scales probed by our observations ($\sim10^2-10^4\au$). However, we calculate a partial $\tau_{\rm kendall}$ (whilst controlling for distance) of 0.337 and corresponding $p$-value of 0.009, which hints at the presence of correlation but is not conclusive (i.e. $p<0.001$). Possible inaccuracies in the measurement of $\theta_{\rm maj}$ resulting from deconvolution errors or non-constant mass ejection may have affected the quality of results. Further, more sensitive and multi-epoch radio observations would be therefore be required to more thoroughly establish, or dismiss, the tentative relationship seen above.

\section{Summary and Conclusions}
\label{sec4:conclusions}
Presented here, our radio observations towards forming massive stars at a variety of evolutionary stages represents the largest radio survey of jets associated with massive protostars to date. It has resulted in the detection of 14 (confirmed) ionised jets coincident with the MYSOs' infrared positions, of which 10 are determined to be associated with shock-ionised, radio lobes. Including those radio sources which hold jet-candidacy status this increases to a total of 38. Within $60\arcsec$ of the pointing centres and linked to other sites of star formation, a further 22 jets or candidates (6 of which are associated with lobes) are found. Analysis of the radio properties of this new, northern-hemisphere sample of ionised jets, as well as that from \citetalias{Purser2016}, were conducted to determine ionised jets' role in massive star formation, as well as their properties, resulting in the following conclusions:
\begin{enumerate}
\item Towards our IRDC subsample, $5.8\GHz$ radio emission is not detected to a level of $\sim1\mJy\kpc^2$ in cores with a luminosity-to-mass ratio of $\nicefrac{L_{\rm C}}{M_{\rm C}}\leq40$. Combined with the detection of masers towards pre-\textsc{Hii} region cores, this agrees with the standard evolutionary picture of molecular cores.
\item In agreement with the previous statistical study of \citetalias{Purser2016}, jet radio luminosities are found to scale with MYSO bolometric luminosity as $S_{\rm 5.8 \GHz} D^2 \propto \Lbol^{0.71\pm0.24}$, the same as for low mass jets. This indicates a common mechanism for the launch of ionised jets across all masses.
\item From comparison with the `dust-free' studies of \citetalias{Purser2016}, our work shows that dust emission accounts for an average of $46\pm10\%$ of an ionised jet's observed, Q-band ($44\GHz$) flux. This highlights the importance of well sampled cm/mm/sub-mm SEDs in the accurate deduction of ionised jet properties.
\item Non-detection of shock-ionised lobes towards lower-luminosity radio jets is primarily due to the sensitivity limit of our observations. This does not preclude the possibility that lower-luminosity radio jets are less likely to produce these lobes, for which observations with increased sensitivity are required.
\item Through calculation of the jets' mass-loss rates we observe the a correlation with bolometric luminosity of $\jml\propto\Lbol^{0.91\pm0.19}$. For radiative line-driving of jets it is predicted that $\jml\propto\Lbol^{\nicefrac{3}{2}}$ and therefore we conclude that this can not be the dominant launching mechanism of jets in the high-mass regime.
\item Comparing empirically-determined accretion rates with our calculated jet mass loss rates gives a typical value for $\nicefrac{\jml}{\dot{m}_{\rm acc}} \sim0.19$, consistent with current, magnetohydrodynamic, jet-launching models, yet higher than the low-mass case. Case-by-case measurements of important jet properties, such as velocity or ionisation fraction are required to discriminate between different magnetocentrifugal launching mechanisms.
\item Using the results of a previous study of massive, molecular outflows it has been shown that most ionised jets have larger momenta than the molecular outflows, whereby $\nicefrac{p_{\rm jet}}{p_{\rm outflow}}> 1$. This indicates that the outflows can indeed be powered by the ionised jets through mechanical entrainment.
\item From the maximum physical sizes of the radio emission from `jet-like' sources, an ionised jet is required to explain the presence of ionised gas past the gravitational radius for each MYSO. This rejects the hypothesis that weak, compact radio emission towards MYSOs stems from small, optically-thick \textsc{Hii} regions.
\end{enumerate}

For future works it has been shown that constraining the spectral properties of the jets themselves, at sub-mm, mm and cm wavelengths is crucial in accurately determining the jets' physical parameters. Relationships between the jets and various properties of the MYSOs themselves would be constrained further and ultimately the mechanisms for launch, collimation and general relationship with their environment elucidated. As briefly mentioned, a future e-MERLIN, matching-beam, C-band radio survey of these targets is planned and results will follow in a future publication.

\section*{Acknowledgements}
SJDP gratefully acknowledges the studentship funded by the Science and Technology Facilities Council of the United Kingdom (STFC) and also support from the advanced grant H2020--ERC--2016--ADG--74302 from the European Research Council (ERC) under the European Union's Horizon 2020 Research and Innovation programme. We would also like to acknowledge and thank the referee, whose comments helped to improve this work.

This paper has made use of information from the RMS survey database at \url{http://www.ast.leeds.ac.uk/RMS} which was constructed with support from the Science and Technology Facilities Council of the United Kingdom.

The National Radio Astronomy Observatory is a facility of the National Science Foundation operated under cooperative agreement by Associated Universities, Inc.

Throughout this work we also made use of \texttt{astropy}, a community-developed core python package for astronomy \citep[version 3.0.1,][]{Astropy2013}, and \texttt{uncertainties}, a python package for calculations with uncertainties (version 3.0.1) developed by Eric O. Lebigot, for plotting and error propagation purposes respectively.\\

\section*{Data Availability}
The data underlying this article are available on GitHub at \url{https://github.com/SimonP2207/RadioJetsFromYSOs}, and can be freely accessed.

\bibliographystyle{mnras}
\bibliography{Biblio.bib}

\appendix
\section{Tables}
\label{sec:appendixtables}
\onecolumn
\input{Tables/AppendixA/BeamsNoises}
\input{Tables/AppendixA/CBandTable1}
\input{Tables/AppendixA/QBandTable1}
\input{Tables/AppendixA/DeconvolvedSizes}
\input{Tables/AppendixA/ClassAlphaGamma}
\input{Tables/AppendixA/JMLTable}

\newpage
\section{Figures}
\label{sec:appendixfigures}
All imaging results are presented in this section. Common symbols/representations used throughout this appendix are noted explicitly below:
\begin{itemize}
	\item For the IRDC sample, cyan contours represent ATLASGAL \citep{Urquhart2014} $870\micron$ emission.
	\item Blue and black contours represent C-band and Q-band data, respectively.
	\item In infrared imagery, a white ellipse in the bottom left corner shows the relevant instrument's approximate PSF.
	\item In all panels, a green cross shows the position of $6.7\GHz$ methanol masers detected at C-band.
	\item Scalebars are illustrated in the bottom right of some panels. In panels where they are missing, the physical scale is the same as the one subplot of that band where the scalebar is present.
	\item If image noise levels are indicated in some panels, this is because they deviate from the values given in \autoref{tab:BeamsNoises}. This is either because a different part of the primary beam is shown, or a strong, nearby source is present.
\end{itemize}
\input{CPlotsPDFs}
\clearpage
\section{Calculation of mass loss rates}
\label{sec:appendixJMLs}
\begin{figure} 
	\includegraphics[width=\textwidth]{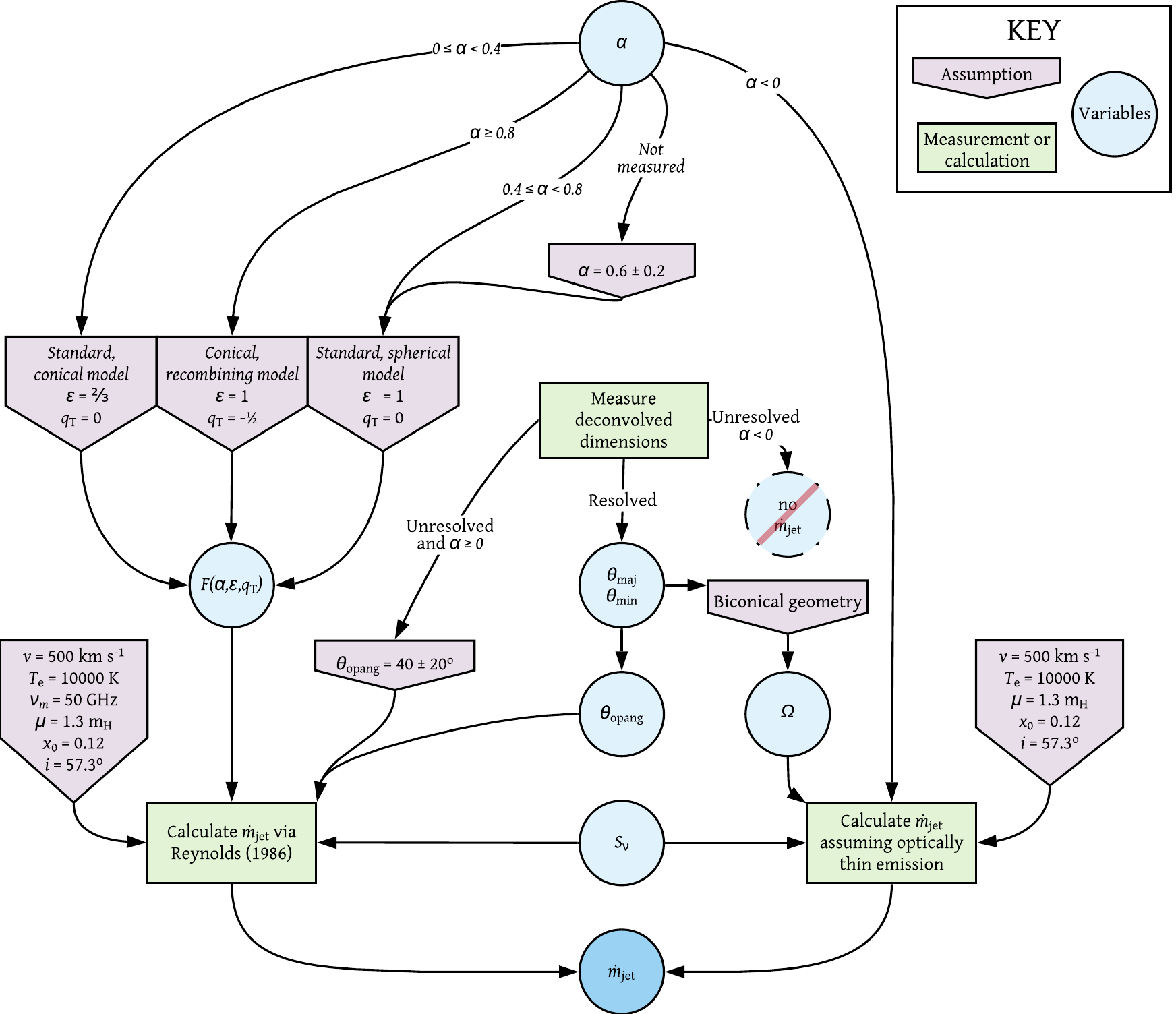}
	\caption{Paradigm used for calculating mass loss rate, $\dot{m}_{\rm jet}$, in ionised jets and disc-winds.}
	\label{fig:jmlflowchart}
\end{figure}

For calculating the mass loss rate of an ionised outflow, one of two approaches is used based upon the thermal emission's spectral index. These approaches are summarised in \autoref{fig:jmlflowchart} and discussed below.

In cases where $\alpha\geq0$, we use the analytical work of \citet{Reynolds1986}. Mass loss rates are calculated using \autoref{eq:jetmassloss} for which one of three physical jet-models is selected to calculate a value for $F\left(\alpha,\epsilon,q_{\rm T}\right)$ (\autoref{eq:Fmassloss}) . Each model differs in its power-law describing jet-width and temperature as functions of distance along the jet's axis, which are $w\left( r \right)\propto r^\epsilon$ and $T\left( r \right) \propto r^{q_{\rm T}}$, respectively. In cases where $0\leq \alpha < 0.4$ we adopt the `standard, conical' model representing jets whose material is still undergoing significant collimation, with $\epsilon=\nicefrac{2}{3}$ and $q_{\rm T}=0$. For $0.4 \leq \alpha < 0.8$ the `standard, spherical' model is used with $\epsilon=1$ and $q_{\rm T}=0$, whereby jet material is ballistic with no significant collimation. Finally, for $\alpha\geq0.8$ the `conical, recombining' model is used with $\epsilon=1$ and $q_{\rm T}=-\nicefrac{1}{2}$ to approximate jets with some degree of recombination/cooling in their flow. Deconvolved dimensions (from CASA's \texttt{imfit} task) are used to calculate opening angle whereby $\theta_{\rm 0}=\tan^{-1}\left( \nicefrac{\theta_{\rm maj}}{\theta_{\rm min}} \right)$. Assumed  values for velocity, temperature, turnover frequency, mean atomic weight, ionisation fraction and inclination are $v = 500\kmps$ (typical of MYSO jets), $T_{\rm e}=10000\K$ (above which collisional cooling with metals is significant), $\nu_{\rm m}=50\GHz$ (a lower limit from observational consensus), $\mu=1.3\, m_{\rm H}$ \citep[typical solar abundances;][]{Lodders2003}, $x_{\rm 0}=12\%$ \citep[from][]{Fedriani2019} and $i=57\fdg3$ (average inclination of a uniformly-distributed, random sample), respectively. In cases where no deconvolved dimensions could be measured, the weighted average opening angle of all jets from \citetalias{Purser2016} and this work is used, $\theta_0=35\pm16^\circ$.

\begin{align}
\left[ \frac{\dot{m}_{\rm jet}}{\rm M_\odot\,\yr^{-1}} \right] &= 1.17\times10^{12} \frac{v_{\rm jet} \, \mu \, S_\nu^{\nicefrac{3}{4}} \, D^{\nicefrac{3}{2}}\, \nu_{\rm m}^{\nicefrac{3\alpha}{4}-\nicefrac{9}{20}} \, \theta_\mathrm{0}^{\nicefrac{3}{4}}}{x_{\rm 0} \, \nu^{\nicefrac{3\alpha}{4}} \, T_e^{\nicefrac{3}{40}}\,\sin\left( i\right)^{\nicefrac{1}{4}}\,F\left(\alpha,\epsilon,q_{\rm T}\right)^{\nicefrac{3}{4}}}\label{eq:jetmassloss}\\
F\left(\alpha,\epsilon,q_{\rm T}\right) &= \frac{4.41}{q_\tau(\alpha-2)(\alpha+0.1)}\label{eq:Fmassloss}\\
q_\tau &= \frac{2.1\left( 1 + \epsilon + q_T \right)}{\alpha - 2}\nonumber
\end{align}

\noindent where all quantities are in SI units, unless explicitly stated.

\begin{figure} 
	\includegraphics[width=\textwidth]{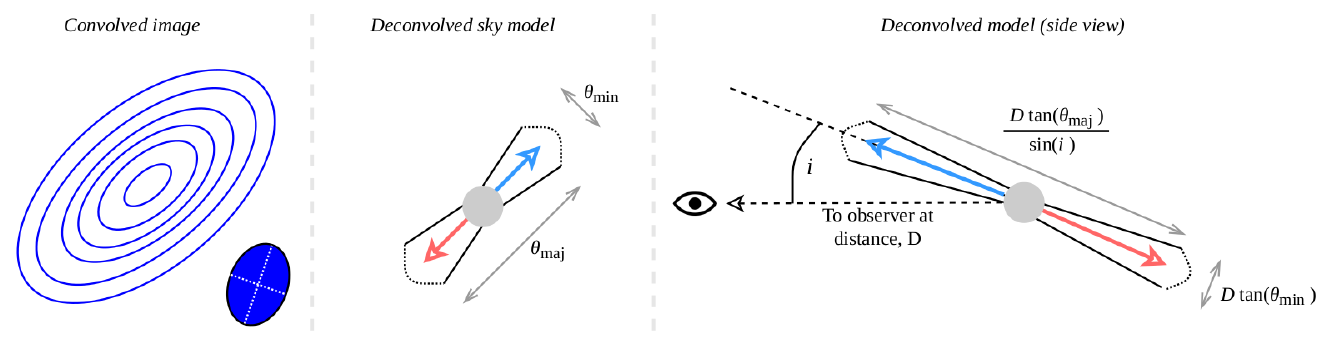}
	\caption{General schematic of an ionised jet, its dimensions and their relation to a clean radio image. Blue and red arrows indicate the jet moving towards and away from the observer, respectively.}
	\label{fig:jetprofile}
\end{figure} 

For those jets with $\alpha<0$, we take a second approach. Since optically thin, radio emission in this case is a measure of the full, ionised content of the emitting plasma. We approximate the jet as having biconical geometry with a total length, $D\,\tan\left( \theta_{\rm maj} \right) \, \sin \left( i \right)^{-1}$  and width, $D\, \tan\left(\theta_{\rm min} \right)$ (see \autoref{fig:jetprofile}). This leads to solid angle and volume as defined in Equations \ref{eq:BiconicalSolidAngle} and \ref{eq:BiconicalVolume}. Since we are in the regime whereby $h\nu\ll kT$ (assuming $T_{\rm e}=10000\K$), the Rayleigh-Jeans approximation can be used in conjunction with $T_{\rm b}=T_{\rm e}\left( 1-{\rm e}^{-\tau_{\rm \nu}} \right)$ to calculate the optical depth of the plasma (\autoref{eq:opticaldepth}). The amount of plasma along the line of sight, or emission measure, can be computed (in units of $\pc\,\cm^{-6}$) using \autoref{eq:emissionmeasure} \citep{MezgerHenderson1967} where the Gaunt factor is $\langle g_{\rm ff}\rangle = \ln ( 4.955\times10^7 \, \nu ) + 1.5 \ln \left( T_{\rm e} \right)$. With a value for emission measure the average electron density, $n_{\rm e}$, within the biconical jet can be calculated since $EM = \int n_{\rm e}^2\, ds$. Constant density is assumed and the average jet thickness, $ds=D\,\tan\left( \theta_{\rm min} \right) \left[ 2 \, \sin\left( i \right) \right]^{-1}$, is used. Mass loss rate is thereby computed using \autoref{eq:opthinjml} assuming a constant velocity over the timescale, $t = D\,\tan\left( \theta_{\rm maj} \right) \, \left( 2\, \sin \left( i \right) v_{\rm jet} \right)^{-1}$. For the other parameters the same values are used as with the alternate method detailed above. In the cases where no deconvolved dimensions could be measured, no mass loss rate is derived.

\begin{align}
	\Omega &= \frac{ \theta_{\rm maj} \, \theta_{\rm min}}{2} \label{eq:BiconicalSolidAngle}\\
	V &= \frac{\pi \, D^3 \,\tan \left( \theta_{\rm maj} \right) \, \tan^2 \left( \theta_{\rm min} \right) }{12\sin\left(i\right)} \label{eq:BiconicalVolume}\\
	\tau_{\rm \nu} &= - \ln \left( 1 -  \frac{S_{\rm \nu}\, c^2}{2\,k_{\rm B}\, \nu^2\, \Omega \, T_{\rm e}} \right) \label{eq:opticaldepth}\\
	\left[ \frac{EM}{\rm \pc \cm^{-6}} \right] &= 3.318\times10^{-17} \, \tau_\nu \, T^{\nicefrac{3}{2}}_{\rm e} \, \nu^{2} \langle g_{\rm ff} \rangle^{-1} \label{eq:emissionmeasure} \\
	\left[ \frac{\dot{m}_{\rm jet}}{\rm M_\odot\,\yr^{-1}} \right] &= 3.171\times10^{-23}  \frac{V \, n_{\rm e} \, \mu \, \sin \left( i \right) v_{\rm jet}}{D\,\tan\left( \theta_{\rm maj} \right) \, x_0} \label{eq:opthinjml}
\end{align}

SI units are used unless explicitly stated in the relevant equations.

\twocolumn
\section{Individual Classification/Notes}
\label{sec:appendixnotes}
\input{ObjectNotes}

\bsp	
\label{lastpage}
\end{document}

%% file: Commands.tex
\newcommand{\pc}{{\rm\thinspace pc}}
\newcommand{\cm}{{\rm\thinspace cm}}
\newcommand{\km}{{\rm\thinspace km}}
\newcommand{\mm}{{\rm\thinspace mm}}

\newcommand{\kmps}{\hbox{${\thinspace \rm\km\s^{-1}\,}$}}

\newcommand{\kpc}{{\rm\thinspace kpc}}
\newcommand{\K}{{\rm\thinspace K}}

\newcommand{\s}{{\rm\thinspace s}}
\newcommand{\au}{{\rm\thinspace au}}

\newcommand{\yr}{{\rm\thinspace yr}}

\newcommand{\mJy}{{\rm\thinspace mJy}}
\newcommand{\uJy}{{\rm\thinspace \upmu Jy}}

\newcommand{\Msol}{\hbox{${\rm\thinspace M_{\odot}}$}}
\newcommand{\Lsol}{\hbox{${\rm\thinspace L_{\odot}}$}}
\newcommand{\Lbol}{\hbox{$\thinspace L_{\rm{bol}}$}}

\newcommand{\mas}{{\rm\thinspace mas}}

\newcommand{\rahr}{\hbox{$\rm^h$}}
\newcommand{\ramin}{\hbox{$\rm^m$}}
\newcommand{\rasec}{\hbox{$\rm^s$}}
\newcommand{\GHz}{{\rm\thinspace GHz}}
\newcommand{\MHz}{{\rm\thinspace MHz}}
\def\pccmEM{\rm \pc \cm^{-3}}
\newcommand{\um}{{\rm\thinspace \upmu m}}
\newcommand{\jml}{\hbox{$\dot{m}_{\rm jet}$}}

%% file: Tables/IRDCTargets.tex
\begin{footnotesize}
\begin{table*}
\centering
\caption{All cores within the C-band field of view of our observed IRDC subsample. Core names marked with an asterisk denote the core used as the observational phase centre for its IRDC complex (with the exception of G033.69-00.01; see \autoref{sec:obsinfo}). Distances, bolometric luminosities and IR ratios, and their errors, are derived following the method described in \autoref{sec:distslumsirratios}. Values for $\Lbol$ given as ranges show the range of possible $\Lbol$. Calculations of this range used our derived distances and the ranges in luminosity given by \citet{Rathborne2010}. This uncertainty in luminosity is due to the lack of $24\um$ or $60-100\um$ detections and their constraint upon the SEDs. Mid-infrared flux ratios have superscripts which determine how their flux ratios were calculated, which are $^1$: Direct interpolation and ratio of measured fluxes; $^2$: Derived from fitted SED parameters.}
\begin{tabularx}{\textwidth}{lYYYYYY}
\toprule
IRDC Core & $\alpha$ & $\delta$ & $D$ & $\Lbol$ & $\frac{S_{70\um}}{S_{24\um}}$ & $M_{\rm core}$\\
 & $\left[ \mathrm{J2000} \right] $ & $\left[ \mathrm{J2000} \right] $ & $\left[ \kpc \right] $ & $ \left[ \Lsol \right] $ &  & $\left[ \Msol \right] $\\
\midrule
G018.82-00.28 MM2$^*$ & $18 \rahr 26 \ramin 23 \fs 4$ & $-12 \degr 39 \arcmin 37\arcsec$ & $4.53^{+0.49}_{-0.39}$ & $8920^{+2680}_{-2410}$ & $^1273\pm103$ & $80^{+24}_{-21}$\\
\addlinespace[0.075cm]
G018.82-00.28 MM4 & $18 \rahr 26 \ramin 15 \fs 5$ & $-12 \degr 41 \arcmin 32\arcsec$ & $4.53^{+0.49}_{-0.39}$ & $265^{+116}_{-111}$ & $^1461\pm174$ & $264^{+79}_{-71}$\\
\addlinespace[0.075cm]
G018.82-00.28 MM5 & $18 \rahr 26 \ramin 21 \fs 0$ & $-12 \degr 41 \arcmin 11\arcsec$ & $4.53^{+0.49}_{-0.39}$ & $34-426$ & $-$ & $149^{+57}_{-57}$\\
\addlinespace[0.075cm]
G024.08+00.04 MM1$^*$ & $18 \rahr 34 \ramin 57 \fs 0$ & $-07 \degr 43 \arcmin 26\arcsec$ & $3.53^{+0.41}_{-0.45}$ & $14300^{+7300}_{-7400}$ & $^21127\pm516$ & $68^{+21}_{-22}$\\
\addlinespace[0.075cm]
G024.08+00.04 MM2 & $18 \rahr 34 \ramin 51 \fs 1$ & $-07 \degr 45 \arcmin 32\arcsec$ & $3.53^{+0.41}_{-0.45}$ & $303^{+206}_{-209}$ & $^24866\pm5788$ & $84^{+27}_{-29}$\\
\addlinespace[0.075cm]
G024.08+00.04 MM3 & $18 \rahr 35 \ramin 02 \fs 2$ & $-07 \degr 45 \arcmin 25\arcsec$ & $3.53^{+0.41}_{-0.45}$ & $47-399$ & $-$ & $53^{+18}_{-18}$\\
\addlinespace[0.075cm]
G024.08+00.04 MM4 & $18 \rahr 35 \ramin 02 \fs 6$ & $-07 \degr 45 \arcmin 56\arcsec$ & $3.53^{+0.41}_{-0.45}$ & $58^{+38}_{-38}$ & $^29371\pm11181$ & $49^{+16}_{-17}$\\
\addlinespace[0.075cm]
G024.08+00.04 MM5 & $18 \rahr 35 \ramin 07 \fs 4$ & $-07 \degr 45 \arcmin 46\arcsec$ & $3.53^{+0.41}_{-0.45}$ & $21-289$ & $-$ & $37^{+13}_{-13}$\\
\addlinespace[0.075cm]
G024.33+00.11 MM1$^*$ & $18 \rahr 35 \ramin 07 \fs 9$ & $-07 \degr 35 \arcmin 04\arcsec$ & $3.56^{+0.44}_{-0.49}$ & $18600^{+8700}_{-9000}$ & $^21204\pm479$ & $214^{+68}_{-73}$\\
\addlinespace[0.075cm]
G024.33+00.11 MM10 & $18 \rahr 35 \ramin 27 \fs 9$ & $-07 \degr 35 \arcmin 32\arcsec$ & $3.56^{+0.44}_{-0.49}$ & $179^{+177}_{-178}$ & $^2439\pm505$ & $30^{+10}_{-11}$\\
\addlinespace[0.075cm]
G024.33+00.11 MM11 & $18 \rahr 35 \ramin 05 \fs 1$ & $-07 \degr 35 \arcmin 58\arcsec$ & $3.56^{+0.44}_{-0.49}$ & $1-201$ & $-$ & $136^{+50}_{-50}$\\
\addlinespace[0.075cm]
G024.33+00.11 MM4 & $18 \rahr 35 \ramin 19 \fs 4$ & $-07 \degr 37 \arcmin 17\arcsec$ & $3.56^{+0.44}_{-0.49}$ & $26-272$ & $-$ & $178^{+64}_{-64}$\\
\addlinespace[0.075cm]
G024.33+00.11 MM6 & $18 \rahr 35 \ramin 07 \fs 7$ & $-07 \degr 34 \arcmin 33\arcsec$ & $3.56^{+0.44}_{-0.49}$ & $51-391$ & $-$ & $119^{+43}_{-43}$\\
\addlinespace[0.075cm]
G024.33+00.11 MM8 & $18 \rahr 35 \ramin 23 \fs 4$ & $-07 \degr 37 \arcmin 21\arcsec$ & $3.56^{+0.44}_{-0.49}$ & $10-172$ & $-$ & $120^{+43}_{-43}$\\
\addlinespace[0.075cm]
G024.33+00.11 MM9 & $18 \rahr 35 \ramin 26 \fs 5$ & $-07 \degr 36 \arcmin 56\arcsec$ & $3.56^{+0.44}_{-0.49}$ & $650^{+781}_{-785}$ & $^225\pm29$ & $59^{+21}_{-22}$\\
\addlinespace[0.075cm]
G024.60+00.08 MM1 & $18 \rahr 35 \ramin 40 \fs 2$ & $-07 \degr 18 \arcmin 37\arcsec$ & $3.41^{+0.39}_{-0.35}$ & $861^{+499}_{-491}$ & $^1137\pm52$ & $93^{+29}_{-28}$\\
\addlinespace[0.075cm]
G024.60+00.08 MM2 & $18 \rahr 35 \ramin 35 \fs 7$ & $-07 \degr 18 \arcmin 09\arcsec$ & $3.41^{+0.39}_{-0.35}$ & $985^{+418}_{-406}$ & $^12702\pm1021$ & $94^{+29}_{-27}$\\
\addlinespace[0.075cm]
G024.60+00.08 MM3$^*$ & $18 \rahr 35 \ramin 41 \fs 1$ & $-07 \degr 18 \arcmin 30\arcsec$ & $3.41^{+0.39}_{-0.35}$ & $133-473$ & $-$ & $12^{+4}_{-4}$\\
\addlinespace[0.075cm]
G028.28-00.34 MM1$^*$ & $18 \rahr 44 \ramin 15 \fs 0$ & $-04 \degr 17 \arcmin 54\arcsec$ & $3.08^{+0.39}_{-0.43}$ & $4660^{+3950}_{-3980}$ & $^2166\pm143$ & $123^{+41}_{-44}$\\
\addlinespace[0.075cm]
G028.28-00.34 MM2 & $18 \rahr 44 \ramin 21 \fs 3$ & $-04 \degr 17 \arcmin 37\arcsec$ & $3.08^{+0.39}_{-0.43}$ & $3970^{+3470}_{-3500}$ & $^2142\pm128$ & $96^{+33}_{-34}$\\
\addlinespace[0.075cm]
G028.28-00.34 MM3 & $18 \rahr 44 \ramin 13 \fs 4$ & $-04 \degr 18 \arcmin 05\arcsec$ & $3.08^{+0.39}_{-0.43}$ & $3560^{+2820}_{-2850}$ & $^2172\pm124$ & $21^{+7}_{-7}$\\
\addlinespace[0.075cm]
G028.37+00.07 MM1$^*$ & $18 \rahr 42 \ramin 52 \fs 1$ & $-03 \degr 59 \arcmin 45\arcsec$ & $4.56^{+0.70}_{-0.55}$ & $34100^{+16700}_{-15400}$ & $^29811\pm4016$ & $447^{+164}_{-140}$\\
\addlinespace[0.075cm]
G028.37+00.07 MM10 & $18 \rahr 42 \ramin 54 \fs 0$ & $-04 \degr 02 \arcmin 30\arcsec$ & $4.56^{+0.70}_{-0.55}$ & $1080^{+880}_{-850}$ & $^21691\pm1571$ & $131^{+49}_{-42}$\\
\addlinespace[0.075cm]
G028.37+00.07 MM11 & $18 \rahr 42 \ramin 42 \fs 7$ & $-04 \degr 01 \arcmin 44\arcsec$ & $4.56^{+0.70}_{-0.55}$ & $302^{+246}_{-239}$ & $^2871\pm830$ & $143^{+54}_{-47}$\\
\addlinespace[0.075cm]
G028.37+00.07 MM14 & $18 \rahr 42 \ramin 52 \fs 6$ & $-04 \degr 02 \arcmin 44\arcsec$ & $4.56^{+0.70}_{-0.55}$ & $2-176$ & $-$ & $17^{+6}_{-6}$\\
\addlinespace[0.075cm]
G028.37+00.07 MM16 & $18 \rahr 42 \ramin 40 \fs 2$ & $-04 \degr 00 \arcmin 23\arcsec$ & $4.56^{+0.70}_{-0.55}$ & $6-516$ & $-$ & $227^{+90}_{-90}$\\
\addlinespace[0.075cm]
G028.37+00.07 MM17 & $18 \rahr 43 \ramin 00 \fs 0$ & $-04 \degr 01 \arcmin 34\arcsec$ & $4.56^{+0.70}_{-0.55}$ & $2-146$ & $-$ & $77^{+30}_{-30}$\\
\addlinespace[0.075cm]
G028.37+00.07 MM2 & $18 \rahr 42 \ramin 37 \fs 6$ & $-04 \degr 02 \arcmin 05\arcsec$ & $4.56^{+0.70}_{-0.55}$ & $20500^{+13500}_{-12900}$ & $^2411\pm179$ & $171^{+63}_{-54}$\\
\addlinespace[0.075cm]
G028.37+00.07 MM4 & $18 \rahr 42 \ramin 50 \fs 7$ & $-04 \degr 03 \arcmin 15\arcsec$ & $4.56^{+0.70}_{-0.55}$ & $2240^{+1740}_{-1690}$ & $^21725\pm1468$ & $120^{+45}_{-39}$\\
\addlinespace[0.075cm]
G028.37+00.07 MM6 & $18 \rahr 42 \ramin 49 \fs 0$ & $-04 \degr 02 \arcmin 23\arcsec$ & $4.56^{+0.70}_{-0.55}$ & $22-1294$ & $-$ & $127^{+51}_{-51}$\\
\addlinespace[0.075cm]
G028.37+00.07 MM9 & $18 \rahr 42 \ramin 46 \fs 7$ & $-04 \degr 04 \arcmin 08\arcsec$ & $4.56^{+0.70}_{-0.55}$ & $5-507$ & $-$ & $200^{+79}_{-79}$\\
\addlinespace[0.075cm]
G028.67+00.13 MM1$^*$ & $18 \rahr 43 \ramin 03 \fs 1$ & $-03 \degr 41 \arcmin 41\arcsec$ & $4.57^{+0.73}_{-0.49}$ & $10400^{+8400}_{-8000}$ & $^2137\pm91$ & $39^{+15}_{-12}$\\
\addlinespace[0.075cm]
G028.67+00.13 MM2 & $18 \rahr 43 \ramin 07 \fs 1$ & $-03 \degr 44 \arcmin 01\arcsec$ & $4.57^{+0.73}_{-0.49}$ & $147^{+92}_{-85}$ & $^199\pm37$ & $209^{+81}_{-64}$\\
\addlinespace[0.075cm]
G028.67+00.13 MM5 & $18 \rahr 43 \ramin 10 \fs 1$ & $-03 \degr 45 \arcmin 08\arcsec$ & $4.57^{+0.73}_{-0.49}$ & $6-134$ & $-$ & $40^{+16}_{-16}$\\
\addlinespace[0.075cm]
G028.67+00.13 MM6 & $18 \rahr 43 \ramin 12 \fs 2$ & $-03 \degr 45 \arcmin 39\arcsec$ & $4.57^{+0.73}_{-0.49}$ & $3-183$ & $-$ & $43^{+17}_{-17}$\\
\addlinespace[0.075cm]
G033.69-00.01 MM1 & $18 \rahr 52 \ramin 58 \fs 8$ & $+00 \degr 42 \arcmin 37\arcsec$ & $5.95^{+0.53}_{-0.56}$ & $1480^{+870}_{-880}$ & $^1117\pm44$ & $467^{+132}_{-135}$\\
\addlinespace[0.075cm]
G033.69-00.01 MM10 & $18 \rahr 52 \ramin 52 \fs 7$ & $+00 \degr 38 \arcmin 35\arcsec$ & $5.95^{+0.53}_{-0.56}$ & $13-587$ & $-$ & $42^{+13}_{-13}$\\
\addlinespace[0.075cm]
G033.69-00.01 MM11 & $18 \rahr 52 \ramin 56 \fs 2$ & $+00 \degr 41 \arcmin 48\arcsec$ & $5.95^{+0.53}_{-0.56}$ & $16-702$ & $-$ & $33^{+10}_{-10}$\\
\addlinespace[0.075cm]
G033.69-00.01 MM2 & $18 \rahr 52 \ramin 49 \fs 9$ & $+00 \degr 37 \arcmin 57\arcsec$ & $5.95^{+0.53}_{-0.56}$ & $17300^{+14900}_{-15000}$ & $^2131\pm117$ & $330^{+94}_{-96}$\\
\addlinespace[0.075cm]
G033.69-00.01 MM3 & $18 \rahr 52 \ramin 50 \fs 8$ & $+00 \degr 36 \arcmin 43\arcsec$ & $5.95^{+0.53}_{-0.56}$ & $4260^{+3760}_{-3770}$ & $^2104\pm97$ & $64^{+18}_{-19}$\\
\addlinespace[0.075cm]
G033.69-00.01 MM4 & $18 \rahr 52 \ramin 56 \fs 4$ & $+00 \degr 43 \arcmin 08\arcsec$ & $5.95^{+0.53}_{-0.56}$ & $197^{+117}_{-118}$ & $^177\pm29$ & $298^{+85}_{-87}$\\
\addlinespace[0.075cm]
G033.69-00.01 MM5 & $18 \rahr 52 \ramin 47 \fs 8$ & $+00 \degr 36 \arcmin 47\arcsec$ & $5.95^{+0.53}_{-0.56}$ & $14800^{+10500}_{-10600}$ & $^2316\pm183$ & $66^{+18}_{-18}$\\
\addlinespace[0.075cm]
G033.69-00.01 MM6 & $18 \rahr 52 \ramin 48 \fs 7$ & $+00 \degr 35 \arcmin 58\arcsec$ & $5.95^{+0.53}_{-0.56}$ & $35-491$ & $-$ & $145^{+44}_{-44}$\\
\addlinespace[0.075cm]
G033.69-00.01 MM8 & $18 \rahr 52 \ramin 53 \fs 9$ & $+00 \degr 41 \arcmin 16\arcsec$ & $5.95^{+0.53}_{-0.56}$ & $17-733$ & $-$ & $233^{+72}_{-72}$\\
\addlinespace[0.075cm]
G033.69-00.01 MM9 & $18 \rahr 52 \ramin 58 \fs 1$ & $+00 \degr 41 \arcmin 20\arcsec$ & $5.95^{+0.53}_{-0.56}$ & $4720^{+3980}_{-4000}$ & $^2162\pm140$ & $30^{+8}_{-9}$\\
\bottomrule
\end{tabularx}
\label{tab:irdctargets}
\end{table*}
\end{footnotesize}

%% file: Tables/MYSOTargets.tex
\begin{footnotesize}
\begin{table*}
\centering
\caption{All targets observed within our MYSO subsample. Distances, and their errors, are derived following the method described in \autoref{sec:distslumsirratios} unless a superscript number is present on the left showing which reference it is taken from. These are: $^{\color{blue}1}$ : \citet{Fujisawa2012}; $^{\color{blue}2}$ : \citet{Zhang2009}; $^{\color{blue}3}$ : \citet{Wu2014}; $^{\color{blue}4}$ : \citet{Rygl2012}; $^{\color{blue}5}$ : \citet{Choi2014}; $^{\color{blue}6}$ : \citet{Rygl2010}; $^{\color{blue}7}$ : \citet{Moscadelli2009}; $^{\color{blue}8}$ : \citet{Imai2000}; $^{\color{blue}9}$ : \citet{Hachisuka2006}; $^{\color{blue}10}$ : \citet{Kawamura1998}; $^{\color{blue}11}$ : \citet{Burns2017}; $^{\color{blue}12}$ : \citet{Honma2007}. An asterisk indicates that our data for that object was previously presented by \citet{Rosero2019}.}
\begin{tabularx}{\textwidth}{lY>{\centering\hsize=2cm}XYYYYYY}
\toprule
MYSO & IRAS & Alias & $\alpha$ & $\delta$ & $D$ & $\Lbol$ & $\frac{S_{70\um}}{S_{24\um}}$\\
 &  &  & $\left[ \mathrm{J2000} \right] $ & $\left[ \mathrm{J2000} \right] $ & $\left[ \kpc \right] $ & $ \left[ \Lsol \right] $ & \\
\midrule
G033.6437-00.2277 & 18509+0027 &  & $18 \rahr 53 \ramin 32 \fs 60$ & $00 \degr 31 \arcmin 39 \farcs 0$ & $^{\color{blue}1}3.65^{+0.40}_{-0.35}$ & $11600^{+3500}_{-3200}$ & $15.3\pm2.3$\\
\addlinespace[0.075cm]
G035.1979-00.7427 & 18556+0136 & G35.20-0.74N & $18 \rahr 58 \ramin 12 \fs 99$ & $01 \degr 40 \arcmin 31 \farcs 2$ & $^{\color{blue}2}2.19^{+0.24}_{-0.20}$ & $31000^{+7700}_{-6700}$ & $9.4\pm0.7$\\
\addlinespace[0.075cm]
G035.1992-01.7424 & 18592+0108 & W48 & $19 \rahr 01 \ramin 46 \fs 70$ & $01 \degr 13 \arcmin 24 \farcs 0$ & $^{\color{blue}2}3.27^{+0.56}_{-0.42}$ & $169000^{+60000}_{-46000}$ & $12.4\pm1.1$\\
\addlinespace[0.075cm]
G037.4266+01.5183 & 18517+0437 &  & $18 \rahr 54 \ramin 13 \fs 80$ & $04 \degr 41 \arcmin 32 \farcs 0$ & $^{\color{blue}3}1.88^{+0.08}_{-0.08}$ & $5290^{+1650}_{-1650}$ & $40.9\pm3.7$\\
\addlinespace[0.075cm]
G056.3694-00.6333 & 19363+2018 &  & $19 \rahr 38 \ramin 31 \fs 63$ & $20 \degr 25 \arcmin 18 \farcs 7$ & $6.65^{+0.63}_{-0.73}$ & $6990^{+1700}_{-1870}$ & $11.9\pm1.6$\\
\addlinespace[0.075cm]
G077.5671+03.6911 & 20107+4038 &  & $20 \rahr 12 \ramin 33 \fs 70$ & $40 \degr 47 \arcmin 40 \farcs 7$ & $6.13^{+0.75}_{-0.82}$ & $5330^{+1330}_{-1450}$ & $5.3\pm0.4$\\
\addlinespace[0.075cm]
G078.8699+02.7602 & 20187+4111 & V1318 Cygni & $20 \rahr 20 \ramin 30 \fs 60$ & $41 \degr 21 \arcmin 26 \farcs 6$ & $^{\color{blue}4}1.40^{+0.08}_{-0.08}$ & $6520^{+1110}_{-1110}$ & $5.1\pm0.3$\\
\addlinespace[0.075cm]
G079.8855+02.5517B & 20227+4154 &  & $20 \rahr 24 \ramin 31 \fs 68$ & $42 \degr 04 \arcmin 22 \farcs 4$ & $^{\color{blue}4}1.40^{+0.08}_{-0.08}$ & $2170^{+770}_{-770}$ & $9.8\pm1.2$\\
\addlinespace[0.075cm]
G081.8652+00.7800 &  & W75 IRS2 & $20 \rahr 38 \ramin 35 \fs 36$ & $42 \degr 37 \arcmin 13 \farcs 7$ & $^{\color{blue}4}1.30^{+0.07}_{-0.07}$ & $3100^{+2700}_{-1480}$ & $11.9\pm1.4$\\
\addlinespace[0.075cm]
G083.7071+03.2817 &  &  & $20 \rahr 33 \ramin 36 \fs 51$ & $45 \degr 35 \arcmin 44 \farcs 0$ & $^{\color{blue}4}1.40^{+0.08}_{-0.08}$ & $3870^{+1190}_{-1190}$ & $6.6\pm0.0$\\
\addlinespace[0.075cm]
G084.9505-00.6910 &  &  & $20 \rahr 55 \ramin 32 \fs 47$ & $44 \degr 06 \arcmin 10 \farcs 1$ & $5.87^{+0.80}_{-0.80}$ & $14600^{+4500}_{-4500}$ & $8.9\pm0.1$\\
\addlinespace[0.075cm]
G094.2615-00.4116 & 21307+5049 &  & $21 \rahr 32 \ramin 30 \fs 59$ & $51 \degr 02 \arcmin 16 \farcs 0$ & $5.56^{+0.83}_{-0.83}$ & $10500^{+3300}_{-3300}$ & $12.6\pm1.3$\\
\addlinespace[0.075cm]
G094.3228-00.1671 & 21300+5102 &  & $21 \rahr 31 \ramin 45 \fs 11$ & $51 \degr 15 \arcmin 35 \farcs 3$ & $4.83^{+0.76}_{-0.84}$ & $6840^{+2240}_{-2460}$ & $5.0\pm0.5$\\
\addlinespace[0.075cm]
G094.4637-00.8043 & 21334+5039 &  & $21 \rahr 35 \ramin 09 \fs 11$ & $50 \degr 53 \arcmin 09 \farcs 6$ & $5.36^{+0.76}_{-0.91}$ & $24500^{+7700}_{-9000}$ & $8.2\pm0.6$\\
\addlinespace[0.075cm]
G094.6028-01.7966 & 21381+5000 & V645 Cygni & $21 \rahr 39 \ramin 58 \fs 25$ & $50 \degr 14 \arcmin 20 \farcs 9$ & $^{\color{blue}5}3.95^{+0.41}_{-0.34}$ & $28500^{+7600}_{-6800}$ & $1.9\pm0.1$\\
\addlinespace[0.075cm]
G100.3779-03.5784 & 22142+5206 &  & $22 \rahr 16 \ramin 10 \fs 35$ & $52 \degr 21 \arcmin 34 \farcs 7$ & $^{\color{blue}5}3.46^{+0.20}_{-0.18}$ & $15400^{+5100}_{-5100}$ & $1.9\pm0.2$\\
\addlinespace[0.075cm]
G102.8051-00.7184B & 22172+5549 &  & $22 \rahr 19 \ramin 09 \fs 11$ & $56 \degr 05 \arcmin 00 \farcs 3$ & $4.36^{+0.78}_{-0.78}$ & $2770^{+1030}_{-1030}$ & $9.5\pm1.5$\\
\addlinespace[0.075cm]
G103.8744+01.8558 & 22134+5834 &  & $22 \rahr 15 \ramin 09 \fs 08$ & $58 \degr 49 \arcmin 07 \farcs 8$ & $2.27^{+0.59}_{-0.86}$ & $9460^{+5380}_{-7490}$ & $3.9\pm0.2$\\
\addlinespace[0.075cm]
G105.5072+00.2294 & 22305+5803 &  & $22 \rahr 32 \ramin 23 \fs 85$ & $58 \degr 18 \arcmin 59 \farcs 8$ & $4.87^{+0.95}_{-0.80}$ & $11000^{+4900}_{-4300}$ & $8.1\pm0.9$\\
\addlinespace[0.075cm]
G107.6823-02.2423A & 22534+5653 &  & $22 \rahr 55 \ramin 29 \fs 82$ & $57 \degr 09 \arcmin 24 \farcs 9$ & $5.02^{+0.88}_{-0.80}$ & $4780^{+1720}_{-1570}$ & $4.4\pm0.4$\\
\addlinespace[0.075cm]
G108.1844+05.5187 & 22272+6358A & LDN 1206 & $22 \rahr 28 \ramin 51 \fs 40$ & $64 \degr 13 \arcmin 41 \farcs 0$ & $^{\color{blue}6}0.78^{+0.10}_{-0.08}$ & $873^{+736}_{-396}$ & $32.5\pm2.6$\\
\addlinespace[0.075cm]
G108.4714-02.8176 &  &  & $23 \rahr 02 \ramin 32 \fs 07$ & $56 \degr 57 \arcmin 51 \farcs 3$ & $^{\color{blue}5}3.24^{+0.11}_{-0.10}$ & $2650^{+340}_{-330}$ & $12.8\pm0.3$\\
\addlinespace[0.075cm]
G108.5955+00.4935A & 22506+5944 &  & $22 \rahr 52 \ramin 38 \fs 71$ & $60 \degr 00 \arcmin 44 \farcs 5$ & $^{\color{blue}5}2.47^{+0.22}_{-0.19}$ & $1230^{+240}_{-220}$ & $6.8\pm0.8$\\
\addlinespace[0.075cm]
G108.7575-00.9863 & 22566+5828 &  & $22 \rahr 58 \ramin 47 \fs 25$ & $58 \degr 45 \arcmin 01 \farcs 6$ & $4.52^{+0.85}_{-0.78}$ & $15200^{+14300}_{-8800}$ & $8.1\pm0.6$\\
\addlinespace[0.075cm]
G110.0931-00.0641 & 23033+5951 &  & $23 \rahr 05 \ramin 25 \fs 16$ & $60 \degr 08 \arcmin 15 \farcs 4$ & $4.70^{+0.85}_{-0.78}$ & $13900^{+5300}_{-4900}$ & $12.5\pm2.1$\\
\addlinespace[0.075cm]
G111.2348-01.2385 & 23151+5912 &  & $23 \rahr 17 \ramin 21 \fs 01$ & $59 \degr 28 \arcmin 48 \farcs 0$ & $^{\color{blue}5}3.33^{+1.23}_{-0.71}$ & $24000^{+17900}_{-10600}$ & $5.8\pm0.6$\\
\addlinespace[0.075cm]
G111.2552-00.7702 & 23139+5939 &  & $23 \rahr 16 \ramin 10 \fs 39$ & $59 \degr 55 \arcmin 28 \farcs 2$ & $^{\color{blue}5}3.34^{+0.27}_{-0.23}$ & $9200^{+8090}_{-4480}$ & $10.4\pm1.4$\\
\addlinespace[0.075cm]
G111.5671+00.7517$^{\color{blue}*}$ & 23118+6110 & NGC 7538 IRS9 & $23 \rahr 14 \ramin 01 \fs 75$ & $61 \degr 27 \arcmin 19 \farcs 8$ & $^{\color{blue}7}2.65^{+0.12}_{-0.11}$ & $47000^{+40900}_{-22300}$ & $35.1\pm1.0$\\
\addlinespace[0.075cm]
G114.0835+02.8568 & 23262+6401 &  & $23 \rahr 28 \ramin 27 \fs 76$ & $64 \degr 17 \arcmin 38 \farcs 4$ & $4.48^{+0.84}_{-0.77}$ & $8170^{+3860}_{-3660}$ & $4.0\pm0.4$\\
\addlinespace[0.075cm]
G118.6172-01.3312 & 00127+6058 &  & $00 \rahr 15 \ramin 27 \fs 83$ & $61 \degr 14 \arcmin 18 \farcs 9$ & $3.13^{+0.72}_{-0.66}$ & $6010^{+2850}_{-2620}$ & $14.4\pm1.6$\\
\addlinespace[0.075cm]
G126.7144-00.8220 & 01202+6133 &  & $01 \rahr 23 \ramin 33 \fs 17$ & $61 \degr 48 \arcmin 48 \farcs 2$ & $1.14^{+0.39}_{-0.56}$ & $6490^{+4490}_{-6410}$ & $7.0\pm0.4$\\
\addlinespace[0.075cm]
G133.7150+01.2155 & 02219+6152 & W3 IRS5 & $02 \rahr 25 \ramin 40 \fs 77$ & $62 \degr 05 \arcmin 52 \farcs 4$ & $^{\color{blue}8}1.83^{+0.14}_{-0.14}$ & $192000^{+168000}_{-94000}$ & $3.6\pm0.4$\\
\addlinespace[0.075cm]
G134.2792+00.8561 & 02252+6120 &  & $02 \rahr 29 \ramin 01 \fs 93$ & $61 \degr 33 \arcmin 30 \farcs 5$ & $^{\color{blue}9}2.04^{+0.07}_{-0.07}$ & $3300^{+1090}_{-1090}$ & $6.3\pm0.5$\\
\addlinespace[0.075cm]
G136.3833+02.2666A & 02461+6147 &  & $02 \rahr 50 \ramin 08 \fs 57$ & $61 \degr 59 \arcmin 52 \farcs 1$ & $3.39^{+0.86}_{-0.56}$ & $7180^{+3830}_{-2640}$ & $4.2\pm0.4$\\
\addlinespace[0.075cm]
G138.2957+01.5552$^{\color{blue}*}$ & 02575+6017 & AFGL 4029 IRS1 & $03 \rahr 01 \ramin 31 \fs 32$ & $60 \degr 29 \arcmin 13 \farcs 2$ & $3.17^{+0.70}_{-0.64}$ & $17200^{+8100}_{-7500}$ & $4.6\pm0.3$\\
\addlinespace[0.075cm]
G139.9091+00.1969A$^{\color{blue}*}$ & 03035+5819 & AFGL 437S & $03 \rahr 07 \ramin 24 \fs 52$ & $58 \degr 30 \arcmin 43 \farcs 3$ & $3.41^{+0.73}_{-0.73}$ & $12300^{+6000}_{-6000}$ & $3.3\pm0.2$\\
\addlinespace[0.075cm]
G141.9996+01.8202 & 03236+5836 & AFGL 490 & $03 \rahr 27 \ramin 38 \fs 76$ & $58 \degr 47 \arcmin 00 \farcs 1$ & $0.97^{+0.46}_{-0.40}$ & $2140^{+2040}_{-1780}$ & $2.8\pm0.1$\\
\addlinespace[0.075cm]
G143.8118-01.5699 &  &  & $03 \rahr 24 \ramin 50 \fs 95$ & $54 \degr 57 \arcmin 32 \farcs 7$ & $2.58^{+0.76}_{-0.49}$ & $10400^{+6200}_{-4100}$ & $32.0\pm0.4$\\
\addlinespace[0.075cm]
G148.1201+00.2928 & 03523+5343 & AFGL 5107 & $03 \rahr 56 \ramin 15 \fs 36$ & $53 \degr 52 \arcmin 13 \farcs 0$ & $3.22^{+0.89}_{-0.57}$ & $3980^{+2410}_{-1720}$ & $3.7\pm0.4$\\
\addlinespace[0.075cm]
G160.1452+03.1559 & 04579+4703 &  & $05 \rahr 01 \ramin 39 \fs 89$ & $47 \degr 07 \arcmin 21 \farcs 7$ & $2.05^{+0.80}_{-0.46}$ & $2310^{+1870}_{-1140}$ & $4.9\pm0.4$\\
\addlinespace[0.075cm]
G173.4839+02.4317 & 05358+3543 & SH 2-233 IR & $05 \rahr 39 \ramin 09 \fs 92$ & $35 \degr 45 \arcmin 17 \farcs 2$ & $^{\color{blue}10}2.00^{+0.60}_{-0.60}$ & $2890^{+3040}_{-2190}$ & $13.4\pm1.2$\\
\addlinespace[0.075cm]
G174.1974-00.0763 & 05274+3345 & AFGL 5142 IRS2 & $05 \rahr 30 \ramin 46 \fs 06$ & $33 \degr 47 \arcmin 54 \farcs 1$ & $^{\color{blue}11}2.14^{+0.05}_{-0.05}$ & $6760^{+2090}_{-2090}$ & $9.1\pm0.7$\\
\addlinespace[0.075cm]
G177.7291-00.3358 & 05355+3039 & CPM 18 & $05 \rahr 38 \ramin 47 \fs 16$ & $30 \degr 41 \arcmin 18 \farcs 1$ & $^{\color{blue}10}2.00^{+0.60}_{-0.60}$ & $3760^{+2280}_{-2280}$ & $10.0\pm0.8$\\
\addlinespace[0.075cm]
G183.3485-00.5751 & 05480+2545 &  & $05 \rahr 51 \ramin 11 \fs 15$ & $25 \degr 46 \arcmin 16 \farcs 4$ & $^{\color{blue}10}2.00^{+0.60}_{-0.60}$ & $4170^{+2580}_{-2580}$ & $19.7\pm1.7$\\
\addlinespace[0.075cm]
G188.9479+00.8871 & 06058+2138 & AFGL 5180 IRS1 & $06 \rahr 08 \ramin 53 \fs 40$ & $21 \degr 38 \arcmin 28 \farcs 1$ & $^{\color{blue}5}2.10^{+0.03}_{-0.03}$ & $10400^{+1000}_{-1000}$ & $9.2\pm0.8$\\
\addlinespace[0.075cm]
G189.0307+00.7821 & 06056+2131 & AFGL 6366S & $06 \rahr 08 \ramin 40 \fs 52$ & $21 \degr 31 \arcmin 00 \farcs 4$ & $^{\color{blue}10}2.00^{+0.60}_{-0.60}$ & $20000^{+12500}_{-12500}$ & $8.8\pm0.8$\\
\addlinespace[0.075cm]
G192.6005-00.0479 & 06099+1800 & S255IR NIRS3 & $06 \rahr 12 \ramin 54 \fs 01$ & $17 \degr 59 \arcmin 23 \farcs 1$ & $^{\color{blue}11}1.78^{+0.12}_{-0.12}$ & $28300^{+5100}_{-5100}$ & $10.6\pm0.8$\\
\addlinespace[0.075cm]
G196.4542-01.6777 & 06117+1350 & S269 IRS2 & $06 \rahr 14 \ramin 37 \fs 06$ & $13 \degr 49 \arcmin 36 \farcs 4$ & $^{\color{blue}12}5.28^{+0.24}_{-0.22}$ & $93900^{+14300}_{-13900}$ & $11.8\pm0.7$\\
\bottomrule
\end{tabularx}
\label{tab:mysotargets}
\end{table*}
\end{footnotesize}

%% file: Tables/ClassificationNumbers.tex
\begin{table*}
\centering
\caption{Classifications of radio sources associated with each IRDC core \citep[separated by core classification according to][]{Rathborne2010} and MYSO. The final row indicates the number of cores/MYSOs associated with radio detections to those without. Classifications as a jet with a `C' in parentheses indicates candidacy status as an ionised jet, while `L' indicates association with shock-ionised lobes (see penultimate paragraph of \autoref{sec:intro}).}
\begin{tabularx}{\textwidth}{YYYYYY}
\toprule
   \multirow{2}{*}{Type}   &          \multicolumn{4}{c}{IRDCs}           &  \multirow{2}{*}{MYSOs}   \\\cline{2-5}
          &    R     &    A     &    I     &    Q     &          \\
\midrule
    Photo-ionised disc-wind    &    0     &    0     &    0     &    0     &    2     \\
   \textsc{Hii} region   &    5     &    0     &    1     &    0     &    3     \\
   Jet    &    1     &    0     &    0     &    0     &    3     \\
  Jet (C)  &    0     &    1     &    0     &    0     &    22    \\
  Jet (L)  &    0     &    0     &    0     &    0     &    10    \\
 Jet (L,C)  &    0     &    0     &    0     &    0     &    3     \\
 Unknown  &    1     &    0     &    0     &    0     &    2     \\
\midrule
Detection ratio      &   6:5    &   1:6    &  1:7     &   0:19    &   45:3    \\
\bottomrule
\end{tabularx}
\label{tab:classnumbers}
\end{table*}

%% file: Tables/AppendixA/BeamsNoises.tex
\begin{center}
\begin{longtable}{lcccccccc}
\caption{Convolving beams (columns 2 and 3) and noise levels (columns 3 and 4) for each observed field at each observed frequency. Note that these are the noise levels at the pointing centre. The 5th and 6th columns show whether the noise in the image was limited thermally, or by dynamic range/spatial frequency $uv$-coverage limitations. Columns 7 and 8 reference the figure showing the image data and section detailed notes/literature review, respectively.\label{tab:BeamsNoises}}\\
\toprule
Target & $\theta_{\rm beam}^{\rm C}$ & $\theta_{\rm beam}^{\rm Q}$ & $\sigma_{\rm RMS}^{\rm C}$ & $\sigma_{\rm RMS}^{\rm Q}$ & \multicolumn{2}{c}{Thermal Noise?} & Figure & Notes \\
 & & & $\left[ \uJy \right]$ & $\left[ \uJy \right]$ & C-band & Q-band & & \\
\midrule
\endfirsthead
\caption*{\textbf{Table \ref{tab:BeamsNoises}:} Continued}\\
\toprule
Target & $\theta_{\rm beam}^{\rm C}$ & $\theta_{\rm beam}^{\rm Q}$ & $\sigma_{\rm RMS}^{\rm C}$ & $\sigma_{\rm RMS}^{\rm Q}$ & \multicolumn{2}{c}{Thermal Noise?} & Figure & Notes \\
 & & & $\left[ \uJy \right]$ & $\left[ \uJy \right]$ & C-band & Q-band & & \\
\midrule
\endhead
\bottomrule
\endfoot
\multicolumn{9}{c}{\textit{IRDC Sample}} \\
\midrule
G018.82-00.28 & $0\farcs558 \times 0\farcs283;\,-33\fdg1$ & $0\farcs072 \times 0\farcs038;\,-25\fdg9$ & 105 & 57 & N & N & \ref{cplot:G018.82} & \autoref{sec:G018.82} \\
G024.08+00.04 & $0\farcs551 \times 0\farcs313;\,-25\fdg5$ & $0\farcs082 \times 0\farcs047;\,-26\fdg7$ & 9 & 29 & Y & Y & \ref{cplot:G024.08} & \autoref{sec:G024.08} \\
G024.33+00.11 & $0\farcs386 \times 0\farcs297;\,-06\fdg6$ & $0\farcs064 \times 0\farcs038;\,-25\fdg8$ & 6 & 37 & Y & Y & \ref{cplot:G024.33} & \autoref{sec:G024.33} \\
G024.60+00.08 & $0\farcs382 \times 0\farcs297;\,-03\fdg8$ & $0\farcs064 \times 0\farcs038;\,-25\fdg4$ & 6 & 35 & Y & Y & \ref{cplot:G024.60} & \autoref{sec:G024.60} \\
G028.28-00.34 & $0\farcs364 \times 0\farcs303;\,-176\fdg6$ & $0\farcs069 \times 0\farcs047;\,-19\fdg8$ & 82 & 35 & N & Y & \ref{cplot:G028.28} & \autoref{sec:G028.28} \\
G028.37+00.07 & $0\farcs364 \times 0\farcs301;\,-03\fdg2$ & $0\farcs068 \times 0\farcs046;\,-21\fdg2$ & 6 & 31 & Y & Y & \ref{cplot:G028.37} & \autoref{sec:G028.37} \\
G028.67+00.13 & $0\farcs358 \times 0\farcs303;\,+00\fdg9$ & $0\farcs058 \times 0\farcs038;\,-22\fdg0$ & 10 & 33 & N & Y & \ref{cplot:G028.67} & \autoref{sec:G028.67} \\
\end{longtable}
\end{center}

%% file: Tables/AppendixA/CBandTable1.tex
\begin{center}
\begin{longtable}{lccccccc}
\caption[C-band positions and fluxes for all objects detected in the VLA survey]{Positions and fluxes at C-band for all detected sources derived from \textsc{imfit}. A `PP' value in R.A. or Declination error columns indicates that the source is extended and listed positions are those of the pixel of peak-flux. Cases where flux upper-limits are given with no positional entries are for Q-band only detections. Quoted errors do not include the uncertainty in the flux scale.\label{tab:CBandPosFlux}}\\
\toprule
Object & Lobe & Right Ascension & $\delta\,{\rm R.A.}$ & Declination & $\delta\,{\rm Dec.}$ & $S_{5.8}^{\rm Peak}$ & $S_{5.8}$ \\
 &  & $\left[ {\rm J2000} \right]$ & $\left[ \mathrm{mas} \right]$ & $\left[ {\rm J2000} \right]$ & $\left[ \mathrm{mas} \right]$ & $\left[\mJy \, {\rm beam^{-1}}\right]$ & $\left[\mJy\right]$ \\
\midrule
\endfirsthead
\caption*{\textbf{Table \ref{tab:CBandPosFlux}:} Continued}\\
\toprule
Object & Lobe & Right Ascension & $\delta\,{\rm R.A.}$ & Declination & $\delta\,{\rm Dec.}$ & $S_{5.8}^{\rm Peak}$ & $S_{5.8}$ \\
 &  & $\left[ {\rm J2000} \right]$ & $\left[ \mathrm{mas} \right]$ & $\left[ {\rm J2000} \right]$ & $\left[ \mathrm{mas} \right]$ & $\left[\mJy \, {\rm beam^{-1}}\right]$ & $\left[\mJy\right]$ \\
\midrule
\endhead
\bottomrule
\endfoot
\multicolumn{8}{c}{\textit{IRDC Sample}} \\
\midrule
G018.82-00.28 & Core & $18\rahr 26\ramin 23\fs6516$ & 1 & $-12\degr 39\arcmin 37\farcs549$ & 1 & $18.89 \pm 0.11$ & $36.62 \pm 0.25$ \\
G024.08+00.04 & A & $18\rahr 34\ramin 57\fs1881$ & 6 & $-07\degr 43\arcmin 26\farcs234$ & 6 & $1.05 \pm 0.01$ & $4.55 \pm 0.08$ \\
 & B1 & $18\rahr 34\ramin 59\fs6276$ & 7 & $-07\degr 43\arcmin 01\farcs633$ & 9 & $0.77 \pm 0.01$ & $1.71 \pm 0.06$ \\
 & B2 & $18\rahr 34\ramin 59\fs5937$ & 5 & $-07\degr 43\arcmin 00\farcs313$ & 7 & $1.47 \pm 0.01$ & $5.52 \pm 0.06$ \\
 & B3 & $18\rahr 34\ramin 59\fs5918$ & 25 & $-07\degr 42\arcmin 57\farcs355$ & 34 & $0.30 \pm 0.01$ & $2.47 \pm 0.13$ \\
 & B & $18\rahr 34\ramin 59\fs5980$ & PP & $-07\degr 43\arcmin 00\farcs433$ & PP & $1.47 \pm 0.01$ & $9.70\pm0.16$ \\
G024.33+00.11 & A & $18\rahr 35\ramin 08\fs1393$ & 4 & $-07\degr 35\arcmin 04\farcs096$ & 5 & $0.23 \pm 0.01$ & $0.28 \pm 0.01$ \\
\end{longtable}
\end{center}

%% file: Tables/AppendixA/QBandTable1.tex
\begin{center}
\begin{longtable}{lccccccc}
\caption[Q-band positions and fluxes for all objects detected in the VLA survey]{Positions and fluxes at Q-band for all detected sources derived from \textsc{imfit}. Entries with no information at all are outside of the Q-band primary beam. Cases with flux upper-limits but no positional data show sources that were detected at C-band but were undetected within the primary beam at Q-band. Quoted errors do not include the uncertainty in the flux scale.\label{tab:QBandPosFlux}}\\
\toprule
Object & Lobe & Right Ascension & $\delta\,{\rm R.A.}$ & Declination & $\delta\,{\rm Dec.}$ & $S_{44}^{\rm Peak}$ & $S_{44}$ \\
 &  & $\left[ {\rm J2000} \right]$ & $\left[ \mathrm{mas} \right]$ & $\left[ {\rm J2000} \right]$ & $\left[ \mathrm{mas} \right]$ & $\left[\mJy \, {\rm beam^{-1}}\right]$ & $\left[\mJy\right]$ \\
\midrule
\endfirsthead
\caption*{\textbf{Table \ref{tab:QBandPosFlux}:} Continued}\\
\toprule
Object & Lobe & Right Ascension & $\delta\,{\rm R.A.}$ & Declination & $\delta\,{\rm Dec.}$ & $S_{44}^{\rm Peak}$ & $S_{44}$ \\
 &  & $\left[ {\rm J2000} \right]$ & $\left[ \mathrm{mas} \right]$ & $\left[ {\rm J2000} \right]$ & $\left[ \mathrm{mas} \right]$ & $\left[\mJy \, {\rm beam^{-1}}\right]$ & $\left[\mJy\right]$ \\
\midrule
\endhead
\bottomrule
\endfoot
\multicolumn{8}{c}{\textit{IRDC Sample}} \\
\midrule
G018.82-00.28 & Core & $18\rahr 26\ramin 23\fs6570$ & 5 & $-12\degr 39\arcmin 37\farcs400$ & 5 & $2.37 \pm 0.06$ & $51.00 \pm 1.10$ \\
G024.08+00.04 & A & - & - & - & - & $<0.09$ & $<0.09$ \\
 & B1 & - & - & - & - & - & - \\
 & B2 & - & - & - & - & - & - \\
 & B3 & - & - & - & - & - & - \\
 & B & - & - & - & - & - & - \\
G024.33+00.11 & A & $18\rahr 35\ramin 08\fs1391$ & 4 & $-07\degr 35\arcmin 04\farcs122$ & 3 & $0.51 \pm 0.04$ & $1.18 \pm 0.12$ \\
\end{longtable}
\end{center}

%% file: Tables/AppendixA/DeconvolvedSizes.tex
\begin{center}
\begin{longtable}{lccccccc}
\caption[C-band deconvolved sizes and position angles for all sources from the VLA survey]{Deconvolved sizes and position angles at both C and Q-band for all detected sources. All values were derived using \textsc{imfit}.\label{tab:DeconvSizes}}\\
\toprule
Object & Lobe & $\theta_{\rm maj}^{5.8}$ & $\theta_{\rm min}^{5.8}$ & $\theta_{\rm PA}^{5.8}$ & $\theta_{\rm maj}^{44}$ & $\theta_{\rm min}^{44}$ & $\theta_{\rm PA}^{44}$ \\
 &  & $\left[ \mathrm{mas} \right]$ & $\left[ \mathrm{mas} \right]$ & $\left[ \degr \right]$ & $\left[ \mathrm{mas} \right]$ & $\left[ \mathrm{mas} \right]$ & $\left[ \degr \right]$\\
\midrule
\endfirsthead
\caption*{\textbf{Table \ref{tab:DeconvSizes}:} Continued}\\
\toprule
Object & Lobe & $\theta_{\rm maj}^{5.8}$ & $\theta_{\rm min}^{5.8}$ & $\theta_{\rm PA}^{5.8}$ & $\theta_{\rm maj}^{44}$ & $\theta_{\rm min}^{44}$ & $\theta_{\rm PA}^{44}$ \\
 &  & $\left[ \mathrm{mas} \right]$ & $\left[ \mathrm{mas} \right]$ & $\left[ \degr \right]$ & $\left[ \mathrm{mas} \right]$ & $\left[ \mathrm{mas} \right]$ & $\left[ \degr \right]$\\
\midrule
\endhead
\bottomrule
\endfoot
\multicolumn{8}{c}{\textit{IRDC Sample}} \\
\midrule
G018.82-00.28 & Core & $417 \pm 7$ & $324 \pm 3$ & $134 \pm 2$ & $577 \pm 58$ & $417 \pm 42$ & $116 \pm 10$ \\
G024.08+00.04 & A & $817 \pm 25$ & $742 \pm 26$ & $102 \pm 16$ & - & - & - \\
 & B1 & $641 \pm 35$ & $367 \pm 21$ & $137 \pm 4$ & - & - & - \\
 & B2 & $1496 \pm 22$ & $219 \pm 5$ & $144 \pm 0$ & - & - & - \\
 & B3 & $1765 \pm 97$ & $650 \pm 51$ & $36 \pm 2$ & - & - & - \\
 & B & - & - & - & - & - & - \\
G024.33+00.11 & A & $179 \pm 36$ & $111 \pm 92$ & $122 \pm 29$ & $111 \pm 14$ & $26 \pm 13$ & $118 \pm 5$ \\
\end{longtable}
\end{center}

%% file: Tables/AppendixA/ClassAlphaGamma.tex
\begin{center}
\begin{longtable}{lccccc}
\caption[Classifications and calculated values for $\alpha$, $\gamma$ and major axis position angle differences for sources detected in the VLA survey]{Classifications, derived spectral indices for flux ($\alpha$) and major axis length ($\gamma$) and calculated changes in position angles from C to Q-band.\label{tab:AlphaGamma}}\\
\toprule
Object & Lobe & Class. & $\alpha$  & $\gamma$ & $\Delta\theta_{\rm PA}$ \\
 &  &  &  &  & $\left[ \degr \right]$\\
\midrule
\endfirsthead
\caption*{\textbf{Table \ref{tab:AlphaGamma}:} Continued}\\
\toprule
Object & Lobe & Class. & $\alpha$  & $\gamma$ & $\Delta\theta_{\rm PA}$ \\
 &  &  &  &  & $\left[ \degr \right]$\\
\midrule
\endhead
\bottomrule
\endfoot
\multicolumn{6}{c}{\textit{IRDC Sample}} \\
\midrule
G018.82-00.28 & Core & UCHII & $\;\;\;\,+0.16\pm0.06$ & $\;\;\;\,+0.16\pm0.05$ & $18\pm10$ \\
G024.08+00.04 & A & UCHII & $<-1.96\pm0.03$ & - & - \\
 & B1 & EG & - & - & - \\
 & B2 & EG & - & - & - \\
 & B3 & EG & - & - & - \\
G024.33+00.11 & A & Jet & $\;\;\;\,+0.70\pm0.08$ & $\;\;\;\,-0.24\pm0.12$ & $4\pm29$ \\
 & B & UK & - & - & - \\
\end{longtable}
\end{center}

%% file: Tables/AppendixA/JMLTable.tex
\begin{center}
\begin{longtable}{lcccc}
\caption{Classifications, opening angles and mass loss rates for all jet-like radio sources detected. Mass loss rates given are the weighted average of $\jml$ calculated at each frequency available. Calculation of mass loss rates and their errors at each frequency is discussed in \autoref{sec:appendixJMLs}. \label{tab:JMLsOpangs}}\\
\toprule
MYSO & Lobe & $\theta_{\rm OA}^{\rm C}$ & $\theta_{\rm OA}^{\rm Q}$ & $\dot{m}_{\rm jet}$\\
 &  & $\left[ \degr \right]$ & $\left[ \degr \right]$ & $\left[ \times 10^{-6}\Msol\yr^{-1} \right]$\\
\midrule
\endfirsthead
\caption*{\textbf{Table \ref{tab:JMLsOpangs}:} Continued}\\
\toprule
MYSO & Lobe & $\theta_{\rm OA}^{\rm C}$ & $\theta_{\rm OA}^{\rm Q}$ & $\dot{m}_{\rm jet}$\\
 &  & $\left[ \degr \right]$ & $\left[ \degr \right]$ & $\left[ \times 10^{-6}\Msol\yr^{-1} \right]$\\
\midrule
\endhead
\bottomrule
\endfoot
\multicolumn{5}{c}{\textit{IRDC Sample}} \\
\midrule
G024.33+00.11 & A & $64\pm44$ & $26\pm13$ & $25.3\pm13.8$ \\
G028.37+00.07 & A & - & $32\pm17$ & $15.1\pm5.7$ \\
 & A3 & - & $44\pm10$ & $31.9\pm8.4$ \\
\midrule
\multicolumn{5}{c}{\textit{MYSO Sample}} \\
\midrule
G033.6437-00.2277 & A & $68\pm34$ & $43\pm42$ & $6.5\pm3.7$ \\
 & 7 & $52\pm4$ & $31\pm14$ & $12.3\pm4.0$ \\
G037.4266+01.5183 & A & - & - & $0.7\pm0.5$ \\
\end{longtable}
\end{center}

%% file: CPlotsPDFs.tex
\subsection{IRDC Sample}
\label{app:irdcfigures}
\begin{figure*}
\includegraphics[width=\textwidth]{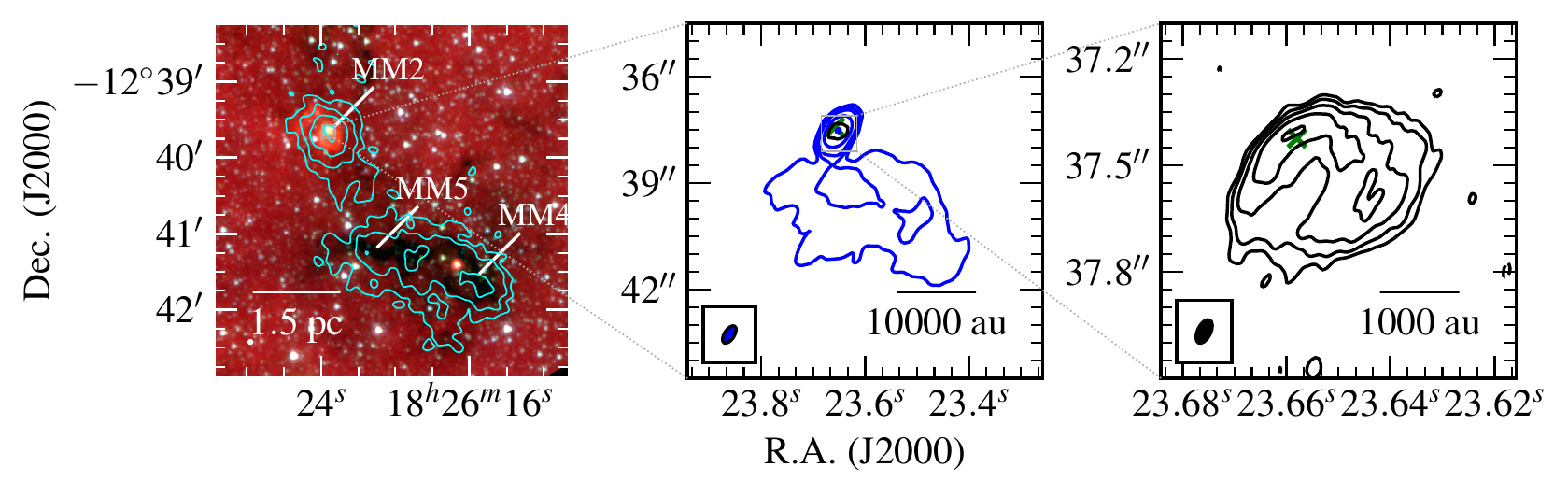}
\caption[Mid-infrared, ATLASGAL and VLA radio images of the IRDC G018.82$-$00.28]{\textbf{G018.82$-$00.28} - Mid-infrared (GLIMPSE $8.0,4.5,3.6\micron$ R, G, B image; left panel), sub-mm (ATLASGAL $870\micron$ cyan contours; left panel)  and radio maps of G018.82$-$00.28 at C-band (blue contours; middle) and Q-band (black contours; right). Restoring beams were $0.558\arcsec\times0.283\arcsec$ at $-33\degr$ and $0.072\arcsec\times0.038\arcsec$ at $-26\degr$, while contour levels are $(-3, 3, 6, 12, 44, 171) \times \sigma$ and $(-3, 3, 6, 11, 21, 40) \times \sigma$ for C and Q-band respectively and ATLASGAL contours are set at $(-3, 3, 5, 9, 15)\times \sigma$ where $\sigma=82\mJy/\mathrm{beam}$. Green crosses show 6.7$\GHz$ methanol maser positions from our data. In the infrared images, a white ellipse is used to denote the approximate PSF of the relevant instrument. All other symbols/values have the usual meaning. If RMS noises are displayed in a subplot, that is because in the location of the source the RMS noise diverges from that quoted in \autoref{tab:BeamsNoises}.}
\label{cplot:G018.82}
\end{figure*}

\begin{figure*} 
\includegraphics[width=\textwidth]{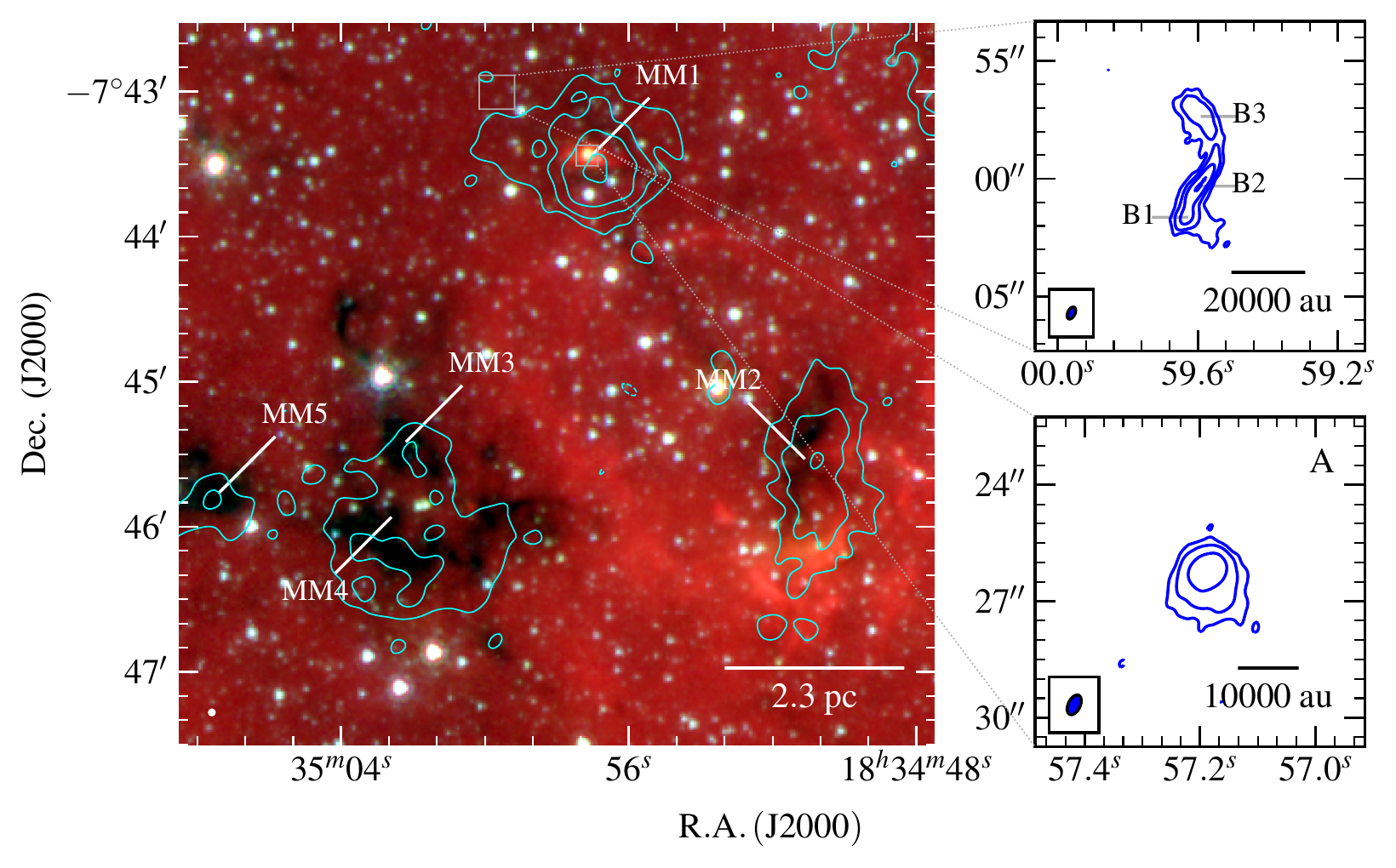}
\caption[Mid-infrared, ATLASGAL and VLA radio images of the IRDC G024.08+00.04]{\textbf{G024.08+00.04} - Mid-infrared (GLIMPSE $8.0,4.5,3.6\micron$ R, G, B image; left panel), sub-mm (ATLASGAL $870\micron$ cyan contours; left panel)  and C-band (blue contours; right panels) images of G024.08+00.04. The C-band restoring beam was $0.551\arcsec\times0.313\arcsec$ at $-25\degr$. The right panels show enlarged C-band maps of components A, B1, B2 and B3. Contour levels are $(-3, 3, 6, 10, 19) \times \sigma$ and ATLASGAL contours are set at $(-3, 3, 5, 9, 15)\times \sigma$ where $\sigma=75\mJy/\mathrm{beam}$.}
\label{cplot:G024.08}
\end{figure*} 
\begin{figure*}
\includegraphics[width=\textwidth]{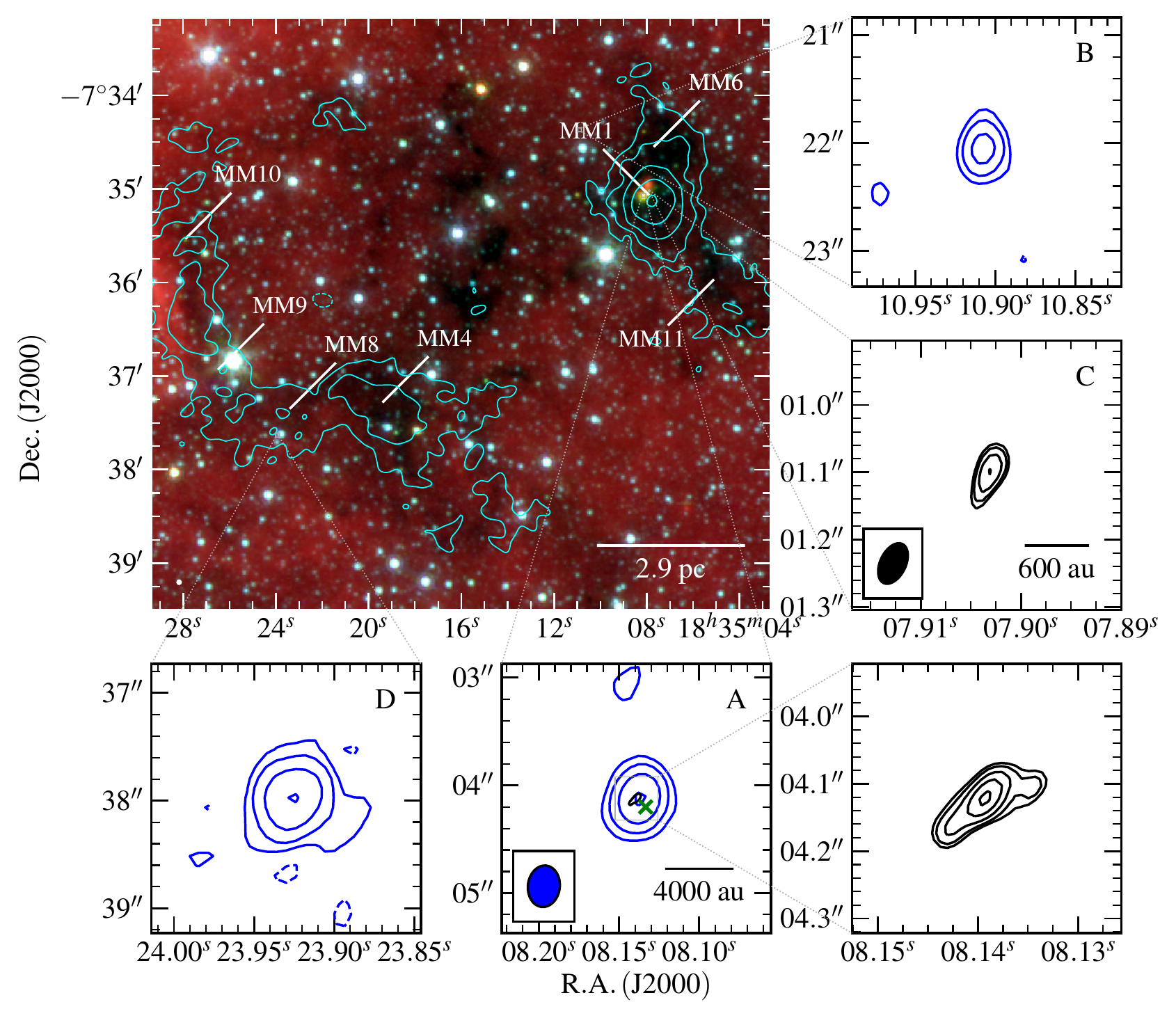}
\caption[Mid-infrared, ATLASGAL and VLA radio images of the IRDC G024.33$+$00.11]{\textbf{G024.33$+$00.11} - Mid-infrared (GLIMPSE $8.0,4.5,3.6\micron$ R, G, B image; top left panel), sub-mm (ATLASGAL $870\micron$ cyan contours; left panel) and radio contour maps of the IRDC, G024.33$+$00.11, at C-band (top and right panels; blue contours) and Q-band (middle and right panels; black contours). Restoring beams were $0.386\arcsec\times0.297\arcsec$ at $-7\degr$ and $0.064\arcsec\times0.038\arcsec$ at $-26\degr$ while contour levels are $(-3, 3, 7, 16, 36) \times \sigma$ and $(-3, 3, 4, 6, 9, 13) \times \sigma$ for C and Q-band respectively and ATLASGAL contours are set at $(-3, 3, 7, 15, 32, 70)\times \sigma$ where $\sigma=80\mJy/\mathrm{beam}$. An exception to the contour levels is for component D which was much brighter than the other C-band sources and so we set its contour levels to $-4, 4, 12, 37, 111)\times\sigma$. All other values have the usual meaning.}
\label{cplot:G024.33}
\end{figure*}

\begin{figure*}
\centering
\includegraphics[width=0.72\textwidth]{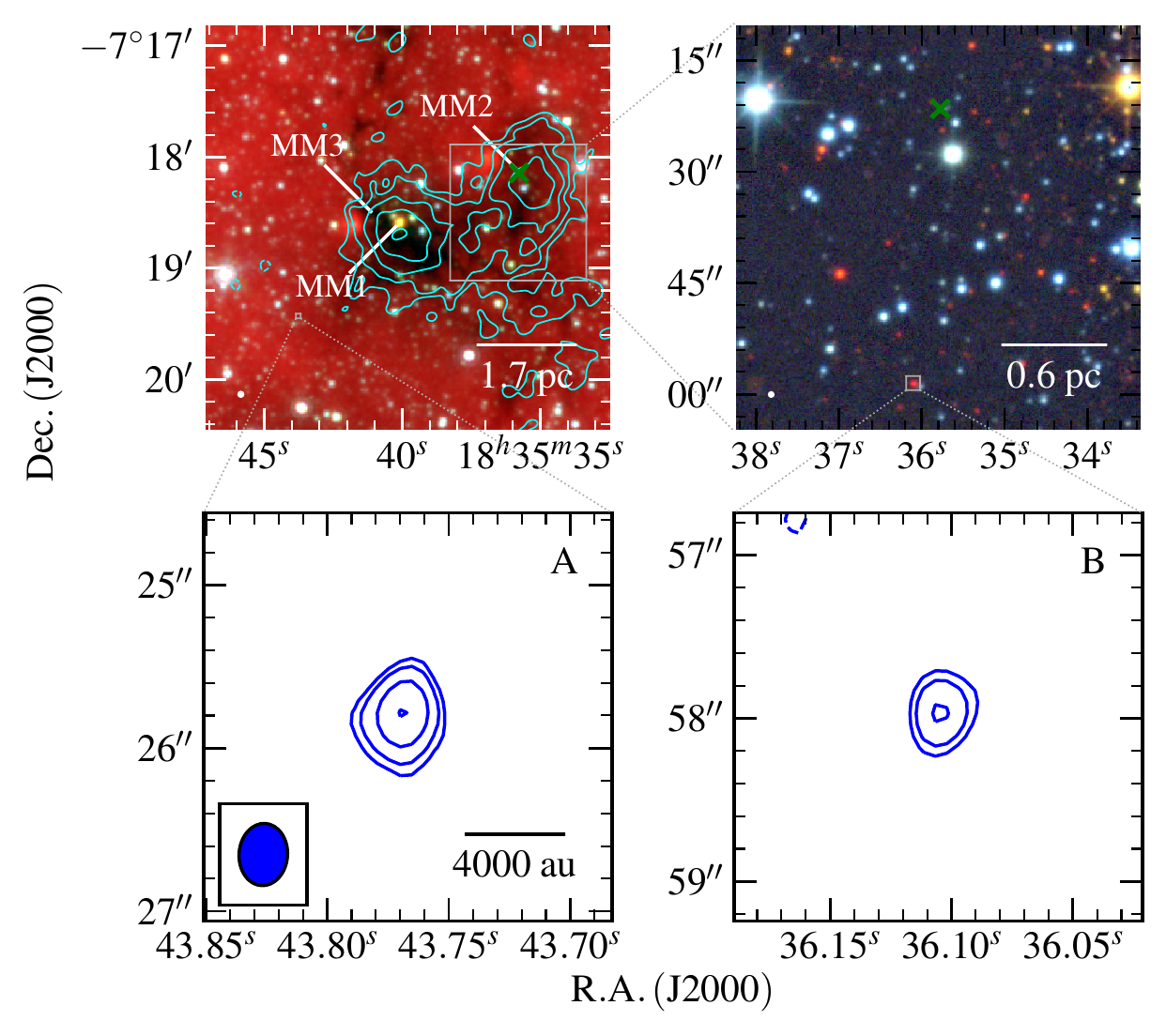}
\caption[Mid-infrared, ATLASGAL and VLA radio images of the IRDC G024.60+00.08]{\textbf{G024.60+00.08} - Mid-infrared (GLIMPSE $8.0,4.5,3.6\micron$ R, G, B image; top left panel), near-infrared (top right panel; UKIDSS, $\mathrm{K,H,J}$ bands in R, G, B colour-scale), sub-mm (ATLASGAL $870\micron$ cyan contours; top left panel) and C-band (blue contours; bottom panels) images of the IRDC, G024.60+00.08. The C-band restoring beam was $0.382\arcsec\times0.297\arcsec$ at $-4\degr$. C-band contour levels are $(-3, 3, 5, 10, 18) \times \sigma$ and ATLASGAL contours are set at $(-3, 3, 5, 8, 14, 24)\times \sigma$ where $\sigma=69\mJy/\mathrm{beam}$. All other values have their usual meaning. Green crosses show 6.7$\GHz$ methanol maser positions from our data.}
\label{cplot:G024.60}
\end{figure*}

\clearpage
\begin{figure*}
\includegraphics[width=\textwidth]{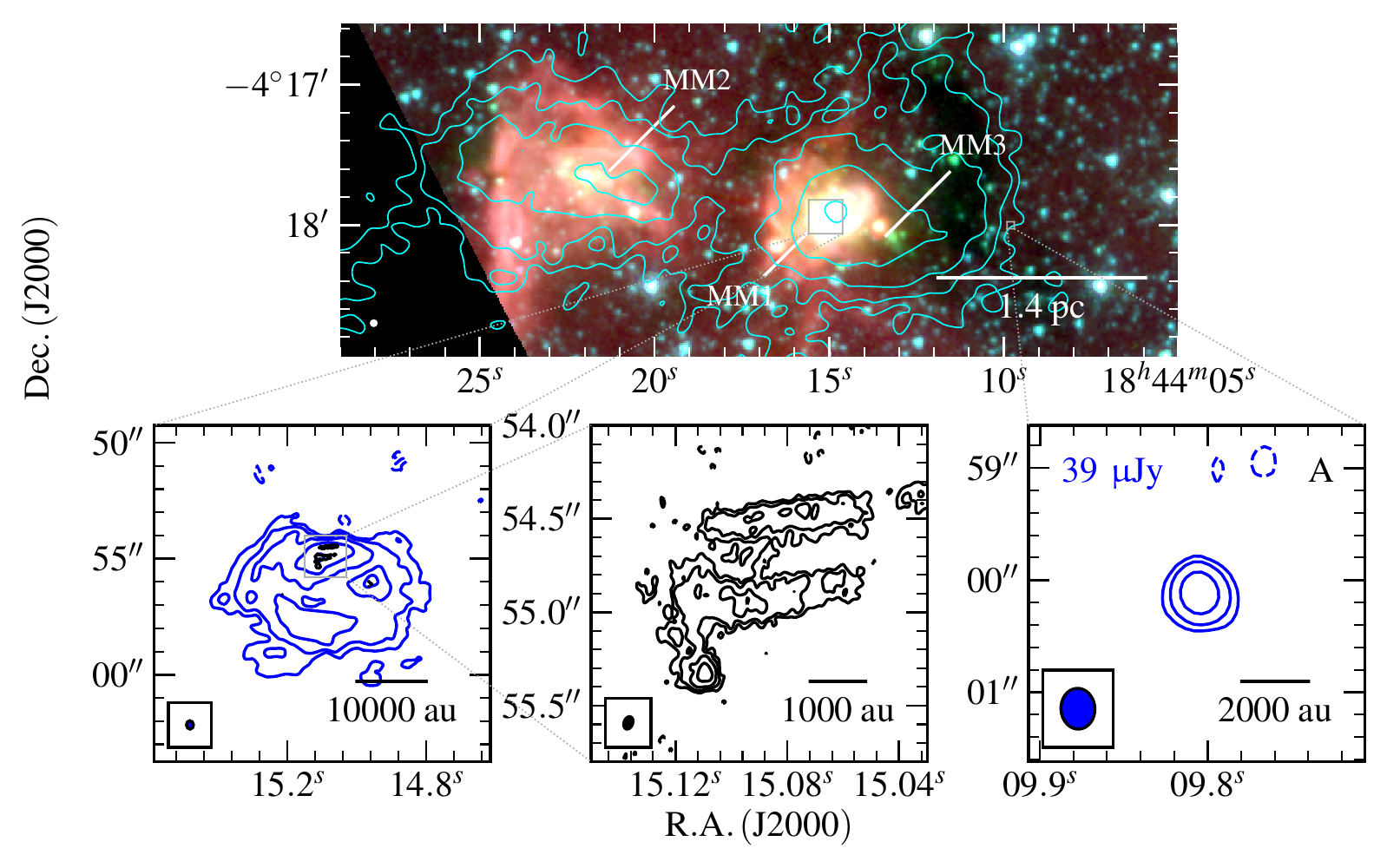}
\caption[Mid-infrared, ATLASGAL and VLA radio images of the IRDC G028.28$-$00.34]{\textbf{G028.28$-$00.34} - Mid-infrared (GLIMPSE $8.0,4.5,3.6\micron$ R, G, B image; top panel), sub-mm (ATLASGAL $870\micron$ cyan contours; top panel)  and radio maps of G028.28$-$00.34 at C-band (blue contours; middle) and Q-band (black contours; bottom). Restoring beams were $0.364\arcsec\times0.303\arcsec$ at $3\degr$ and $0.069\arcsec\times0.047\arcsec$ at $-20\degr$, while contour levels are $(-3, 3, 10, 33, 108) \times \sigma$ and $(-3, 3, 5, 9, 17) \times \sigma$ for C and Q-band respectively. ATLASGAL contours are set at $(-3, 3, 6, 11, 20, 39)\times \sigma$ where $\sigma=68\mJy/\mathrm{beam}$. All other symbols/values have the usual meaning.}
\label{cplot:G028.28}
\end{figure*}

\begin{figure*}
\includegraphics[width=\textwidth]{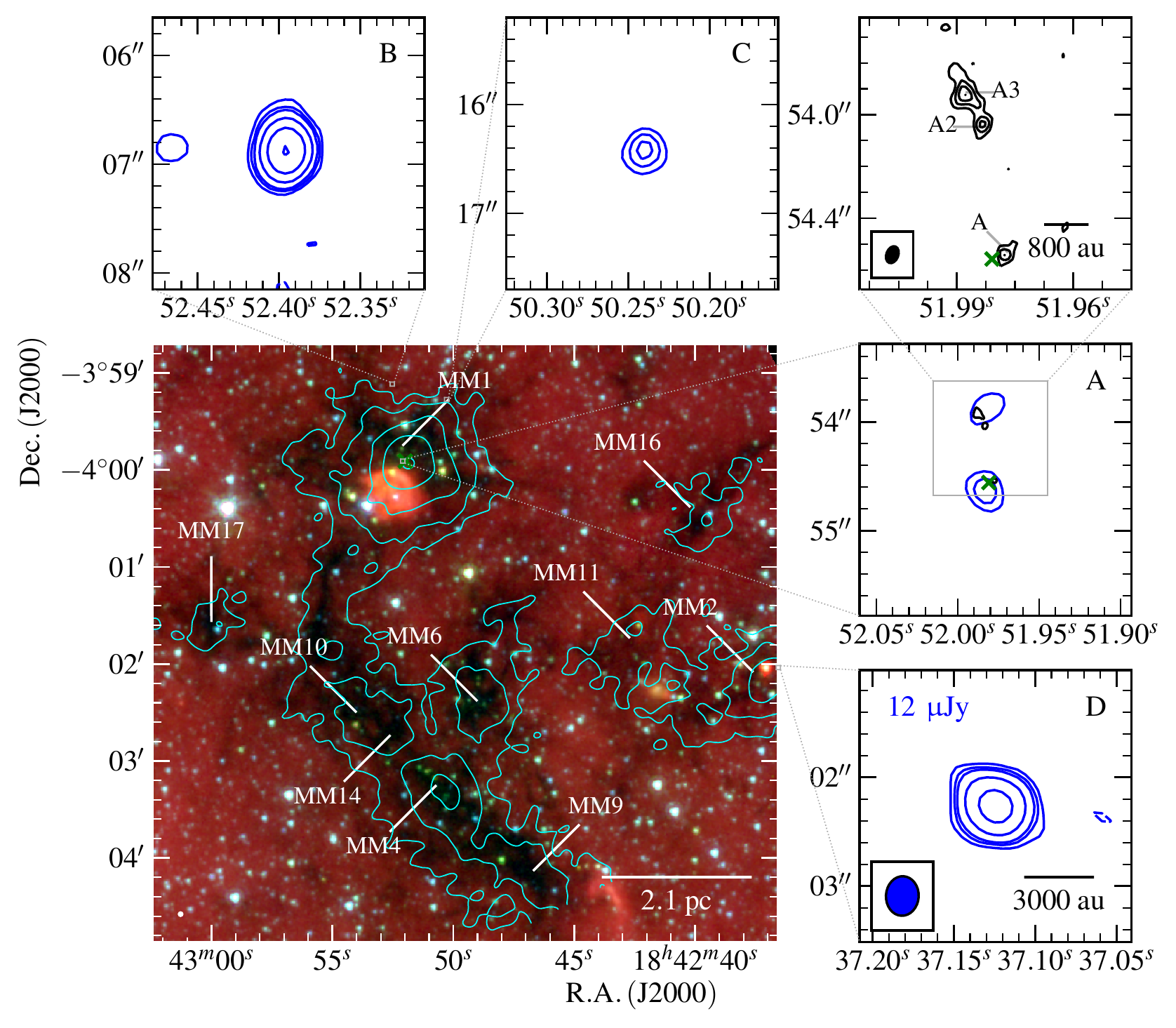}
\caption[Mid-infrared, ATLASGAL and VLA radio images of the IRDC G028.37$+$00.07]{\textbf{G028.37$+$00.07} - Mid-infrared (GLIMPSE $8.0,4.5,3.6\micron$ R, G, B image; left panel), sub-mm (ATLASGAL $870\micron$ cyan contours; left panel) and radio maps of G028.37$+$00.07 at C-band (blue contours; middle panels) and Q-band (black contours; right panel). Restoring beams were $0.364\arcsec\times0.301\arcsec$ at $-3\degr$ and $0.068\arcsec\times0.046\arcsec$ at $-21\degr$, while contour levels are $(-3, 3, 5, 7, 15, 34, 77) \times \sigma$ and $(-3, 3, 5, 7, 9) \times \sigma$ for C and Q-band respectively, while ATLASGAL contours are set at $(-3, 3, 6, 13, 28, 59)\times \sigma$ where $\sigma=87\mJy/\mathrm{beam}$.  All other symbols/values have the usual meaning. Green crosses show 6.7$\GHz$ methanol maser positions from our data.}
\label{cplot:G028.37}
\end{figure*}

\begin{figure*} 
\includegraphics[width=\textwidth]{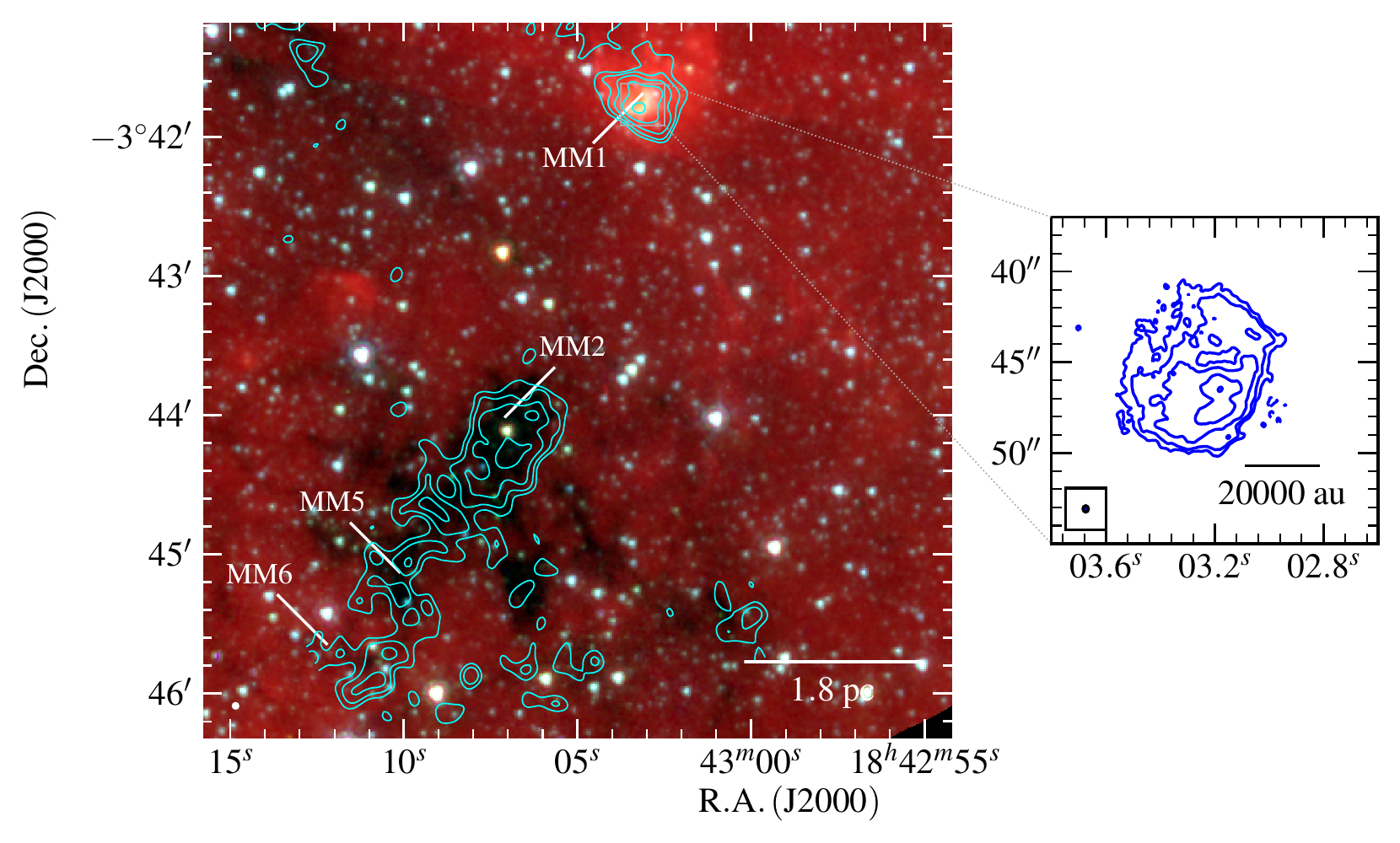}
\caption[Mid-infrared, ATLASGAL and VLA radio images of the IRDC G028.67+00.13]{\textbf{G028.67+00.13} - Mid-infrared (GLIMPSE $8.0,4.5,3.6\micron$ R, G, B image; left panel), sub-mm (ATLASGAL $870\micron$ cyan contours; left panel) and C-band radio map of G028.67+00.13 (right panel). The restoring beams used was $0.358\arcsec\times0.303\arcsec$ at $1\degr$, while contour levels are  set to $(-4, 4, 9, 20, 43, 96) \times \sigma$. All other symbols/values have the usual meaning.}
\label{cplot:G028.67}
\end{figure*} 

\begin{figure*} 
\includegraphics[width=\textwidth]{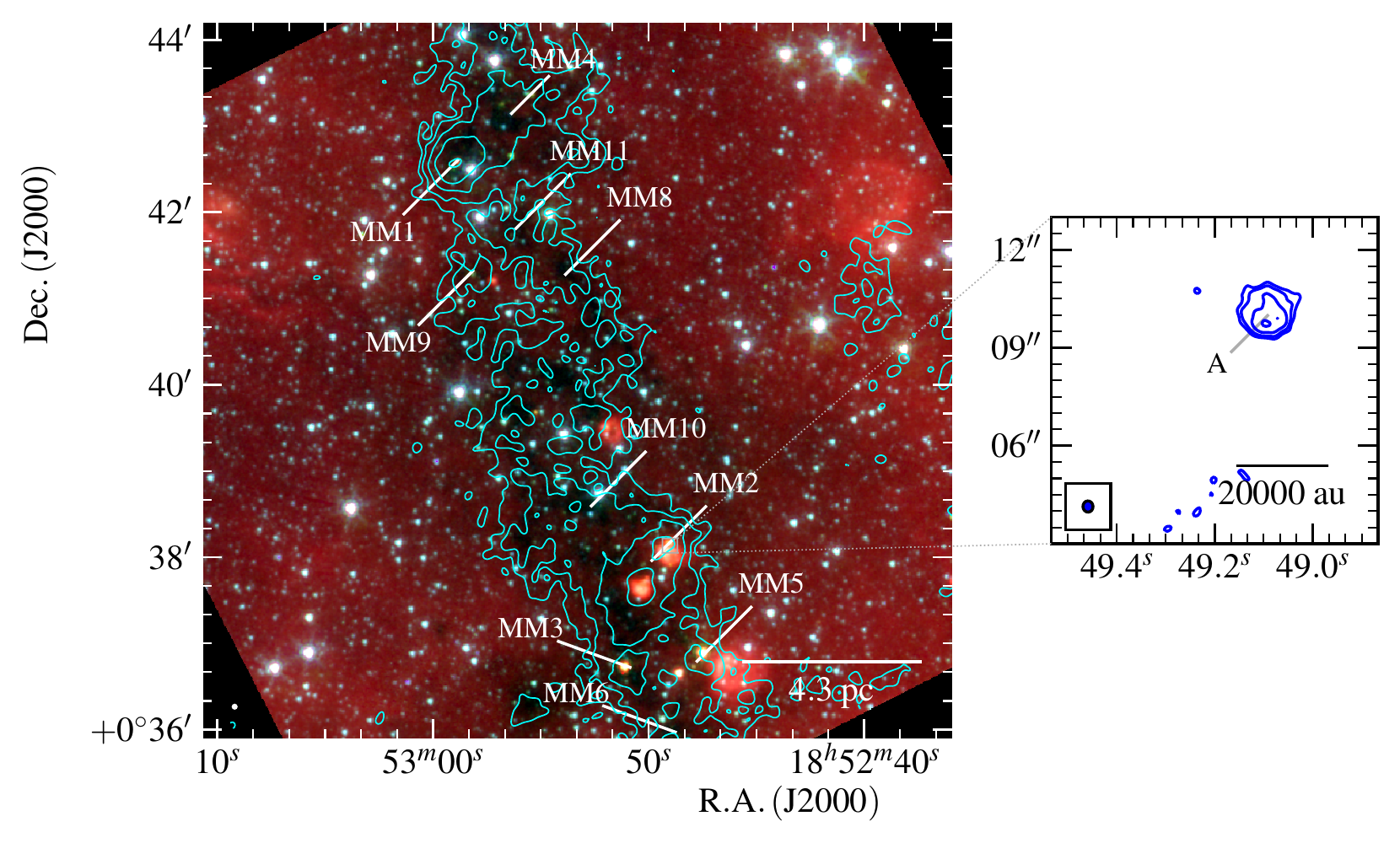}
\caption[Mid-infrared, ATLASGAL and VLA radio images of the IRDC G033.69$-$00.01]{\textbf{G033.69$-$00.01} - Mid-infrared (GLIMPSE $8.0,4.5,3.6\micron$ R, G, B image; left panel), sub-mm (ATLASGAL $870\micron$ cyan contours; left panel) and C-band radio map of G033.69$-$00.01 MM2 (right panel). The restoring beams used was $0.445\arcsec\times0.364\arcsec$ at $-53\degr$, while contour levels are  set to $(-4, 4, 7, 13, 22) \times \sigma$, while ATLASGAL contours are set at $(-3, 3, 5, 9, 15, 25)\times \sigma$ where $\sigma=85\mJy/\mathrm{beam}$. All other symbols/values have the usual meaning.}
\label{cplot:G033.69}
\end{figure*} 

\clearpage
\subsection{MYSO Sample}
\label{app:mysofigures}

\begin{figure*}
	\includegraphics[width=\textwidth]{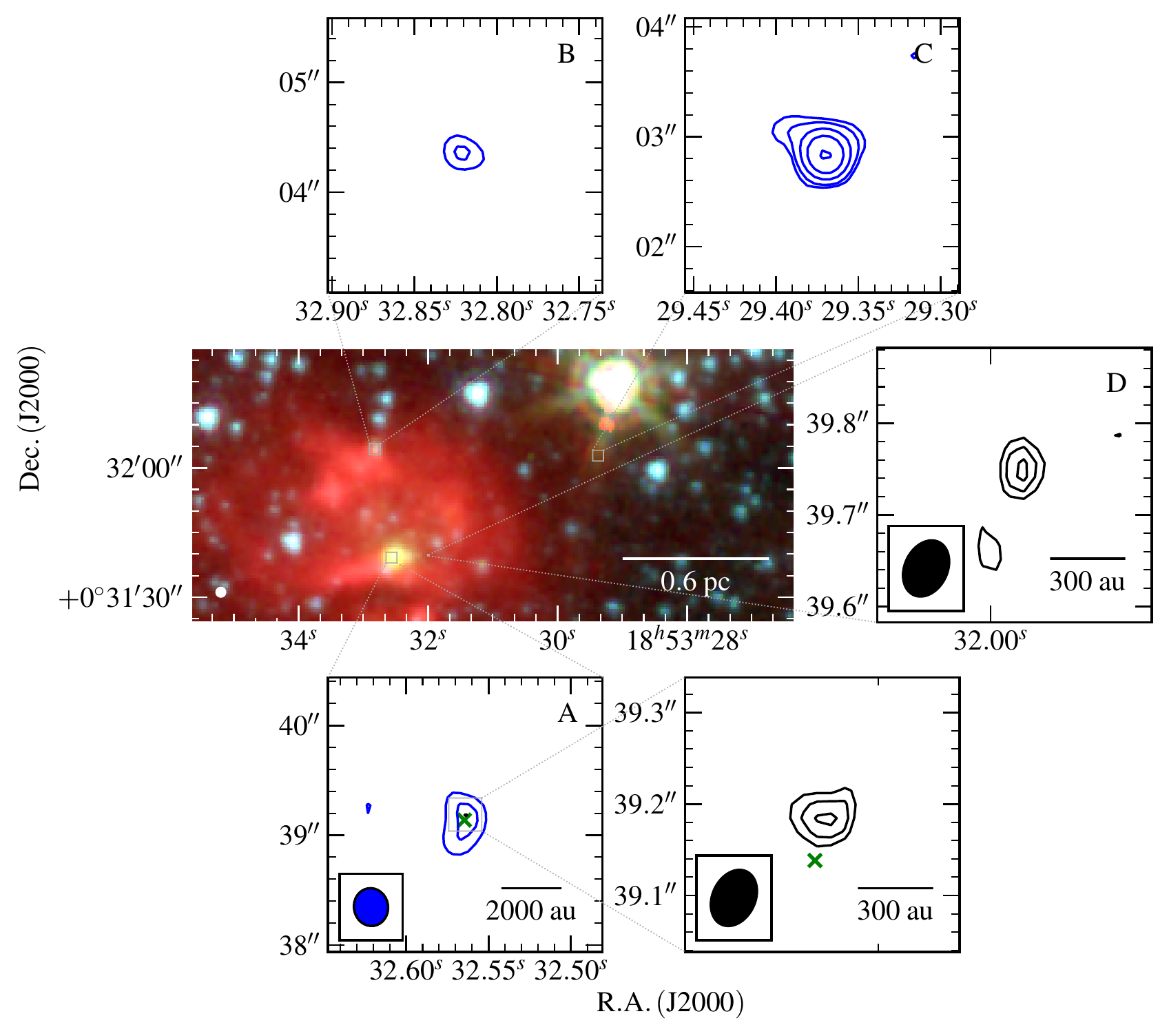}
	\caption[Mid-infrared and VLA radio images of the MYSO G033.6437$-$00.2277]{\textbf{G033.6437$-$00.2277} - Mid-infrared (GLIMPSE $8.0,4.5,3.6\micron$ R, G, B image; left panel) and radio maps of G033.6437$-$00.2277 at C-band (blue contours; top and bottom left panels) and Q-band (black contours; bottom right and right panels). Restoring beams were $0.346\arcsec\times0.311\arcsec$ at $9\degr$ and $0.064\arcsec\times0.047\arcsec$ at $-26\degr$, while contour levels are $(-3, 3, 5, 10, 18, 33) \times \sigma$ and $(-3, 3, 4, 5) \times \sigma$ for C and Q-band respectively. All other symbols/values have the usual meaning. Green crosses show 6.7$\GHz$ methanol maser positions from our data.}
	\label{cplot:G033.6437}
\end{figure*}

\begin{figure*}
\includegraphics[width=\textwidth]{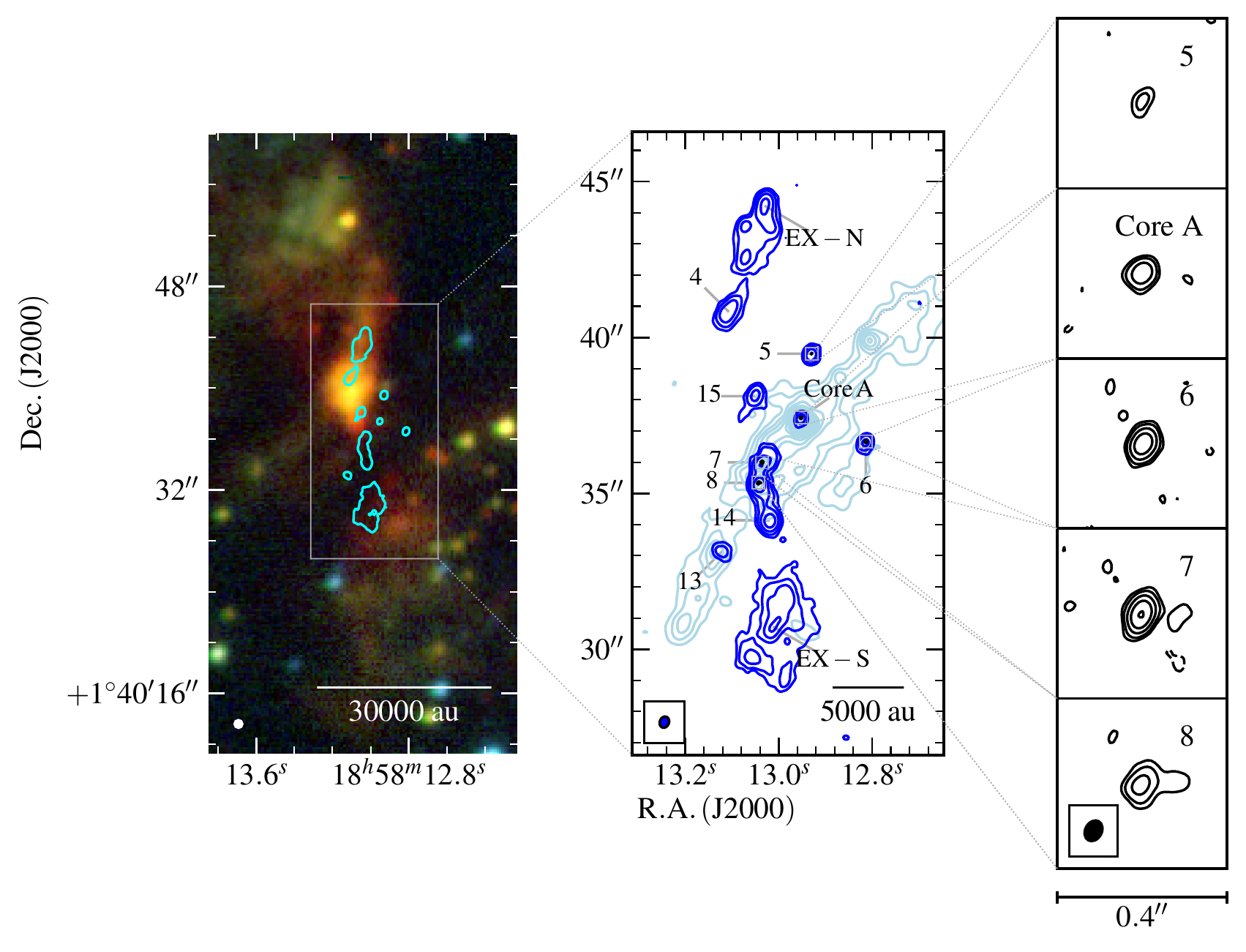}
\caption[Mid-infrared and VLA radio images of the MYSO IRAS 18556+0136]{\textbf{G035.1979$-$00.7427} - Mid-infrared (GLIMPSE $8.0,4.5,3.6\micron$ R, G, B image; left panel), ALMA $343\GHz$ \citep[][light blue contours; middle panel]{SanchezMonge2014}, C-band (dark blue contours; middle panel) and Q-band (black contours; middle/right panels) images of G035.1979$-$00.7427 (or IRAS 18556+0136). Restoring beams were $0.367\arcsec\times0.293\arcsec$ at $-21\degr$ and $0.051\arcsec\times0.039\arcsec$ at $-29\degr$ for the C and Q-band data respectively. The right panels show enlarged maps of components 5, 6, 7, 8 and Core A at Q-band. Contour levels are $(-4, 4, 11, 29, 78) \times \sigma$ and $(-3, 3, 6, 14, 29, 63) \times \sigma$ for C and Q-band respectively where $\sigma=35.1\,\uJy\,{\rm beam}^{-1}$ for the Q-band image. All other values have the usual meaning.}
\label{cplot:G035.1979}
\end{figure*}

\begin{figure*}
	\includegraphics[width=\textwidth]{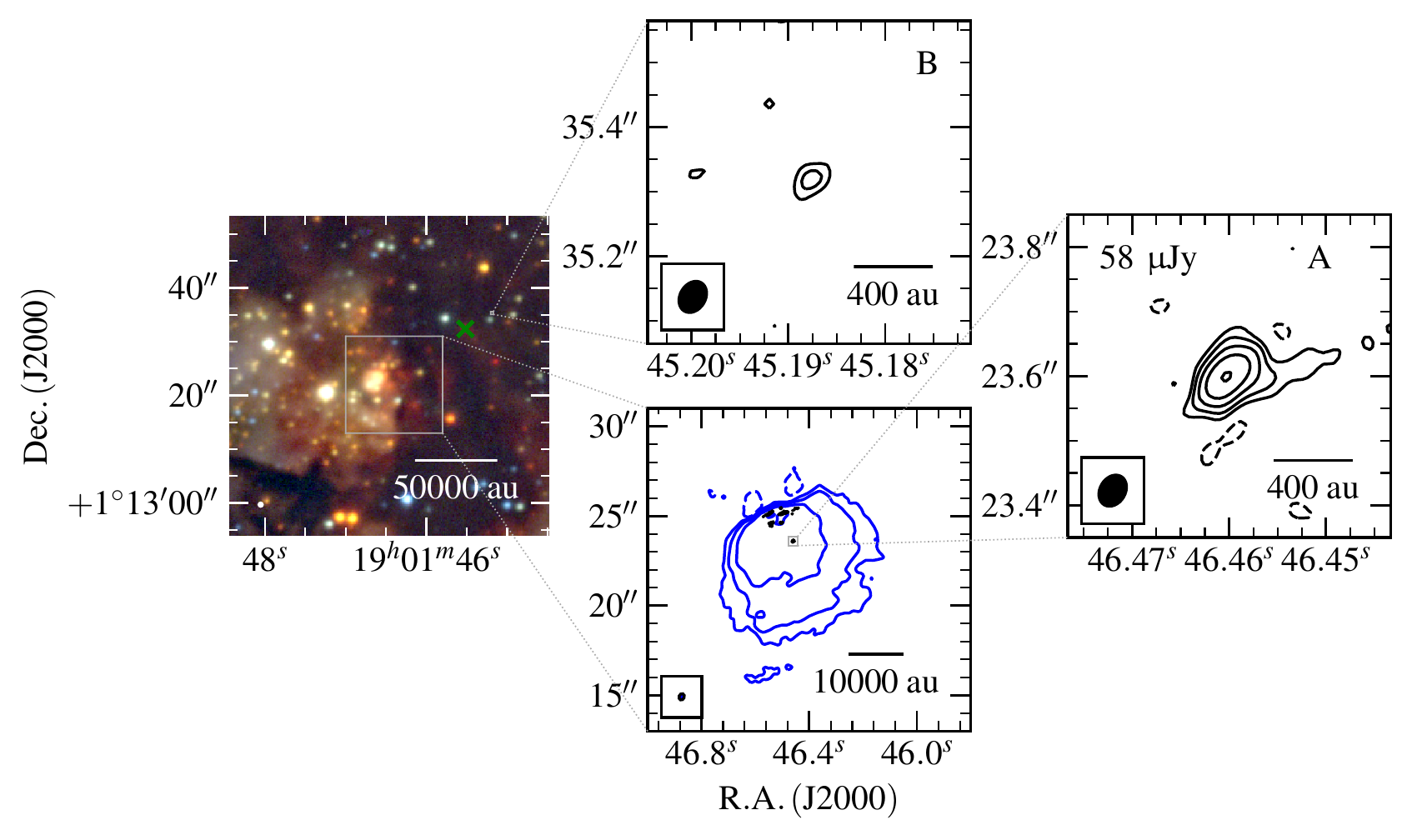}
	\caption[$\,\,\,$Near-infrared and VLA radio images of the MYSO/\textsc{Hii} region W48, or G035.1992$-$01.7424]{\textbf{G035.1992$-$01.7424} - Near-infrared (R, G, B colour-scale, top left panel; UKIDSS, $\mathrm{K,H,J}$ bands), C-band (blue contours; bottom panel) and Q-band (black contours; top and right panels) images of W48 (G035.1992$-$01.7424). The C-band restoring beam was $0.364\arcsec\times0.289\arcsec$ at $-21\degr$ and $0.051\arcsec\times0.039\arcsec$ at $-31\degr$, while contour levels are $(-3, 3, 11, 40, 147) \times \sigma$ and $(-3, 3, 6, 11, 20, 38) \times \sigma$ for the C and Q-band data respectively. All other values have their usual meaning.}
	\label{cplot:G035.1992}
\end{figure*}

\begin{figure*}
	\includegraphics[width=\textwidth]{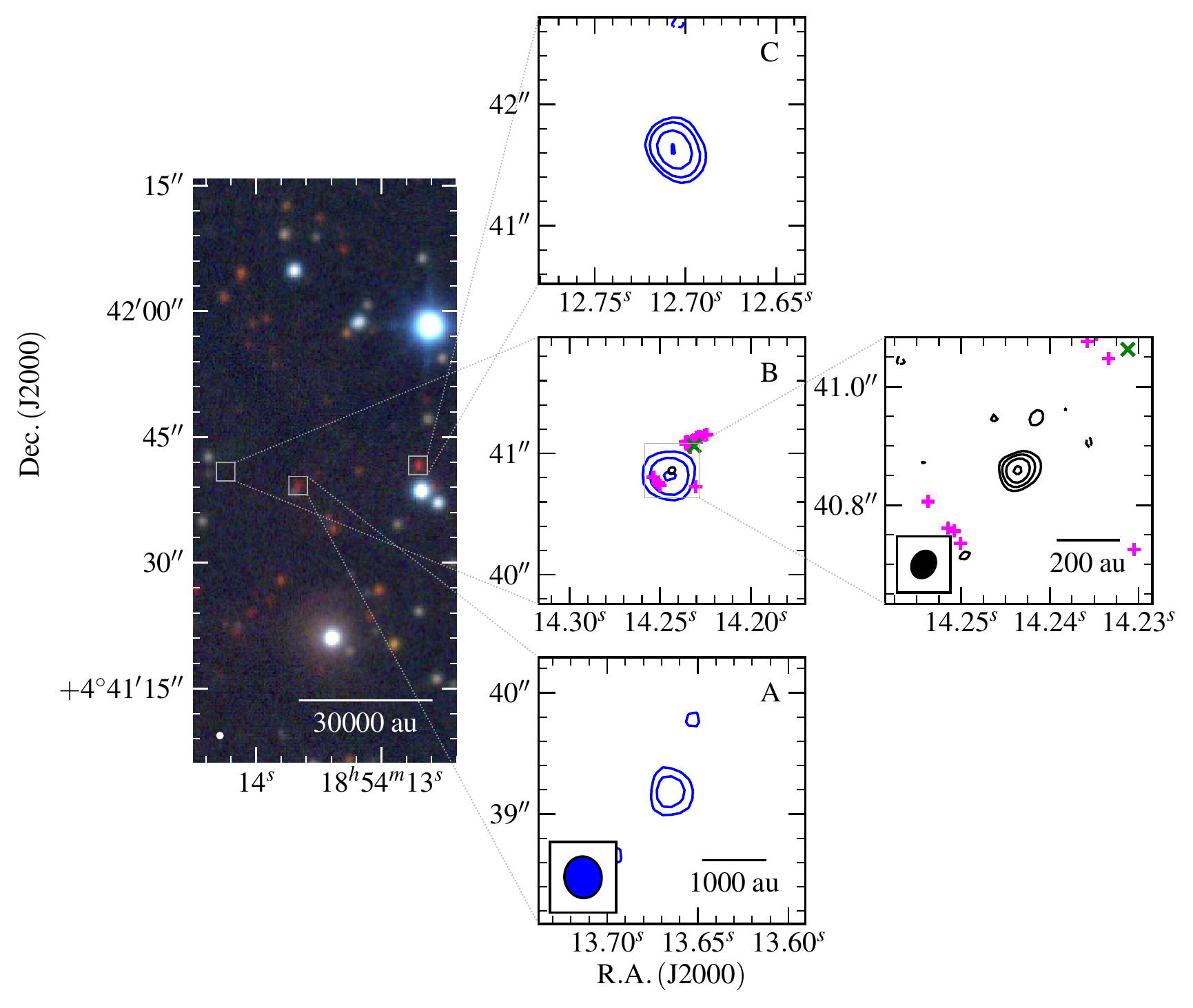}
	\caption[Near-infrared and VLA radio images of the MYSO IRAS 18517+0437]{\textbf{G037.4266+01.5183} - Near-infrared (R, G, B colour-scale, left panel; 2MASS, $\mathrm{K,H,J}$ bands) and radio contour maps of G037.4266+01.5183 (IRAS 18517+0437) at C-band (bottom, blue contours) and Q-band (bottom, black contours). Restoring beams were $0.344\arcsec\times0.307\arcsec$ at $8\degr$ and $0.048\arcsec\times0.039\arcsec$ at $-31\degr$ for the C and Q-band data respectively. The bottom panels show enlarged maps of components A, B and C at both bands. Contour levels are $(-3, 3, 5, 9, 15) \times \sigma$ and $(-3, 3, 5, 8, 13) \times \sigma$ for C and Q-band respectively. All other values have the usual meaning. Pink '+' markers represent the methanol masers detected by \citet{Surcis2015}. Green crosses show 6.7$\GHz$ methanol maser positions from our data.}
	\label{cplot:G037.4266}
\end{figure*}

\begin{figure*}
\centering
\includegraphics[width=0.72\textwidth]{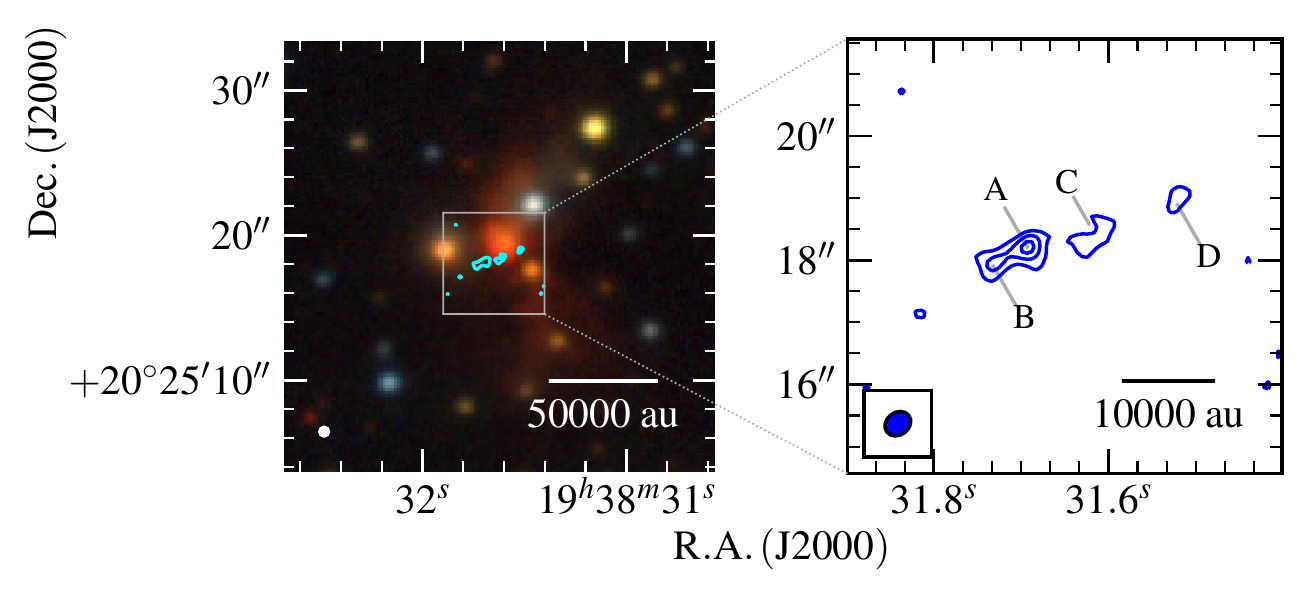}
\caption[Near-infrared and VLA radio images of the MYSO G056.3694$-$00.6333]{\textbf{G056.3694$-$00.6333} - Near-infrared (left panel; UKIDSS, $\mathrm{K,H,J}$ bands in R, G, B colour-scale) and C-band radio map (robustness of 2) of G056.3694$-$00.6333 (right panel). The restoring beams used was $0.445\arcsec\times0.364\arcsec$ at $-53\degr$, while contour levels are  set to $(-3, 3, 5, 7) \times \sigma$. All other symbols/values have the usual meaning. To highlight the slight offset between the reddened UKIDSS source and radio lobes, the radio $3\sigma$ contour is overlayed upon the NIR image (cyan).}
\label{cplot:G056.3694}
\end{figure*} 

\begin{figure*}
\centering
\includegraphics[width=0.72\textwidth]{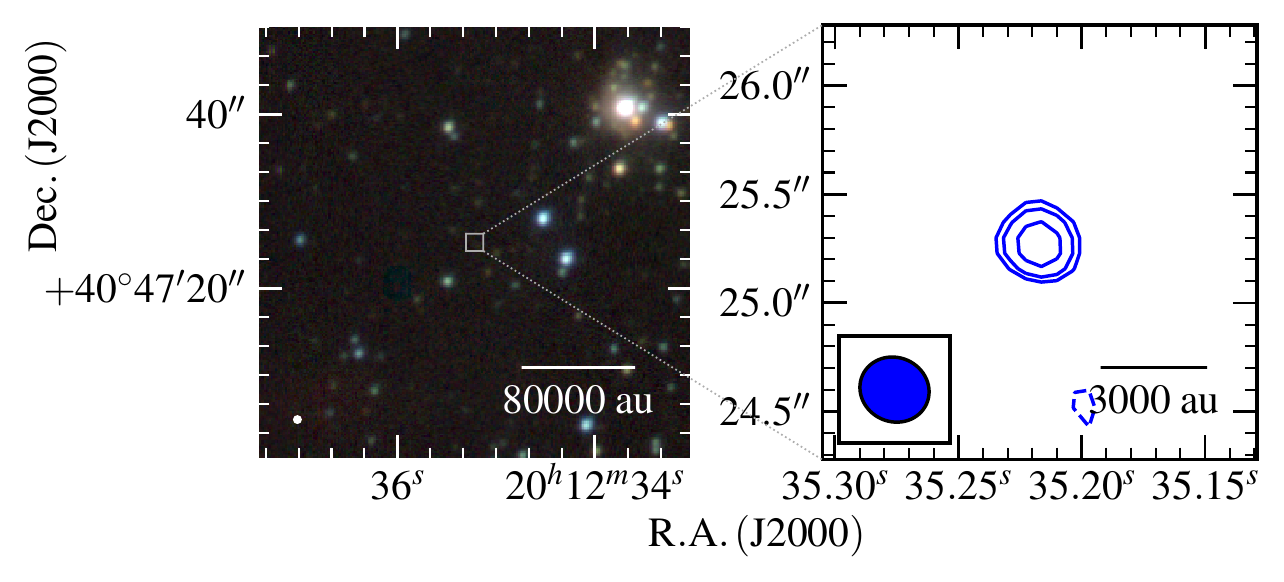}
\caption[Near-infrared and VLA radio images of the MYSO G077.5671+03.6911]{\textbf{G077.5671+03.6911} - Near-infrared (left panel; UKIDSS, $\mathrm{K,H,J}$ bands in R, G, B colour-scale) and C-band radio map of G077.5671+03.6911 (right panel). The restoring beams used was $0.323\arcsec\times0.295\arcsec$ at $66\degr$. Contour levels are $(-3, 3, 5, 7) \times \sigma$. All other symbols/values have the usual meaning.}
\label{cplot:G077.5671}
\end{figure*} 

\begin{figure*}
\includegraphics[width=\textwidth]{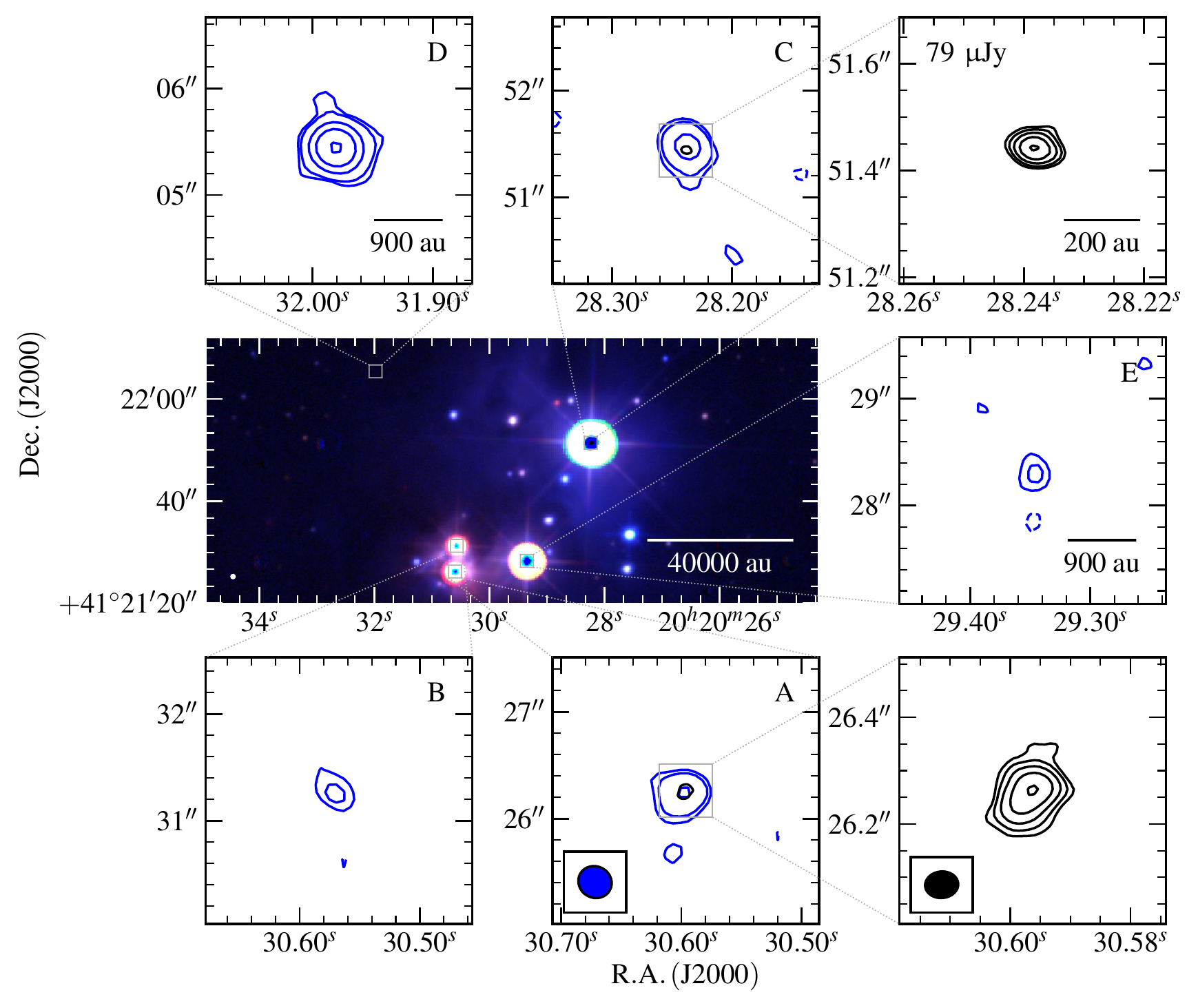}
\caption[Near-infrared and VLA radio images of the MYSO G078.8699+02.7602]{\textbf{G078.8699+02.7602} - Near-infrared (middle row, left panel; UKIDSS, $\mathrm{K,H,J}$ R, G, B colour-scale) and radio maps of G078.8699+02.7602 at C-band (blue contours; middle row and top right panels) and Q-band (black contours; bottom panels). Restoring beams were $0.319\arcsec\times0.295\arcsec$ at $61\degr$ and $0.062\arcsec\times0.050\arcsec$ at $-85\degr$ for the C and Q-band data respectively. Contour levels are $(-3, 3, 6, 16, 35, 80) \times \sigma$ for C-band images, $(-3, 3, 5, 10, 20, 37) \times \sigma$ for the Q-band image of A and $(-3, 3, 5, 8, 12, 19) \times \sigma$ for the Q-band image of C (due to varying noise across the primary beam). All other symbols/values have the usual meaning.}
\label{cplot:G078.8699}
\end{figure*}

\begin{figure*}
\includegraphics[width=\textwidth]{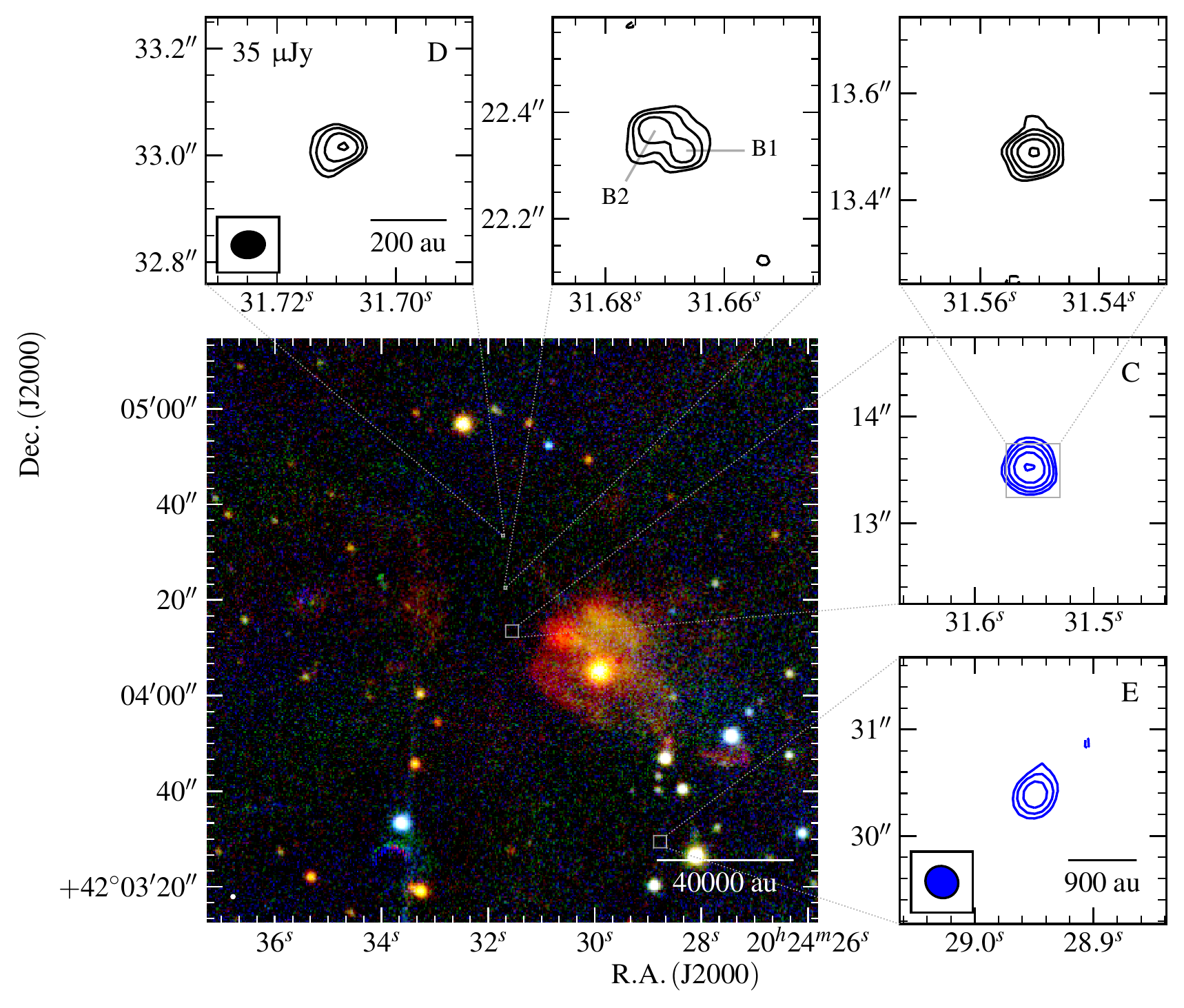}
\caption[Near-infrared and VLA radio images of the MYSO G079.8855+02.5517]{\textbf{G079.8855+02.5517} - Near-infrared (bottom left panel; UKIDSS, $\mathrm{K,H,J}$ bands in R, G, B colour-scale) and radio maps of G079.8855+02.5517 at C-band (blue contours; top left, top middle, bottom left, bottom middle and middle right panels) and Q-band (black contours; top right/ bottom left panels). Restoring beams were $0.316\arcsec\times0.293\arcsec$ at $54\degr$ and $0.062\arcsec\times0.050\arcsec$ at $-84\degr$ for the C and Q-band data respectively. Contour levels are $(-3, 3, 6, 11, 22) \times \sigma$ for C-band and Q-band images of lobes A, B and C1/C2, whilst being $(-3, 3, 4, 7, 10) \times \sigma$ for the Q-band image of D (due to varying noise across the primary beam). All other symbols/values have the usual meaning.}
\label{cplot:G079.8855}
\end{figure*}

\begin{figure*}
\includegraphics[width=\textwidth]{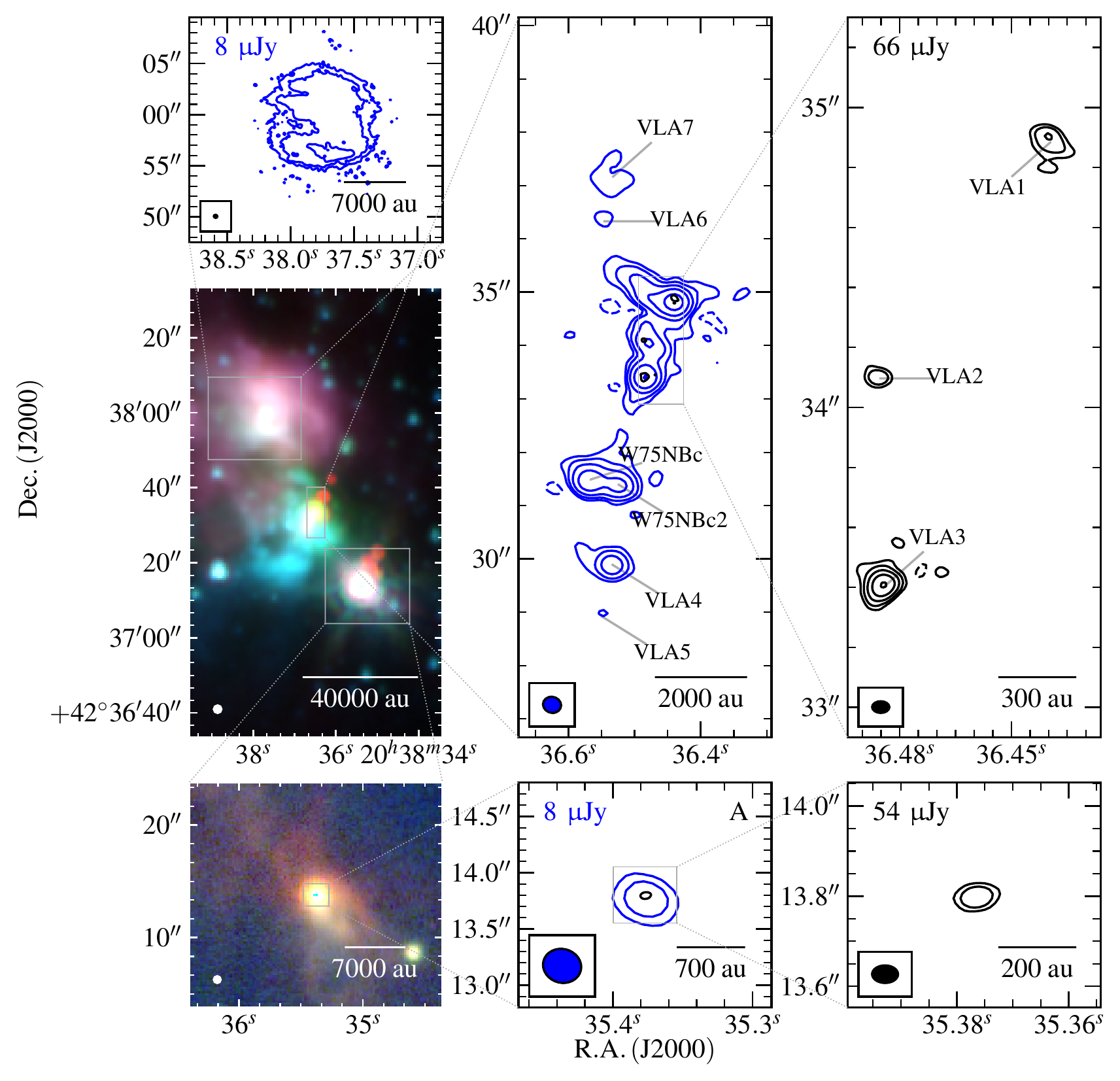}
\caption[Near-infrared, mid-infrared and VLA radio images of the MYSO G081.8652+00.7800]{\textbf{G081.8652+00.7800} - Mid-infrared (middle, left panel; GLIMPSE, $8.0,4.5,3.6\micron$ R, G, B colour-scale), near-infrared (bottom left panel; UKIDSS, $\mathrm{K,H,J}$ bands in R, G, B colour-scale) and radio maps of G081.8652+00.7800 at C-band (blue contours; middle, central and bottom-middle) and Q-band (black contours; right and bottom right). Restoring beams were $0.313\arcsec\times0.293\arcsec$ at $43\degr$ and $0.051\arcsec\times0.039\arcsec$ at $89\degr$, while contour levels are set at $(-4, 4, 9, 21, 49, 112)$ and $(-4, 4, 10, 25, 61, 153) \times \sigma$ for the C and Q-band data respectively. Varying noise levels are the result of dynamic range limitations leading to non-Gaussian noise as well as standard primary beam effects.}
\label{cplot:G081.8652}
\end{figure*}

\begin{figure*}
\includegraphics[width=\textwidth]{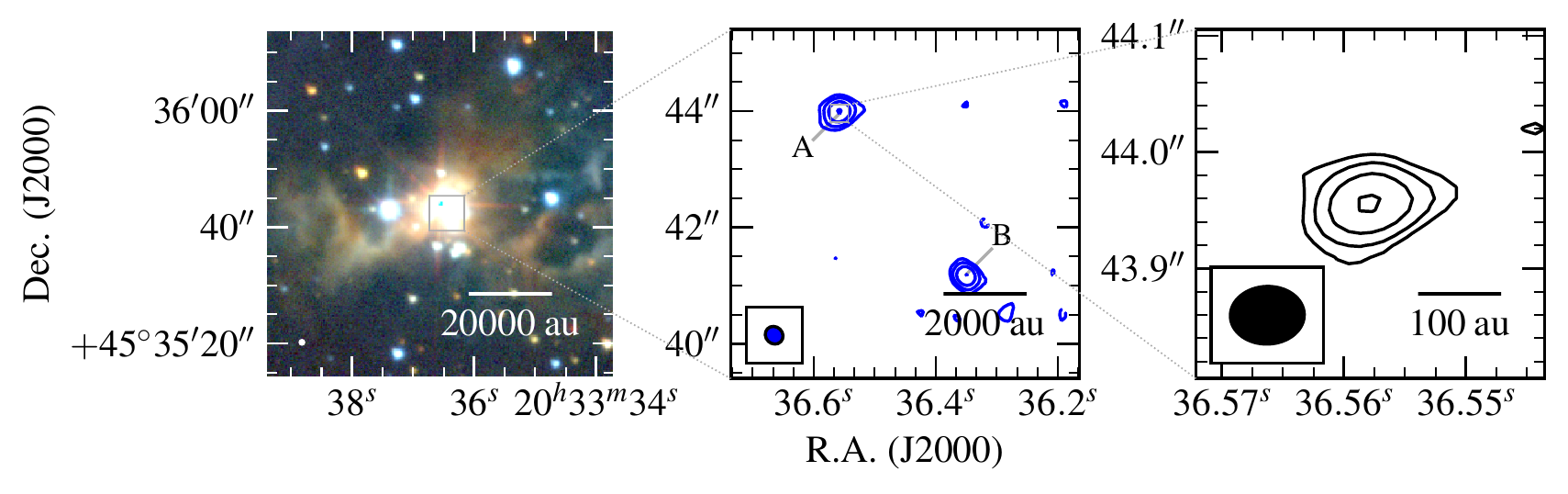}
\caption[Near-infrared and VLA radio images of the MYSO G083.7071+03.2817]{\textbf{G083.7071+03.2817} - Near-infrared (left panel; UKIDSS, $\mathrm{K,H,J}$ bands in R, G, B colour-scale) and radio maps of G083.7071+03.2817 at C-band (blue contours; middle) and Q-band (black contours; bottom). Restoring beams were $0.319\arcsec\times0.290\arcsec$ at $46\degr$ and $0.064\arcsec\times0.050\arcsec$ at $-88\degr$ for the C and Q-band data respectively. Contour levels are $(-3, 3, 6, 13, 27) \times \sigma$ and $(-3, 3, 5, 10, 18) \times \sigma$ for C and Q-band respectively. All other symbols/values have the usual meaning.}
\label{cplot:G083.7071}
\end{figure*}

\begin{figure*}
\centering
\includegraphics[width=0.72\textwidth]{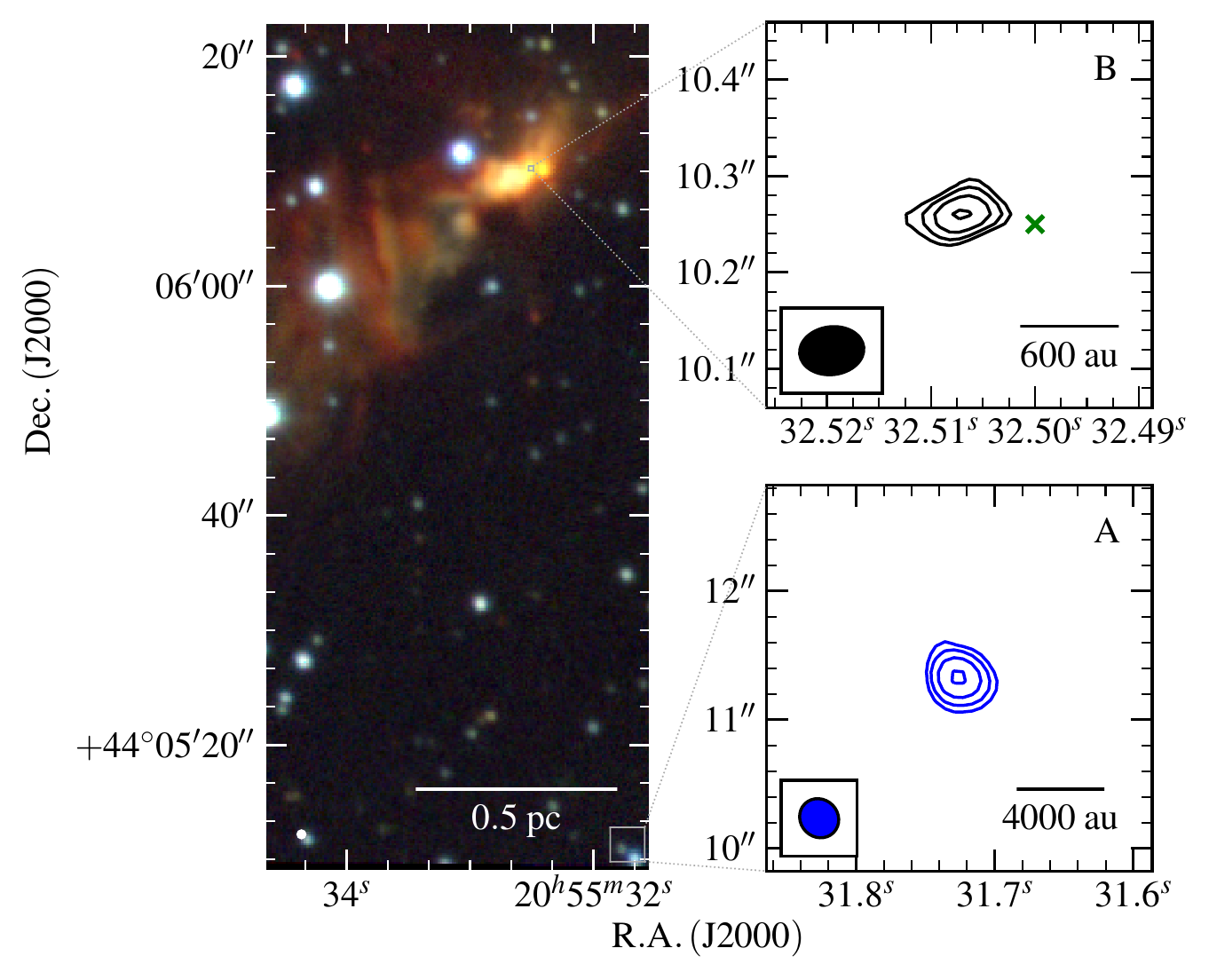}
\caption[Near-infrared and VLA radio images of the MYSO G084.9505$-$00.6910]{\textbf{G084.9505$-$00.6910} - Near-infrared (left panel; UKIDSS, $\mathrm{K,H,J}$ bands in R, G, B colour-scale) and radio maps of the G084.9505$-$00.6910 field at C-band (blue contours; bottom right) and Q-band (black contours; top right). Restoring beams were $0.316\arcsec\times0.292\arcsec$ at $45\degr$ and $0.067\arcsec\times0.050\arcsec$ at $-83\degr$ for the C and Q-band (robustness of 2) data respectively. Contour levels are $(-3, 3, 5, 9, 16) \times \sigma$ and $(-3, 3, 4, 6, 8) \times \sigma$ for C and Q-band respectively. All other symbols/values have the usual meaning. Green crosses show 6.7$\GHz$ methanol maser positions from our data.}
\label{cplot:G084.9505}
\end{figure*} 

\clearpage
\begin{figure*}
\includegraphics[width=\textwidth]{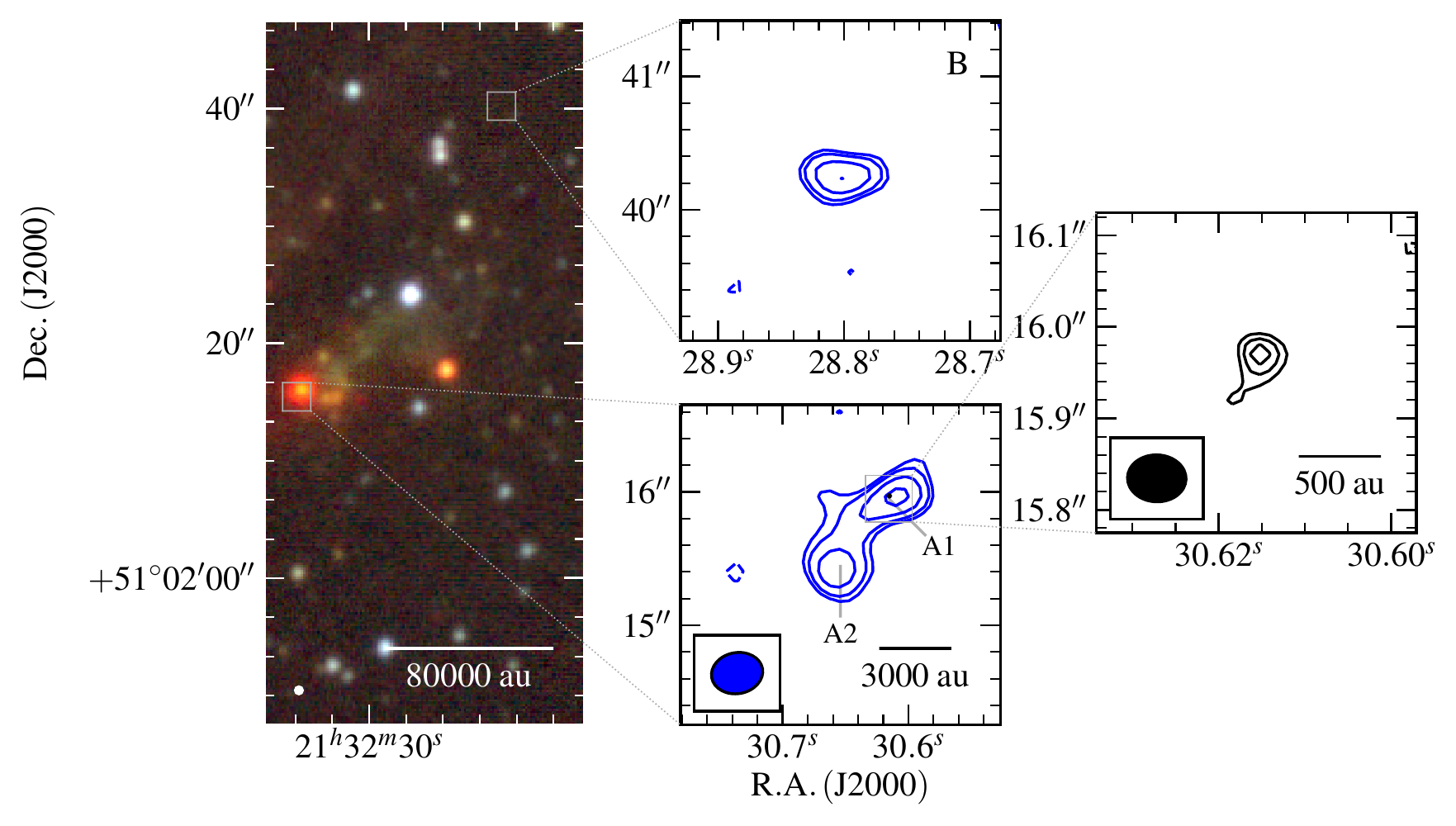}
\caption[Near-infrared and VLA radio images of the MYSO G094.2615$-$00.4116]{\textbf{G094.2615$-$00.4116} - Near-infrared (left panel; UKIDSS, $\mathrm{K,H,J}$ bands in R, G, B colour-scale) and radio maps of the G094.2615$-$00.4116 field at C-band (blue contours; middle) and Q-band (black contours; right). Restoring beams were $0.387\arcsec\times0.308\arcsec$ at $-80\degr$ and $0.064\arcsec\times0.051\arcsec$ at $88\degr$ for the C and Q-band data respectively. Contour levels are $(-3, 3, 4, 6, 9) \times \sigma$ and $(-3, 3, 4, 5) \times \sigma$ for C and Q-band (robustness of 2) respectively. All other symbols/values have the usual meaning.}
\label{cplot:G094.2615}
\end{figure*}

\begin{figure*}
\includegraphics[width=\textwidth]{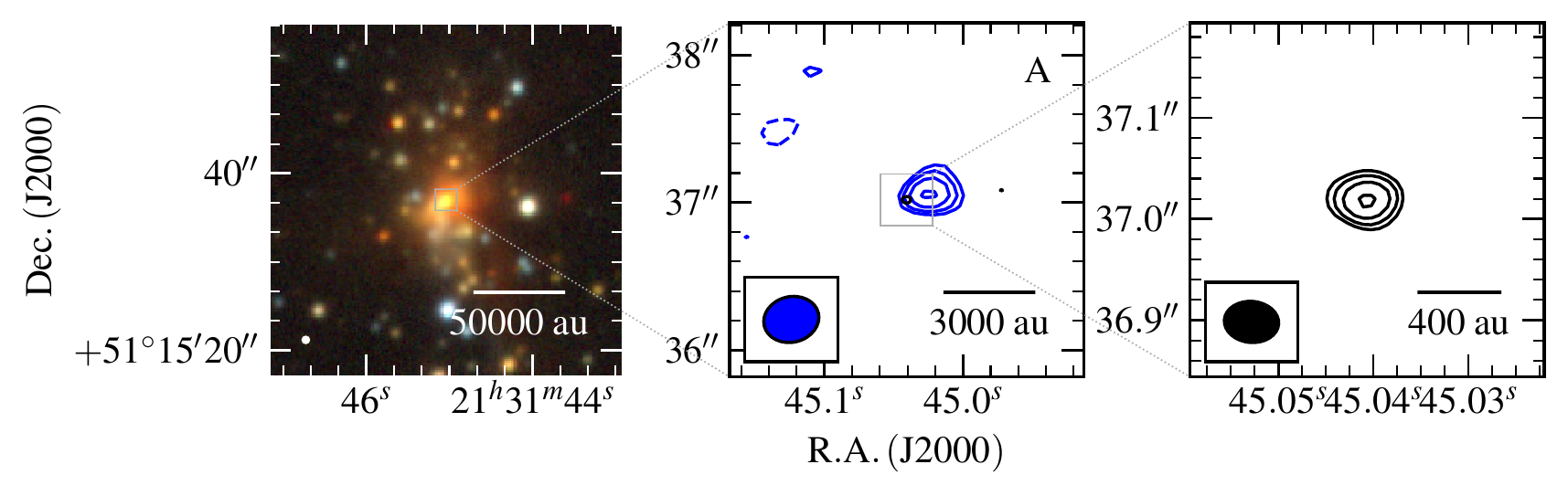}
\caption[Near-infrared and VLA radio images of the MYSO G094.3228$-$00.1671]{\textbf{G094.3228$-$00.1671} - Near-infrared (left panel; UKIDSS, $\mathrm{K,H,J}$ bands in R, G, B colour-scale) and radio maps of the G094.3228$-$00.1671 field at C-band (blue contours; middle) and Q-band (black contours; right). Restoring beams were $0.377\arcsec\times0.309\arcsec$ at $-77\degr$ and $0.054\arcsec\times0.041\arcsec$ at $-85\degr$ for the C and Q-band data respectively. Contour levels are $(-3, 3, 4, 6, 8) \times \sigma$ and $(-3, 3, 4, 7, 10) \times \sigma$ for C and Q-band respectively. All other symbols/values have the usual meaning.}
\label{cplot:G094.3228}
\end{figure*}

\begin{figure*}
\includegraphics[width=\textwidth]{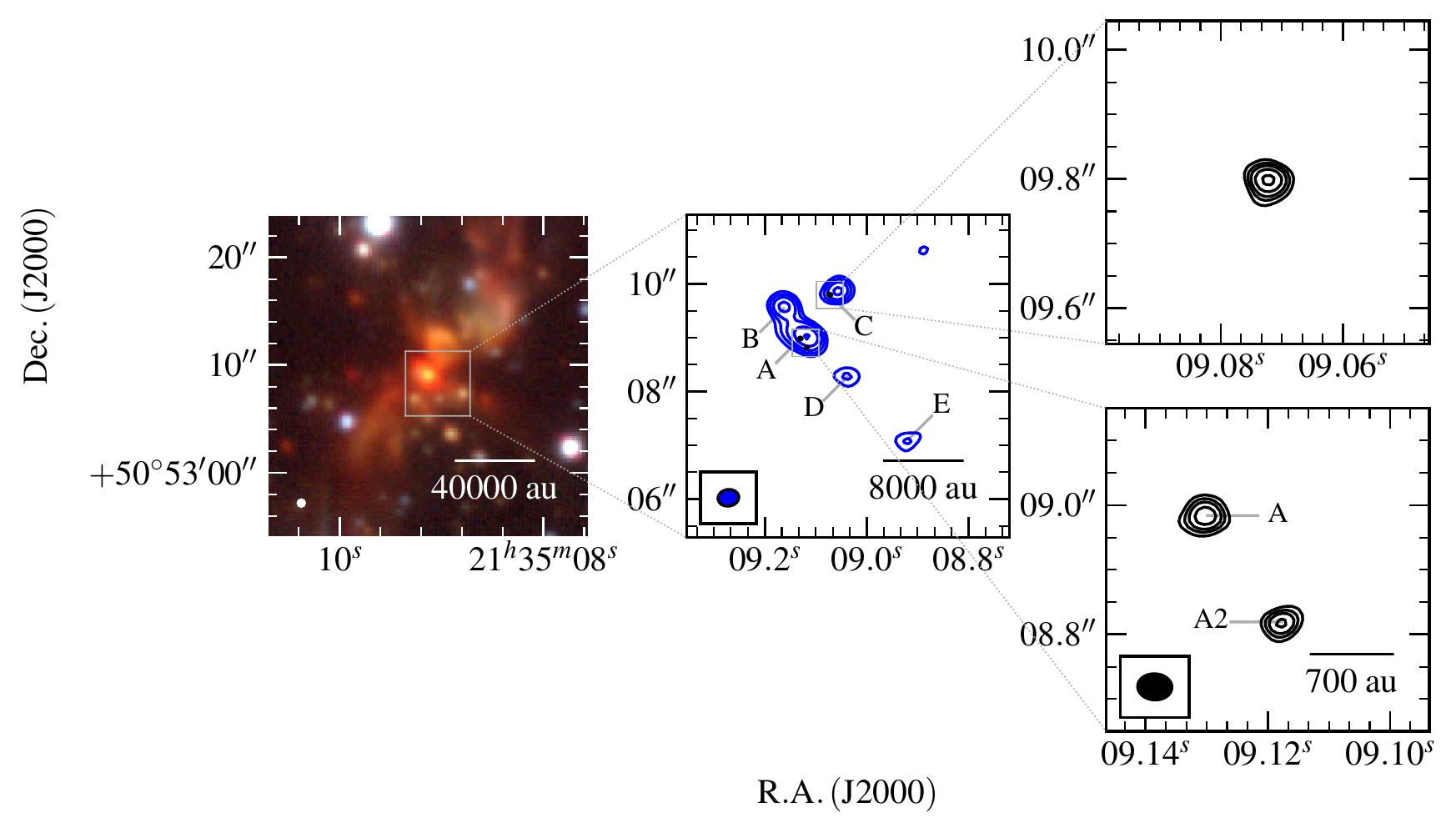}
\caption[Near-infrared and VLA radio images of the MYSO G094.4637$-$00.8043]{\textbf{G094.4637$-$00.8043} - Near-infrared (left panel; UKIDSS, $\mathrm{K,H,J}$ bands in R, G, B colour-scale) and radio maps of the G094.4637$-$00.8043 field at C-band (blue contours; middle) and Q-band (black contours; right). Restoring beams were $0.392\arcsec\times0.312\arcsec$ at $-82\degr$ and $0.053\arcsec\times0.041\arcsec$ at $87\degr$ for the C and Q-band data respectively. Contour levels are $(-4, 4, 8, 15, 29) \times \sigma$ and $(-4, 4, 6, 8, 11, 15) \times \sigma$ for C and Q-band respectively. All other symbols/values have the usual meaning.}
\label{cplot:G094.4637}
\end{figure*}

\begin{figure*}
\includegraphics[width=\textwidth]{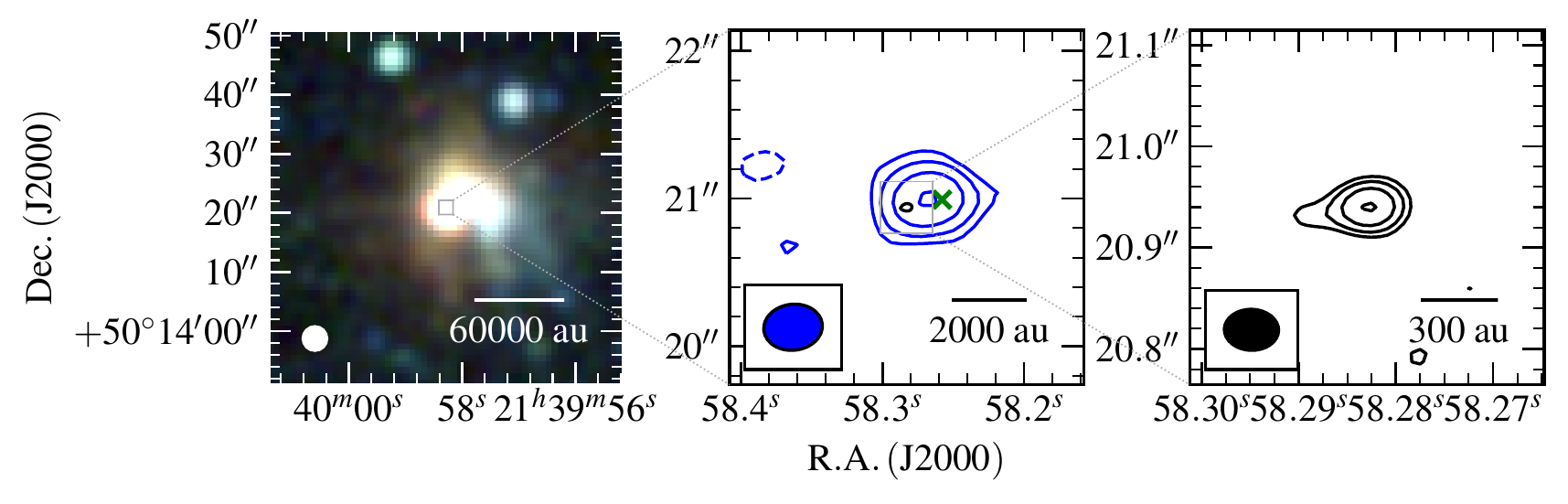}
\caption[Near-infrared and VLA radio images of the MYSO G094.6028$-$01.7966]{\textbf{G094.6028$-$01.7966} - Near-infrared (left panel; 2MASS, $\mathrm{K,H,J}$ bands in R, G, B colour-scale) and radio maps of G094.6028$-$01.7966 at C-band (blue contours; middle) and Q-band (black contours; bottom). Restoring beams were $0.399\arcsec\times0.311\arcsec$ at $-84\degr$ and $0.054\arcsec\times0.041\arcsec$ at $88\degr$ for the C and Q-band data respectively. Contour levels are $(-3, 3, 7, 15, 32) \times \sigma$ and $(-3, 3, 5, 9, 15) \times \sigma$ for C and Q-band respectively. All other symbols/values have the usual meaning. Green crosses show 6.7$\GHz$ methanol maser positions from our data.}
\label{cplot:G094.6028}
\end{figure*}

\begin{figure*}
\includegraphics[width=\textwidth]{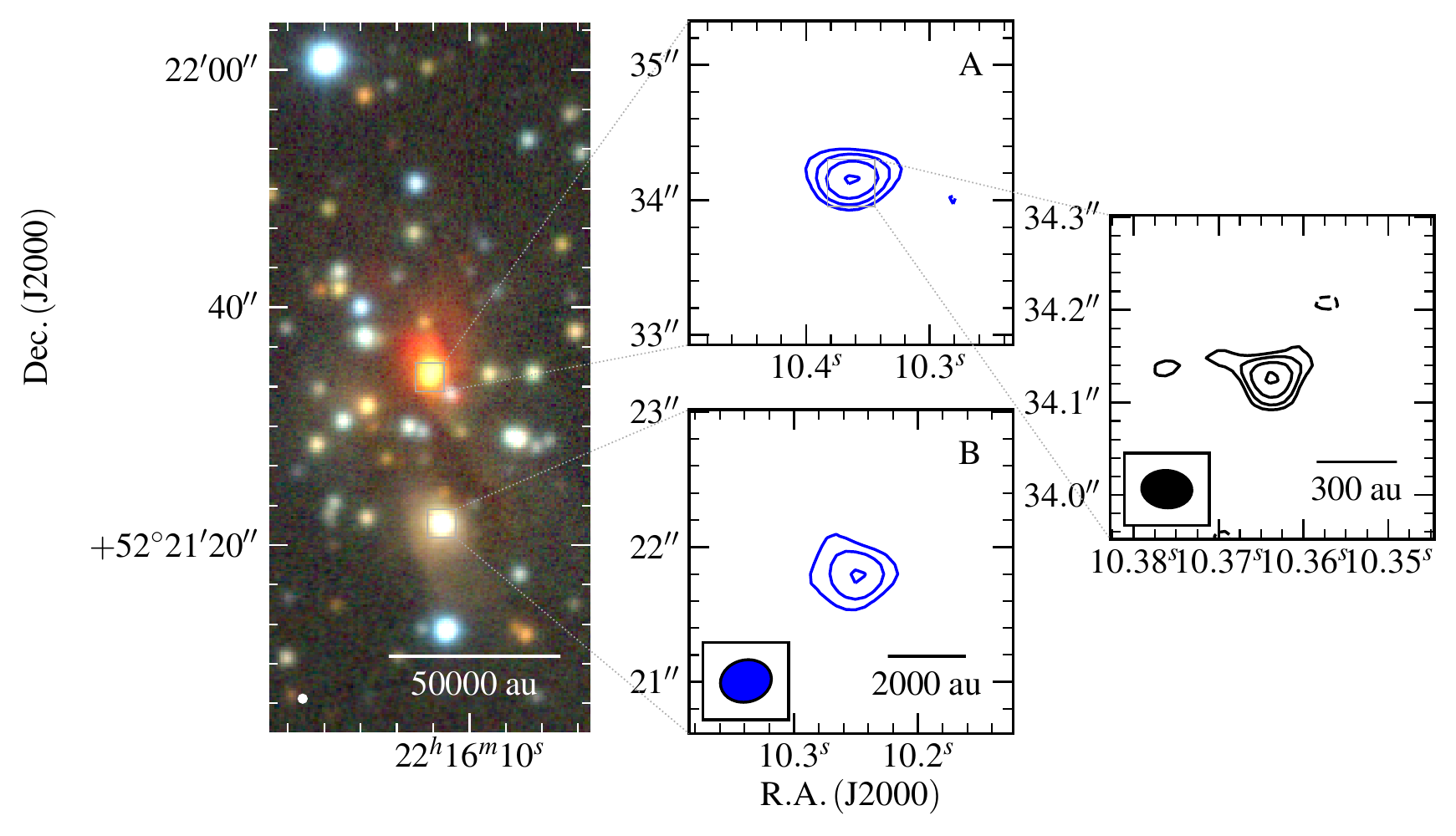}
\caption[Near-infrared and VLA radio images of the MYSO G100.3779$-$03.5784]{\textbf{G100.3779$-$03.5784} - Near-infrared (left panel; UKIDSS, $\mathrm{K,H,J}$ bands in R, G, B colour-scale) and radio maps of the G100.3779$-$03.5784 field at C-band (blue contours; middle) and Q-band (black contours; right). Restoring beams were $0.381\arcsec\times0.310\arcsec$ at $-76\degr$ and $0.054\arcsec\times0.041\arcsec$ at $87\degr$ for the C and Q-band data respectively. Contour levels are $(-3, 3, 5, 8, 13) \times \sigma$ and $(-3, 3, 4, 7, 10) \times \sigma$ for C and Q-band (robustness of 2) respectively. All other symbols/values have the usual meaning.}
\label{cplot:G100.3779}
\end{figure*}

\begin{figure*}
\includegraphics[width=\textwidth]{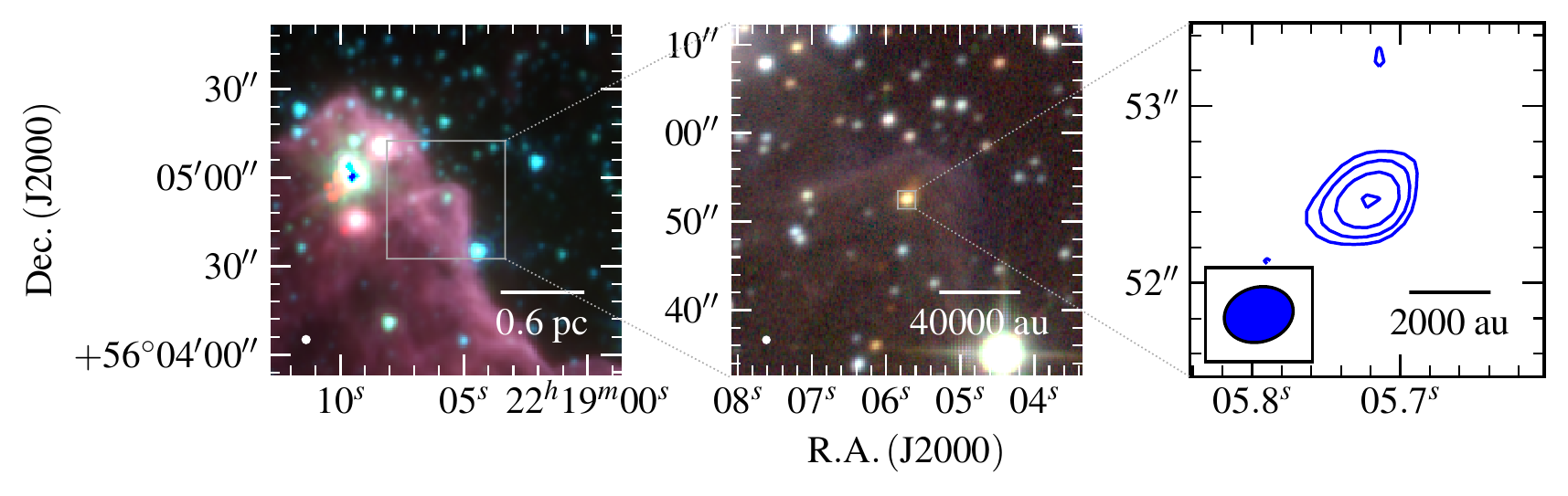}
\caption[Near-infrared, mid-infrared and VLA radio images of the MYSO G102.8051$-$00.7184]{\textbf{G102.8051$-$00.7184} - Mid-infrared (left panel; GLIMPSE, $8.0,4.5,3.6\micron$ R, G, B colour-scale), near-infrared (middle panel; UKIDSS, $\mathrm{K,H,J}$ bands in R, G, B colour-scale) and C-band radio map of G102.8051$-$00.7184 (right panel). The restoring beams used was $0.395\arcsec\times0.309\arcsec$ at $-75\degr$. Contour levels are $(-3, 3, 5, 9, 14) \times \sigma$. All other symbols/values have the usual meaning.}
\label{cplot:G102.8051}
\end{figure*} 

\begin{figure*}
\includegraphics[width=\textwidth]{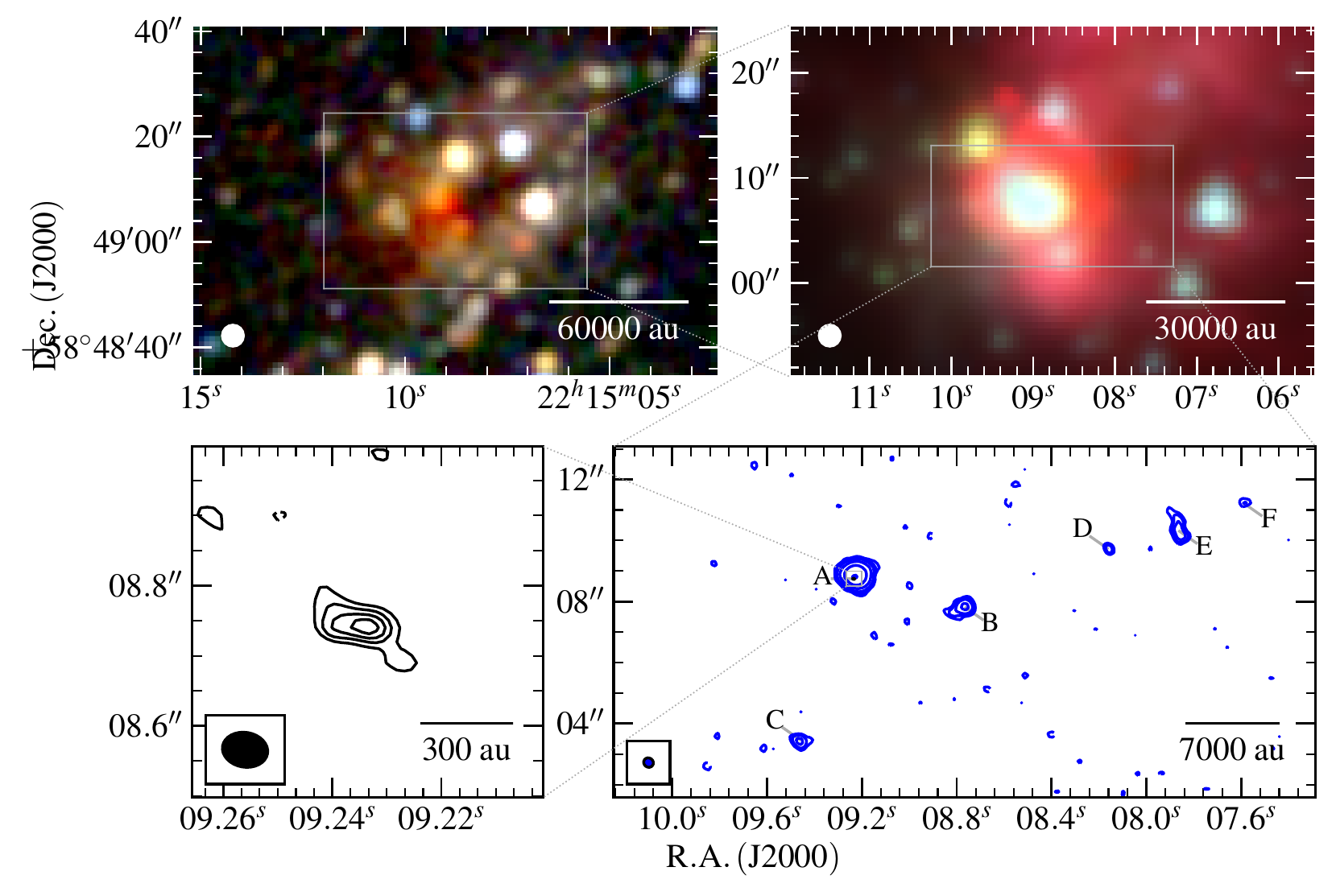}
\caption[Near-infrared, mid-infrared and VLA radio images of the MYSO G103.8744+01.8558]{\textbf{G103.8744+01.8558} - Mid-infrared (top right panel; GLIMPSE, $8.0,4.5,3.6\micron$ R, G, B colour-scale), near-infrared (top left panel; 2MASS, $\mathrm{K,H,J}$ R, G, B colour-scale) and radio maps of G103.8744+01.8558 at C-band (blue contours; bottom-right) and Q-band (black contours; bottom left). Restoring beams were $0.327\arcsec\times0.302\arcsec$ at $54\degr$ and $0.065\arcsec\times0.050\arcsec$ at $80\degr$, while contour levels are $(-3, 3, 5, 9, 24, 70, 199) \times \sigma$ and $(-3, 3, 4, 5, 6) \times \sigma$ for C and Q-band (robustness of 2) respectively. All other symbols have their usual meanings.}
\label{cplot:G103.8744}
\end{figure*}

\begin{figure*}
\includegraphics[width=\textwidth]{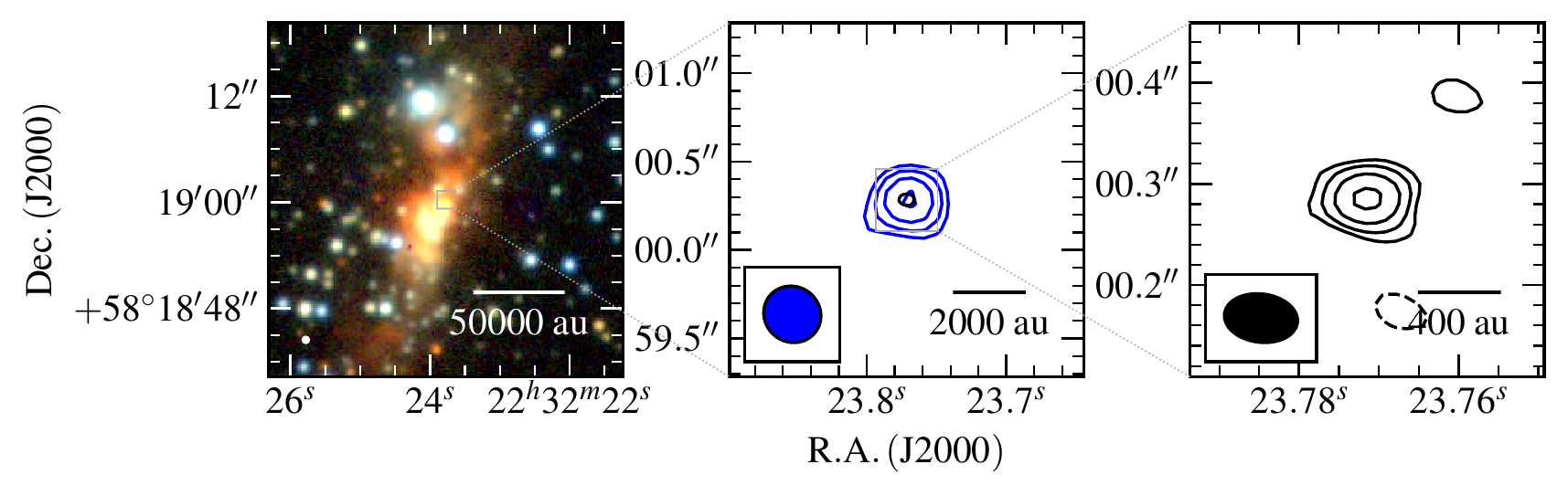}
\caption[Near-infrared and VLA radio images of the MYSO G105.5072+00.2294]{\textbf{G105.5072+00.2294} - Near-infrared (R, G, B colour-scale, left panel; UKIDSS, $\mathrm{K,H,J}$ bands) and radio maps of the G105.5072+00.2294 field at C-band (blue contours; middle) and Q-band (black contours; right). Restoring beams were $0.330\arcsec\times0.312\arcsec$ at $51\degr$ and $0.073\arcsec\times0.048\arcsec$ at $83\degr$, while contour levels are $(-3, 3, 4, 6, 9) \times \sigma$ and $(-3, 3, 5, 7, 12) \times \sigma$ for C and Q-band (robustness of 2) respectively. All other symbols/values have the usual meaning.}
\label{cplot:G105.5072}
\end{figure*}

\begin{figure*}
\includegraphics[width=\textwidth]{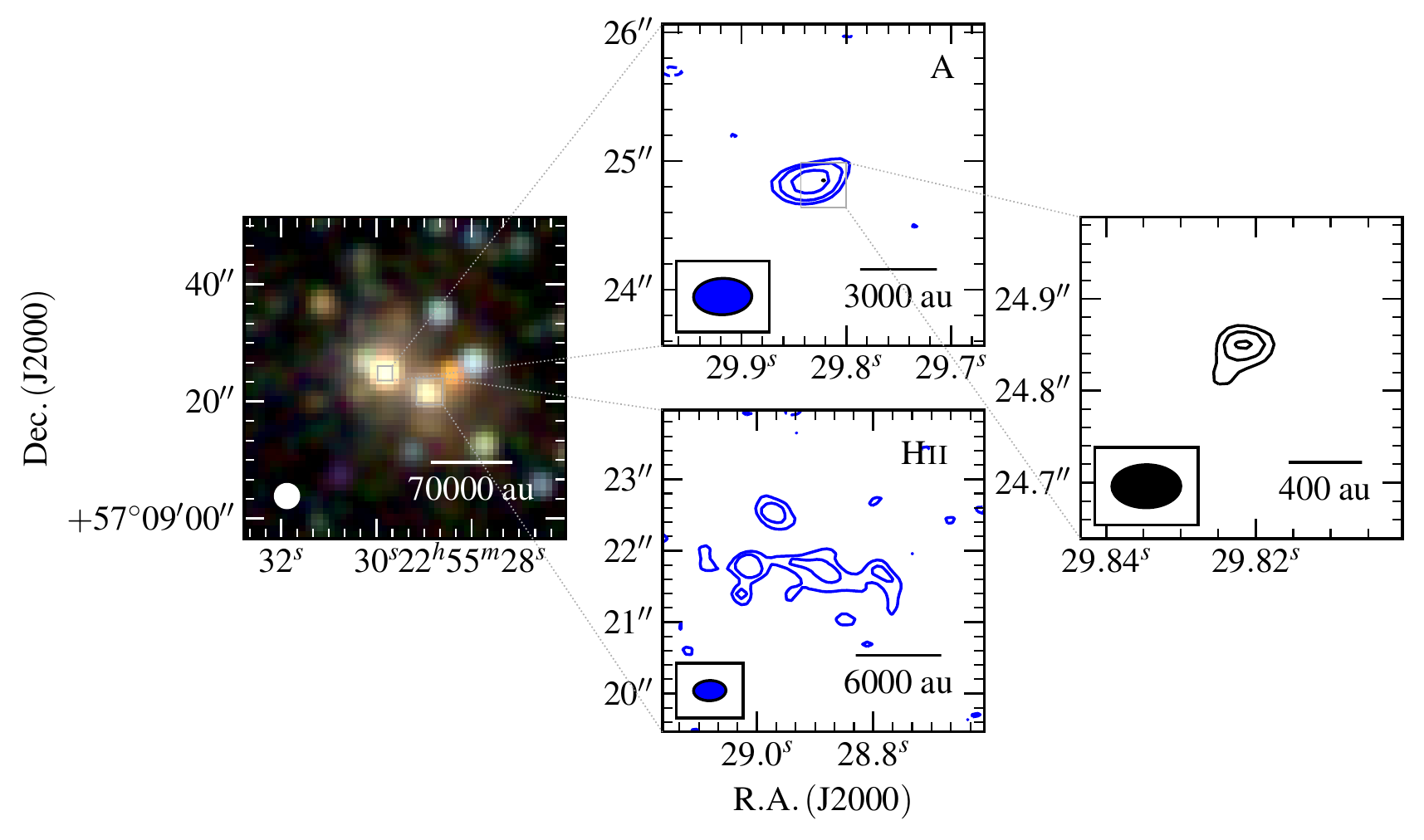}
\caption[Near-infrared and VLA radio images of the MYSO G107.6823$-$02.2423A]{\textbf{G107.6823$-$02.2423A} - Near-infrared (R, G, B colour-scale, left panel; 2MASS, $\mathrm{K,H,J}$ bands) and radio contour maps of G107.6823$-$02.2423A at C-band (top and bottom panels; blue contours) and Q-band (right panel; black contours). Restoring beams were $0.564\arcsec\times0.352\arcsec$ at $-88\degr$ and $0.075\arcsec\times0.047\arcsec$ at $-90\degr$ while contour levels are $(-3, 3, 4, 6, 8) \times \sigma$ and $(-3, 3, 4, 5) \times \sigma$ for C and Q-band respectively. A robustness of 2 was utilised for both sets of radio data. All other values have the usual meaning.}
\label{cplot:G107.6823}
\end{figure*}

\begin{figure*}
\includegraphics[width=\textwidth]{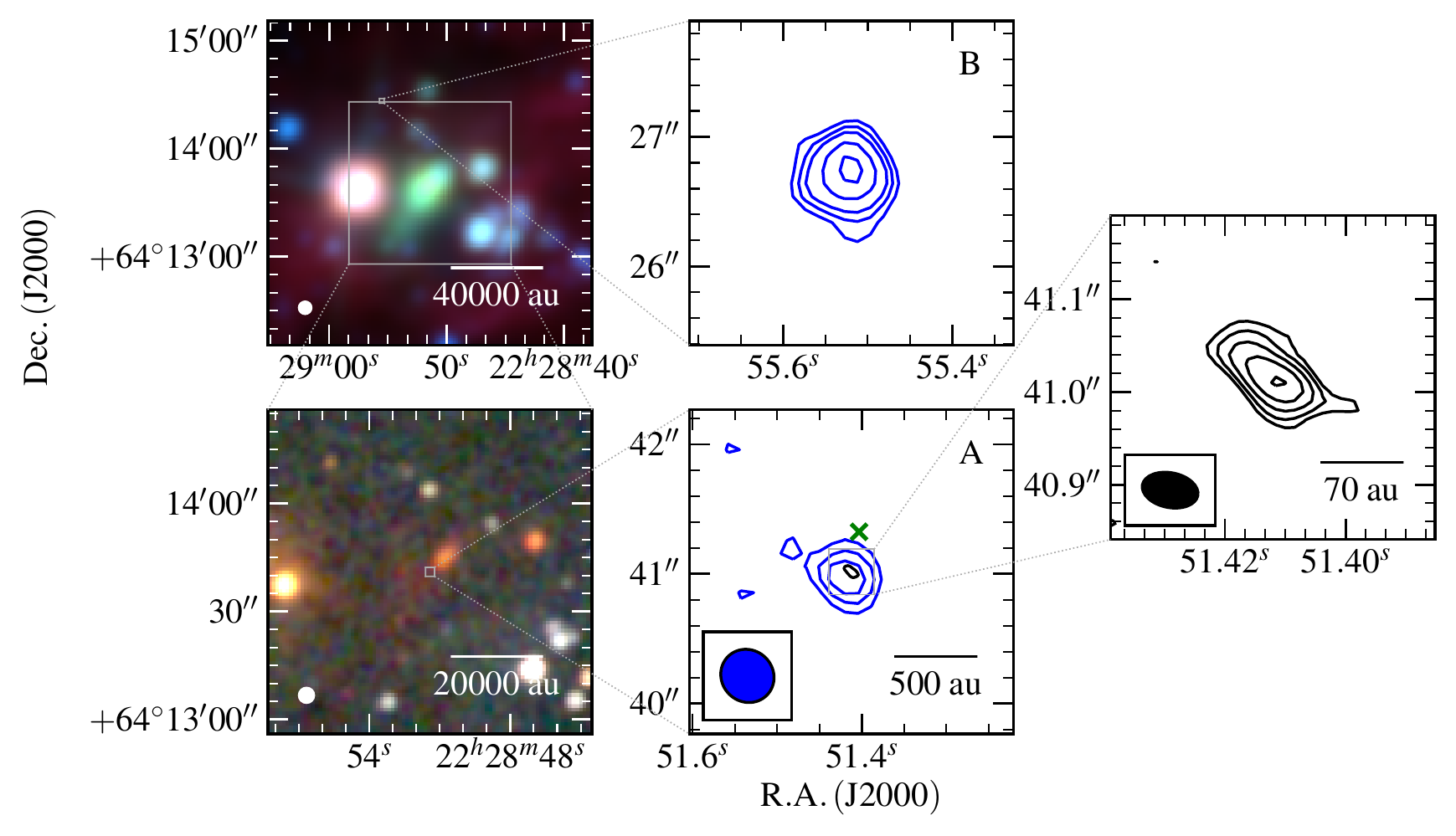}
\caption[Near-infrared, mid-infrared and VLA radio images of the MYSO G108.1844+05.5187]{\textbf{G108.1844+05.5187} - Mid-infrared (R, G, B colour-scale, top-left panel; WISE, $12.0,4.6,3.4\micron$), near-infrared (R, G, B colour-scale, bottom-left panel; 2MASS, $\mathrm{K,H,J}$ bands) and radio contour maps of G108.1844+05.5187 at C-band (top and bottom panels; blue contours) and Q-band (right panel; black contours). Restoring beams were $0.423\arcsec\times0.398\arcsec$ at $45\degr$ and $0.061\arcsec\times0.038\arcsec$ at $79\degr$ while contour levels are $(-3, 3, 4, 5, 7, 9) \times \sigma$ and $(-3, 3, 5, 7, 11, 16) \times \sigma$ for C and Q-band respectively. All other values have the usual meaning. Green crosses show 6.7$\GHz$ methanol maser positions from our data.}
\label{cplot:G108.1844}
\end{figure*}

\begin{figure*}
\includegraphics[width=\textwidth]{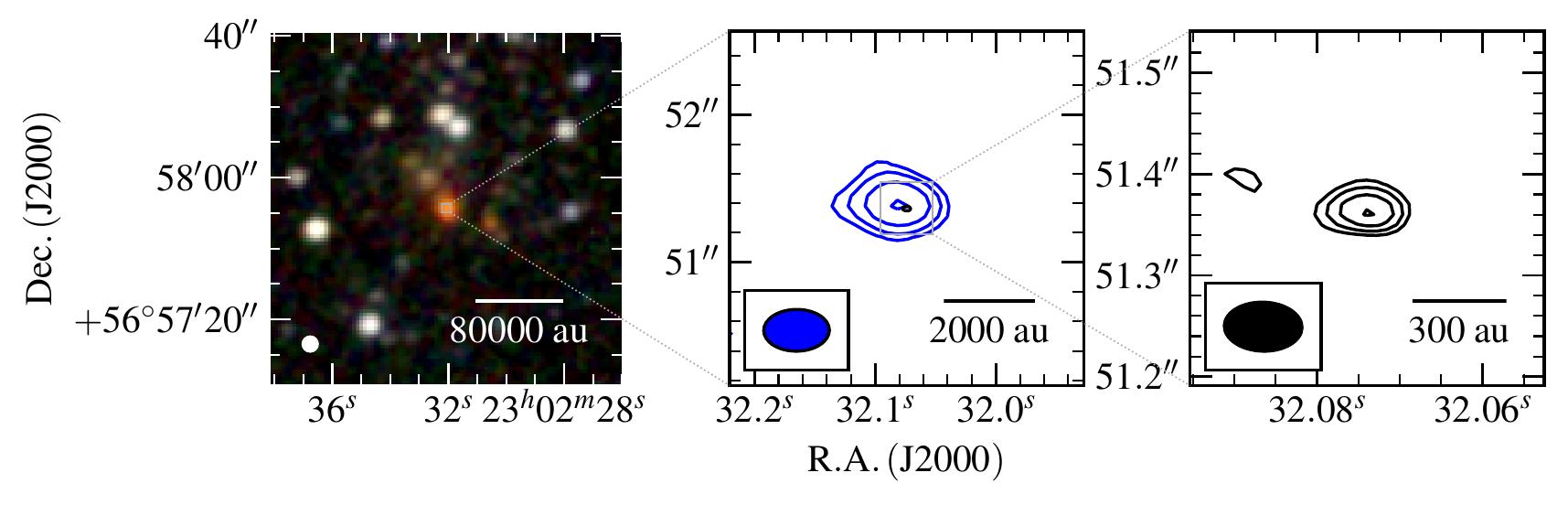}
\caption[Near-infrared and VLA radio images of the MYSO G108.4714$-$02.8176]{\textbf{G108.4714$-$02.8176} - Near-infrared (left panel; 2MASS, $\mathrm{K,H,J}$ bands R, G, B colour-scale) and radio maps of the G108.4714$-$02.8176 field at C-band (blue contours; middle) and Q-band (black contours; right). Restoring beams were $0.445\arcsec\times0.282\arcsec$ at $-90\degr$ and $0.063\arcsec\times0.038\arcsec$ at $88\degr$ for the C and Q-band data respectively. Contour levels are $(-3, 3, 5, 10, 17) \times \sigma$ and $(-3, 3, 4, 5, 7) \times \sigma$ for C and Q-band respectively. All other symbols/values have the usual meaning.}
\label{cplot:G108.4714}
\end{figure*}

\begin{figure*}
\includegraphics[width=\textwidth]{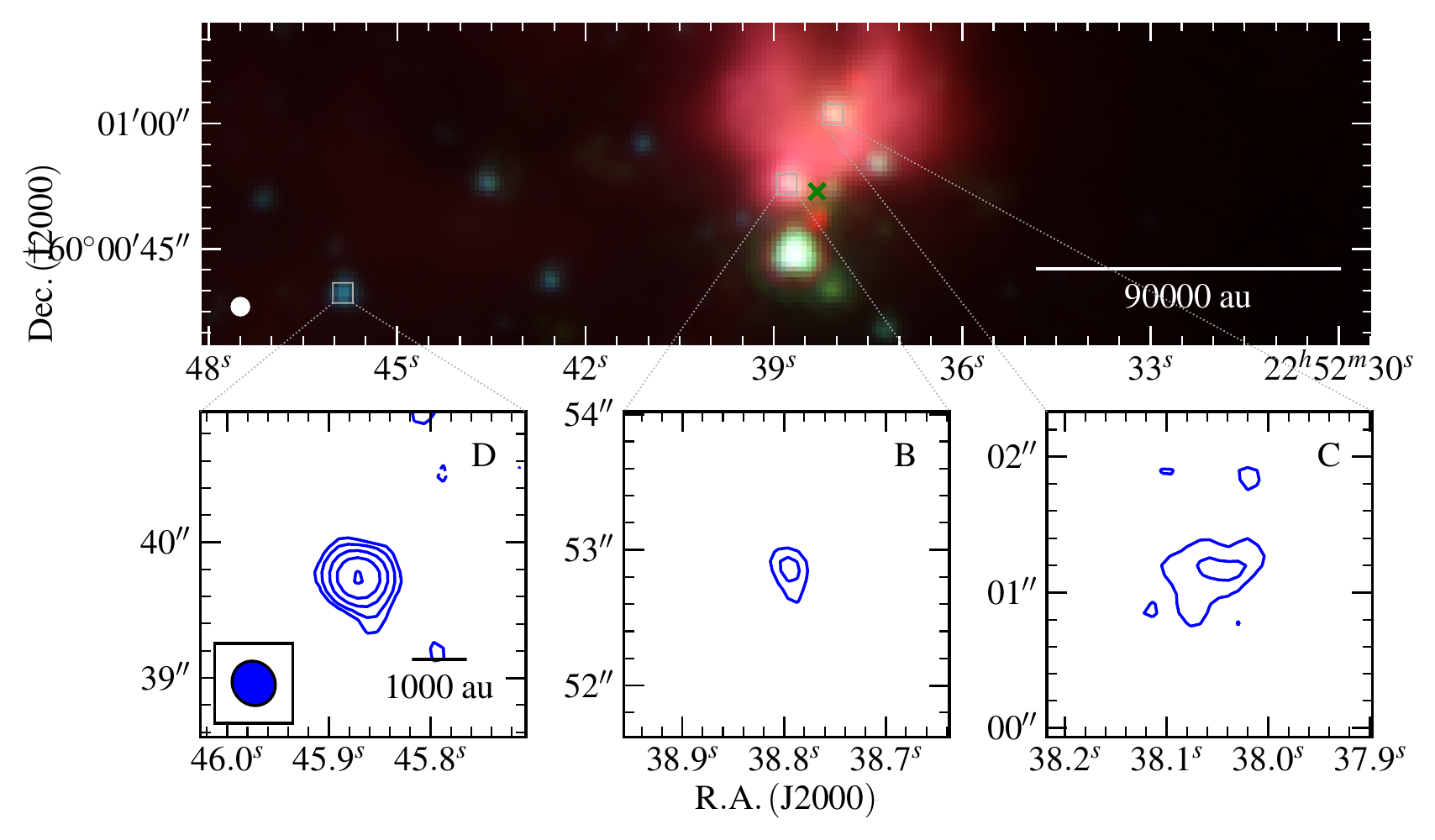}
\caption[Mid-infrared and VLA radio images of the MYSO G108.5955+00.4935C]{\textbf{G108.5955+00.4935A} - Mid-infrared (R, G, B colour-scale, top panel; GLIMPSE $8.0,4.5,3.6\micron$ bands) and C-band (blue contours; bottom panels) images of G108.5955+00.4935B and G108.5955+00.4935C. The C-band restoring beam was $0.337\arcsec\times0.311\arcsec$ at $35\degr$. Contour levels are $(-3, 3, 5, 9, 15, 26) \times \sigma$ and all other values have their usual meaning. It is important to note that the methanol maser indicated on the MIR plot (green cross) isn't represented on radio images due to non-detection of continuum emission.}
\label{cplot:G108.5955}
\end{figure*}

\begin{figure*}
\includegraphics[width=\textwidth]{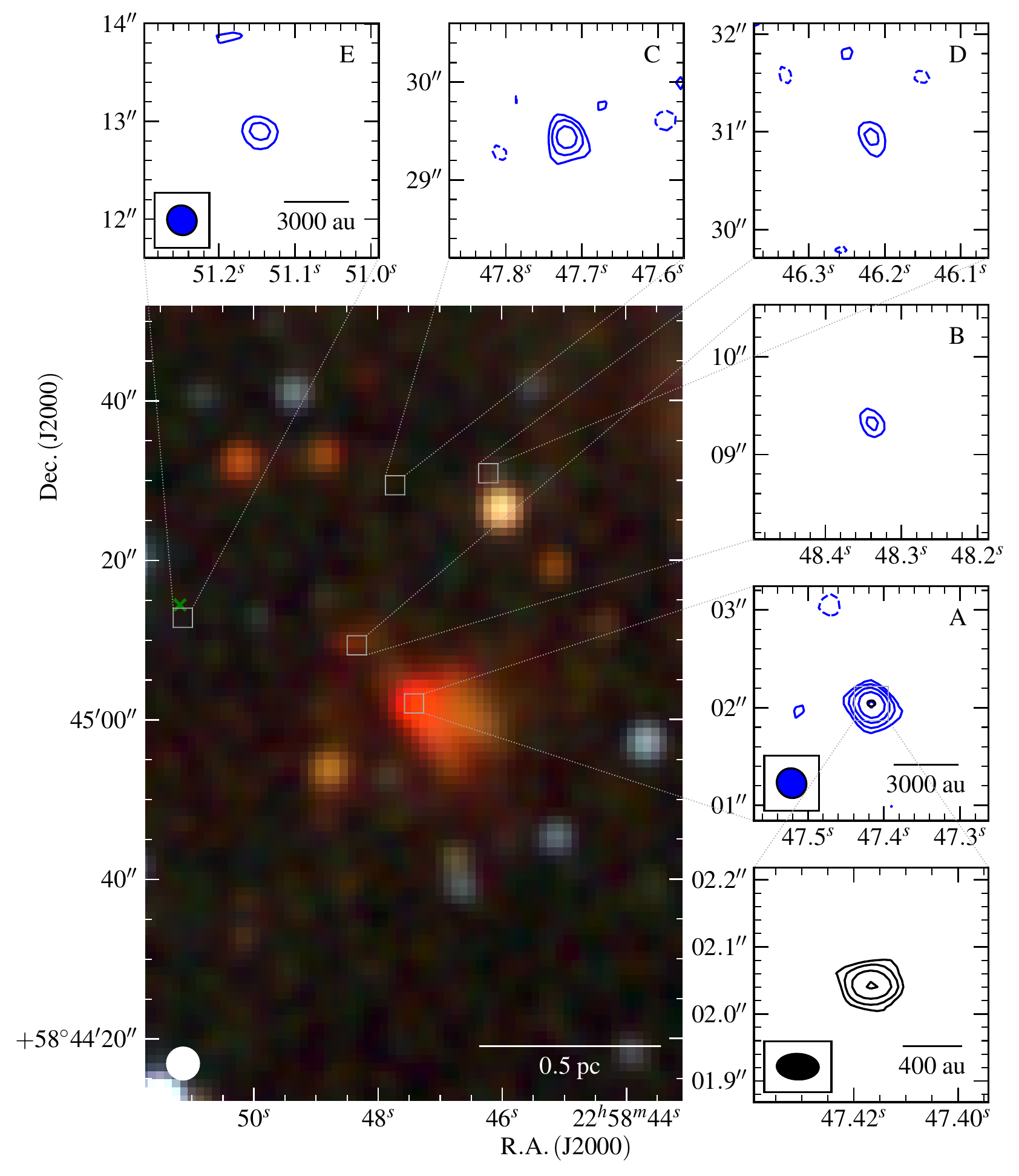}
\caption[Mid-infrared and VLA radio images of the MYSO G108.7575$-$00.9863]{\textbf{G108.7575$-$00.9863} - Mid-infrared (R, G, B colour-scale, left panel; 2MASS, $\mathrm{K,H,J}$ bands R, G, B colour-scale) and radio maps of G108.7575$-$00.9863 at C-band (blue contours; top and middle-right panels) and Q-band (black contours; bottom right). Restoring beams were $0.312\arcsec\times0.295\arcsec$ at $44\degr$ and $0.063\arcsec\times0.038\arcsec$ at $88\degr$, while contour levels were set at $(-3, 3, 5, 8, 12, 19)$ and $(-3, 3, 5, 9, 15) \times \sigma$ for the C and Q-band (robustness of 2) data respectively. Varying noise levels are the result of missing spatial scale information leading to non-Gaussian noise and primary beam effects.  Green crosses show 6.7$\GHz$ methanol maser positions from our data.}
\label{cplot:G108.7575}
\end{figure*}

\begin{figure*}
\includegraphics[width=\textwidth]{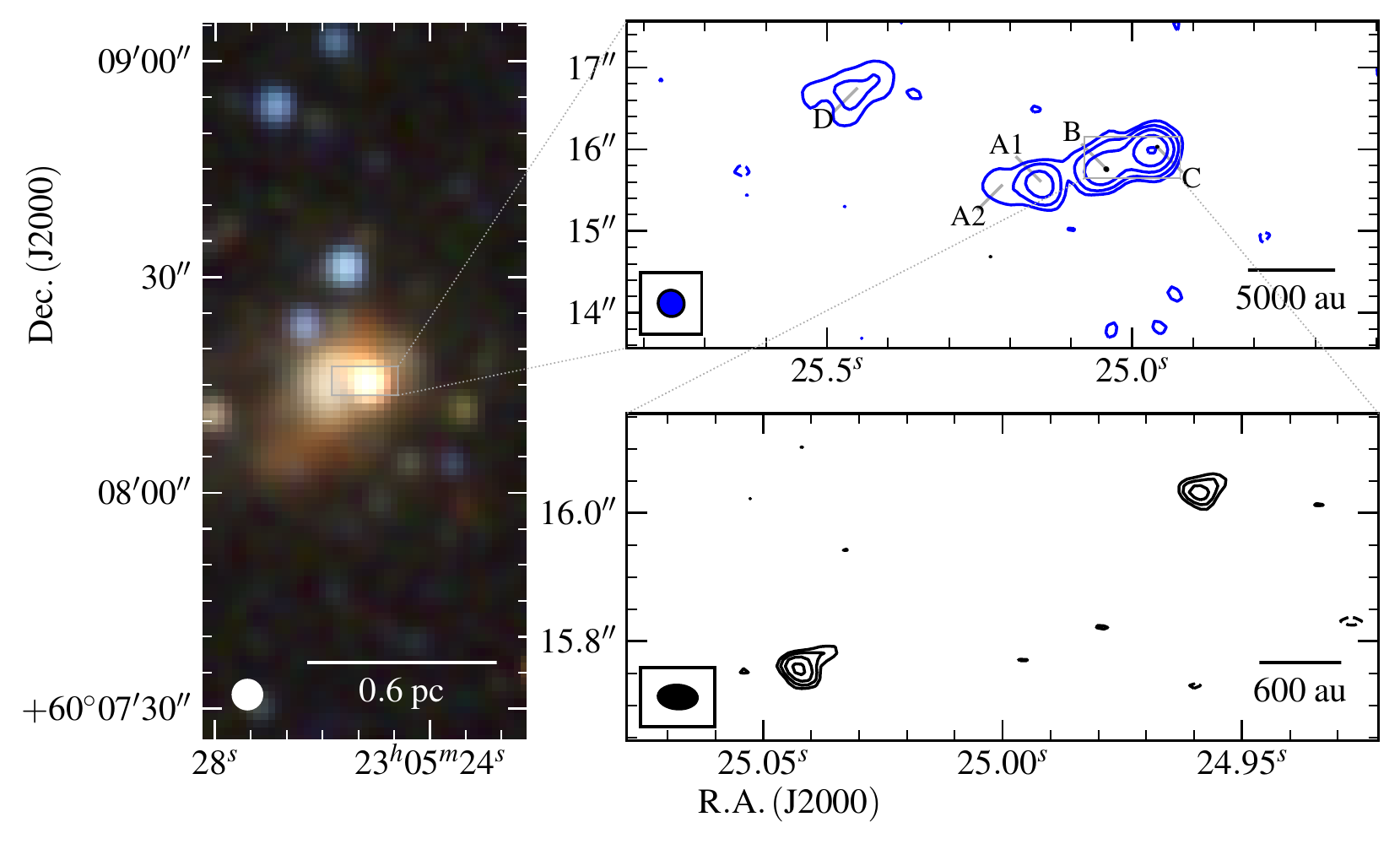}
\caption[Near-infrared and VLA radio images of the MYSO G110.0931$-$00.0641]{\textbf{G110.0931$-$00.0641} - Near-infrared (R, G, B colour-scale, left panel; 2MASS, $\mathrm{K,H,J}$ bands ) and radio maps of the G110.0931$-$00.0641 field at C-band (blue contours; middle) and Q-band (black contours; right). Restoring beams were $0.326\arcsec\times0.314\arcsec$ at $36\degr$ and $0.063\arcsec\times0.038\arcsec$ at $86\degr$, while contour levels are $(-3, 3, 7, 14, 32, 69) \times \sigma$ and $(-3, 3, 4, 5, 6) \times \sigma$ for C and Q-band respectively. All other symbols/values have the usual meaning.}
\label{cplot:G110.0931}
\end{figure*}

\clearpage
\begin{figure*}
\includegraphics[width=\textwidth]{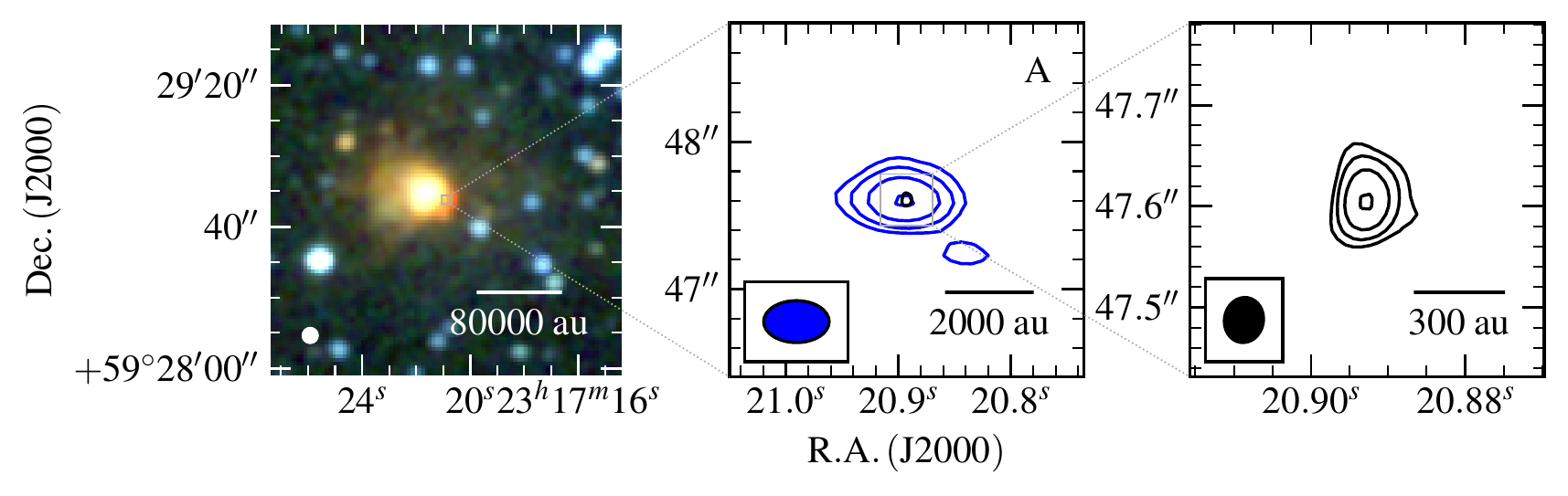}
\caption[Near-infrared and VLA radio images of the MYSO G111.2348$-$01.2385]{\textbf{G111.2348$-$01.2385} - Near-infrared (left panel; 2MASS, $\mathrm{K,H,J}$ bands R, G, B colour-scale) and radio maps of the G111.2348$-$01.2385 field at C-band (blue contours; middle) and Q-band (black contours; right). Restoring beams were $0.443\arcsec\times0.284\arcsec$ at $90\degr$ and $0.045\arcsec\times0.039\arcsec$ at $-10\degr$ while contour levels are set at $(-3, 3, 6, 12, 24) \times \sigma$ and $(-3, 3, 6, 12, 25) \times \sigma$ for C and Q-band respectively. All other symbols/values have the usual meaning.}
\label{cplot:G111.2348}
\end{figure*}

\begin{figure*}
\includegraphics[width=\textwidth]{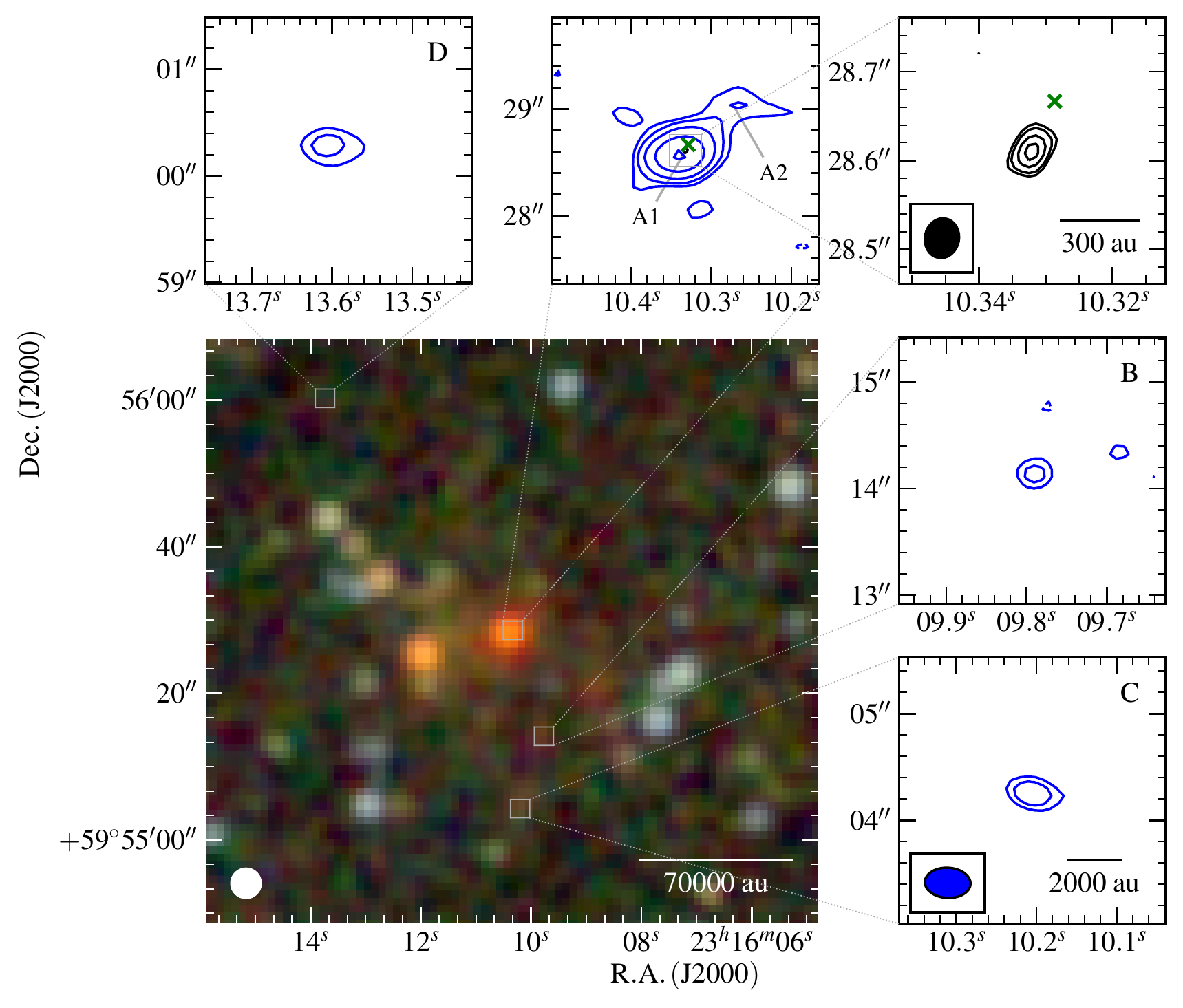}
\caption[Near-infrared and VLA radio images of the MYSO G111.2552$-$00.7702]{\textbf{G111.2552$-$00.7702} - Near-infrared (bottom left panel; 2MASS, $\mathrm{K,H,J}$ R, G, B colour-scale) and radio maps of G111.2552$-$00.7702 at C-band (blue contours; top left/middle and right middle/bottom panels) and Q-band (black contours; top right panel). Restoring beams were $0.430\arcsec\times0.283\arcsec$ at $87\degr$ and $0.045\arcsec\times0.038\arcsec$ at $-11\degr$ while contour levels are set at $(-3, 3, 5, 9, 17, 30) \times \sigma$ and $(-3, 3, 4, 6, 8) \times \sigma$ for the C-band and Q-band images respectively. All other symbols/values have the usual meaning. Green crosses show 6.7$\GHz$ methanol maser positions from our data.}
\label{cplot:G111.2552}
\end{figure*}

\begin{figure*}
\includegraphics[width=\textwidth]{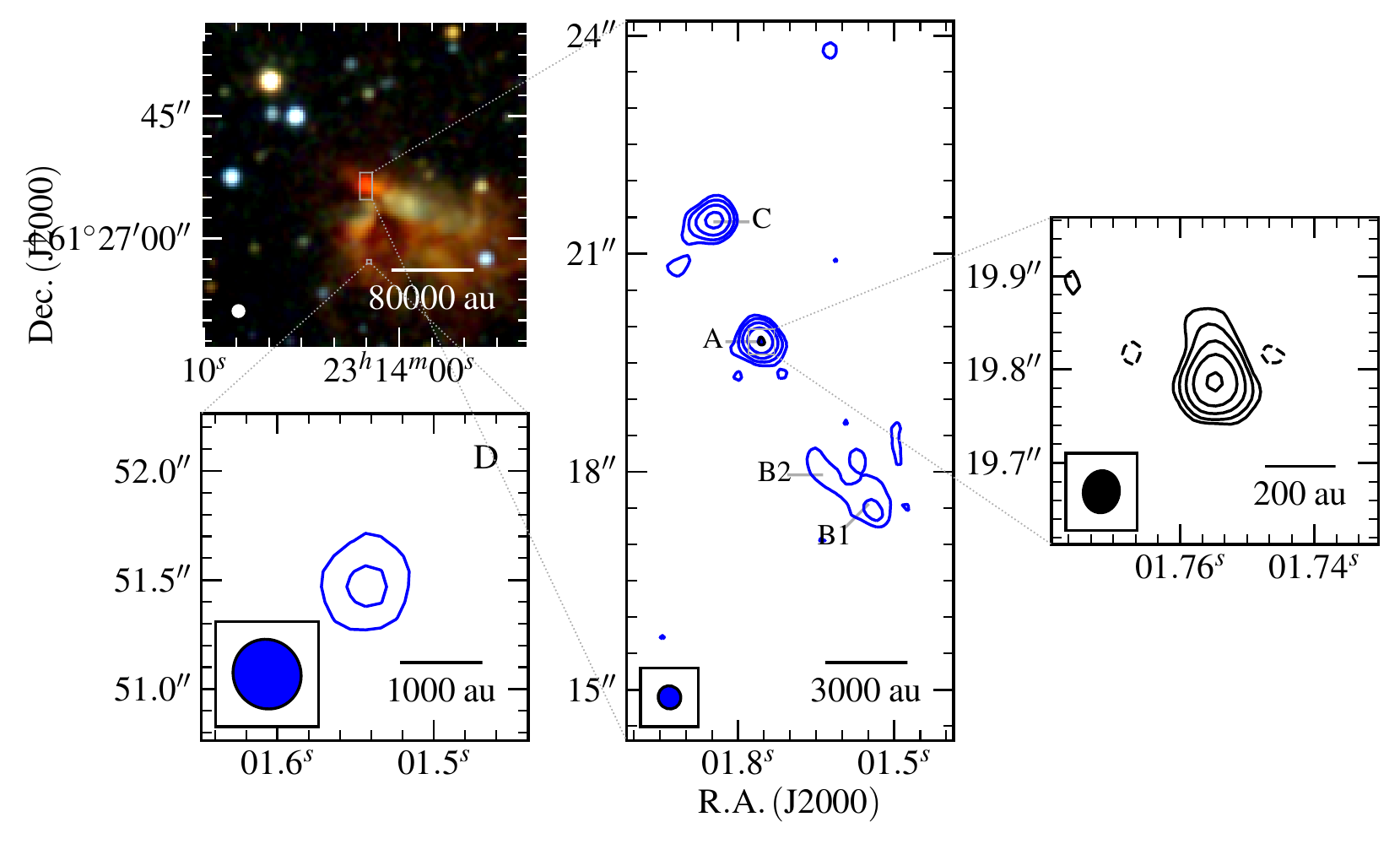}
\caption[Near-infrared and VLA radio images of the MYSO G111.5671+00.7517]{\textbf{G111.5671+00.7517} - Near-infrared (R, G, B colour-scale, left panel; 2MASS, $\mathrm{K,H,J}$ bands) and radio contour maps of G111.5671+00.7517 at C-band (middle and bottom left panels; blue contours) and Q-band (right panel; black contours). Restoring beams were $0.324\arcsec\times0.308\arcsec$ at $31\degr$ and $0.045\arcsec\times0.038\arcsec$ at $-9\degr$ while contour levels are $(-3, 3, 6, 12, 24, 48) \times \sigma$ and $(-3, 3, 6, 12, 23, 46) \times \sigma$ for C and Q-band respectively. All other values have the usual meaning.}
\label{cplot:G111.5671}
\end{figure*}

\begin{figure*}
\includegraphics[width=\textwidth]{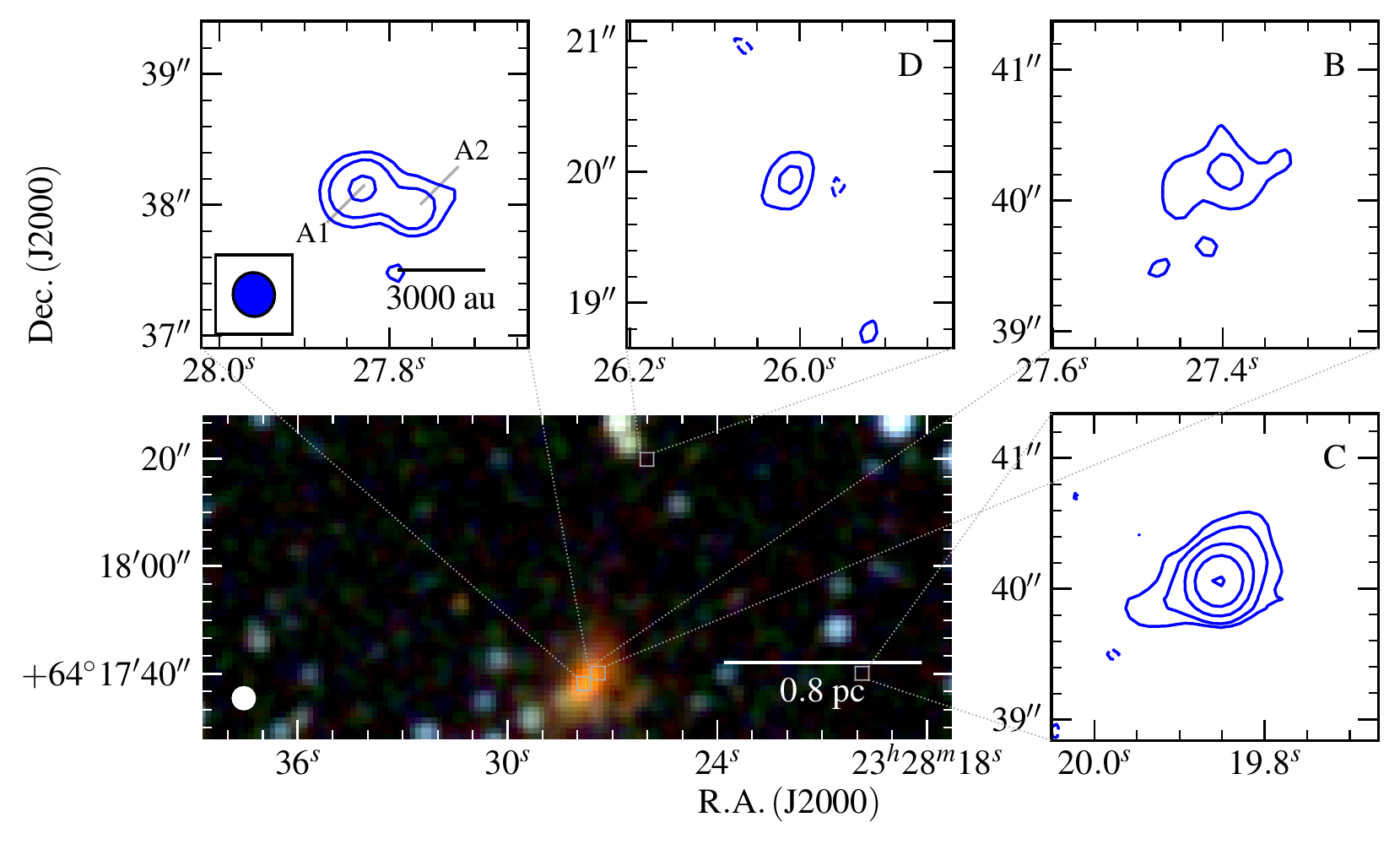}
\caption[Near-infrared and VLA radio images of the MYSO G114.0835+02.8568]{\textbf{G114.0835+02.8568} - Near-infrared colour-scale (R, G, B colour-scale, left panel; 2MASS, $\mathrm{K,H,J}$ bands) and radio contour maps of G114.0835+02.8568 at C-band (blue contours). The restoring beam and contour levels used were $0.337\arcsec\times0.317\arcsec$ at $13\degr$ and $(-3, 3, 6, 14, 30, 64) \times \sigma$. All other values have the usual meaning.}
\label{cplot:G114.0835}
\end{figure*}

\begin{figure*}
\includegraphics[width=5in]{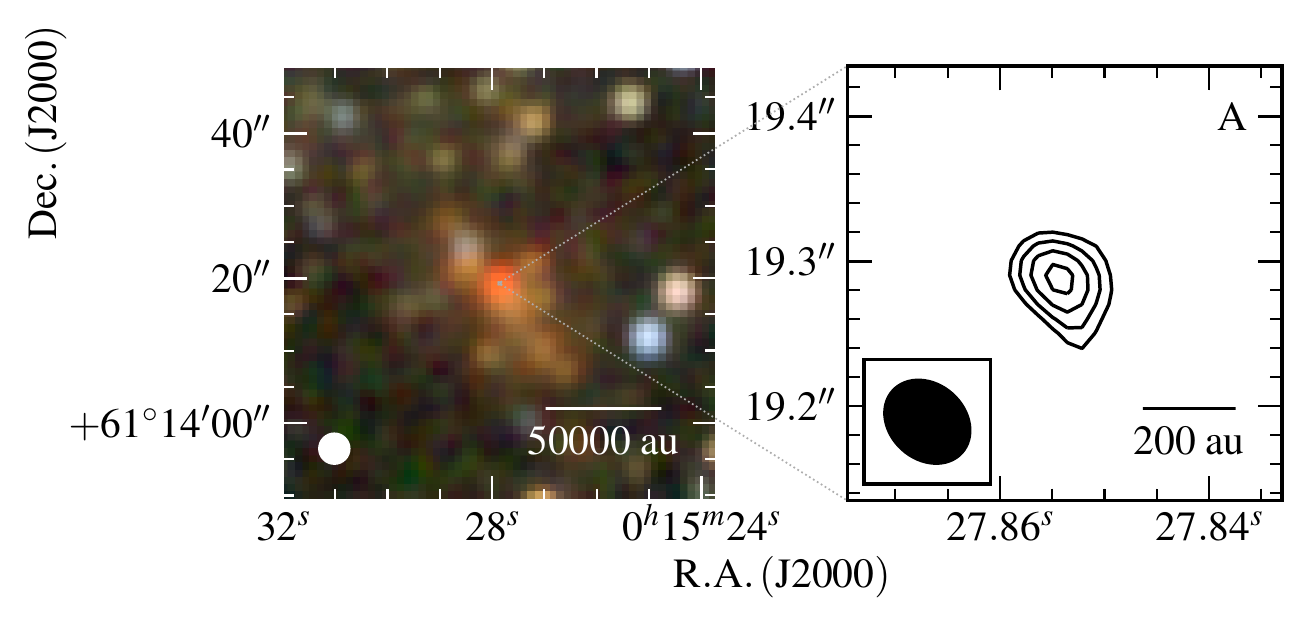}
\caption[Near-infrared and VLA radio images of the MYSO G118.6172$-$01.3312]{\textbf{G118.6172$-$01.3312} - Near-infrared colour-scale (R, G, B colour-scale, left panel; 2MASS, $\mathrm{K,H,J}$ bands) and radio contour map of G118.6172$-$01.3312 at Q-band (right panel). The restoring beam and contour levels used were $0.064\arcsec\times0.051\arcsec$ at $48\degr$ and $(-3, 3, 4, 5, 6) \times \sigma$ respectively. All other values have the usual meaning.}
\label{cplot:G118.6172}
\end{figure*}

\begin{figure*}
\includegraphics[width=\textwidth]{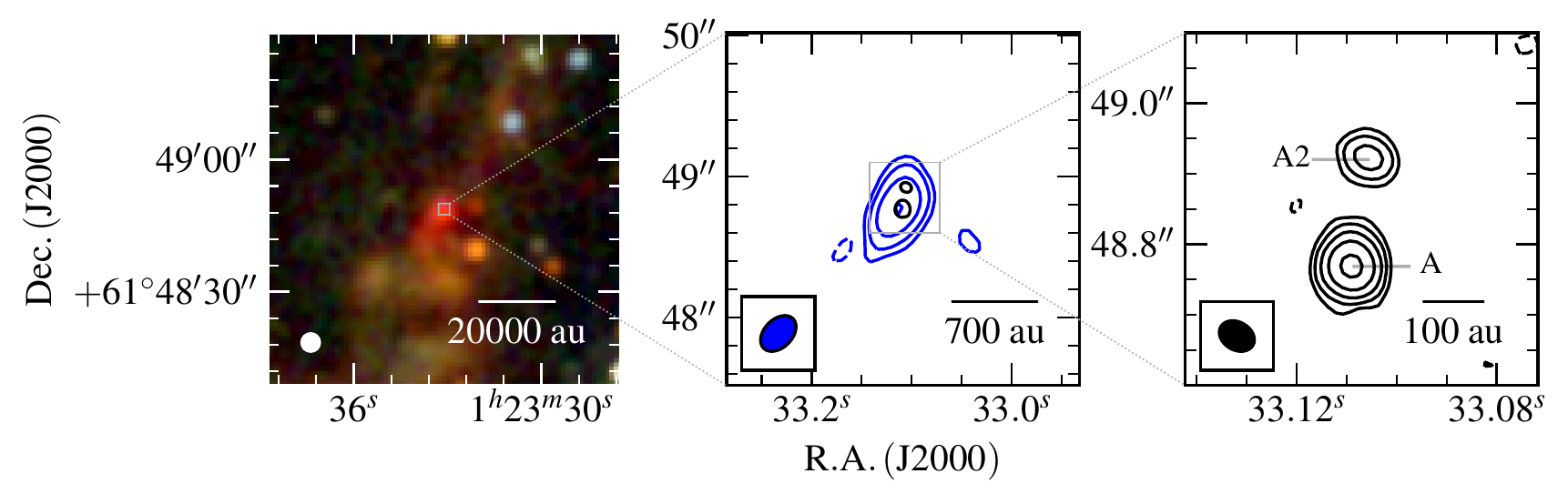}
\caption[Near-infrared and VLA radio images of the MYSO G126.7144$-$00.8220]{\textbf{G126.7144$-$00.8220} - Near-infrared (left panel; 2MASS, $\mathrm{K,H,J}$ bands R, G, B colour-scale) and radio maps of the G126.7144$-$00.8220 field at C-band (blue contours; middle) and Q-band (black contours; right). Restoring beams were $0.295\arcsec\times0.200\arcsec$ at $-44\degr$ and $0.053\arcsec\times0.040\arcsec$ at $60\degr$, while contour levels are $(-3, 3, 7, 18, 43) \times \sigma$ and $(-3, 3, 7, 16, 37, 86) \times \sigma$ for C (robustness of -1) and Q-band respectively. All other symbols/values have the usual meaning.}
\label{cplot:G126.7144}
\end{figure*}

\begin{figure*}
\includegraphics[width=\textwidth]{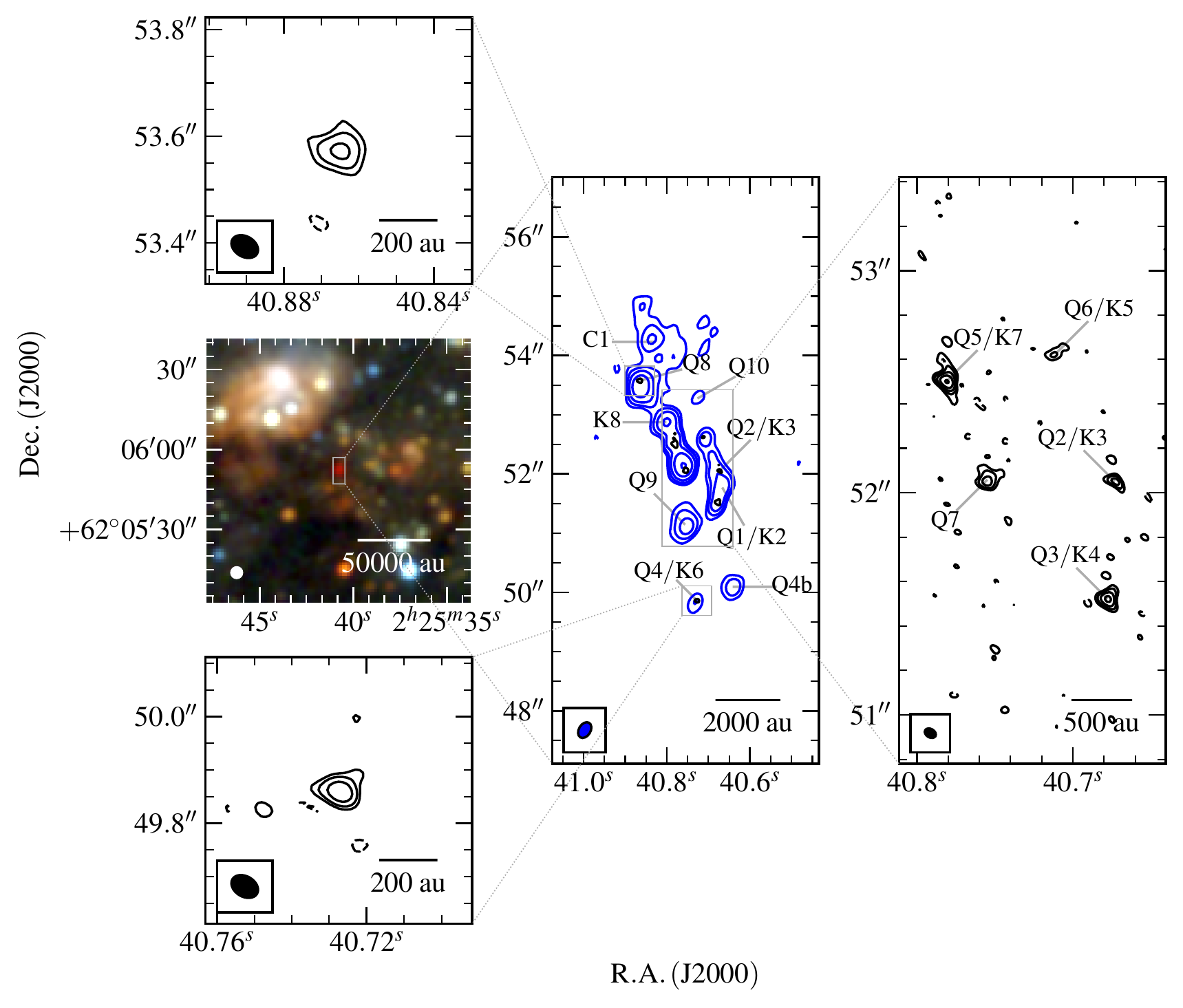}
\caption[Near-infrared and VLA radio images of the MYSO G133.7150+01.2155]{\textbf{G133.7150+01.2155} - Near-infrared (R, G, B colour-scale, middle left panel; 2MASS, $\mathrm{K,H,J}$ bands) and radio contour maps of G133.7150+01.2155 at C-band (middle, blue contours) and Q-band (top left, bottom right and right panels; black contours). Restoring beams were $0.287\arcsec\times0.197\arcsec$ at $-31\degr$ and $0.054\arcsec\times0.038\arcsec$ at $59\degr$ with contour levels set to $(-4, 4, 9, 19, 42, 93) \times \sigma$ and $(-3, 3, 6, 12, 24, 48) \times \sigma$ for C (robust$=-1$) and Q-bands respectively. All other values have the usual meaning.}
\label{cplot:G133.7150}
\end{figure*}

\begin{figure*}
\includegraphics[width=\textwidth]{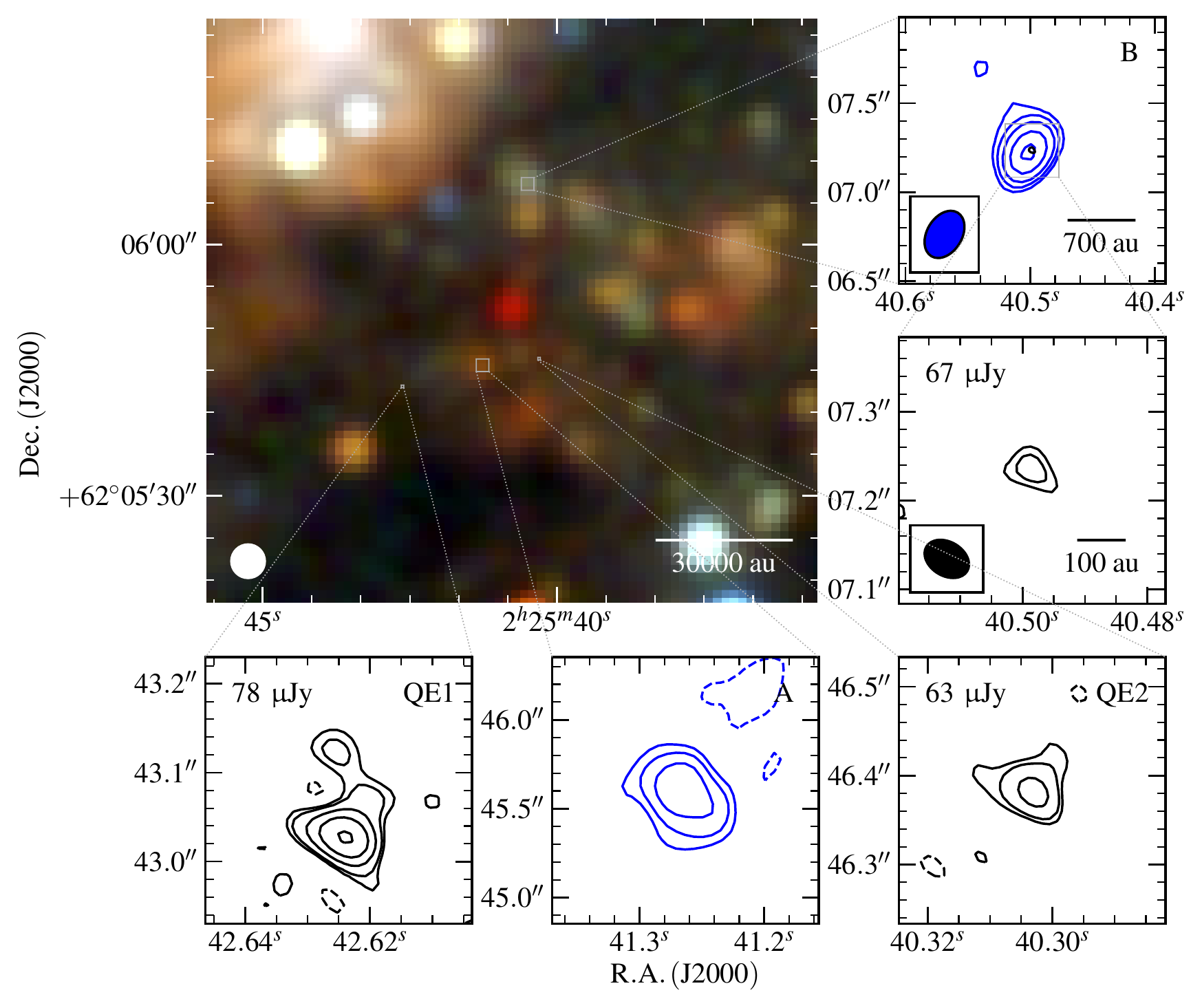}
\caption[Near-infrared and VLA radio images of the MYSO G133.7150+01.2155]{\textbf{G133.7150+01.2155 (Other sources)} - Near-infrared (R, G, B colour-scale, middle left panel; 2MASS, $\mathrm{K,H,J}$ bands) and radio contour maps of the other detected sources in the G133.7150+01.2155 field. Restoring beams and contour colours are the same as in Figure \ref{cplot:G133.7150}, with contour levels set to $(-3, 3, 5, 8, 14, 22) \times \sigma$ and $(-3, 3, 5, 12, 24, 47) \times \sigma$ for C (robust$=-1$) and Q-bands respectively.}
\label{cplot:G133.7150extrasources}
\end{figure*}

\begin{figure*}
\centering
\includegraphics[width=0.72\textwidth]{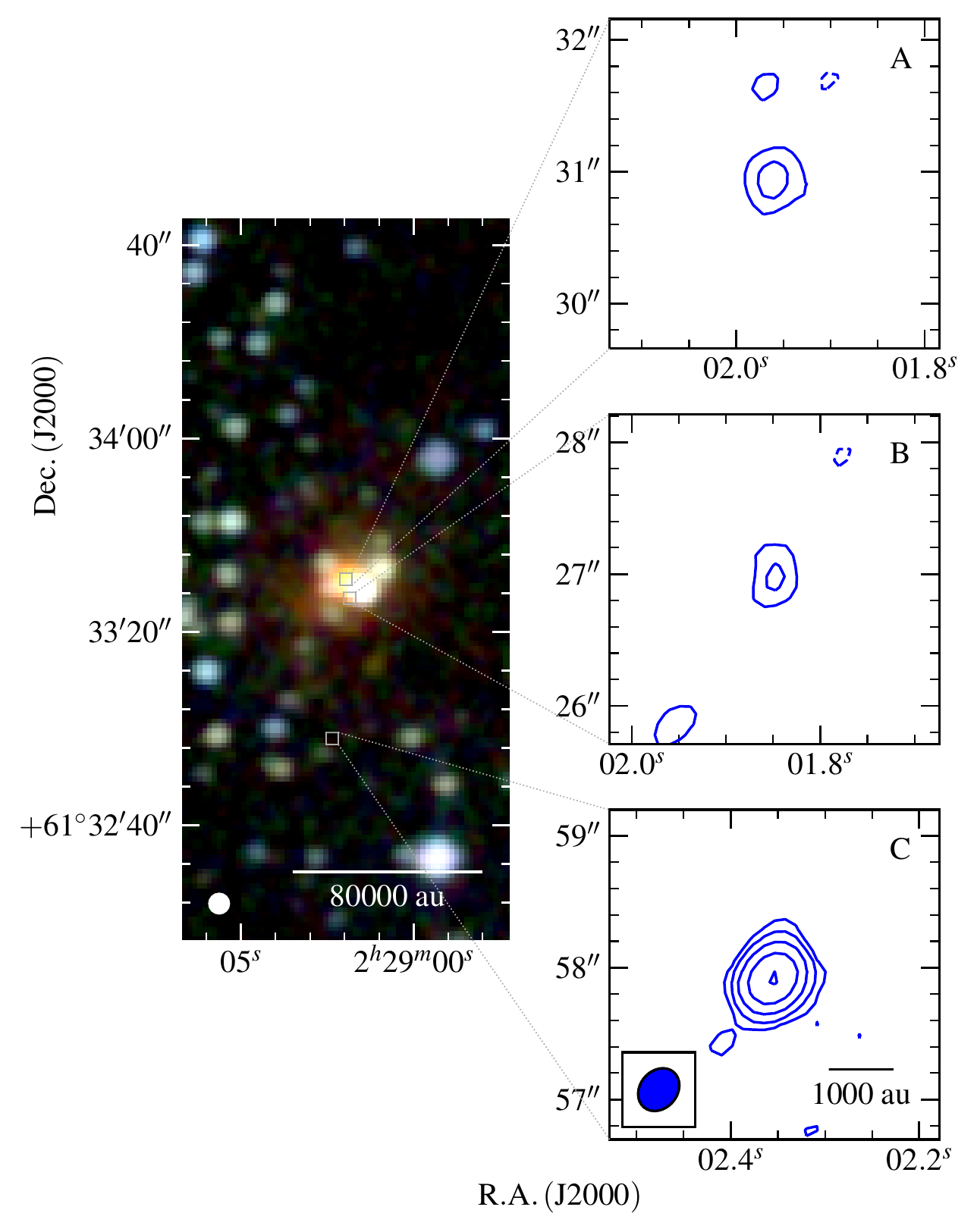}
\caption[Near-infrared and VLA radio images of the MYSO G134.2792+00.8561]{\textbf{G134.2792+00.8561} - Near-infrared (R, G, B colour-scale, left panel; 2MASS, $\mathrm{K,H,J}$ bands) and radio contour maps of G134.2792+00.8561 at C-band (right panels, blue contours). The restoring beam used was $0.350\arcsec\times0.284\arcsec$ at $-40\degr$, while contour levels are $(-3, 3, 7, 17, 41, 97) \times \sigma$. All other values have the usual meaning.}
\label{cplot:G134.2792}
\end{figure*} 
\clearpage
\begin{figure*}
\includegraphics[width=\textwidth]{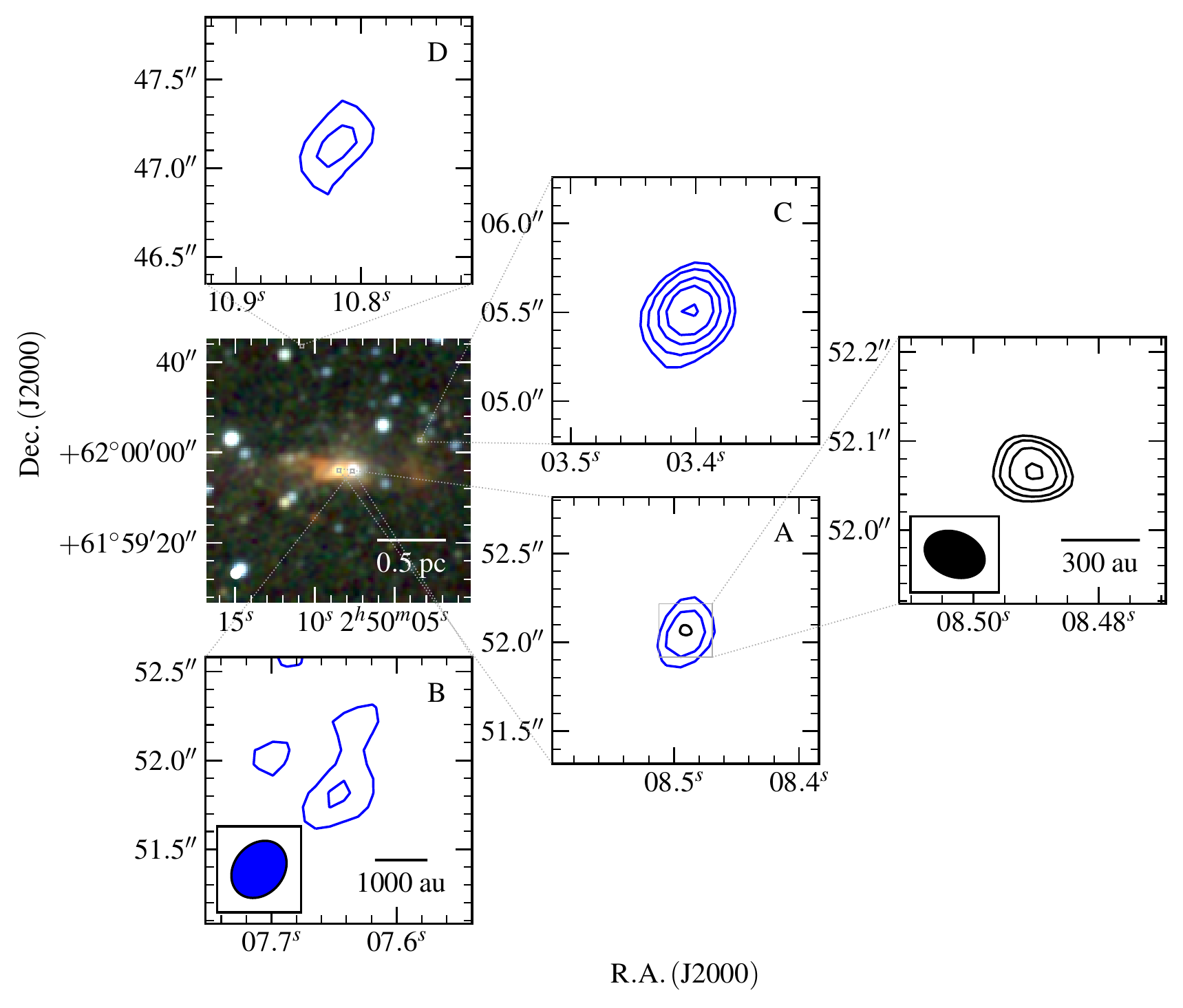}
\caption[Near-infrared and VLA radio images of the MYSO G136.3833+02.2666]{\textbf{G136.3833+02.2666} - Near-infrared (R, G, B colour-scale, left panel; 2MASS, $\mathrm{K,H,J}$ bands) and radio contour maps of G136.3833+02.2666 at C-band (top left, bottom left and middle panels, blue contours) and Q-band (right panel, black contours). Restoring beams were $0.349\arcsec\times0.279\arcsec$ at $-40\degr$ and $0.070\arcsec\times0.050\arcsec$ at $69\degr$ utilising a robustness of 2 for the Q-band data. Contour levels are $(-3, 3, 5, 9, 13, 20) \times \sigma$ and $(-3, 3, 4, 6, 9) \times \sigma$ for C and Q-band respectively. Varying noises across the C-band primary beam are represented. All other values have the usual meaning.}
\label{cplot:G136.3833}
\end{figure*}

\begin{figure*}
\includegraphics[width=\textwidth]{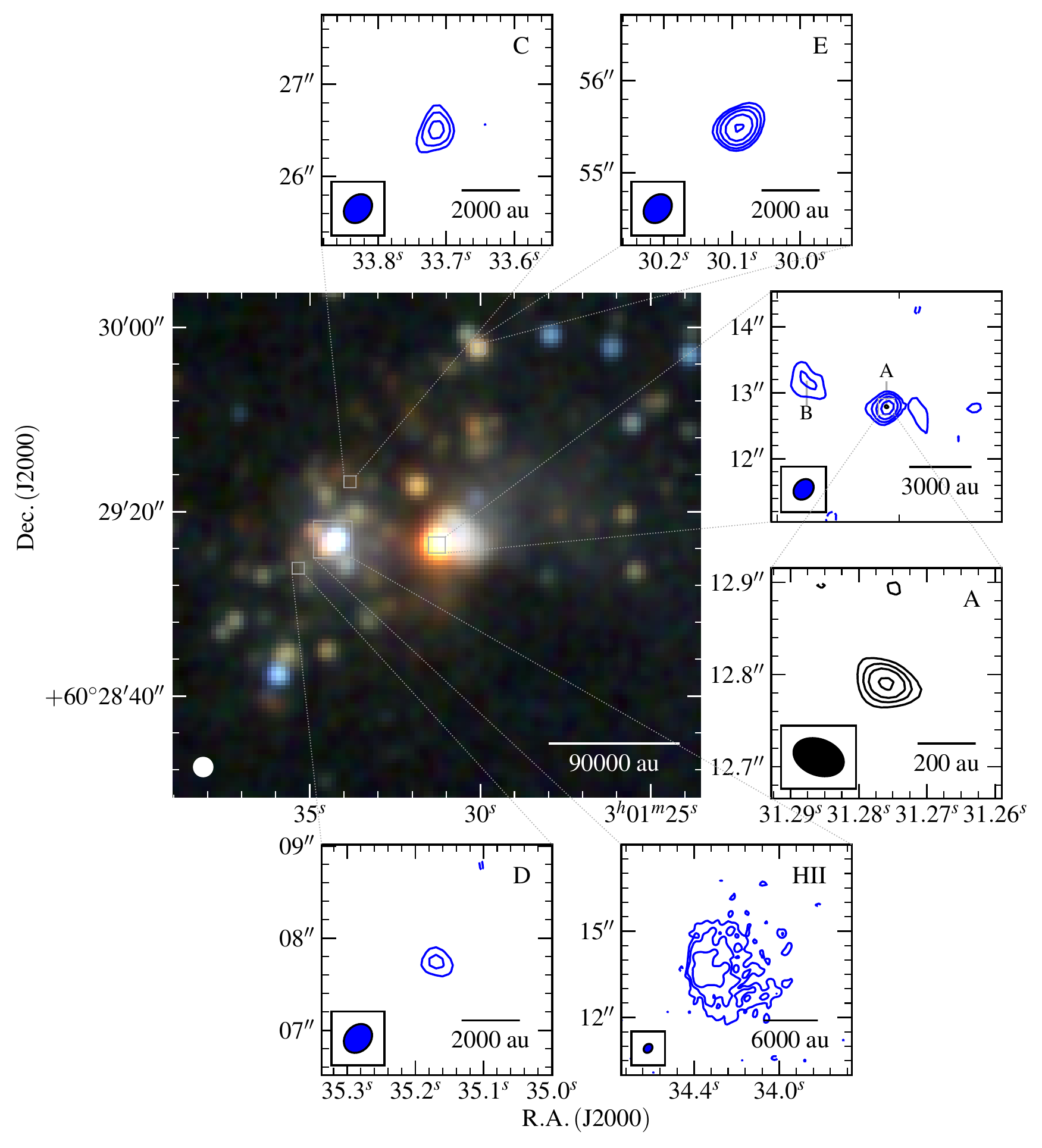}
\caption[Near-infrared and VLA radio images of the MYSO G138.2957+01.5552]{\textbf{G138.2957+01.5552} - Near-infrared (R, G, B colour-scale, central panel; 2MASS, $\mathrm{K,H,J}$ bands) and radio contour maps of G138.2957+01.5552 at C-band (top, bottom and top-right panels, blue contours) and Q-band (bottom-right panel, black contours). Restoring beams were $0.342\arcsec\times0.273\arcsec$ at $-42\degr$ and $0.056\arcsec\times0.039\arcsec$ at $69\degr$. Contour levels are $(-3, 3, 5, 8, 13, 22) \times \sigma$ and $(-3, 3, 5, 7, 10) \times \sigma$ for C and Q-band respectively. All other values have the usual meaning.}
\label{cplot:G138.2957}
\end{figure*}

\begin{figure*}
\includegraphics[width=\textwidth]{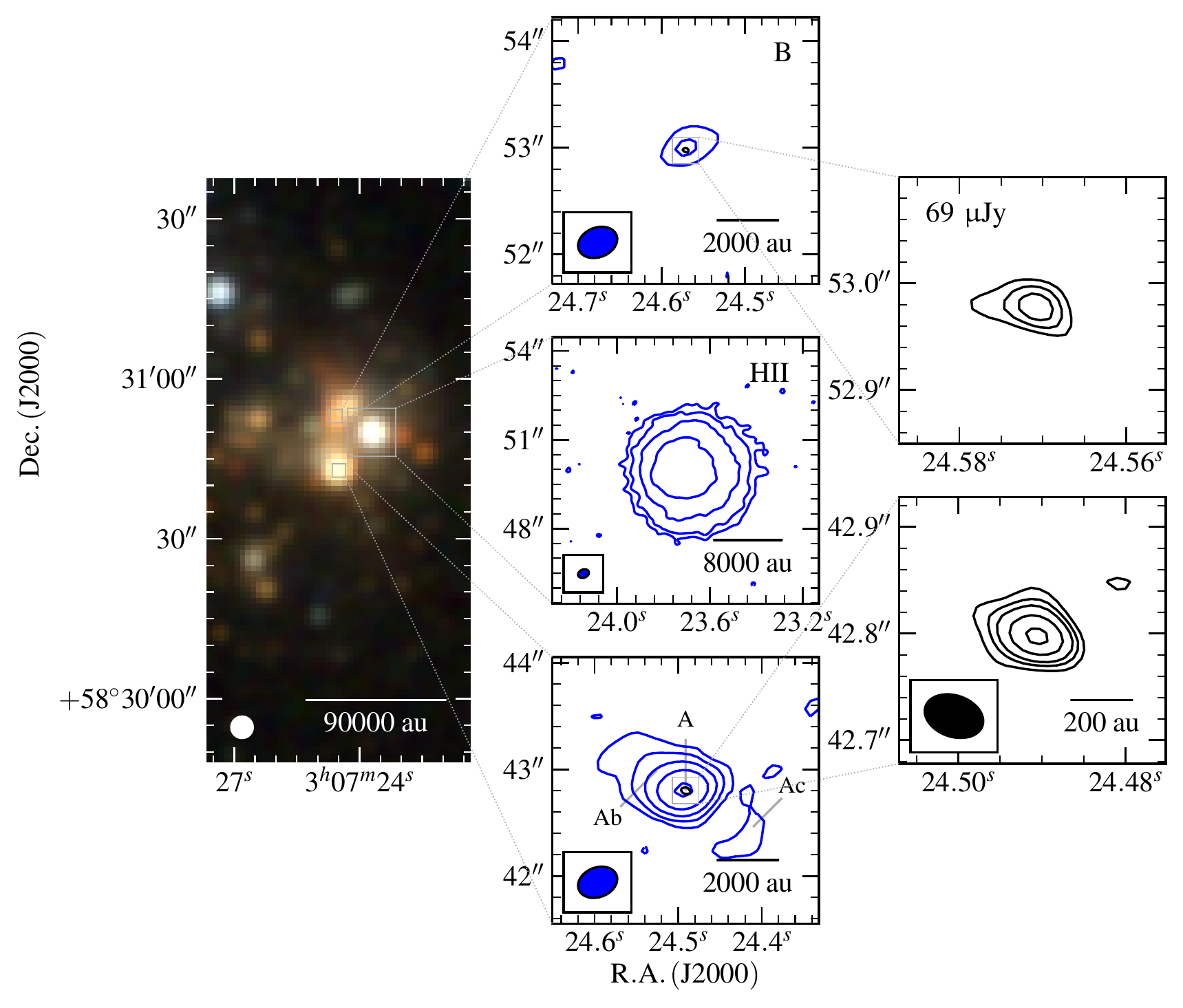}
\caption[Near-infrared and VLA radio images of the MYSO G139.9091+00.1969A]{\textbf{G139.9091+00.1969A} - Near-infrared (R, G, B colour-scale, left panel; 2MASS, $\mathrm{K,H,J}$ bands) and radio contour maps of G139.9091+00.1969A at C-band (centre-column panels, blue contours) and Q-band (right panels, black contours). Restoring beams were $0.379\arcsec\times0.278\arcsec$ at $-69\degr$ and $0.056\arcsec\times0.039\arcsec$ at $71\degr$. Contour levels are $(-3, 3, 7, 15, 32, 71) \times \sigma$ and $(-3, 3, 5, 7, 11, 16) \times \sigma$ for C and Q-band respectively. All other values have the usual meaning.}
\label{cplot:G139.9091}
\end{figure*}

\begin{figure*}
\includegraphics[width=\textwidth]{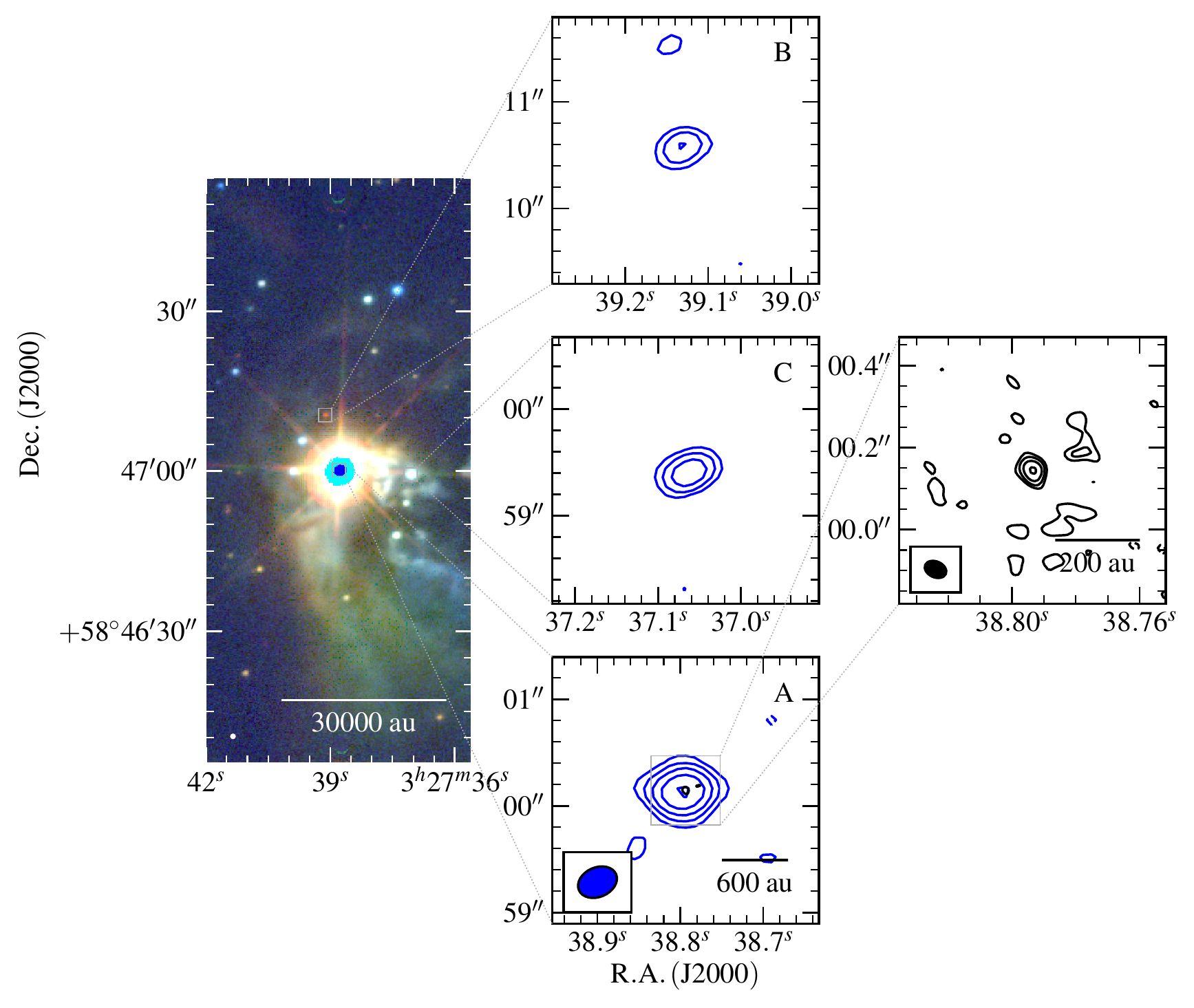}
\caption[Near-infrared and VLA radio images of the MYSO G141.9996+01.8202]{\textbf{G141.9996+01.8202} - Near-infrared (R, G, B colour-scale, left panel; UKIDSS, $\mathrm{K,H,J}$ bands) and radio contour maps of G141.9996+01.8202 at C-band (centre-column panels, blue contours) and Q-band (right panel, black contours). Restoring beams were $0.379\arcsec\times0.277\arcsec$ at $-66\degr$ and $0.055\arcsec\times0.039\arcsec$ at $66\degr$. Contour levels are $(-3, 3, 6, 11, 20, 37) \times \sigma$ and $(-3, 3, 5, 7, 12) \times \sigma$ for C and Q-band respectively. All other values have the usual meaning.}
\label{cplot:G141.9996}
\end{figure*}

\begin{figure*}
\includegraphics[width=\textwidth]{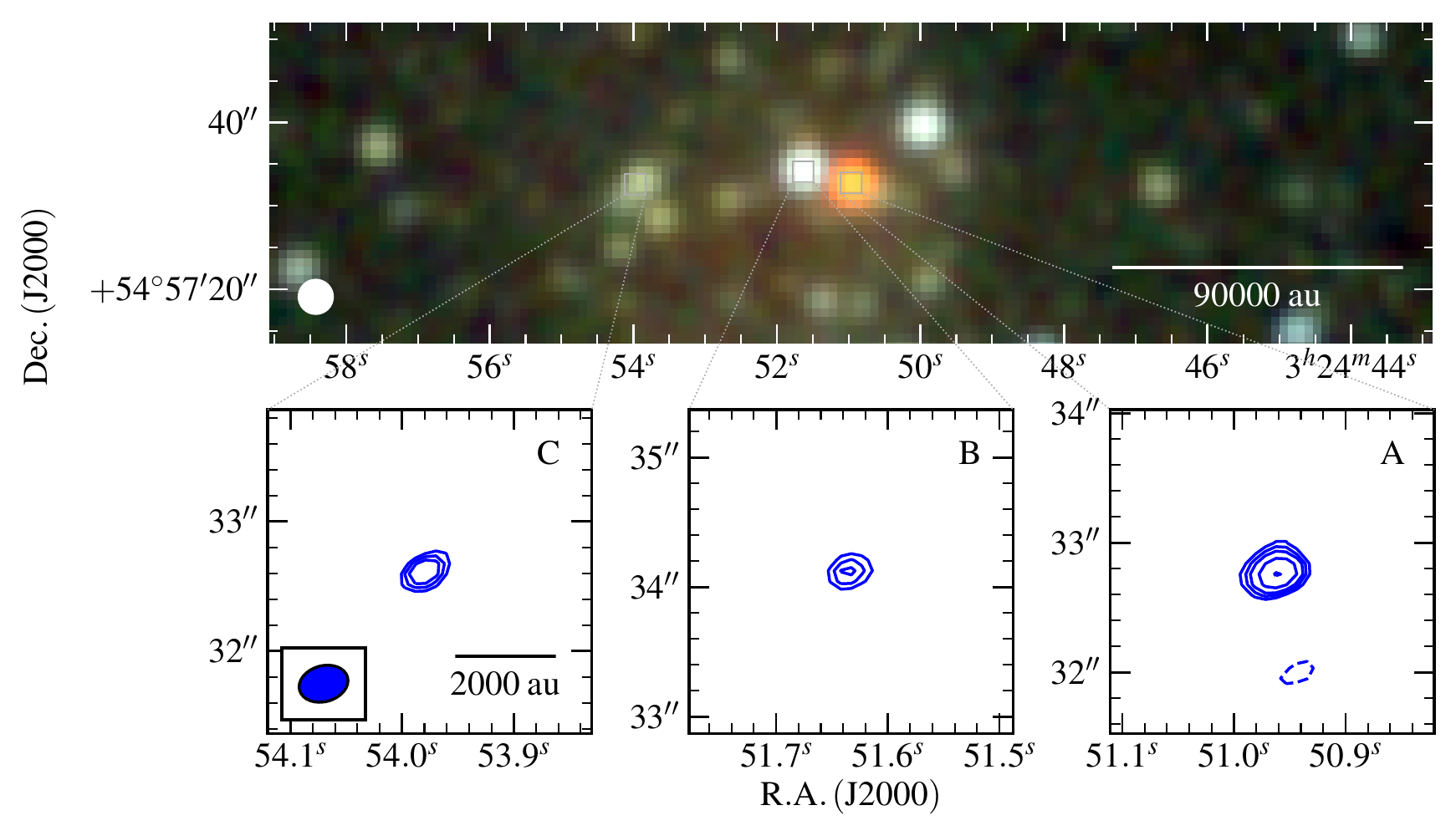}
\caption[Near-infrared and VLA radio images of the MYSO G143.8118$-$01.5699]{ \textbf{G143.8118$-$01.5699} - Near-infrared (R, G, B colour-scale, top panel; 2MASS, $\mathrm{K,H,J}$ bands) and C-band contour maps of G143.8118$-$01.5699 (bottom panels; blue contours). The restoring beams used was $0.384\arcsec\times0.277\arcsec$ at $-75\degr$ and contour levels are $(-3, 3, 4, 5, 7, 10) \times \sigma$. All other values have the usual meaning.}
\label{cplot:G143.8118}
\end{figure*}

\begin{figure*}
\includegraphics[width=\textwidth]{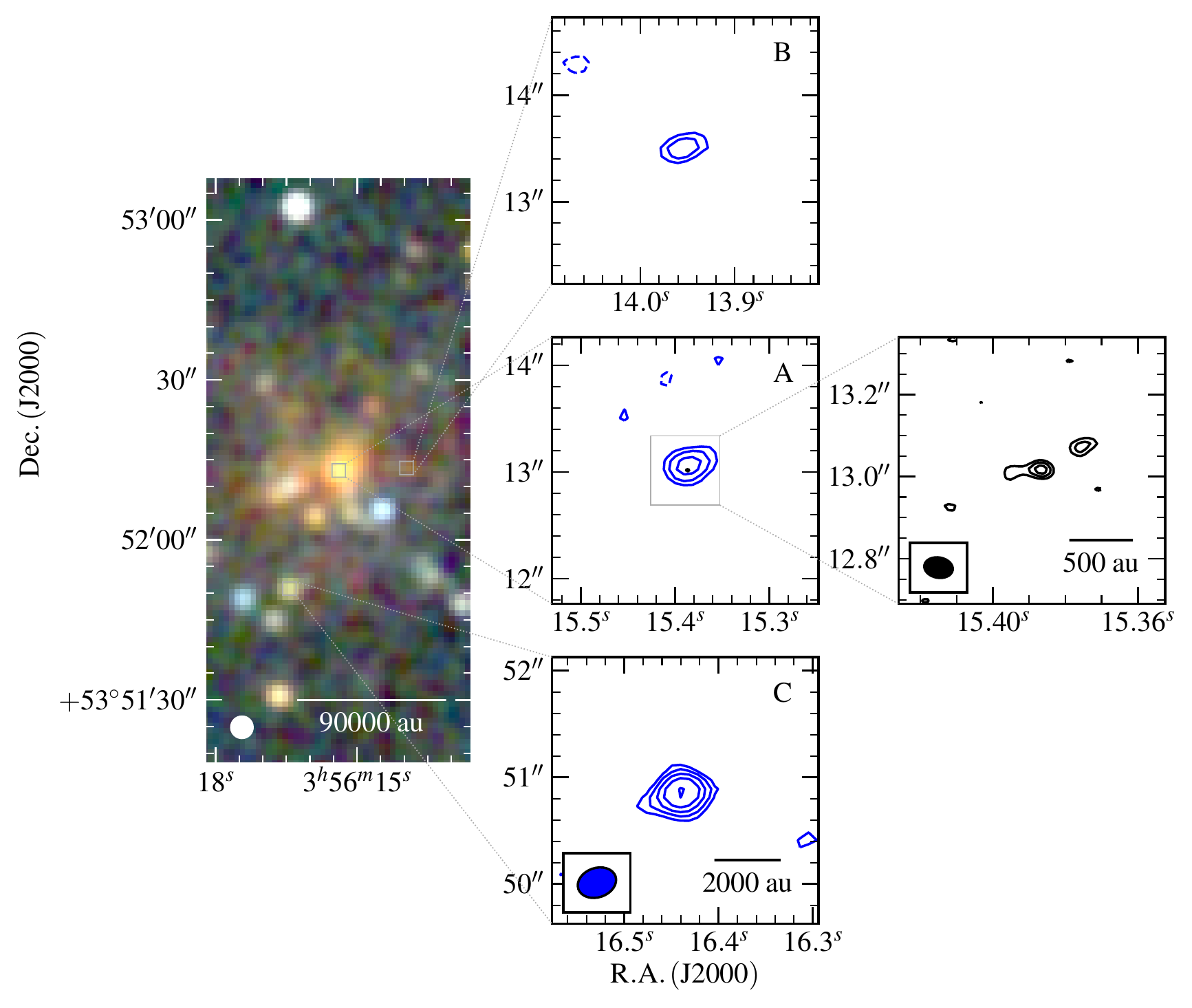}
\caption[Near-infrared and VLA radio images of the MYSO G148.1201+00.2928]{\textbf{G148.1201+00.2928} - Near-infrared (R, G, B colour-scale, left panel; 2MASS, $\mathrm{K,H,J}$ bands) and radio contour maps of G148.1201+00.2928 at C-band (centre-column panels, blue contours) and Q-band (right panel, black contours). Restoring beams were $0.368\arcsec\times0.275\arcsec$ at $-71\degr$ and $0.070\arcsec\times0.050\arcsec$ at $80\degr$. Contour levels are $(-3, 3, 5, 8, 12, 20) \times \sigma$ and $(-3, 3, 4, 5) \times \sigma$ for C and Q-band (robustness of 2) respectively. All other values have the usual meaning.}
\label{cplot:G148.1201}
\end{figure*}

\begin{figure*}
\centering
\includegraphics[width=0.72\textwidth]{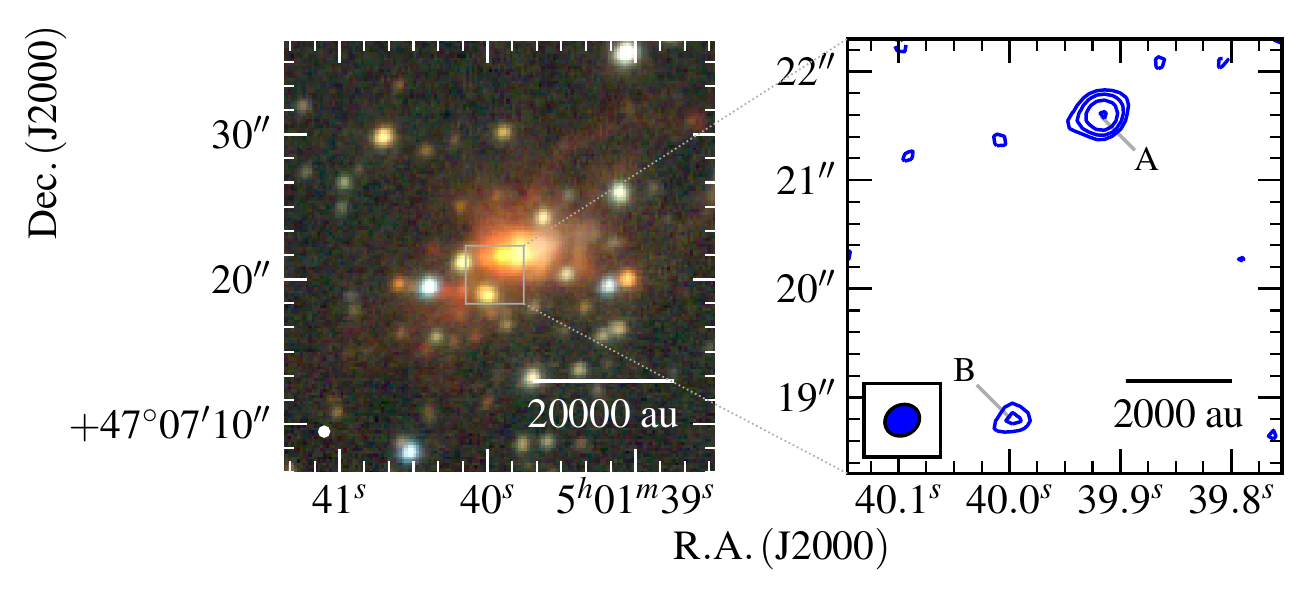}
\caption[Near-infrared and VLA radio images of the MYSO G160.1452+03.1559]{\textbf{G160.1452+03.1559} - Near-infrared (R, G, B colour-scale, left panel; UKIDSS, $\mathrm{K,H,J}$ bands) and radio contour maps of G160.1452+03.1559 at C-band (right panel, blue contours). The restoring beam for the C-band data was $0.328\arcsec\times0.280\arcsec$ at $-62\degr$, while contour levels are set at $(-3, 3, 5, 8, 13) \times \sigma$. All other values have the usual meaning.}
\label{cplot:G160.1452}
\end{figure*}

\begin{figure*}
\includegraphics[width=\textwidth]{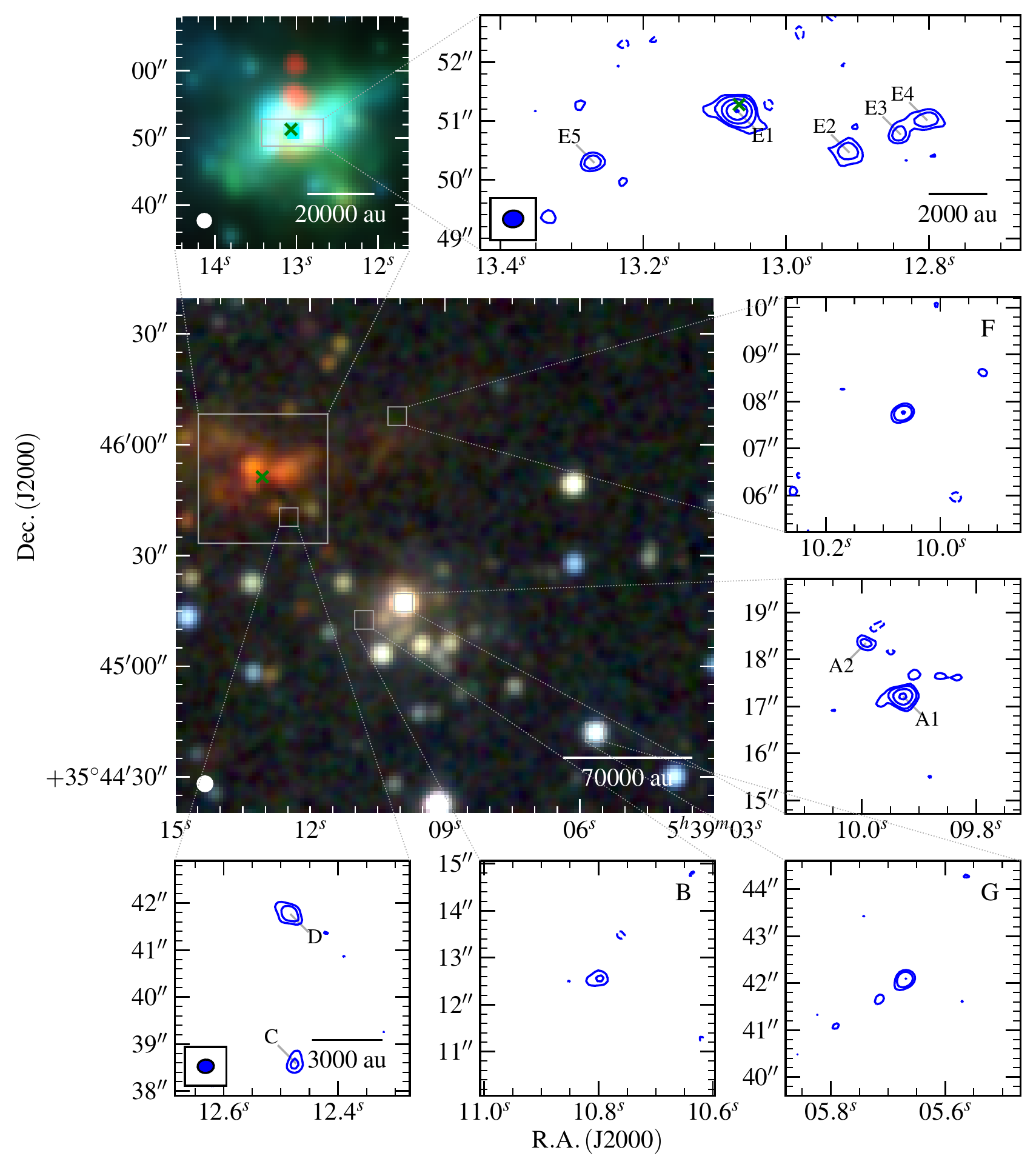}
\caption[Near-infrared, mid-infrared and VLA radio images of the MYSO G173.4839+02.4317]{\textbf{G173.4839+02.4317} - Near-infrared (R, G, B colour-scale, central, large panel; 2MASS, $\mathrm{K,H,J}$ bands), mid-infrared (R, G, B colour-scale, top left panel; GLIMPSE $8.0,4.5,3.6\micron$ bands) and radio contour maps of G173.4839+02.4317 at C-band (other panels, blue contours). The restoring beam for the C-band data was $0.347\arcsec\times0.286\arcsec$ at $-84\degr$, while contour levels are set at $(-3, 3, 5, 8, 14, 24) \times \sigma$. All other values have the usual meaning. Green crosses show 6.7$\GHz$ methanol maser positions from our data.}
\label{cplot:G173.4839}
\end{figure*}

\begin{figure*}
\includegraphics[width=\textwidth]{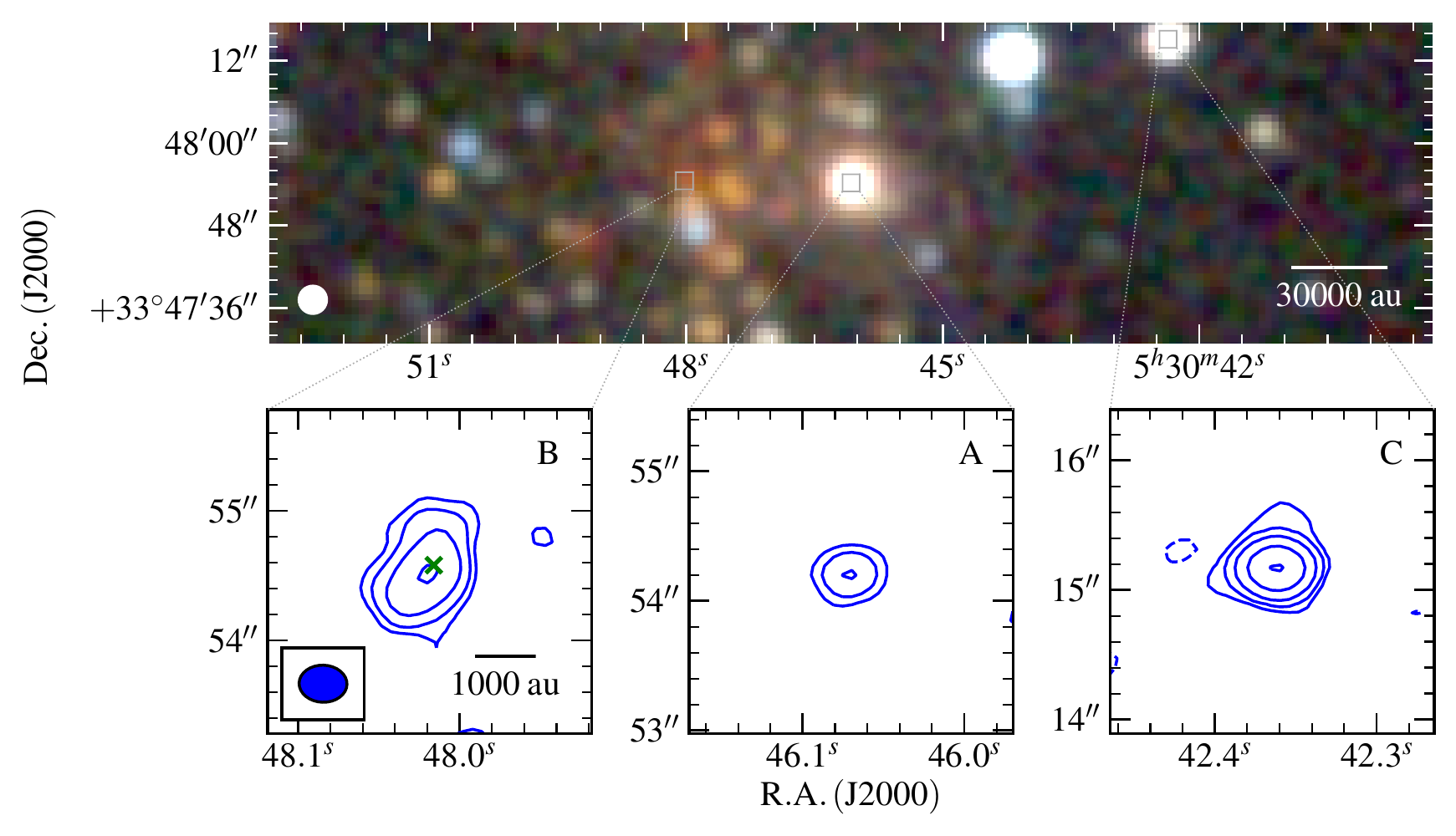}
\caption[Near-infrared and VLA radio images of the MYSO G174.1974$-$00.0763]{\textbf{G174.1974$-$00.0763} - Near-infrared (R, G, B colour-scale, central, large panel; 2MASS, $\mathrm{K,H,J}$ bands) and radio contour maps of G174.1974$-$00.0763 at C-band (other panels, blue contours). The restoring beam for the C-band data was $0.369\arcsec\times0.286\arcsec$ at $88\degr$, while contour levels are set at $(-3, 3, 7, 17, 46, 114) \times \sigma$. All other values have the usual meaning. Green crosses show 6.7$\GHz$ methanol maser positions from our data.}
\label{cplot:G174.1974}
\end{figure*}

\begin{figure*}
\centering
\includegraphics[width=0.72\textwidth]{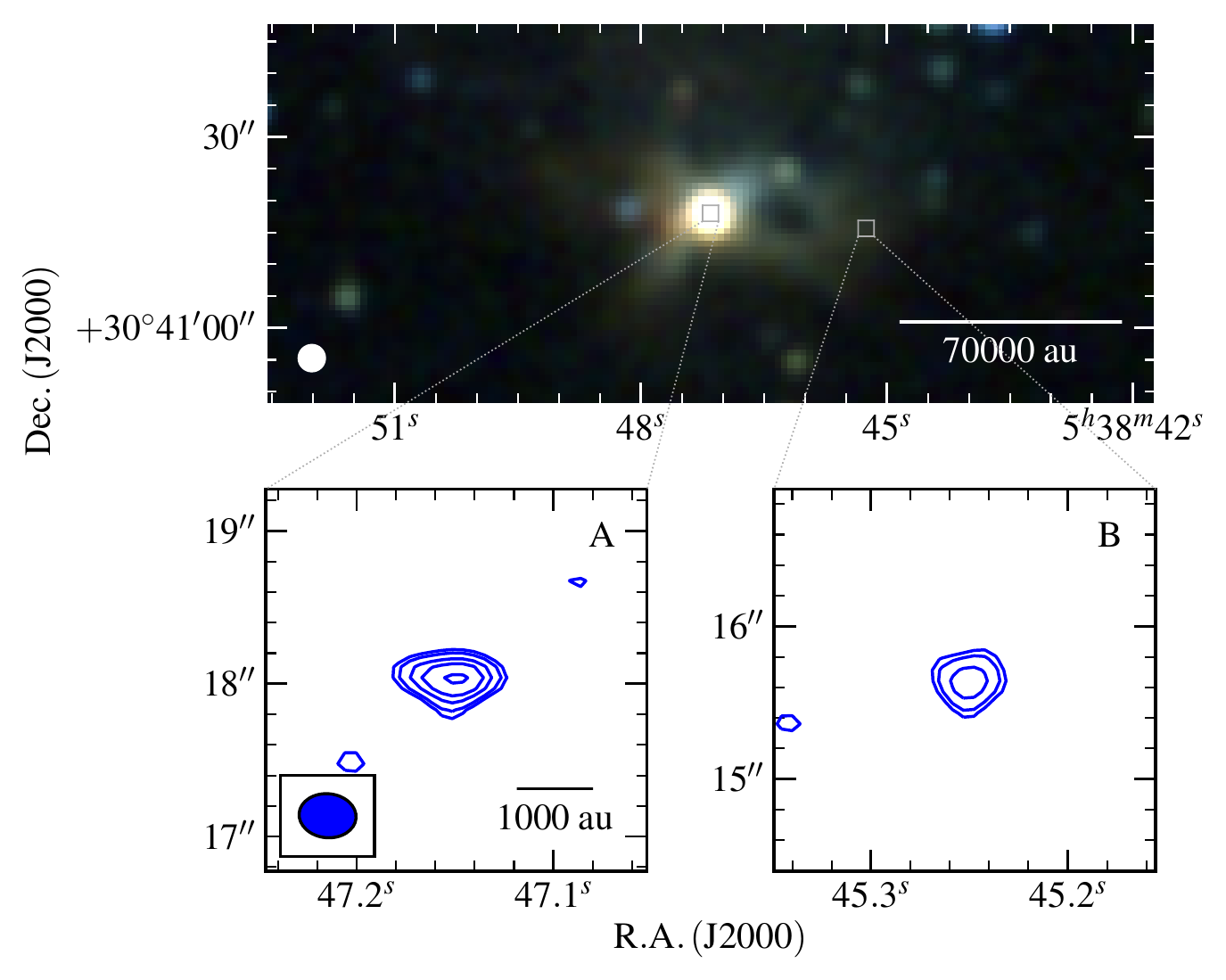}
\caption[Near-infrared and VLA radio images of the MYSO G177.7291$-$00.3358]{\textbf{G177.7291$-$00.3358} - Near-infrared (R, G, B colour-scale, central, large panel; 2MASS, $\mathrm{K,H,J}$ bands) and C-band (blue contours; bottom panels) images of G177.7291$-$00.3358. The C-band restoring beam was $0.374\arcsec\times0.288\arcsec$ at $84\degr$. Contour levels are $(-3, 3, 4, 6, 8, 11) \times \sigma$ and all other values have their usual meaning.}
\label{cplot:G177.7291}
\end{figure*}

\begin{figure*}
\centering
\includegraphics[width=0.72\textwidth]{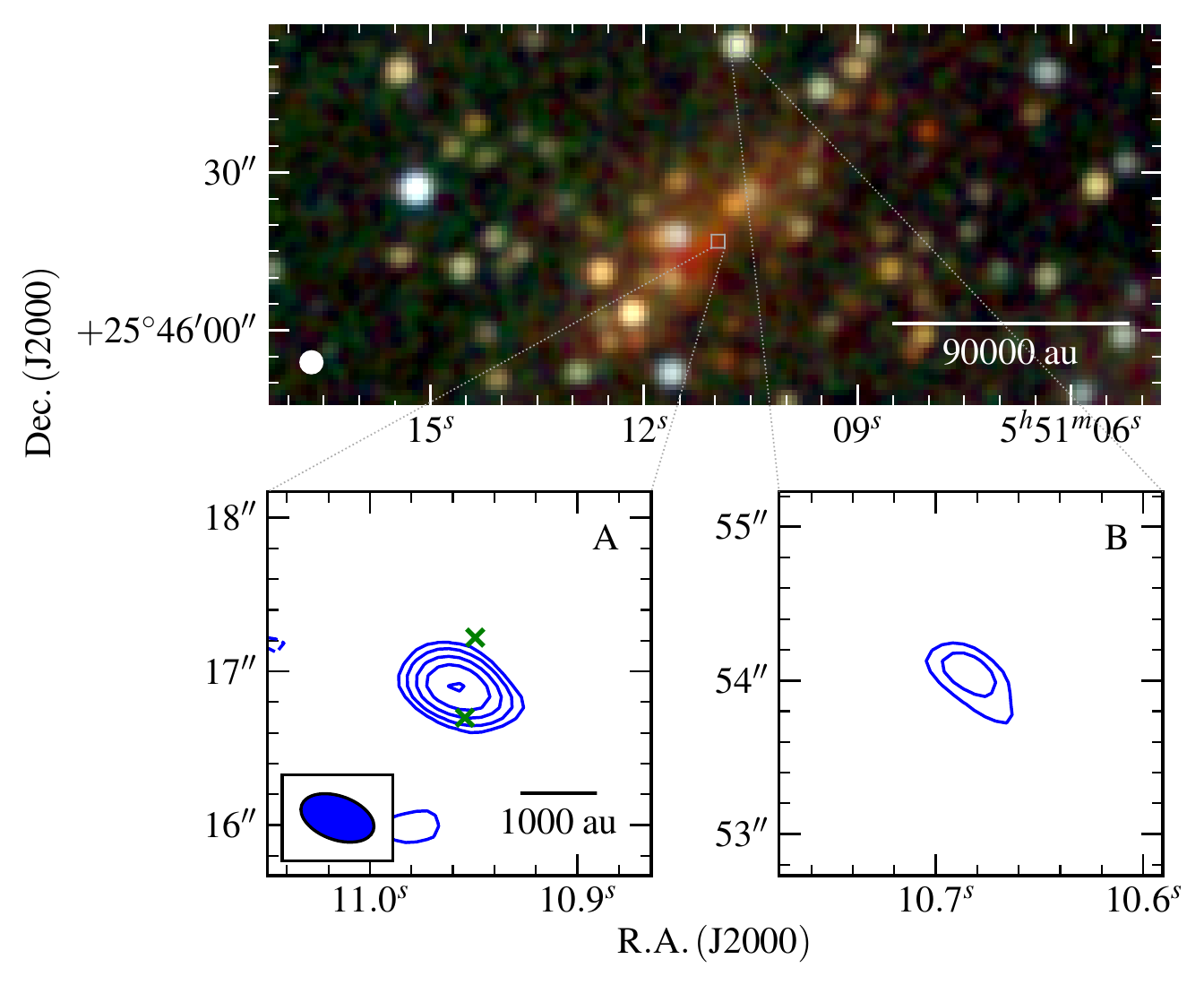}
\caption[Near-infrared and VLA radio images of the MYSO G183.3485$-$00.5751]{\textbf{G183.3485$-$00.5751} - Near-infrared (R, G, B colour-scale, central, large panel; 2MASS, $\mathrm{K,H,J}$ bands) and C-band (blue contours; bottom panels) images of G183.3485$-$00.5751. The C-band restoring beam was $0.495\arcsec\times0.285\arcsec$ at $70\degr$. Contour levels are $(-3, 3, 5, 8, 12, 19) \times \sigma$ and all other values have their usual meaning. Green crosses show 6.7$\GHz$ methanol maser positions from our data.}
\label{cplot:G183.3485}
\end{figure*}

\begin{figure*}
\includegraphics[width=\textwidth]{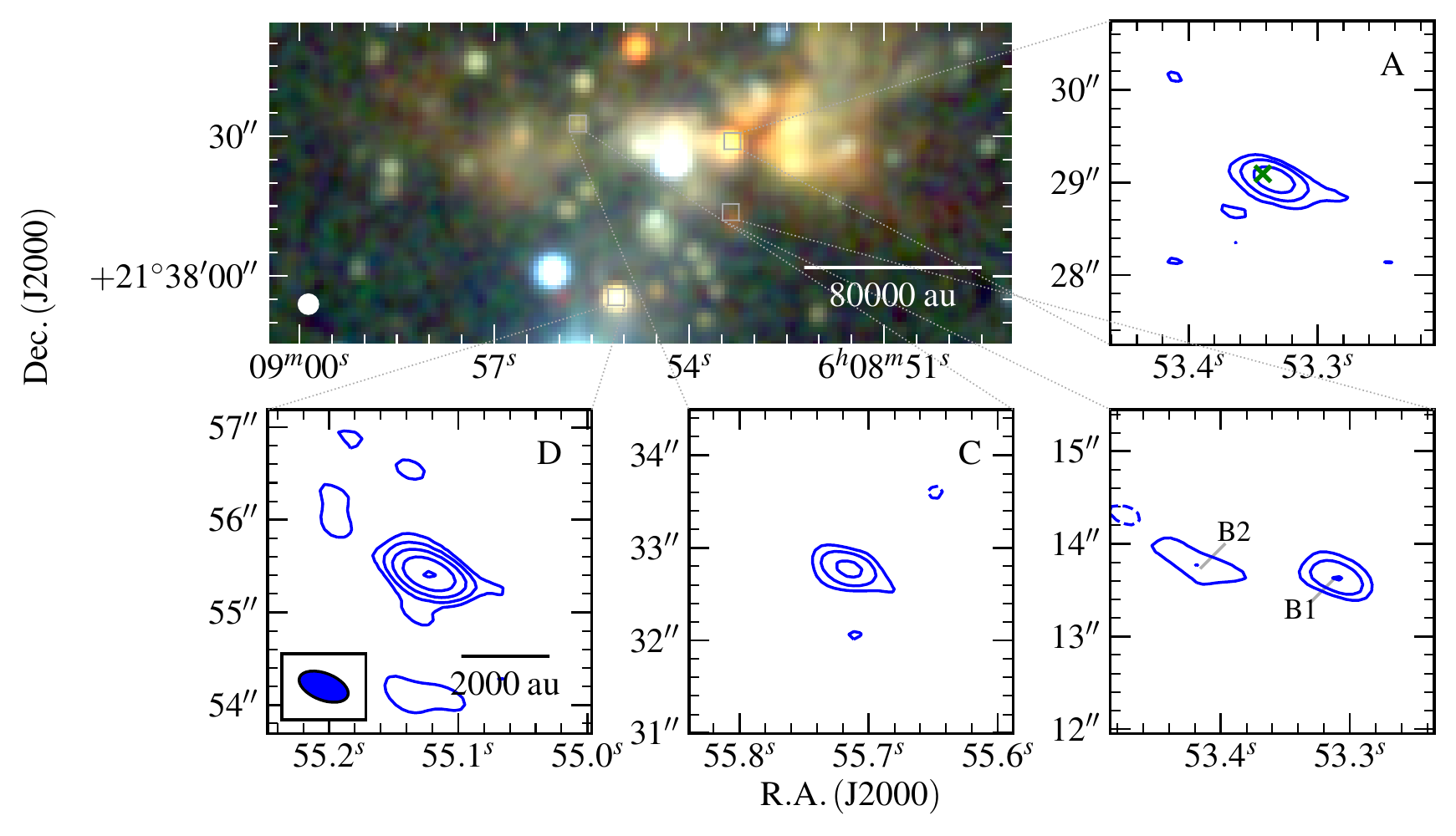}
\caption[Near-infrared and VLA radio images of the MYSO G188.9479+00.8871]{\textbf{G188.9479+00.8871} - Near-infrared (R, G, B colour-scale, top left panel; 2MASS, $\mathrm{K,H,J}$ bands) and C-band (blue contours; bottom and top right panels) images of G188.9479+00.8871. The C-band restoring beam was $0.558\arcsec\times0.290\arcsec$ at $69\degr$. Contour levels are $(-3, 3, 6, 12, 24, 48) \times \sigma$ and all other values have their usual meaning. Green crosses show 6.7$\GHz$ methanol maser positions from our data.}
\label{cplot:G188.9479}
\end{figure*}

\begin{figure*}
\includegraphics[width=\textwidth]{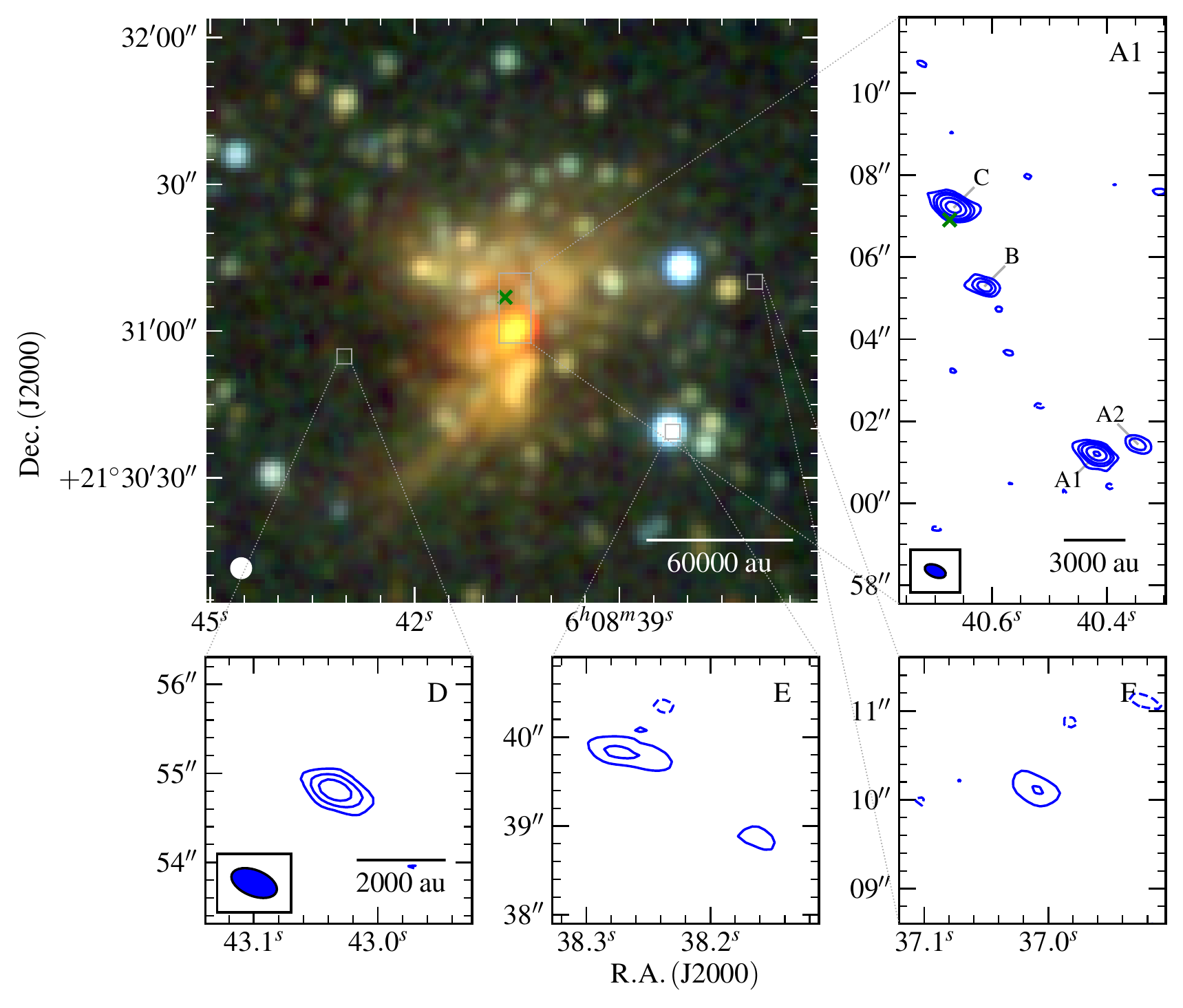}
\caption[Near-infrared and VLA radio images of the MYSO G189.0307+00.7821]{\textbf{G189.0307+00.7821} - Near-infrared (R, G, B colour-scale, top left panel; 2MASS, $\mathrm{K,H,J}$ bands) and C-band (blue contours; bottom and top right panels) images of G189.0307+00.7821. The C-band restoring beam was $0.539\arcsec\times0.291\arcsec$ at $68\degr$. Contour levels are $(-3, 3, 6, 10, 19, 34) \times \sigma$ and all other values have their usual meaning. Green crosses show 6.7$\GHz$ methanol maser positions from our data.}
\label{cplot:G189.0307}
\end{figure*}

\begin{figure*}
\includegraphics[width=\textwidth]{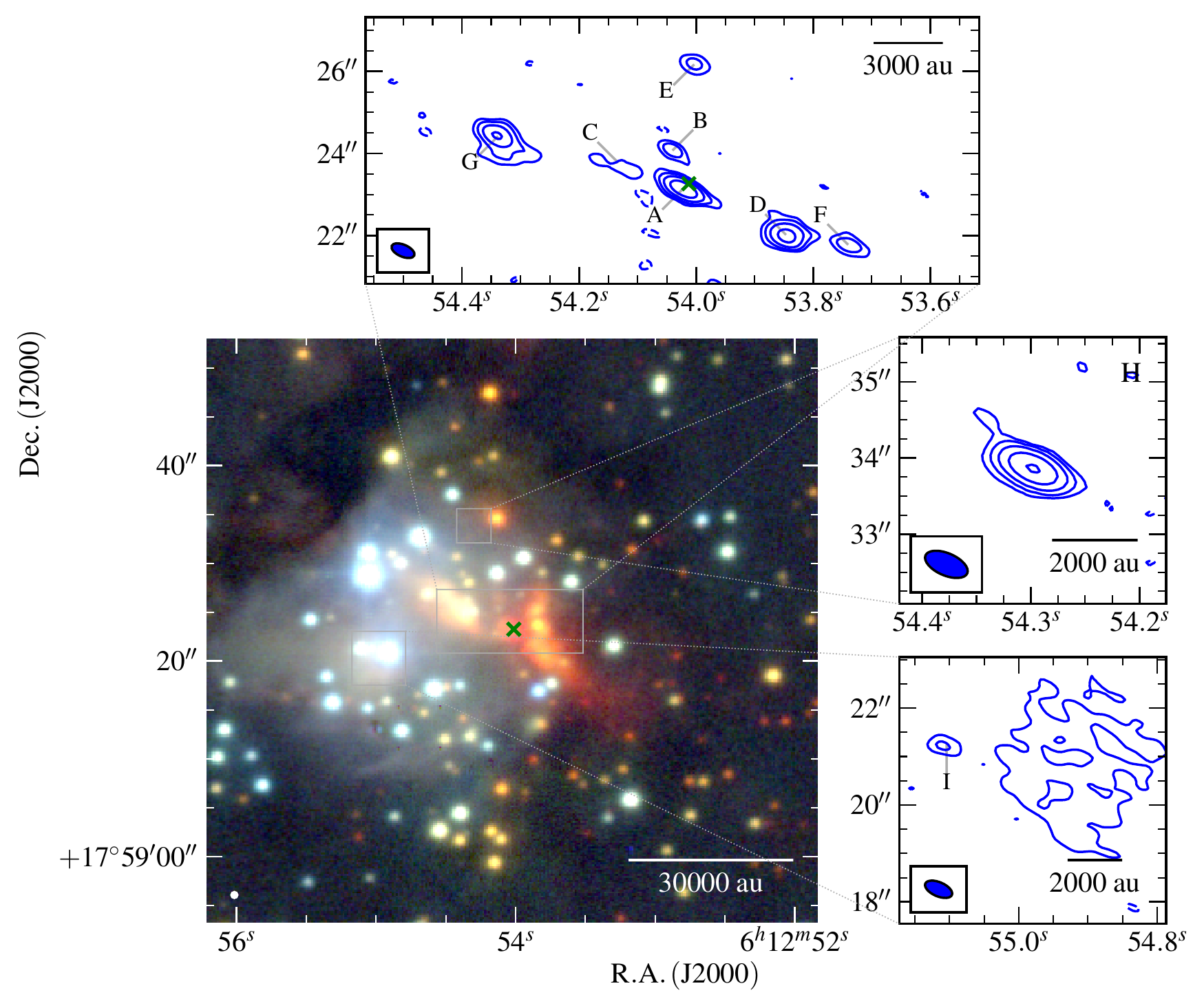}
\caption[$\,\,\,$Near-infrared and VLA radio images of the MYSO G192.6005$-$00.0479]{\textbf{G192.6005$-$00.0479} - Near-infrared (R, G, B colour-scale, bottom left panel; UKIDSS, $\mathrm{K,H,J}$ bands) and C-band (blue contours; other panels) images of G192.6005$-$00.0479. The C-band restoring beam was $0.597\arcsec\times0.292\arcsec$ at $67\degr$. Contour levels are $(-3, 3, 7, 16, 38, 87) \times \sigma$ and all other values have their usual meaning.}
\label{cplot:G192.6005}
\end{figure*}

\begin{figure*}
\includegraphics[width=\textwidth]{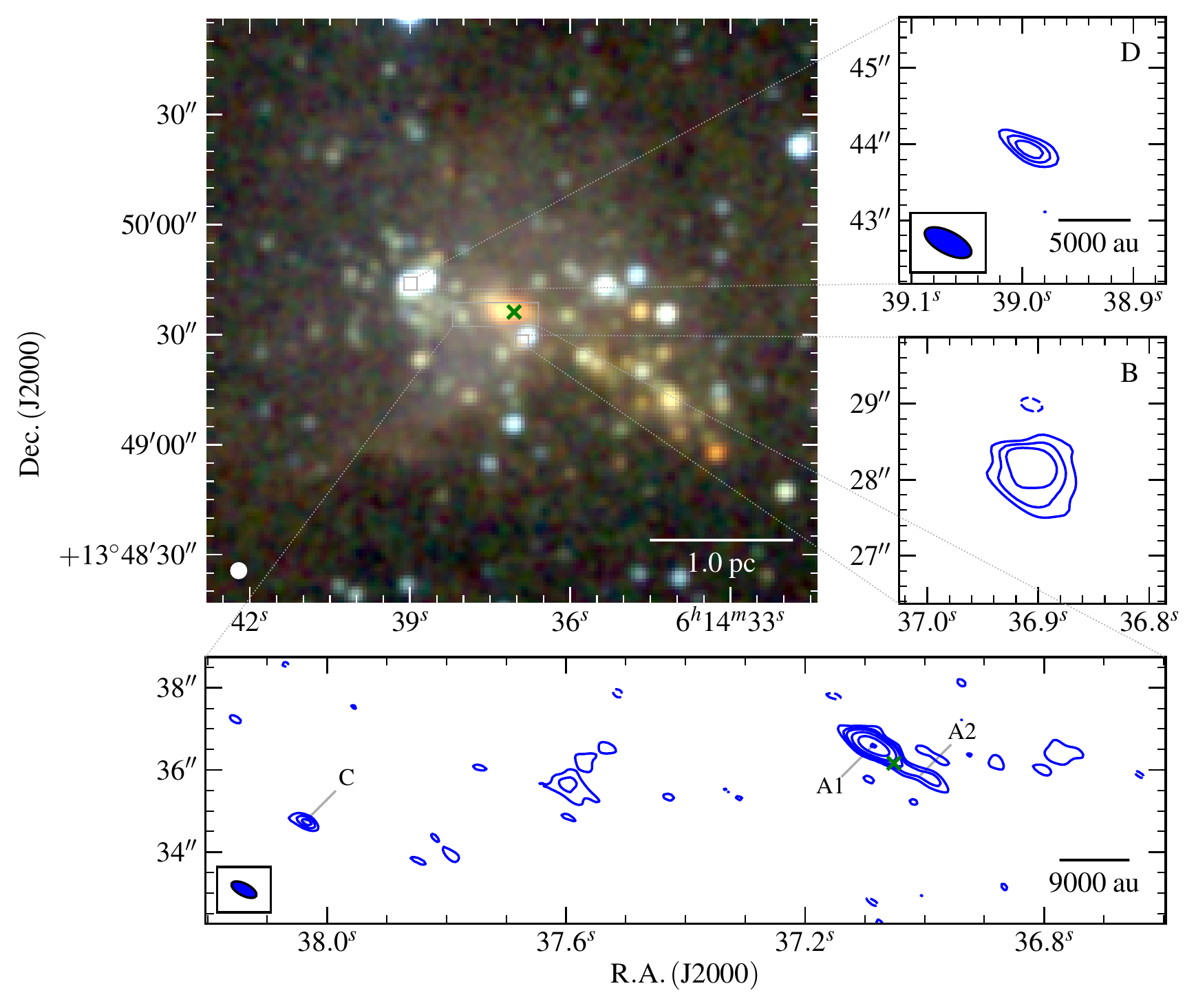}
\caption[$\,\,\,$Near-infrared and VLA radio images of the MYSO G196.4542$-$01.6777]{\textbf{G196.4542$-$01.6777} - Near-infrared (R, G, B colour-scale, top left panel; UKIDSS, $\mathrm{K,H,J}$ bands) and C-band (blue contours; other panels) images of G196.4542$-$01.6777. The C-band restoring beam was $0.674\arcsec\times0.293\arcsec$ at $63\degr$. Contour levels are $(-3, 3, 5, 7, 13, 27, 55) \times \sigma$ and all other values have their usual meaning. Green crosses show 6.7$\GHz$ methanol maser positions from our data.}
\label{cplot:G196.4542}
\end{figure*}

%% file: ObjectNotes.tex
Throughout this section we discuss each field with detected emission in both/either C-band or Q-band imagery. Subsequently the detected radio emission is then compared to previous observations in the literature across all wavelengths, to both decide upon the classification of each individual radio source and also to discern the overall star formation picture towards each core/clump.
\subsection{IRDC Sample}
\label{sec:irdcnotes}

\subsubsection{G018.82$-$00.28}
\label{sec:G018.82}
Within the C-band primary beam, there are three, classified cores from \citet{Rathborne2010}, MM2, MM4 and MM5. Another core, MM6 lies approximately halfway between MM4 and MM5, however it remain unclassified. Only MM2 is classified as containing a red MIR source, which is identified as IRAS 18236$-$1241 and also as G018.8330$-$00.3004 from the RMS survey where it is classified as a \textsc{Hii} region. Follow up radio observations with the VLA as part of the CORNISH survey \citep{Purcell2013} detect an extended \textsc{Hii} region possessing an integrated flux of $131.4\mJy$ which is that of G018.8330$-$00.3004/IRAS 18236$-$1241. There is debate as to whether the near or far kinematic distance applies in this case. Many sources quote the far kinematic distance of $12.6\kpc$, yet MIR images show that the \textsc{Hii} region sits in front of the IRDC sitting at the near kinematic distance of $4.6\kpc$ (see left hand panel of \autoref{cplot:G018.82}). In light of the mid-infrared images, we adopt the near kinematic distance.

Approximately $3\arcsec$ east of the pointing centre, coincident with the MM2 core, the mid-infrared source and G018.8330$-$00.3004, we detect a compact source with an extended `tail' at C-band. Q-band imagery does not detect the tail (likely resolved out) but resolves the compact source into a cometary \textsc{Hii} region approximately $0.6\arcsec\times0.4\arcsec$ in size with a calculated physical size of $\sim0.013\pc\times0.010\pc$. Since the full \textsc{Hii} region is much larger in extent than the compact core (see CORNISH imagery) it is therefore classified as an ultracompact \textsc{Hii} region (UC\textsc{Hii}). Methanol maser emission is detected at the head of the cometary morphology where flux is highest. Using our C and Q-band integrated fluxes of $37\pm2\mJy$ and $51\pm5\mJy$ we calculate an optically-thin spectral index of $\alpha_{\rm CQ}=0.16\pm0.06$ for the `core'. A C-band flux of $100\pm5\mJy$ (integrated over the area enclosed by the $3\sigma$ contours) is calculated for both the `core' and `tail' with the latter thereby possessing a flux of $63\pm5\mJy$. However, comparing our C-band image and measured flux with that of the CORNISH survey, it is clear that some flux has been resolved out at C-band ($\sim31\mJy$). With $\Lbol=11883^{+4249}_{-3953}$ (using the distance to MM2 of $4.53^{+0.49}_{-0.39}\kpc$), we would expect the UC\textsc{Hii} region to have a radio luminosity (from its inferred Lyman flux of $2.6\times10^{46}\,{\rm s}^{-1}$) of  $\sim280\mJy\kpc^2$. This is below its measured value of $S_{5.8}D^2\sim 2700\mJy\kpc^2$ by an order of magnitude. However, if we assume a distance of $12.6\kpc$ and therefore $\Lbol=91938\pm29965\Lsol$, radio luminosity inferred from  the Lyman flux is $\sim45200\mJy\kpc^2$, which is only larger by a factor of two than the measured $S_{5.8}D^2\sim 20800\mJy\kpc^2$. This lends support to the \textsc{Hii} region being located at the far distance, rather than the near and since it is seen in front of the molecular core, the same can be said for the IRDC complex. Due to the strong flux and spatial scale range of the \textsc{HCHii} region, our images are dynamic range-limited and no other emission is detected.

\subsubsection{G024.08+00.04}
\label{sec:G024.08}
\citet{Rathborne2010} observed five, $1.2\mm$ cores (MM$1-5$) all of which are quiescent apart from MM1 which contains a reddened MIR source and is hosted by the brightest sub-mm clump of the complex.  \citet{Simon2006} catalogued the IRDC complex at a distance of $3.8\kpc$ using a $v_{\rm lsr}$ of $52.5\pm5.1\kmps$. Using that velocity we calculate near and far kinematic distances of $3.53^{+0.41}_{-0.45}\kpc$ and $11.71^{+0.44}_{-0.50}\kpc$, respectively. The RMS survey identifies the reddened MIR source as the \textsc{Hii} region G024.1838+00.1198, which was allocated a kinematic distance of $7.7\kpc$ by the RMS survey ($v_{\rm lsr}=113.5\kmps$) placing it behind MM1 with which it is coincident. 


At C-band we detect a cometary \textsc{Hii} region (designated A) $3\arcsec$ to the east of the pointing centre (coincident with MM1 and G024.1838+00.1198) and a bright, elongated source $45\arcsec$ to the NE comprised of at least 3 separate components (B1, B2 and B3). At Q-band we do not detect any significant flux in the primary beam. With a physical, FWHM size of $0.030\pc\times0.024\pc$ for the RMS distance of $7.7\kpc$, A is a typical UC\textsc{Hii} region. Collectively we term B1, B2 and B3 as `B' which has upper-limits to its radio spectral indices of $\alpha_{\rm CQ}<-0.4$ (from a wide-field Q-band image). However this calculated upper limit may be affected by loss of flux at Q-band considering its extended nature in C-band images. Since B is not coincident with any IR source, or sub-mm, continuum emission, and is located away from MIR extinction patches, we classify it as extragalactic in origin. No radio emission was detected towards any of the other cores within the C-band primary beam.

\subsubsection{G024.33+00.11}
\label{sec:G024.33}
VLA, $3.6\cm$ observations showed a $0.33\mJy$ radio source (VLA1) at $\alpha \mathrm{(J2000)} = 18\rahr 35\ramin 08.1\rasec$, $\delta \mathrm{(J2000)} = -07\degr 35\arcmin 04\arcsec$ and a $0.29\mJy$ source (VLA2) at $\alpha \mathrm{(J2000)} = 18\rahr 35\ramin 24.0\rasec$, $\delta \mathrm{(J2000)} = -07\degr 37\arcmin 38\arcsec$ \citep{Battersby2010}. VLA1 and VLA2 are coincident with the mm-cores, MM1 ($15\arcsec$ in diameter) and MM8 ($40\arcsec$ in diameter) respectively, observed by \citet{Rathborne2006}. Unusually, MM1 is determined to be at the kinematic far-distance ($6.3\kpc$) unlike the other mm-cores of \citet{Rathborne2010} (which are at distances of $3.7\kpc$). Near-infrared H$_2$, $2.122\um$ images show two lobes of shock H$_2$ emission (B and C) aligned along a position angle of $125\degr$ and separated by $\sim6\arcsec$, \citep[described as a bipolar H$_2$ outflow by][]{Lee2013}. \citet{LopezSepulcre2010} detect a wide angle, bipolar outflow centred on MM1 whose outflow axis is not well defined in channel intensity maps.

At C-band we detect three radio sources, A, B and D. A \citep[VLA 1 of][]{Battersby2010} is located at the centre of MM1 coincident with an MIR `yellowball', whilst B is distinctly positioned away from the patch of mid-infrared extinction. No mid-infrared source is seen coincident with B, however in UKIDSS imagery it is possibly co-located with a faint red source in close proximity/confusion with a bright blue star. Source D \citep[VLA 2 of][]{Battersby2010} is located within the MM8 core ($18\arcsec$ SE of its peak position), has a relatively large flux ($3.4\pm0.18\mJy$) and may be coincident with a blue, GLIMPSE source. D is also variable over the 4 years between the observations of \citet{Battersby2010} and those here, increasing in flux by $\sim 3 \mJy$ in that period. Comparing our C-band flux for A with the $8.3\GHz$ flux obtained by \citet{Battersby2010} of $0.33\mJy$ (their VLA1), we derive $\alpha=-0.2\pm0.3$ making its emission mechanism ambiguous. Only source A is detected at Q-band, however another compact Q-band source, C, is detected $4.6\arcsec$ to the NW of A. Source C does not have a C-band counterpart, giving a lower limit on its spectral index (assuming no variability) of $\alpha>1.2$. Both A and C are resolved at Q-band, while only A is resolved at C-band. As mentioned previously, a bipolar H$_2$ outflow is present which is associated to our source A and aligned with A's major axis. In combination with its spectral index of $\alpha=0.70\pm0.11$, we determine this to be an ionised jet, especially given its elongated radio morphology, with C and Q-band major axes that are parallel, showing minimal dust contributions at Q-band. Interestingly A seems to have a western extension to its Q-band emission, whose exact nature remains undetermined.  Source B, C and D remain unclassified.

\subsubsection{G024.60+00.08}
\label{sec:G024.60}
\citet{Rathborne2010} detect four mm-cores in total towards this IRDC (MM1$-$4) of which MM4 remains unclassified. An earlier work \citep{Rathborne2006} determined that the complex itself was comprised of two separate complexes at different distances from us. The furthest distance, $6.3\kpc$, was obtained for MM2 while for the rest of the cores, a distance of $3.7\kpc$ was adopted. Interestingly, the same work used PdBI observations at $1.2$ and $3\mm$ to show that MM1 and MM2 are actually composed of 3 and 5 separate condensations respectively. Towards both MM1 and MM2 \citet{LopezSepulcre2010} detect bipolar outflows in $\mathrm{HCO^+(1-0)}$ whose red lobes are oriented at position angles of $\sim280\degr$ and $\sim10\degr$ respectively. Previous radio observations \citep{Battersby2010} at $8.3\GHz$ revealed two point-like sources, one distinctly offset from the IRDC ($\sim4\arcmin$ to the south of MM1) and the other closer but still outside the obvious mid-infrared extinction.

Two C-band continuum sources are detected within the primary beam. Source A is not coincident with any near, or mid, infrared source and is distinctly located away from the mid-infrared extinction of the IRDC, and therefore we believe it to be extragalactic. In contrast, Source B is located on a western filament of the cloud and is coincident with a reddened, UKIDSS source and white, GLIMPSE source. It is determined that B is a star which has evolved past the YSO stage and is located on the near-side of the IRDC on account of its mid-infrared colours. At Q-band no significant radio emission was detected across the primary beam. A methanol maser was also observed and located within the NW armature of the IRDC's extinction coincident with both a GLIMPSE point source, the mm-core MM2 and one of the bipolar outflows, but with no associated, C-band, continuum emission ($3\sigma$ upper limit of $18\uJy$). 

\subsubsection{G028.28$-$00.34}
\label{sec:G028.28}
A deep radio survey by \citet{Cyganowski2011} detected a $4\sigma$, unresolved, radio-continuum source (EGO G28.28$-$0.36$-$CM1) with a peak flux at $8.3\GHz$ of $0.21\mJy$, centred on the EGO at $18\rahr44\ramin13.33\rasec$, $-04\degr18\arcmin04.3\arcsec$. Three other sources were detected (F G28.28$-$0.36$-$CM1, CM2 and CM3) in their field of view, two of which (CM1 and CM3) were coincident with extended, reddened, GLIMPSE sources and the other isolated from any infrared emission. No evidence for a molecular outflow \citep{LopezSepulcre2010} or H$_2$ line emission \citep{Lee2013} was found in the region however, suggesting minimal outflow or jet activity. At $1.2\mm$ \citet{Rathborne2010} detected 4 cores, of which 3 have classifications (all `R'), which are contained within 2 ATLASGAL clumps at $870\um$. Core MM1 harbours the RMS survey's \textsc{Hii} region, G028.2875$-$00.3639 which was determined to lie at the far, kinematic distance of $11.6\kpc$.

An obvious \textsc{Hii} region (G028.2875$-$00.3639) is detected at C-band and Q-band with an extent of $\sim0.4\pc$ which is therefore classified as a compact \textsc{Hii} (C\textsc{Hii} region). In the higher frequency image much of the extended emission is resolved out, leaving a compact source with an \textsc{imfit} derived size of $1900\au\times1600\au$. Since the extended emission was not resolved out in the C-band images, the Q-band data was split in to two sub-bands (with central frequencies of $42\GHz$ and $46\GHz$), reimaged and fluxes for the compact source recorded. Using the peak fluxes in both sub-bands derives a spectral index of $\alpha\sim-0.1$, which likely indicates optically thin free-free emission, therefore ruling out a stellar or disc wind. Integrated fluxes (over $3\sigma$ contours) of $\sim440\mJy$ are measured at $5.8\GHz$. We believe this to be an underestimate of the optically-thin flux considering the value of $\alpha=0.6\pm0.2$ between $5$ and $9\GHz$ \citep[from the data of][]{Kurtz1994,Purcell2013}. Comparing the near and far distance estimates, with which we derive values for $\Lbol$ of $23700\Lsol$ and $340000\Lsol$ (from the RMS survey's values), we believe the near distance to be incorrect since the radio luminosity of the \textsc{Hii} region is too large for $\Lbol=23700\Lsol$. Therefore, the core may also lie at the far kinematic-distance considering the \textsc{Hii} region lies in front of it.

Another C-band source (outside the Q-band primary beam) was detected (Source A) coincident with G28.28$-$0.36$-$CM2 from \citet{Cyganowski2011}, but has no near, or mid-infrared counterpart and lies outside the IRDC's extinction and ATLASGAL, sub-mm, continuum emission. Deconvolved dimensions at $8.3\GHz$ show A to be resolved with a size of $0.69\arcsec\times0.54\arcsec$, whereas in our C-band data only upper limits on A's physical size could be determined. This tends to suggest that our A-configuration observations have likely resolved out any extended flux detected in \citet{Cyganowski2011}. Therefore, comparing their peak flux to our peak flux to determine a rough spectral index yields $\alpha=-0.2\pm0.2$, supporting our extragalactic classification. At a robustness of 0.5 without removing short baselines, no source is detected coincident with the EGO, however this is most likely due to the anticipated, weak cm-emission and dynamic range-limited image.

\subsubsection{G028.37+00.07}
\label{sec:G028.37}
Within our C-band primary beam \citet{Rathborne2010} identifies 11 cores with a wide range of evolutionary classes (Q, I and A). Their MM1 and MM2 were determined to possess luminosities of $\sim25000\Lsol$ with both being members of clumps containing EGOs and \textsc{Hii} regions from inspection of GLIMPSE imagery. At $870\um$ MM1's parental clump is the most luminous, has the widest NH$_3\,(1,1)$ line widths ($4.3\kmps$) and highest rotational temperatures of the entire complex \citep[clump P2 of][]{Wang2008}, likely a consequence of internal heating. No previous cm-emission, excluding the \textsc{Hii} regions detected by \citet{Battersby2010}, has been detected towards MM1. Towards MM2, however, a point-like (with a $2.4\arcsec$ C-configuration, VLA, synthesised beam) radio source with an $8.3\GHz$ flux of $1.7\pm0.2\mJy$ was observed to be coincident with a possibly extended, reddened, GLIMPSE source \citep[VLA4 of][]{Battersby2010}. No sub-mm, line observations are found in the literature targeting molecular outflow activity and therefore no comment can be made on the presence of large-scale outflows.

Radio images in \autoref{cplot:G028.37} show the detection of four C-band sources, designated A, B, C and D, and three Q-band sources, A (co-located with the C-band source), A2 and A3. A CH$_3$OH maser is detected and coincident with A, which possesses a spectral index of $\alpha=1.38\pm0.19$. Source B is resolved in C-band and coincident with a $\sim2.6\,\mJy$ source from archival, $20\cm$, GPS images, indicating a non-thermal spectral index of $\alpha\simeq-1$ (assuming minimal resolving out effects). Comparison with the observations of \citet{Battersby2010}, show that B is registered as their point-like VLA6 with a $3.6\cm$ flux of $380\pm40\uJy$ yielding a value for $\alpha$ of $-1.0\pm0.5$, agreeing with the GPS comparison. In conjunction with its non-detection at Q-band ($4\sigma$ upper limit of $360\uJy$), location away from cold-dust, sub-mm emission (ATLASGAL $870\um$) and lack of a MIR-counterpart, this confirms its extragalactic nature. From our data alone, component C's nature can not be determined due to its high upper limit for spectral index ($\alpha<1.1$) and ambiguous positioning on the edge of sub-mm continuum emission. Source A3 from Q-band is coincident with a $3\sigma$, C-band source which looks slightly extended at a position angle of $\sim-45\degr$. Using an \textsc{imfit} derived flux of $47\pm15\uJy$ for the C-band, $3\sigma$ emission, a spectral index of $\alpha=1.60\pm0.18$ is calculated. A Q-band, \textsc{imfit}-derived position angle for A3 show it to be elongated at an angle of $25\pm7\degr$, while the C-band, \textsc{imfit} derived centroid of emission is offset to the Q-band's at a position angle of $\sim-55\degr$ (i.e. roughly perpendicular to the Q-band elongation axis). This suggests that A3 may be dominated by thermal, free-free emission from an ionised jet at C-band (ejection axis at $\theta_{PA}\sim-50\degr$), but by dust from an accretion disc ($\sim800\au$ in diameter) at Q-band. Source A2, which is offset by $750\au$ at a position angle of $-154\degr$ from A3, is likely to be the dust emission from a close-by, coeval YSO/core. We therefore believe A, A2 and A3 represent a triple system of YSOs. Component D is the same source as VLA4 from \citet{Battersby2010} and in conjunction with their flux at X-band we derive a spectral index of $1.7\pm0.5$ indicating optically thick, free-free emission. From its MIR colour/morphology, deconvolved size ($0.007\pc\times0.003\pc$) and radio spectral index, D is likely a HC\textsc{Hii} for which we calculate a powering star of ZAMS type B2, corresponding to a mass of $11\Msol$ and luminosity of $6600\Lsol$. This estimate is based upon the assumption that the emission is optically thin at $3.6\cm$ however it does provide a lower limit for the bolometric luminosity which is in agreement with that derived by \citet{Rathborne2010} of $23329\Lsol$.

\subsubsection{G028.67+00.13}
\label{sec:G028.67}
At C-band we detect one source, being the \textsc{Hii} region coincident with the mm-core MM1 from \citet{Rathborne2010}. We measure an integrated flux of $62\mJy$ and deconvolved size of $4.0\arcsec\times3.2\arcsec$ (or $0.09\pc\times0.07\pc$ at the distance of $4.8\kpc$), typical of a compact \textsc{Hii} region. From the flux, and assuming optically thin emission at $5.8\GHz$, we infer a powering star of ZAMS type B1 equivalent to a mass of $15\Msol$ or luminosity of $20000\Lsol$ \citep{Davies2011}, in rough agreement with $12000\Lsol$ derived for MM1. We establish a $3\sigma$ upper flux limit for C-band emission towards MM2 of $24\uJy$.


\subsubsection{G033.69$-$0.01}
\label{sec:G033.69}
Mid-IR images show a filamentary structure extending along a rough N-S axis along which \citet{Rathborne2010} classify 10 cores with a wide range in evolutionary status. The most luminous of these cores are MM2 and MM5, which possess red and active classifications respectively. From inspection of GLIMPSE images, the reddened MIR source associated to MM2 is likely an \textsc{Hii} region on account of its extended morphology. Comparison with ATLASGAL images show that MM2 lies approximately halfway between two $870\um$ emission peaks, both of which are likely associated to extended \textsc{Hii} regions. The northern part of the overall filamentary complex appears less evolved and quiescent, but harbours the highest mass core (MM1 at $750\Msol$) which itself harbours a reddened MIR source.

Only a cometary UC\textsc{Hii} region (A) $1.8\arcsec$ ($0.05\pc$) across is detected $18\arcsec$ to NW of MM2's given position with a flux of $\sim1.5\mJy$ at C-band, indicative of a ZAMS B1 powering star with a mass between $12-15\Msol$ and equivalent to a luminosity of $9-18\times10^3\Lsol$ \citep{Davies2011}. No Q-band emission is detected across any of the 3 pointings towards this complex. Source A is also located $8\arcsec$ to the north and $37\arcsec$ to the NW of more expanded, resolved out \textsc{Hii} regions which show prominently in the GLIMPSE, RGB images of \autoref{cplot:G033.69}. Core MM2 is offset from A by $17\arcsec$ to the SE but still encompasses the detected \textsc{Hii} region due to its large diameter \citep[$41\arcsec$][]{Rathborne2007}. Considering its radio-derived bolometric luminosity, it is likely to be the major contributor to the IR derived luminosity of $22373\Lsol$.



\subsection{MYSO Sample}
\label{sec:mysonotes}

\subsubsection{G033.6437$-$00.2277}
\label{sec:G033.6437}
Classified as a diffuse \textsc{Hii} region in the RMS survey, from inspection of GLIMPSE imagery a `yellowball' coincides with the diffuse PAH, $8\um$ emission from the extended \textsc{Hii} region (G033.6437$-$00.2277 in the RMS survey). Towards the MYSO, methanol maser emission is detected \citep{Bartkiewicz2009} in an arc along a rough position angle of $70\degr$. These masers are known to periodically flare \citep{Fujisawa2012} on short timescales of $\sim1$ day.

At C-band we detect 3 sources names A, B and C of which only A has a corresponding Q-band detection associated to it. Approximately $8.5\arcsec$ to the west of A, a Q-band only radio object is detected (D) to a $5\sigma$ level. Methanol maser emission is also detected at A's position, which itself is situated at the centre of the \textsc{Hii} region's diffuse $8\um$ emission with a mid-infrared `yellowball' (see RGB image of \autoref{cplot:G033.6437}). Without any further information in terms of previously detected outflows or other indicators of jet activity, we classify A to be a jet candidate. Source B is coincident with a blue source in UKIDSS, with an upper limit to its radio spectral index of $<0.53\pm0.14$. We believe it to be more evolved than the YSO stage on account of its non-reddened NIR profile and therefore classify it with an evolved status. Due to its isolated nature and absence of a mid-infrared detection, C is classified to be extragalactic. Unusually D is only detected at Q-band, and a spectral index lower limit of $\alpha>1.52\pm0.17$ is deduced. We believe this to be a YSO in the vicinity of G033.6437$-$00.2277, and due to its high spectral index, may display variability or be dominated by dust emission at $44\GHz$.

\subsubsection{G035.1979$-$00.7427}
\label{sec:G035.1979}
Better known as G35.20$-$0.74N, \citet{SanchezMonge2014} used ALMA observations at $350\GHz$ (light blue contours of \autoref{cplot:G035.1979}) to detect 6 evenly-spaced, dense cores, labelled A through F, aligned along a position angle of $\sim140\degr$ over an extent of $\sim15\arcsec$, or $0.15\pc$ at $2.19\kpc$. Cores A and B showed coherent velocity structures indicative of Keplerian discs in rotation around central objects of mass $4-18\Msol$, as well as harbouring precessing jets perpendicular to the suspected discs. Previous to that work, \citet{GibbHoare2003} analysed $5$ and $8.5\GHz$ VLA observations, detecting 11 radio sources at either/both frequencies. Of these 11 sources, those named $1-4$ and $7-11$ appeared to be aligned along a N-S axis, with 7 (or G35.2N) and 8 coincident with core B from \citet{SanchezMonge2014}. Another radio work by \citet{Beltran2016} analysed VLA, B-configuration observations at $15\GHz$, $23\GHz$ and $43\GHz$ and concluded that their radio source 8 \citep[named 7 by][]{GibbHoare2003} was in fact a hypercompact \textsc{Hii} region. As a note, it is the naming convention of \citet{GibbHoare2003} that we follow in this work. Fits to the SED yielded ranges in bolometric luminosity from $(0.7-2.2)\times10^5\Lsol$, dependent upon cavity opening angle and inclination \citep{Zhang2013}. Most recently, \citet{Fedriani2019} used the C-band observations presented here to derive the ionisation fraction of the ionised jet launched from core B at various distances from the core. From their results, we calculate that the ionisation fraction drops with distance from the launching point as $x_{\rm i}\propto r^{-0.5\pm0.3}$.

Q-band images show 5 compact radio sources in the field of view, all of which coincide with, and maintain the naming of, the previously detected sources in \citet{GibbHoare2003} apart from the source we refer to as `Core A' (see \autoref{cplot:G035.1979}) which inherits its name from the sub-mm source by the same name \citep{SanchezMonge2014}. Sources 7 \citep[`Core B' of][]{SanchezMonge2014} and 8 both show lobes of faint radio emission $\sim0.09\arcsec$ (or $200\au$) to the west of their peaks, which are recorded as 7b and 8b respectively. At C-band we detect all Q-band sources, and five more designated 4, 13, 14, 15, EX$-$S and EX$-$N (i.e. extended south/north). Lobes 13, 14 and 15 are new detections, while EX$-$S and EX$-$N are comprised of sources 9, 10, 11 and 1, 2, 3 from \citet{GibbHoare2003} respectively. 

\begin{figure}
\centering
\includegraphics[width=84mm]{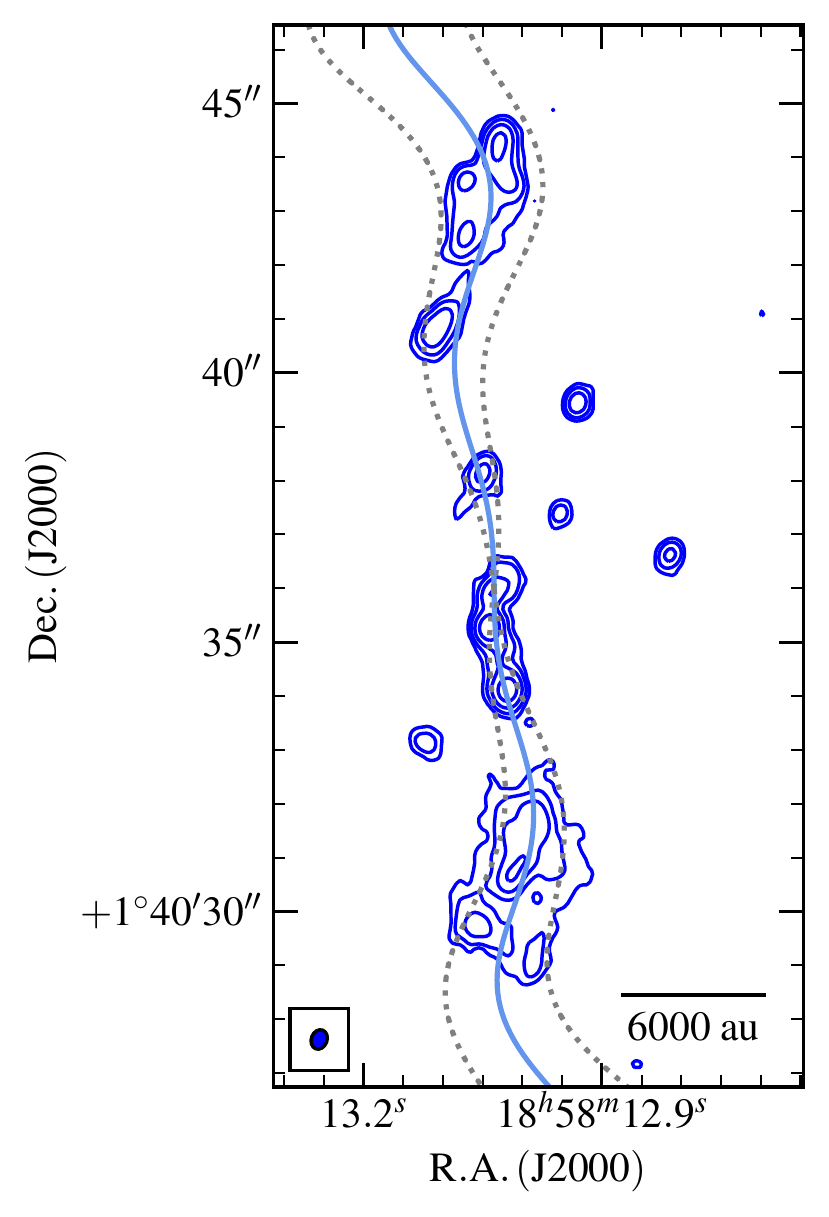}
\caption[Fitted precessing jet model of G35.20$-$0.74N overlaid on radio flux contour map]{A radio contour map of G35.20$-$0.74N with a fitted jet model overlaid (light blue line), with the derived, minimum jet opening angle defined by the grey, dotted lines. Contour levels are the same as in \autoref{cplot:G035.1979}.}
\label{fig:18556precession}
\end{figure}

Comparison with the ALMA image of \citet{SanchezMonge2014} (central panel of \autoref{cplot:G035.1979}) show 4, 5, 14, 15, EX$-$S and EX$-$N are not associated with any sub-mm continuum sources, but are with the diffuse emission from cavity walls in both near and mid-infrared images. From UKIDSS imagery, 5 is possibly associated with a NIR source and, like 6, is probably more evolved than the radio sources associated to strong sub-mm emission. Considering their flat spectral indices and non-association with sub-mm emission, 5 and 6 are likely small, compact \textsc{Hii} regions. Due to its jet-like spectral index of $0.64\pm0.06$ and perfect alignment with 14 and 15, we determine that 7 is the driving jet behind the precessing axis defined by 4, 7, 8, 14, 15, EX$-$S and EX$-$N. This disagrees with the interpretation of \citet{Beltran2016} that source 7 (their source 8) is an HC\textsc{Hii} region. We attribute this discrepancy between the two works to our better resolution, and thus measurement of spectral index, allowing us to be sure that source 7 is elongated and jet-like. Assuming a velocity of $500\kmps$ for the jet, we determine a rough precession period of $150\pm50\yr$ from fitting a simple jet model to the lobe positions (\autoref{fig:18556precession}). Assuming this motion has a simple relation to the period of a possible, binary companion, we determine (for a $19\Msol$ central object) that the companion should orbit at a distance of $80\pm20\au$. Although the nature of 8 may appear ambiguous as a \textsc{Hii} region, jet or optically thin lobe of shocked emission ($\alpha=-0.11\pm0.09$), its position relative to 7 changes significantly ($\sim0.24\arcsec$) in comparison with the results of \citet{GibbHoare2003}, indicating a more transient phenomena. Therefore, we split the bolometric luminosity of $150000\Lsol$ evenly between Core A and 7 for any further analysis.

\subsubsection{G035.1992$-$01.7424}
\label{sec:G035.1992}
Our RMS source is also known as the UC\textsc{Hii} region, W48A, located in the W48 massive-star forming complex. \citet{Rygl2014} detected 3 FIR clumps towards the W48 region, one (H1) centred on the UC\textsc{Hii} region and the other two, H2 and H3, located $16\arcsec$ and $63\arcsec$ to the west respectively. Both H1 and H2 have one $2.3\mm$-core each, named H1a and H2a respectively, while H3 has two $2.3\mm$-cores (H3a and H3b). Clump H2 was associated with methanol, hydroxyl and water maser activity, while clump H3 displayed a roughly north-east to south-west outflow traced by the CO ($3-2$) transition. Ultimately this work concluded that sequential star formation from (in order of most to least evolved) H1, to H2 and lastly to H3 explained their observations, with H2 bordering the UC\textsc{Hii}'s photo-dissociation region (PDR) and therefore being pressured into collapse by the expanding \textsc{Hii} region. Most recently, $22\GHz$ radio images employing $uv$-cuts of $>320\mathrm{k}\lambda$ by \citet{Masque2017} resolved out much of the emission and detected a compact source at $\alpha \, (\mathrm{J2000})=19\rasec 01\ramin 46.46\rasec$, $\delta \, (\mathrm{J2000}) = 01\degr 13\arcmin 23.6\arcsec$ (centre of W48A) with a flux of $5.24\pm0.14\mJy$ and deconvolved size of $(67\pm3\mas)\times(39\pm3\mas)$, at a position angle of $131\pm5\degr$.

At C-band we detect a cometary UC\textsc{Hii}, W48A, which limits the image in dynamic range due to its extended, bright emission. When moving to Q-band, the lack of short $uv$-spacings filtered out much of the extended structure and we detect partially resolved out emission at the location of the cometary \textsc{Hii} region's bow shock. Further to this, we detect a compact source roughly at the \textsc{Hii} region's geometric centre \citep[i.e.\ at the location of the compact source detected by][]{Masque2017}. From the K-band, radio flux previously recorded by \citet{Masque2017}, we derive a spectral index of $-0.11\pm0.17$ with no change in deconvolved position angle or size (which makes the hypothesis of an ionized jet unlikely). From the models of a spherical, ionized wind \citep{Panagia1975}, this looks like an unlikely candidate for an ionized, stellar wind. To investigate this possibility however, from the measured flux at $44\GHz$ of $4.84\pm0.19\mJy$, we infer a mass loss rate of (assuming $v = 3000\kmps$) of $5.6 \times 10^{-5} \Msol \yr^{-1}$ \citep{WrightBarlow1975}. From the models of \citet{Vink2001}, we can infer a mass loss rate expected for a source of given luminosity/mass. For W48's luminosity of $1.5 \times 10^5 \Lsol$ we expect a mass loss rate of $\sim2.5\times 10^{-6}\Msol \yr^{-1}$ at best, which is around 20 times too small, compared to our measured flux at $44\GHz$, which consequently rules out the stellar wind picture. With the calculated, optically-thin, spectral index, we can accurately calculate an emission measure of $(1.32\pm0.14)\times 10^{9} \pc\cm^{-6}$, an average electron density of $(1.24\pm0.14)\times10^6 \cm^{-3}$ and a B1 type ($7600\Lsol$, $11\Msol$) powering star, assuming it is a small \textsc{Hii} region. Considering W48A is powered by a star (or cluster) with a bolometric luminosity of $170000\Lsol$, we can say that A is not powering the overall region. Therefore we deduce A to be a candidate, gravitationally-trapped \textsc{Hii} region, since half the major axis size ($106\pm20\au$) is smaller than the gravitational radius ($r_g=122\au$ for $M_\star=11.3\Msol$).

Source B has no obvious K-band, or WISE $3/4\um$ counterpart though we can not pass comment for longer wavelengths due to saturation of the WISE images at $12\um$ and $22\um$. Although not detected at C-band, we establish a dynamic range limited, $4\sigma$, upper limit on its flux of $147\um$ and therefore a lower limit on its spectral index of $\alpha>-0.16\pm0.11$. Assuming no variability in the source, this would suggest a thermal object, likely a deeply embedded, extremely young YSO. With the limitations to radio interferometric imaging from W48A, K or Ku-band observations would likely be needed to adequately establish the spectral index of B and therefore provide a firmer classification.

A methanol maser was detected which is co-located with a reddened source in UKIDSS NIR imagery, however no continuum emission is detected in C (dynamic range limited) or Q-band (with $4\sigma$ upper flux limits of $158\uJy$ and $112\uJy$ respectively) images. This methanol maser is located within clump H2/core H2a from \citet{Rygl2014} and likely points to a young MYSO on account of its non-detection at radio wavelengths.

\subsubsection{G037.4266+01.5183}
\label{sec:G037.4266}
\citet{LopezSepulcre2010} detect a N-S oriented molecular outflow in C$^{18}$O$(2-1)$. Subsequently, \citet{Surcis2015} used VLBI to identify 19 CH$_3$OH maser spots divided into two groups, A and B, to the NW and SE of G37.43+1.51's position. No clear velocity gradients were observed in either group. However, group A shared their velocity with the outflow's blue lobe, thereby associating the maser emission with the N-S molecular outflow of \citet{LopezSepulcre2010}.

We detect one Q-band, $\sim0.5\mJy$, point-like source offset $\sim11\arcsec$ to the north east of the pointing centre, which coincides with both a mid-infrared (in WISE imagery) source and B at C-band. A CH$_3$OH maser spot is detected $\sim0\farcs3$ NW of B, itself coincident maser group A from \citet{Surcis2015}. Both maser groups from \citet{Surcis2015} are plotted (pink crosses) in \autoref{cplot:G037.4266} for reference. Two more C-band sources are detected, A and C, which both have red, NIR counterparts. While B possesses a thermal spectral index of $1.16\pm0.12$, A and C were not detected at $44\GHz$ and we derive upper limits of $\alpha<0.59\pm0.11$ and $\alpha<0.29\pm0.06$, respectively. No physical dimensions could be deconvolved using \textsc{imfit} and therefore we can only classify these three sources as jet candidates.

\subsubsection{G056.3694$-$00.6333}
\label{sec:G056.3694}
Near-infrared UKIDSS imagery shows a reddened point source at the centre of some nebulosity towards the south west and north of the point source.

At C-band we see a string-like morphology slightly offset to the reddened, UKIDSS point source. Considering this morphology, this is likely an ionized jet with lobes, however a lack of results from other wavelengths in the literature means we must classify this as a candidate. Due to its proximity to the near-infrared source, C is assumed to be the YSO where the base of the jet is located, while A, B and D are shock-ionized lobes associated to the jet's flow. Ideally high-resolution mm-observations would help to establish which components are affiliated with cores to clarify the picture further. In the case that C is indeed a jet, while A, B and D are shock-ionized radio lobes, mm sources should only be associated to C.

\subsubsection{G077.5671+03.6911}
\label{sec:G077.5671}
Near infrared 2MASS images show this object to be at the centre of a possible cluster. Previous ammonia and water maser observations failed to detect significant emission \citep{Urquhart2011}, unusual for our sample of MYSOs.

One source, unrelated to the MYSO, was detected $23\arcsec$ to the SE of the pointing centre in the C-band clean images, however nothing was detected at Q-band. Considering that it has no associated near or mid-infrared source, it is most likely extragalactic in origin though its non-thermal or thermal nature can not be constrained ($\alpha<0.25\pm0.11$).

\subsubsection{G078.8699+02.7602}
\label{sec:G078.8699}
Another well studied region in our sample containing 3 Herbig Ae/Be stars named BD$+40\degr4124$, V$1686$ Cygni and V$1318$ Cygni. Observed in both optical and NIR regimes, it was determined that V1318 Cygni was composed of 3 separate optical sources dubbed north, middle and south \citep{Aspin1994}, of which `south' was concluded to be a very young, intermediate-mass, Herbig Ae/Be star with a low-mass binary companion, `north', separated by $\sim5000\au$ and connected via the diffuse optical source `middle'. At sub-mm wavelengths, V$1318$ Cygni is by far the dominant source \citep[see Figure 22 of][]{Sandell2011}. \citet{Navarete2015} observed this region in both K-band continuum ($2.2\um$) and $\mathrm{H}_2$ line ($2.122\um$) NIR filters. They detected a bipolar outflow in shocked $\mathrm{H}_2$ emission consisting of 4 lobes (sources 1a, 1b, 1c and 1d) roughly aligned at a position angle of $\sim150\degr$ and extending over a length of $0.78\pc$. The approximate midpoint of these lobes is located at $\alpha\mathrm{(J2000)}=20\rahr20\ramin30.6\rasec, \,\delta\mathrm{(J2000)}=41\degr22\arcmin06\arcsec$. Two more lobes (2 and 3) of shock emission were detected $\sim2\arcmin$ from their pointing centre (which coincides that of our observations) at a position angle of $10\degr$.

UKIDSS imagery shows 4 bright sources within $1\arcmin$ of our field centre (\autoref{cplot:G078.8699}), the most easterly two being separated from eachother by $\sim5\arcsec$, displaying red NIR colours and coincident with V$1318$ Cygni's `north' and `south' source from \citet{Aspin1994} respectively. A less reddened source can be seen $\sim14\arcsec$ to their east (V$1686$ Cygni) and the brightest source, blue in colour, is located $\sim40\arcsec$ to the NE (BD$+40\degr4124$). At C-band we detect radio emission from all of these NIR sources (A, B, C and E), as well as from a NIR dark source located $47\arcsec$ NNW of the pointing centre with profiterole-like morphology (D). Only sources A, B and C are detected at Q-band.

Source A is clearly resolved at Q-band with dimensions of $(99\pm6\times50\pm7)\au$ for the listed distance of $1.4\kpc$, while possessing values for $\alpha$ and $\gamma$ of $1.41\pm0.08$ and $>-0.54\pm0.03$ respectively. Its deconvolved position angle of $-31\pm6\degr$ aligns it with the shocked emission lobes, 1c and 1d, of \citet{Navarete2015}. Source B is extremely low signal to noise ($3\sigma$ level at Q-band) and is listed as a Q-band detection on the basis of its coincidence with the corresponding C-band source but does remain consistent with the standard jet model of \citet{Reynolds1986} whereby $\alpha=0.64\pm0.23$ and $\gamma=-0.91\pm0.26$, which also describes a stellar wind. Without more supporting evidence however, we must classify is as a jet candidate. Source C remains unresolved at both frequencies and we calculate a spectral index of $\alpha=1.12\pm0.08$ for it, but given its bright, blue NIR colours it has clearly evolved past the MYSO stage. For D we determine it to be extragalactic on account of its IR dark nature mixed with a limit to its spectral index of $\alpha<0.41\pm0.19$. Our radio object E is not detected at Q-band and we establish an upper limit on $\alpha$ of $<0.50\pm0.13$ and although definitely galactic, its more specific nature can not be determined. Considering the fact that optical counterparts exist to these objects (apart from D), it is highly likely that some are Herbig Ae/Be stars. As for the bolometric luminosities of each object, based on correction for the more accurate distance of \citet{Rygl2012}, we adopt $5972\Lsol$ and $528\Lsol$ for A and B respectively \citep[from][]{Aspin1994}, while for C we adopt a luminosity of $1080\Lsol$ \citep{Lumsden2012}.

\subsubsection{G079.8855+02.5517}
\label{sec:G079.8855}
Previous $\,^{12}\mathrm{CO}\left(3-2\right)$ maps detect an outflow with red and blue lobes along a position angle of $\sim120\degr$ and roughly centred on the RMS source G079.8855+02.5517B \citep{Gottschalk2012}.

Within the RMS database, our C-band and Q-band detected source C is designated as G079.8855+02.5517C, while our Q-band-only source B1/B2 corresponds to RMS source G079.8855+02.5517B (the more luminous of the two by an order of magnitude). None of our sources are detected in UKIDSS, but C and B1/B2 are strongly saturated in GLIMPSE images apart from at $3.5\um$. With a separation of just $97\pm8\au$ ($69\pm4\mas$ at a distance of $1400\pm80\pc$), B1 and B2 are likely a young binary system. Considering that B2 is elongated along a position angle of $131\pm15\degr$, parallel with the molecular outflow observed by \citet{Gottschalk2012}, we believe that it is an ionized jet whose flux is contributed to by dust emission (hence the steep spectral index of $\alpha > 1.6\pm0.1$). 

Other detected emission in the field is the C-band source E (outside the Q-band primary beam) and Q-band source D. Since E has no near or mid-infrared counterpart, its most likely classification is extragalactic, while source D, with its steep ($\alpha>1.6\pm0.1$) thermal spectral index and mid IR-dark appearance, is most probably a YSO at the beginning of its evolution.

\subsubsection{G081.8652+00.7800 (W75N)}
\label{sec:G081.8652}
W75N is arguably the most well-studied region of our sample. Our C-band field of view encompasses 3 significant objects, one being an \textsc{Hii} region (RMS designation G081.8789+00.7822), another being the string of radio lobes \citep{CarrascoGonzalez2010} with confusing outflow morphology \citep{Shepherd2003} and the last being a reddened MIR source (RMS designation G081.8652+00.7800) to the SW of the well studied radio object. The string of 5 radio lobes is oriented at a position angle of $\sim170\degr$ from the clean maps of \citet{CarrascoGonzalez2010} who determined their source VLA3 to be a thermal jet ($\alpha=0.6\pm0.1$) powering the Herbig-Haro object, Bc, with the natures of VLA1, VLA2 and VLA4 uncertain. Significant radio flux variability was detected towards Bc over the period encapsulating all of their studied, archival observations ($\sim13\yr$). 

At both C and Q-bands we detect an unresolved source dubbed `A' which is coincident with the RMS source G081.8652+00.7800 and possesses a spectral index of $\alpha=1.00\pm0.08$ indicative of thermal free-free emission. Upper limits on A's physical dimensions of $(<50\times<24)\au$ are established from the higher frequency images indicating an extremely compact source. As a result, we can only classify this object to be a jet candidate as both its spectral index and extremely small dimensions lend themselves to interpretation as a small HC\textsc{Hii}.

Approximately $24\arcsec$ to the NE of A we detect at least 9 sources in the C-band maps and 3 in the Q-band images. These form the well-studied jet/lobes of the W75N region and we report VLA5, VLA6 and VLA7 to be new sources previously undetected. VLA 1 is elongated at a position angle of $\sim80\degr$, which is parallel with the large CO outflow reported in \citet{Shepherd2003}, however \textsc{imfit} requires a two-component model at C-band, one for the central `core' and the other for the more elongated emission to the north-east. VLA1's `core' is also detected at Q-band and has a spectral index of $\alpha=0.19\pm0.06$. For the elongated emission we calculate $\alpha<-1.21\pm0.07$. The accuracy of this result is debatable considering resolving out effect at Q-band and accurate deconvolution of VLA1 into two sources at C-band. Due to its strong alignment with an outflow, but uncertain spectral indices we assign in the classification of jet candidate.

VLA2 possesses values for $\alpha$ and $\gamma$ of $0.11\pm0.07$ and $-1.46\pm0.12$ respectively, with deconvolved position angles of $\sim20\degr$ at both frequencies. From our analysis alone, and given the confusing picture of outflow and low signal to noise, we would classify VLA2 as a jet candidate. However considering the work by \citet{CarrascoGonzalez2015}, who observed the `onset of collimation' for the thermal jet associated to VLA2, this radio object is definitively classified as a jet.

VLA4 is seen to possibly be another lobe of shocked emission by \citet{CarrascoGonzalez2010} who required longer time baselines to see if proper motions were significant or not. Taking advantage of the larger time baselines between our $5.8\GHz$ and their 2006, $8.46\GHz$ observations we derive proper motions of $0.13\pm0.04\arcsec$, or $126\pm42\kmps$, at a position angle of $-178\pm24\degr$. We believe that due to the $>3\sigma$ detection of this proper motion, this source is another radio lobe induced by jet shocks.

VLA5, VLA6 and VLA7 all fall well below the detection thresholds of previous radio observations and are both unresolved (VLA5) and distinctly extended (VLA6 and VLA7) in their morphology. Considering the overall morphology of the region, we assume these sources to be radio lobes, much like VLA4, Bc and Bc2, as the result of jet shocks.

The source W75NBc was detected to have proper motions of $\sim220\pm70\kmps$ by \citet{CarrascoGonzalez2010}, however the positions they catalogue for source Bc assume that our sources W75NBc and W75NBc2 (see \autoref{cplot:G081.8652}) are one and the same. Therefore to recalculate proper motions using our data, we used a deconvolved position derived from \textsc{imfit} assuming Bc and Bc2 to be the same source. It is important to note that we use the same phase calibrator ($\mathrm{J}2007+4029$) as that of \citet{CarrascoGonzalez2010}. At $5.8\GHz$, this gives a 2012 epoch position of $\alpha\mathrm{(J2000)}:20\rahr38\ramin36.5525\rasec,\,\delta\mathrm{(J2000)}:42\degr37\arcmin31.4585\arcsec$ (with uncertainties in position of $24.9$ and $9.8\mas$ in $\alpha\mathrm{(J2000)}$ and $\delta\mathrm{(J2000)}$ respectively) translating to a proper motion of $0.12\pm0.05\arcsec$ ($\theta_\mathrm{PA}=161\pm27\degr$) or $114\pm46\kmps$ for the more accurate distance of $1.32\pm0.11\kpc$  from \citet{Rygl2010} ($173\pm69\kmps$ using the distance of $2\kpc$ adopted by \citet{CarrascoGonzalez2010}). Considering this evidence for proper motions in both current and previous data, Bc2/Bc are classified as lobes representing the shocked surfaces of an ionized jet impinging upon its surroundings.

We also detect (C-band only) the \textsc{Hii} region (RMS designation G081.8789 +00.7822) whose material stretches across a region approximately $9.5\times8.6\arcsec$ in angular size, or $0.064\times0.058\pc$, classifying it as an ultracompact \textsc{Hii} region. Considering its flux, assuming optically thin emission and using the models of \citet{Davies2011} we estimate it to have a bolometric luminosity (from its derived Lyman flux) of $6600\Lsol$, mass of $11\Msol$ and therefore equivalent ZAMS type of B1. However, it must be noted that this is a lower-limit on both its flux and spatial extent due to potential spatial-filtering on these scales.

\subsubsection{G083.7071+03.2817}
\label{sec:G083.7071}
Apart from galactic plane surveys, no previous significant observations targeting this MYSO exist in the literature. UKIDSS NIR images show extended nebulosity in a bipolar configuration, characteristic of cavities, at a position angle of $\sim120\degr$ and more diffuse emission to the south west. 

Towards this source we detect two C-band radio objects, A and B, and one Q-band source coincident with A. Source A is resolved with dimensions of $(0.237\pm0.029\times0.045\pm0.061)\arcsec$ at $118\pm9\degr$ and $(0.044\pm0.01\times0.012\pm0.01)\arcsec$ at $111\pm18\degr$ at C and Q-bands respectively. In turn we calculate spectral index values of $\alpha=0.47\pm0.07$ and $\gamma=-0.83\pm0.13$, in line with the models of \citet{Reynolds1986} and A is therefore a thermal jet, especially given its alignment with NIR reflection nebulae in UKIDSS images (see left panel of \autoref{cplot:G083.7071}). Source B has an upper limit to its spectral index of $\alpha<-0.43\pm0.04$ and, considering its unresolved dimensions at C-band and position angle with respect to A of $\sim-140\degr$ (i.e.\ perpendicular to A's major axis), is most likely extragalactic in origin.

\subsubsection{G084.9505$-$00.6910}
\label{sec:G084.9505}
Near infrared RGB images from UKIDSS show a well defined example of a bipolar reflection nebulae at a position angle of $\sim 120\degr$, centred on the MYSO, however no other previous targeted observations of this object are present in the literature. 

At Q-band we detect one source, B, which is not resolved using \textsc{imfit} but does appear slightly elongated along a position angle of $\sim120\degr$, parallel to what appears to be reflection nebulae (likely cavity walls) in UKIDSS imagery (left panel, \autoref{cplot:G084.9505}). At C-band no continuum source is detected coincident with the B, but a methanol maser is detected whose \textsc{imfit} derived position is offset to B's Q-band peak by $\sim0.8\arcsec$. Without further information we classify this to be a jet candidate, especially given its unusually steep \citep[but still consistent with the models of][]{Reynolds1986} spectral index of $\alpha>1.25$ suggesting flux contribution from dust emission. One source is detected only at C-band, A, which is outside of the Q-band primary beam and so no further analysis can be conducted leaving it with an unknown classification. However, it does not have either a UKIDSS (NIR) or WISE (MIR) counterpart meaning it is likely extragalactic.

\subsubsection{G094.2615$-$00.4116}
\label{sec:G094.2615}
A knot of $\mathrm{H}_2$ $2.122\um$ emission is detected to the SE of our source A1 \citep[source 1 of A from][]{Varricatt2010} and associated to a molecular outflow oriented NW to SE at an angle of $141\degr$ \citep{Fontani2004}. 

We detect 3 sources within $50\arcsec$ of the pointing centre, A1, A2 and B. Source B is only detected at C-band, does not have an NIR counterpart and, from WISE imagery, probably does not have a MIR counterpart either (its position is on the limit of source confusion). Therefore B is classified as extragalactic in origin. A1 has both a C-band and Q-band detection inferring a spectral index of $\alpha=0.47\pm0.19$, however no value for $\gamma$ could be obtained due to the relatively low ($5\sigma$) SNR at Q-band. Considering A1's obvious elongation at C-band, apparent NIR cavities oriented at the same position angle and a parallel outflow, we classify it as a thermal jet. A2 is likely to be a lobe associated to A1's jet, produced via shock emission. However its extended morphology ($1800\pm400\au$ by $1000\pm800\au$) and unknown MIR status make it a possibility that it is a small \textsc{Hii} region of a B3, ZAMS-type star \citep[mass of $\sim8\Msol$, for the calculated value $\log(\mathrm{N_i})=44.23\pm0.09$, or $\mathrm{L}_\star=2300\pm180\Lsol$, from the models of][]{Davies2011}.

\subsubsection{G094.3228$-$00.1671}
\label{sec:G094.3228}
Knots of H$_2$, $2.122\um$ emission were detected by \citet{Navarete2015} and classified to be bipolar in distribution. Their orientation with respect to the RMS source are at position angles of $355\degr$, $185\degr$, $290\degr$ and $165\degr$ for H$_2$ lobes 1a, 1b, 1c and 1d respectively.

At both frequencies we detect an unresolved radio source designated as A $\sim2\arcsec$ to the NW of the pointing centre and coincident with the reddened NIR source apparent in the left panel of \autoref{cplot:G094.3228}. At Q-band, A's position is separated from that at C-band by $0.18\arcsec$ at a position angle of $107\degr$, however the positional accuracy code of the Q-band phase calibrator is `C' and therefore we can not say if this offset is real. We derive a spectral index of $\alpha=1.19\pm0.14$ between C and Q-bands, with an upper limit on the physical size of the emission at Q-band of $<(114\times40)\au$. Due to the unresolved nature of the radio source, it can not be determined if the emission is parallel, or perpendicular, to the previously recorded outflow axes from \citet{Navarete2015}. We therefore classify A as a jet candidate.

\subsubsection{G094.4637$-$00.8043}
\label{sec:G094.4637}
\citet{Smith1992} used near-infrared line and continuum observations to show that activity from multiple jets is likely present in this region considering the wide range in outflow axes seen in their H$_2$, $2.12\um$ images. More recent H$_2$ line observations \citep{Navarete2015} showed 5 distinct regions of emission, with what appears to be 2 bipolar outflows at position angles of $355/180\degr$ (1a, 1b) and $210/30\degr$ (2a, 2b), and a monopolar outflow (3a) at a position angle of $-60\degr$. The monopolar outflow is also aligned with the obvious NE/SW bipolar reflection nebula in NIR images, and the 2a/2b outflow was aligned with another, fainter, NIR reflection nebula (see left panel of \autoref{cplot:G094.4637}).

In standard images with a robustness of $0.5$ (\autoref{cplot:G094.4637}), source A at C-band is coincident with A and A2 at Q-band, with an average spectral index (combining the two Q-band fluxes) of $\bar{\alpha}=0.65\pm0.10$, typical of ionised jets. However considering their separation of $1090\pm170\au$ ($202\pm2\mas$ at $5.36\pm0.84\kpc$) and current theoretical ranges of jet launching radii between $10-100\au$, we believe that A1 and A2 are close binaries. Imaging the C-band data with a more uniform robustness of $-2$ (increasing the effective resolution) shows that C-band source A is most likely the same source as Q-band source A, while A2 appears more separate from the C-band lobe. Using this approach, A's spectral index is derived to be $0.39\pm0.09$, while for A2 it is calculated to be $\alpha>1.61\pm0.09$ (calculated with C-band flux's $3\sigma$ upper-limit). Both A and A2 are roughly aligned with components B, D and E at a position angle of $\sim45\degr$, for which we calculate spectral index upper limits of $<0.0$, $<0.4$ and $<0.5$ respectively. Component C is detected at both bands and has a jet-like spectral index of $0.86\pm0.09$, and although \textsc{imfit} was unable to deconvolve any sizes, at Q-band the emission looks slightly elongated along a position angle of $\sim10\degr$. 

Comparing our results to those of \citet{Navarete2015} who detect three distinct directions of collimated outflow, we detect 3 heavily thermal components (A, A2 and C). Although no definitive spectral indices could be deduced for B, D or E, we believe that due to their alignment with A, they represent shocked surfaces of optically thin and/or non-thermal emission from a thermal jet ejected at a position angle of $\sim45\degr$ from A. The H$_2$ bipolar outflow components 2a and 2b are aligned along this axis and therefore likely more distant jet-shock features. A2 is likely a close binary to A, though its possible association with an outflow is unknown. The thermal jet at C also has a $4\sigma$ component to its WNW, in line with the monopolar outflow 3a, however without more information, we can not be sure as to C's exact nature. We therefore classify A to be a jet with lobes and both A2 and C as jet candidates.

\subsubsection{G094.6028$-$01.7966}
\label{sec:G094.6028}
More popularly known as V645 Cygni, this object has a rich observational history. Notably \citet{Murakawa2013} detected Br$\gamma$ emission with a P-Cygni profile and a blue absorption feature in $\mathrm{HeI}$ strongly shifted in velocity by $-800\kmps$. This implies fast moving gas in the line of sight towards us. Modelling CO bandhead emission detected during their K-band spectroscopic analysis showed that the accretion disc in the system was almost pole-on, reinforcing the theory that a stellar wind/ionized jet was oriented towards the observer. Previous to this, a bipolar CO $J=3-2$ outflow was detected by \citet{Schulz1989}, centred on V645 Cygni.

At both C-band and Q-band we detect a compact source, `A', at the pointing centre. Methanol maser emission is also seen coincident with A.  Although no physical dimensions could be deconvolved, we derive a spectral index for the emission of $0.49\pm0.08$. Considering the observational history and compatibility with either a disc-wind, or a thermal jet, we classify this object as a jet candidate.

\subsubsection{G100.3779$-$03.5784}
\label{sec:G100.3779}
This YSO was observed by \citet{Moscadelli2016} at frequencies of $6.2$, $13.1$ and $21.7\GHz$ with the VLA in its A-configuration. They detect a compact source which is slightly resolved at the upper two frequency bands with a spectral index of $\alpha=0.84\pm0.25$. Previously \citet{Anglada2002} also observed the same object (their VLA2, which also had a NW extension to its emission), as well as another radio lobe $\sim10\arcsec$ to the south (VLA1). In UKIDSS imagery, a red source is coincident with the previous radio detection, which appears extended in the N-S direction. 

At C-band we detect two sources, which we designate A and B. Source A coincides with VLA2 from \citet{Anglada2002} and therefore the same radio source from \citet{Moscadelli2016}, for which we derive a spectral index of $\alpha=0.79\pm0.12$, in line with that estimated previously. Combining our data with that from \citet{Moscadelli2016}, we derive a more accurate spectral index of $\alpha=0.80\pm0.10$ (neglecting their C-band result due to poor image quality). Although no deconvolved dimensions could be established, source A looks extended in an E-W direction in both C and Q-band images in agreement with $3.6\cm$ morphology seen by \citet{Anglada2002}. Source B, which is extended at C-band, is not detected likely owing to an optically thin spectral index and/or the loss of flux with the smaller synthesised beam. Due to the near IR colours presented in UKIDSS RGB images, we classify B as an \textsc{Hii} region. Although jet-like, without morphological information source A is classified as a jet candidate.

\subsubsection{G102.8051$-$00.7184}
\label{sec:G102.8051}
The RMS survey catalogues three red MSX sources, G102.8051$-$00.7184A, B and C, within $15\arcsec$ of eachother, with bolometric luminosities of $2300\Lsol$, $2300\Lsol$ and $1300\Lsol$ respectively. Four millimetric sources were found by \citet{Palau2013}, with their MM2 being the only possibly massive ($2-10\Msol$) core in the region ($22\rahr19\ramin08.974\rasec$, $56\degr05\arcmin02.97\arcsec$) and the only one driving a CO$(2-1)$ bipolar outflow (position angle of $-20\degr$). \citet{Fontani2004} detected a CO$(1-0)$ outflow at a position angle of $\sim10\degr$ with the NE blue lobe coincident with both [FeII] and H$_2$ line emission indicative of shocks from protostellar outflows. GLIMPSE imagery shows a reddened, extended source with all the typical characteristics of an \textsc{Hii} region approximately $15\arcsec$ NE of G102.8051$-$00.7184B and with a luminosity of $660\Lsol$ (from the RMS survey).

At C-band we detect one source, however it is not coincident with any reddened, MIR source in GLIMPSE, with colours (in near and mid-infrared images) more attributable to a more evolved phase. Further to this, no corresponding source was detected at Q-band, establishing an upper limit to the spectral index of $<0.61\pm0.15$.

\subsubsection{G103.8744+01.8558}
\label{sec:G103.8744}
Previous PdBI (2mm) and VLA observations (multiple frequencies) detect 6 mm-cores associated to the IRAS source (22134+5834), one of which (MM2) is coincident with a detected UC\textsc{Hii} region, VLA1 \citep{Wang2016} which was calculated to be powered by a B1 ZAMS type star. Of the 6 detected cores, MM1 was both the brightest and heaviest with a 2mm flux of $9.3\pm2.0\mJy$ and inferred mass of $6.1\pm1.3\Msol$.

At C-band, we detect 5 sources which we label A, B, C, D and E, of which only A is detected at Q-band. In terms of C-band radio flux, A is the brightest and a spectral index of $\alpha=-0.42\pm0.12$ is derived for A, however we believe that this is due to resolving out effects at Q-band. This is supported by the decrease in deconvolved major axis length with frequency (\textsc{Hii} regions should not change in size). From the optically thin \citep[verified by the previous observations of][]{Wang2016} $5.8\GHz$ emission, the models of \citet{Davies2011} predict a bolometric luminosity of $4200\Lsol$. In comparison to the infrared derived bolometric luminosity of $6800\Lsol$, we believe that the difference of $2600\Lsol$ is supplied by the other 5 cores in the vicinity, most of which comes from MM1, which is coincident with our source B. Our only other detected radio source with a corresponding mm-core is C which is associated to MM4 of \citet{Wang2016}. From the general morphology, we believe D, E and F (which have no IR or mm counterparts) are shock sites whereby ejected material from B is impacting the surrounding dust/gas. We classify B as a jet with lobes on the basis of its elongated morphology along a position angle of $114\degr$ which is aligned with the string of lobes, D, E and F. Although source C may be a YSO, however its exact classification is unknown and we therefore classify it as a jet candidate. 

\subsubsection{G105.5072+00.2294}
\label{sec:G105.5072}
\citet{Molinari2002} detected a $0.12\pm0.03\mJy$ $3.6\cm$ radio source (VLA1) coincident with a `ring' of HCO$^+(1-0)$ line emission towards the MSX source. A CO $J=2-1$ molecular outflow has also been observed whose emission peaks $\sim10\arcsec$ to the NW of the MSX position \citep{Zhang2005}. A definitive position angle for the outflow is difficult to establish with the red lobe elongated at a position angle of $\sim90\degr$ and the blue lobe at an angle of $-160\degr$. H$_2$ 2.122$\um$ observations by \citet{Varricatt2010} showed three distinct patches of shock emission aligned along a position angle of $\sim15\degr$ and separated from the infrared source by $20\arcsec$ to the NNW.

One, elongated radio source is detected at both frequencies (A) with values for $\alpha$ and $\gamma$ of $1.02\pm0.13$ and $-1.05\pm0.44$ respectively. Deconvolved position angles for the major axis agree at both frequencies with a value of $\sim110\degr$ parallel with the NIR reflection nebulae of the outflow cavities apparent in the left panel of \autoref{cplot:G105.5072}. Confusingly this position angle does not align well with the H$_2$ emission or blue CO $J=2-1$ lobe, but is parallel with the elongated red CO $J=2-1$ emission. Assuming the radio source is a jet, it is possible that the outflow axis has precessed towards the west considering the trail of H$_2$ emission and the elongation of the radio source at $110\degr$. However, further observations will be needed to clarify this picture and A is assigned a classification of jet candidate.

\subsubsection{G107.6823$-$02.2423A}
\label{sec:G107.6823}
Previous near-infrared spectroscopic observations detected both Br$\gamma$ and [Fe\textsc{ii}] emission \citep{Cooper2013} with a slit positioned over both G107.6823$-$02.2423A and its neighbouring \textsc{Hii} region G107.6823$-$02.2423B. While the strong Br$\gamma$ is attributable to the \textsc{Hii} region, the $1.64\um$ [Fe\textsc{ii}] emission is a consequence of shocked material, possibly attributable to jet activity. Diffuse $2.122\um$ H$_2$ emission is detected by \citet{Navarete2015} over ranges in position angle from the central source of $210-240\degr$ and $250-360\degr$.

Coincident with two NIR 2MASS sources are two C-band detections, one of which is extended and resolved out (\textsc{Hii}, coincident with G107.6823$-$02.2423B) and the other (A) which is located at the pointing centre and is also detected at Q-band. For the latter we derive a spectral index of $\alpha=1.15\pm0.21$, however the source is point-like at all frequencies (which may be due to the low SNR or it possessing a true, unresolved nature). Due to the lack of further information, this source is classified as a jet candidate.

\subsubsection{G108.1844+05.5187}
\label{sec:G108.1844}
One of the nearest objects in our sample at a distance of $0.776^{+0.104}_{-0.083}\pc$ \citep{Rygl2010}, \citet{Beltran2006} detected a $\mathrm{CO}$ outflow at a position angle of $140\degr$ centred on their source, OVRO 2 (mass of $14.2\Msol$), with the (weak) red lobe towards the SE and (strong) blue to the NW. \citet{Surcis2013} used polarimetric, VLBI observations to detect 29 methanol masers aligned along the same position angle ($145\pm11\degr$) as the molecular outflow and derived a magnetic field position angle of $9\pm15\degr$. For this source we adopt a bolometric luminosity of $873\Lsol$ based on the luminosity found by \citet{Sugitani1989} of $1100\Lsol$, corrected for the more recent distance found by maser parallax \citep{Rygl2010}.

Towards the mm-emission of previous observations, we detect source A to be coincident with OVRO 2 from \citet{Beltran2006}. The emission is elongated along a position angle of $42\pm18\degr$ and $42\pm7\degr$ at C (which is embedded in diffuse emission) and Q-band respectively, giving an offset to the magnetic field position angles of $33\pm18\degr$. Considering that the radio emission is almost perpendicular ($82\pm18\degr$) to the outflow, which itself is significantly offset to the magnetic field direction, a confusing picture is established. However, given the overwhelming evidence for an outflow from source A (and the elongation of a reddened 2MASS source along the outflow axis), we believe that the Q-band emission is tracing a disc of dimensions $(61\pm12)\times(12\pm8)\au$. We therefore conclude that source A traces an ionized jet (candidate) at C-band, but is dominated by disc emission at Q. 

Source B is not coincident with any near or mid-infrared emission, was not detected at Q-band (though was located far out in the primary beam) and does not display any mm-emission at all. Considering these facts, we conclude it to be extragalactic in origin.

\subsubsection{G108.4714$-$02.8176}
\label{sec:G108.4714}
\citet{Navarete2015} detect a bipolar H$_2$ outflow at a position angle of $15\degr$ centred on a reddened 2MASS point source (left panel of \autoref{cplot:G108.4714}). 

We detect one component at both bands, centred on the red 2MASS source, with a spectral index of $\alpha=0.55\pm0.16$. While not at C-band, it is resolved at Q-band ($\gamma>-0.77\pm0.35$), with a major axis aligned along a position angle of $\theta_\mathrm{PA}=101\pm25\degr$, perpendicular to the H$_2$ outflow's direction. With the spectral index indicative of a typical ionised jet or disc wind, we classify this as a disc wind on the basis of its Q-band elongation perpendicular to the established outflow. It is worth noting that the C-band image shows a slight elongation at a position angle parallel with the H$_2$ bipolar outflow.

\subsubsection{G108.5955+00.4925A}
\label{sec:G108.5955}
Associated to the infrared source, IRAS 22506+5944, the RMS survey lists two more MYSOs within $60\arcsec$ of G108.5955+00.4925A, being G108.5955+00.4925B and G108.5955+00.4925C, owing to this being a cluster of at least 15 members \citep{Kumar2006}. GLIMPSE imagery shows diffuse $8\um$ emission centred on G108.5955+00.4925C, characteristic of a \textsc{Hii} region, however it is categorised as an MYSO based on NIR spectral features \citep{Cooper2013}. G108.5955+00.4925A itself is centred on a green (but not extended) MIR object and G108.5955+00.4925B is centred on a reddened MIR compact source. Previous $\mm$ observations \citep{Su2004} detected a $3\mm$ core, which appeared to be driving a bipolar CO outflow at a position angle of $\sim90\degr$ but was not positioned over any of the three RMS MYSOs. 

Three C-band radio sources were identified from the radio clean maps, however we did not detect any at Q-band. In light of the RMS survey's naming schemes, we dub them B, C (G108.5955+00.4925B and G108.5955+00.4925C respectively) and D (not in the RMS database). All are coincident with mid-infrared sources, however D's source is not reddened and therefore is likely of a more evolved evolutionary status. Our source B has a very low flux ($30\pm8\uJy$), and therefore a non-restrictive upper limit to its spectral index of $\alpha<0.73\pm0.14$. Without more information, we classify B to be a jet candidate. On the other hand, C is classified as a \textsc{Hii} region due to its extended morphology, coincidence with a diffuse MIR source and radio flux which matches that expected of a \textsc{Hii} region with a bolometric luminosity of $2700\Lsol$ \citep{Davies2011}, which agrees with the $3000\Lsol$ derived from infrared SED fitting. A CH$_3$OH maser was detected at $\alpha\,(\mathrm{J2000})=22\rahr52\ramin38.3110\rasec, \delta\,(\mathrm{J2000})=60\degr00\arcmin51.885\arcsec$ (with positional uncertainties of $7$ and $9\mas$ in $\alpha$ and $\delta$ respectively) but was not coincident with any RMS source, or infrared/radio continuum source. However, it was coincident with the $3\mm$ core/CO outflow detected by \citet{Su2004}. It is therefore likely that this maser reveals the position of a deeply embedded, relatively unevolved MYSO, considering the already clustered environment towards this source.

\subsubsection{G108.7575$-$00.9863}
\label{sec:G108.7575}
While GLIMPSE images are completely saturated, 2MASS (left panel of \autoref{cplot:G108.7575}) shows a reflection nebula centred on G108.7575$-$00.9863 with its diffuse emission extending over a position angle range of $225-280\degr$, with respect to the MYSO which lies at the heart of a cluster of $38$ members \citep{Chen2009}. NIR observations \citep{Cooper2013} show a relatively featureless spectra, with only a weak Br$\gamma$ line present. Approximately $130\arcsec$ to the NNW is the classical \textsc{Hii} region, Sh2-152, which harbours a cluster and is a well studied object. \citet{Navarete2015} detect H$_2$ 2.122$\um$ emission dispersed widely over a wide area, including diffuse/knotted emission towards Sh2-152 and bipolar outflows centred on G108.7575$-$00.9863 and to the W and ESE of it. No obvious driving force can be determined for the source of the shocking material, though the obvious candidate is G108.7575$-$00.9863, especially due to its coincidence with the bipolar H$_2$ emission, BP1. 

We report the detection of 5, compact C-band sources within $1\arcmin$ of the pointing centre (i.e.\ G108.7575$-$00.9863). These are labelled A$\rightarrow$E, of which A and B have near-infrared counterparts \citep[NIRS 172 and NIRS 182 from][respectively]{Chen2009}. Source A is also situated on a heavily saturated GLIMPSE source, is the only detection at Q-band within the primary beam, possesses a spectral index of $0.94\pm0.09$ and remains unresolved at all frequencies. Without further information, we consequently classify it to be a jet candidate. Due to B's reddened near infrared colours, it is determined to be a cluster member though its evolutionary status remains ambiguous without further information. Source C through E all suffer from the same classification issues of non-restrictive, thermal upper limits to spectral indices and lack of information at other wavelengths and we therefore classify them all to be of unknown nature. A methanol maser coincident with small 3$\sigma$ source $1.5\arcsec$ to the north of E is also detected suggestive of a well-embedded YSO. 

\subsubsection{G110.0931$-$00.0641}
\label{sec:G110.0931}
K-band images from 2MASS show diffuse emission to the south-east, east and north-east of a bright point source centred on G110.0931$-$00.0641's position. A bipolar H$_2$ $2.122\um$ outflow is detected at a position angle of $125\degr$ and centred on the MYSO \citep{Navarete2015}. Radio observations conducted in 2007 at $3.6\cm$ ($8.33\GHz$) using the VLA in its A-configuration detected 3 lobes of emission arranged along an axis at a position angle of $\sim110\degr$ \citep{Rodriguez2012b}. \citet{LopezSepulcre2010} detect a HCO$^+(1-0)$ outflow at a rough position angle of $45\degr$ and elongated emission in C$^{18}$O$(2-1)$ (tracing dense material) perpendicular to it. Both types of emission were centred on G110.0931$-$00.0641.

At C-band we detect 5 radio components named A1 \citep[VLA3 from][]{Rodriguez2012b}, A2, B \citep[VLA2 from][]{Rodriguez2012b}, C \citep[VLA1 from][]{Rodriguez2012b} and D, all of which are within $3\arcsec$ of the pointing centre. Both B and C are detected at Q-band and B's morphology is elongated in the direction of C. Spectral indices were therefore only established for B and C, and were calculated to be $0.34\pm0.14$ and $-0.08\pm0.16$ respectively, indicative of optically thin free-free emission. Components A1, A2, B and C are all approximately aligned on a position angle of $110\degr$, while D is located $2.5\arcsec$ to the NE of A1. 

Under the assumption that A1, A2, B and C are all \textsc{Hii} regions, the sum of the radio flux-inferred bolometric luminosities of $(1.2\pm0.1)\times10^4\Lsol$, is slightly under-luminous for that derived from SED fits of infrared data ($1.7\times10^4\Lsol$). However, B is optically thick and therefore under-luminosity is expected, meaning that a quadruple system of neighbouring \textsc{Hii} regions is still possible at this point. Considering the alignment with previously established outflows, proper motions and changes in physical size from C to Q-bands, an alternative explanation is that of a radio jet/lobe system. Considering the high-positional accuracy quality code of the phase calibrators used at C-band (J$2230+6946$) and Q-band (J$2250+5550$), for the optically thin radio lobe, C, any positional change between frequencies should be solely due to proper motions (which we do not expect to see in the \textsc{Hii} region case). For C, an angular shift of $67\pm3\mas$, at a position angle of $283\pm5\degr$ (along the axis joining the lobes), is deduced from C to Q-bands corresponding to a proper motion of $1092\pm226\kmps$ (using a distance of $4.70\pm0.82\kpc$ and $\Delta t=508\pm52$ days). This is typical of observed velocities for radio lobes detected towards MYSOs and serves to strengthen the case that C is an optically-thin, shock-ionized lobe. On the basis of the excellent agreement of spectral indices with the models of \citet{Reynolds1986}, proper motions observed towards C and the general alignment of the radio lobes with outflows, we classify B as a radio jet with lobes (A1, A2 and C). As for D, in light of the \textsc{Hii} region-like infrared K-band morphology, we believe it is an extended \textsc{Hii} region whereby resolving out is starting to affect recovered flux/morphology at C-band and resulted in a non-detection at Q. 

\subsubsection{G111.2348$-$01.2385}
\label{sec:G111.2348}
Near-infared (2MASS) images show a point source embedded in a reflection nebula extending to the east of it. \citet{Beuther2002} detected a CO $(2-1)$ molecular outflow oriented east to west. Previously, this source was observed by the VLA at both $8.44\GHz$ and $43.4\GHz$ by \citet{Garay2007b}, who observed a compact source centred on the MYSO's location, with a spectral index of $1.1\pm0.2$, and a large, cometary \textsc{Hii} region whose peak is located $15\arcsec$ to the north west. That work concluded the compact source to be a HC\textsc{Hii} region around the MYSO, which was still undergoing accretion. Later $3.6\cm$ ($8.33\GHz$) radio observations using the VLA in its A-configuration detected one radio component elongated at a position angle of $97\degr$ with a line of 8 H$_2$O maser spots roughly perpendicular to it \citep{RodriguezEsnard2014}. Sub-mm observations show two cores in the continuum ($875\um$), one centred on G111.2348$-$01.2385 and the other located $\sim2\arcsec$ to its south west, with an east-west SiO$(8-7)$ outflow centred on G111.2348$-$01.2385 \citep{Beuther2007} and another in the north east-south west direction, presumably from the other core. 

At both frequencies we detect one radio lobe coincident with the NIR point source. A spectral index of $1.10\pm0.07$ \citep[in agreement with that calculated by][]{Garay2007b} is calculated, and interestingly the deconvolved dimensions of the emission show it to be elongated at a position angle of $173\pm9\degr$ at Q-band (unresolved at C-band). This position angle is perpendicular to that derived by previous $8.33\GHz$ observations of $97\degr$ \citep{RodriguezEsnard2014} which, coupled with the steep spectral index, suggests that at Q-band, dust emission starts to dominate over the (perpendicular) ionized component. However, contrary to this, if we take fluxes from the variety of frequencies present in the literature, the spectral index remains steady across all bands. Considering the history of previous observations, an ionized jet may be located at the single radio component we detect, however it is equally likely the emission stems from a HC\textsc{Hii} region. Without further radio information, we can not establish the dust's flux contribution and therefore can not elucidate the nature of the radio object further. Consequently, we assign the classification of jet candidate to this object.

\subsubsection{G111.2552$-$00.7702}
\label{sec:G111.2552}
Radio observations at $8.66\GHz$ \citep{Tofani1995,Sridharan2002} imaged a compact source at the position of G111.2552$-$00.7702 with time-variable flux. Later, higher resolution observations \citep{Trinidad2006} showed the previous compact source (labelled I23139) and an unresolved $3.5\cm$ source $0.5\arcsec$ to its SSW. I23139 was deduced to be an ionized jet on the basis of its spectral index ($0.64\pm0.36$) and masers tracing an outflow \citep[][who derived maser proper motions along a position angle of $\sim-70\degr$]{Goddi2005}. \citet{Varricatt2010} showed that the MYSO is centred on a K-band source which is resolved as a binary (separation of $0.4\arcsec$) in their images. Further to this they also detected a H$_2$ $2.122\um$ knot at a position angle of $65\degr$ from the MYSO. A molecular outflow was detected in CO $(2-1)$ but with no discernible outflow axis since the red and blue lobes lay on top of eachother \citep{Beuther2002}, suggesting a head on orientation. 

Within $1\arcmin$ of the pointing centre, we detect 5 compact, C-band radio objects, which we label A1, A2, B, C and D. Component A1 has a Q-band counterpart with a spectral index, between the two frequencies, of $\alpha=0.16\pm0.12$. The spectral index may be steeper however considering A2's close proximity at C-band which, when using \textsc{imfit}, can lead to flux being wrongly allocated from A2 to A1. Considering the spectral index previously found \citep[$\alpha=0.64\pm0.36$,][]{Trinidad2006}, this may indeed be the case. Because A1's major axis and A2's $\theta_\mathrm{PA}$, with respect to A1, is oriented at a position angle parallel to that of maser proper motions \citep{Goddi2005} resulting from an outflow, we suggest A2 is a shock-excited lobe of emission, with the jet source located at A1. We therefore classify A1 as a thermal jet with lobes.

A CH$_3$OH maser is also detected offset from the continuum at C-band by $0.13\arcsec$ at a position angle of $-48\degr$. Components B and C appear to have near-infrared counterparts from 2MASS images, while D does not. However, because the spectral indices of these three components is not constrained, we must assign them an unknown classification. Further to those sources reported above, a $4\sigma$ source is detected $0.6\arcsec$ SW of A1 and coincident with I23139S (indicated to the south west of A1 by a $3\sigma$ contour in \autoref{cplot:G111.2552}). With an \textsc{imfit} derived flux of $15.5\pm7.6\uJy$ and in combination with the flux at $8.6\GHz$ derived by \citet{Trinidad2006}, we calculate a spectral index of $6.8\pm1.5$, which is clearly not realistic. We therefore propose that I$23139$S is another time variable (over the $11 \yr$ period between their observations and ours) object.

\subsubsection{G111.5671+00.7517}
\label{sec:G111.5671}
One of the best examples of a NIR reflection nebulae is seen in 2MASS images of this source (also known as NGC7548$-$IRS9), which extends towards the south west of the redenned MYSO. Previous $3.6$ and $1.3\cm$ radio observations by \citet{SanchezMonge2008} detected two compact radio components at $1.3\cm$ associated to NGC7548$-$IRS9. While their VLA3 is coincident with G111.5671+00.7517, VLA2 was offset approximately $5\arcsec$ to the west, yet both sources were coincident with K-band NIR (2MASS) point sources. At $3.6\cm$, their images were affected by a bright, extended \textsc{Hii} region's sibelobes from the NW resulting in a non-detection. \citet{Navarete2015} resolved a bipolar H$_2$, $2.122\um$ outflow at a position angle of $\sim155\degr$ with respect to the MYSO, and another in an east-west direction. Sub-mm observations of HCO$^+$ and CO $(1-0)$ have shown two outflows, one coincident with the H$_2$ jet and driven by VLA2 \citep{SanchezMonge2008} and the other associated to VLA3 and driven along a position angle of $\sim70\degr$.

In our C-band images, we detect 5 radio components within $60\arcsec$ of G111.5671+00.7517, labelled, in order of distance from the centre, A (coincident with VLA3), C, B2, B1 and D. Of these 5 components, only A has a NIR or Q-band counterpart. The Q-band counterpart of A is distinctly elongated at a position angle of $19\degr$, pointing towards C, and has a spectral index of $0.87\pm0.06$. Since there are no NIR source associated with C, B1 or B2 by inspection of the high-resolution ($\theta=0.4\arcsec$) RGB, J, H and K-band images of \citet{Mallick2014}, we believe them to be shock-ionized lobes, especially given their optically thin spectral indices. From the morphology of the emission, we believe the jet is launched and collimated at A. On this basis we therefore classify A to be a jet with lobes. Also, considering the position angles of lobes B1, B2 and C with respect to A (in order of separation) of $214\degr$, $205\degr$ and $21\degr$ respectively, in conjunction with the wide angle reflection nebula at K-band, the molecular outflow at $70\degr$ and the current jet's major axis defined by the Q-band data ($19\degr$), we believe A's jet axis to be undergoing rapid, clockwise (on the plane of the sky) precession. Assuming a jet velocity of $500\kmps$ and inclination of $90\degr$ (i.e.\ jet lies in the plane of the sky) this indicates a shift of $12.5\degr$ over $23.2\yr$, or a precession rate of $\sim0.5\degr\,\yr^{-1}$. This would also explain why the molecular outflow being driven from A is outflowing at a position angle of $70\degr$. As for source D, because of its non-detection and location on the edge of the Q-band's primary beam, a non-restrictive upper limit of the spectral index was deduced ($\alpha<0.26\pm0.11$) and therefore we classify as unknown in nature.

\subsubsection{G114.0835+02.8568}
\label{sec:G114.0835}
Being a relatively unstudied object within our sample, the only relevant observations are those of \citet{Navarete2015} who observed a bipolar H$_2$ outflow along a rough north-south axis and centred on the MYSO's position. In near infrared RGB images, the MYSO is centred on an extended (in the north west-south east direction), reddened source. 

Only C-band observations of this field were observed in which we detect 5 radio sources, labelled A1, A2, B, C and D, of which A1 and A2 are spatially coincident with G114.0835+02.8568 and the only sources associated to a NIR point source. Component B looks spatially extended/resolved out and lies within the K-band nebulosity seen in 2MASS images, whereas C (the strongest source in the field) and D are located distinctly away from the central MYSO. For A1/A2, because of the lack of Q-band images, spectral indices can not be established and therefore either component could be a YSO, or a lobe of shocked emission. Without further information, we can not constrain the natures of the radio objects and therefore assign them all with an unknown classification.

\subsubsection{G118.6172$-$01.3312}
\label{sec:G118.6172}
Aside from the usual surveys of the whole galactic plane, no observations towards this source are present in the literature. In 2MASS near-infrared images, the MYSO presents itself as a red source in a cluster of other, slightly less reddened, point sources. GLIMPSE images are verging on saturation at $8\um$ over the MYSO's position, with diffuse $8\um$ (presumably PAH emission) to the north and north east spread across a wide area.

Unusually we only detect one source at Q-band which is not detected at C-band, therefore we establish a lower limit to the spectral index of $>1.59\pm0.14$, which is unusually high for the sample in general. The radio source is just resolved with dimensions of $48\times21\mas$, corresponding to $134\times56\au$, with the major axis oriented at a position angle of $15\degr$. However, the errors on these derived quantities are large due to the low SNR of this source. Considering the lack of any data at other wavelength regimes, as well as at C-band, we classify this as a jet candidate.

\subsubsection{G126.7144$-$00.8220}
\label{sec:G126.7144}
G126.7144$-$00.8220 is situated at the centre of a bipolar, near-infrared nebula aligned at a rough position angle of $-10\degr$, typical of outflow cavities. A $2.122\um$ H$_2$ bipolar outflow centred on the MYSO and driven along a north-south axis $\theta_\mathrm{PA}\sim-10\degr$ was observed by \citet{Navarete2015}. The same work also detected knots of H$_2$ emission to the west and north west. No other relevant, high-resolution observations exist in the literature for this object.

Centred on the reddened 2MASS object, we detect an elongated source at C-band, which splits into two Q-band sources (A and A2) separated by $0.15\arcsec$ ($\sim110\au$ at the RMS survey's distance of $0.7\pc$). We calculate an overall spectral index of $0.83\pm0.05$ using the C-band flux for A and the combined fluxes of both sources at Q-band. For this object, two scenarios fit the radio data we have obtained. The first scenario is that the Q-band sources are a close binary of two YSOs, one of which is driving the previously detected $2.122\um$ H$_2$ outflow. On the other hand, due to their combined, jet-like spectral index and small separation, this could in fact be a biconical jet, whereby we are seeing the direct emission from the launching site of the jet and counter-jet, with a launching radius of $55\au$ (assuming symmetry). This second scenario is supported by the alignment of the two sources with the H$_2$ outflow, however it must be conceded that this may be a chance alignment. With the equal likelihood of both scenarios, we classify both objects as jet candidates.

\subsubsection{G133.7150+01.2155}
\label{sec:G133.7150}
More commonly associated with the alias W3 IRS5, this is an extremely well studied object. Previous radio observations at K and Q-bands by \citet{VanDerTak2005b} detected 5 sources (labelled Q$1\rightarrow$Q$5$) at Q-band and 8 sources (labelled K$1\rightarrow$K$8$) at K-band. Of these, Q1/K2, Q2/K3, Q3/K4, Q4/K6, Q5/K7 and Q6/K5 were determined to be the same Q-band/K-band sources, with ranges in thermal spectral indices of $0.5 \leq \alpha \leq 1.4$ determining them to by YSOs and therefore highlighting the tight clustered environment. Using speckle MIR imaging, the same work determined Q3/K4, Q4/K6 and Q5/K7 to have mid-infrared counterparts. In comparison with previous radio observations \citep[][the latter of which also detected proper motions of $\sim130\kmps$ for K8 and Q2/K3]{Tieftrunk1997,Wilson2003}, Q2/K3, Q3/K4, Q5/K7 and K8 showed flux variability over time. PdBI $1.3/3.4\mm$ imaging by \citet{Rodon2008} detected 6 mm-cores towards W3 IRS5, separated by $<2\arcsec$, MM1$\rightarrow$6, of which MM1 was coincident with Q5/K7, MM2 with Q3/K4, MM3 with Q1/K2 (or possibly Q2/K3) and MM6 with Q4/K6. Simultaneous SiO ($2-1$) and ($5-4$) line imaging was conducted identifying 5 outflows, one driven by MM1 (SIO-a) in a north-east/south-west direction, one by MM2 (SIO-b) which is head on and one by MM4 (SIO-c) aligned east to west, with the other outflows not having any identified driving source. Line emission from SO$_2$ ($22_{2,20}-22_{1,21}$) showed overall rotation with a velocity gradient over the whole cluster at a position angle of $\sim130\degr$, with the blue shifted gas towards the south-east. Diffuse 2.122$\um$ H$_2$ emission containing knots was also detected towards this region \citep{Navarete2015}, though no specific position angle for this emission was given.

Imaging at C-band was dominated by 4 large, extended and bright \textsc{Hii} regions to the West, East and South of the pointing centre. Consequently a uv-range of $>60\mathrm{k\lambda}$ and robustness of -1 were employed to minimise the detrimental effects of bright, extended emission close to the science target.

\begin{figure}
\centering
\includegraphics[width=84mm]{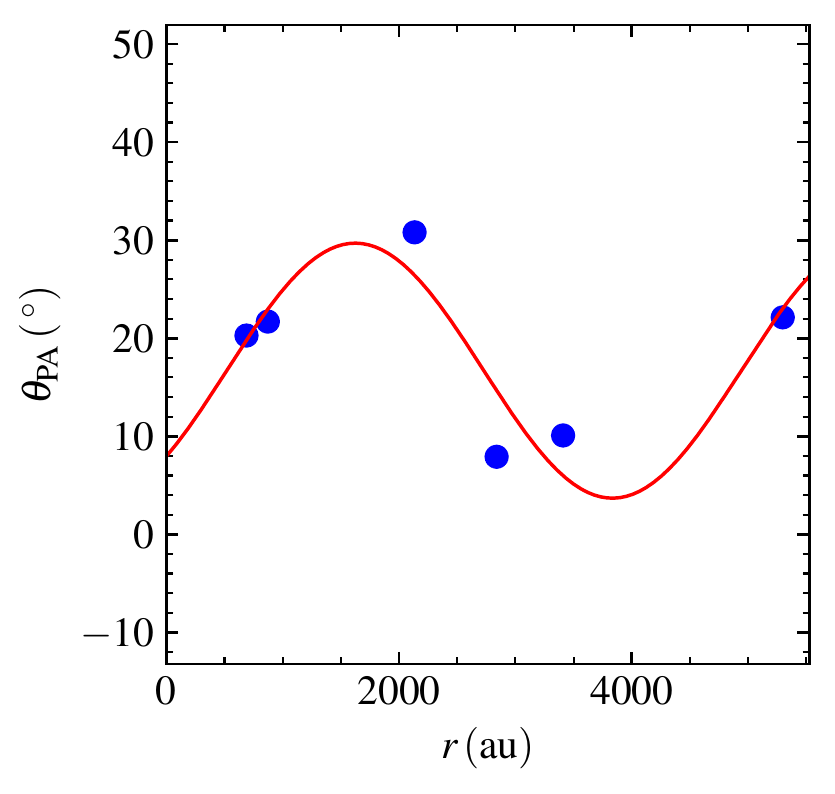}
\caption[Relative separations of shock-ionised lobes from the central thermal jet of G133.7150+01.2155]{A plot of the radii and position angles of lobes Q4b, Q7, Q8, Q9, K8 and C1, with respect to Q5/K7, for the MYSO G133.7150+01.2155. Least squares fitting of these points is shown in red, which represents a simple precession model.}
\label{fig:g133.7150rvspa}
\end{figure}

At C-band we detect emission (left panel of \autoref{cplot:G133.7150}) from 7 previously observed lobes \citep{VanDerTak2005b} labelled Q1/K2, Q2/K3, Q3/K4, Q4/K6, Q5/K7 (MYSO), Q6/K5 and K8, as well as 5 new lobes labelled, Q4b, Q8, Q9, Q10 and C1 (in accordance with previous naming systems) in the immediate vicinity (within $3\arcsec$) of Q5/K7. At Q-band (right panel of \autoref{cplot:G133.7150}) we detect emission coincident with C-band positions for Q2/K3, Q3/K4, Q4, Q5/K7, Q6/K5, Q7 and Q8. 

Due to its elongated morphology, jet-like spectral index ($\alpha=0.85\pm0.07$) and positioning relative to its associated lobes parallel to the SIO-a outflow, Q5/K7 is determined to be a thermal jet with lobes. Shock emission as a result of the jet from Q5/K7 impacting on surrounding material are present in lobes Q4b, Q7, Q8, Q9, K8 and C1. Due to a range in position angle with respect to Q5/K7 we estimate the precession angle, and period, using least squares fitting of a simple, sinusoidal, precession model to the positions (with respect to Q5/K7) for Q4b, Q7, Q8, Q9, K8 and C1 (\autoref{fig:g133.7150rvspa}). The inclination and velocity were assumed to be $90\degr$ (i.e. jet is perpendicular to the line of sight) and $500\kmps$ respectively. This resulted in a derived precession period of $43\pm6\yr$ and precession angle of $37\pm15\degr$, around an axis oriented at a position angle of $16\pm5\degr$. Our derived period should be taken as a lower limit if an inclination of the precession axis towards the observer exists. Interestingly the \textsc{imfit}-derived positions for both Q7 and Q8 change from C to Q band, which is likely due to proper motions if both lobes are optically thin shocks as discussed above. Considering that the C and Q-band observations were taken $\sim 2.7\yr$ apart and assuming these to be shock surfaces of the jet (from Q5/K7) upon ambient material, proper motions (in the plane of the sky) are derived to be $303\pm26\kmps$ at $154\pm1\degr$ and $353\pm30\kmps$ at $14\pm1\degr$ for Q7 and Q8 respectively \citep[adopting a distance of $1.83\pm0.14\kpc$,][]{Imai2000}, consistent with the derived precession model. These velocity magnitudes are often seen in ionised jets towards MYSOs adding to the case that these changes are proper motion based. It must be conceded that due to the extended nature of these lobes at both C and Q-bands, theses motions may in fact be due to resolving-out effects, however the facts that the two proper motions are equal in magnitude but opposite in direction, the ionized gas is optically thin and that the precession model is consistent with these positional changes adds weight to the proper motion interpretation.

From the spectral index of Q3/K4 ($0.76\pm0.08$), and the position angle of its major axis at $\theta^{44}_\mathrm{PA}=161\pm15\degr$ which points at the non-thermal lobe Q1/K2, we classify it as a jet with lobes. As for its northern neighbours, Q2/K3 and Q6/K5, we believe one of them to be driving a jet which is powering lobe Q10 due to positional alignment of the C-band emission. From the C-band images, emission appears to be present between Q2/K3 and Q6/K5, which was associated to Q2/K3 by \textsc{imfit}. We therefore believe Q6/K5 to be another jet candidate in the vicinity, while Q2/K3 is classified as a candidate jet with associated lobes, Q10 and the emission between Q2/K3 and Q6/K5. Q4/K6 possesses a steep spectral index ($\alpha=0.93\pm0.12$) and is coincident with MM6 from \citet{Rodon2008}, and we therefore classify it as a jet candidate, due to its ambiguous properties as either a jet or small HC\textsc{Hii} region.

With the exception of Q6/K5, all the Q-band sources are within a beam-width ($\sim0\farcs04$) of their Q-band counterparts from Van der Tak et al. (2005), with a mean separation of $0\farcs026$. Q6/K5 displays a proper motion of $185\pm7 \mas$, or $100\pm9 \kmps$, along a position angle of $-5\pm2\degr$. However, this latter position comparison is between K and Q-bands. This adds to the complexity of the classification for this source, reinforcing the candidate status of Q6/K5.

Outside the immediate vicinity of G133.7150+01.2155, we also detect two more C-band sources, A and B (the latter of which is also detected at Q), and two sources solely detected at Q-band, QE1 and QE2. These 4 sources have been illustrated separately in \autoref{cplot:G133.7150extrasources}. Both A and B have near-infrared counterparts in 2MASS imagery, and we determine spectral indices of $-0.56\pm0.06$ and $-0.13\pm0.14$ respectively. Considering the \textsc{imfit} derived C-band dimensions of A, this spectral index results from the loss of extended flux at Q-band. Due to its extended nature, we classify it as a UC\textsc{Hii} region powered by a B3 ZAMS type star ($\Lbol=2500\Lsol$, EM$=(5.5\pm3.1)\times10^6\pccmEM$). Source B remains point-like at all frequencies, and due to its optically thin spectral index and unreddened NIR colour we classify it as a small HC\textsc{Hii} region powered by a B3 type ZAMS star ($\Lbol=2300\Lsol$, EM$>7.4\times10^7\pccmEM$) or later. As for the Q-band only sources, QE1 and QE2, we derived steep, thermal spectral indices of $>2.05\pm0.05$ and $>1.43\pm0.07$ respectively. We believe them to be deeply embedded YSOs on account of their non-detection at near-infrared wavelengths and therefore members of the cluster associated to W3 IRS5.

\subsubsection{G134.2792+00.8561}
\label{sec:G134.2792}
In the RGB, 2MASS images of \autoref{cplot:G134.2792}, G134.2792+00.8561 is centred on a reddened, elongated (along a position angle of $\sim110\degr$) source, which itself is spatially confused with a bright, white source. At $11.6\um$, MICHELLE images show the 2MASS source to break up into 2, possibly 3, sources oriented east to west. \citet{Ogura2002} detect a HH object (HH 586) in images of H$\alpha$ emission at coordinates, $\alpha\,(\mathrm{J2000})=02\rahr29\ramin 01.1\rasec$, $\delta\,(\mathrm{J2000})=61\degr 33\arcmin 33\arcsec$ offset from the MYSO's coordinates by $6.3\arcsec$ at a position angle, $\theta_\mathrm{PA}=103\degr$.  A bipolar CO molecular outflow is also detected, centred on G134.2792+00.8561, with an outflow position angle and dynamical timescale of $-30\degr$ and $15000\yr$ respectively \citep{Lefloch1997}. Images of continuum-subtracted, $2.122\um$, H$_2$ emission, show diffuse, knotted and bipolar morphologies predominantly to the west and south-east of the central source \citep{Navarete2015}.

At the MYSO's position, we detect a barely resolved source ($167\times119\mas$ at $\theta_\mathrm{PA}= 109\pm63\degr$), labelled as A, with no corresponding detection at Q-band thereby giving an upper limit to its spectral index of $<0.34\pm0.09$. Due to the alignment of A's major axis with the Herbig-Haro object, HH 586, it is most likely a thermal jet, however without further information we classify it to be a jet candidate. Approximately $4\arcsec$ to the south of A is an unresolved, $7\sigma$ source at C-band whose established upper limit of $\alpha<0.45\pm0.11$ fails to constrain it nature. Similarly, the brightest C-band radio source is located $33\arcsec$ south of the MYSO, named C, which has an upper limit to its spectral index of $<-0.12\pm0.13$. Considering the fact that C does not have a NIR or MIR counterpart, we deduce it to be extragalactic in nature. As for B however, mid-infrared saturation of GLIMPSE images by G134.2792+00.8561 prohibits us from ascertaining its MIR profile and therefore we classify it as unknown in nature. No source was detected over the primary beam in the Q-band image.

\subsubsection{G136.3833+02.2666}
\label{sec:G136.3833}
Relevant previous observations towards this source are limited in the literature, with only millimetric and low-resolution outflow studies being prevalent. \citet{Saito2006} used high resolution mm-studies at $100\GHz$ to resolve three mm-cores within $20\arcsec$ of eachother. The most massive core, core A, is associated to the MYSO, with the other two cores located $15\arcsec$ and $20\arcsec$ to its north east. In the near-infrared, an elongated reflection nebula emanating from the central, reddened source is seen at K-band along an approximate east-west axis. A non-reddened 2MASS source is located $\sim6\arcsec$ to the west of the MYSO.

At C-band we detect four sources, A, B, C and D, of which only D is not associated to a near-infrared source. Q-band images only detect one source at the same position as the C-band source A, yielding a spectral index of $\alpha=0.92\pm0.14$. Unfortunately A remains unresolved at both frequencies and without further studies, is classified as a jet candidate. Due to B's coincidence with a unreddened 2MASS source, and absence of mm-emission \citep{Saito2006} we classify it to be a partially resolved out \textsc{Hii} region. For both C and D we can not make a definite classification and therefore assign them an unknown status.

\subsubsection{G138.2957+01.5552}
\label{sec:G138.2957}
Also known as AFGL 4029-IRS1, this MYSO's parental clump harbours a dense cluster of at least 30 B$-$type stars \citep{Deharveng1997}. Remarkably, a previous optical study of this object exists and established velocities of $500\kmps$ for the highly-inclined [\textsc{Sii}] optical jet emanating from this MYSO \citep{Ray1990}. Previous A-configuration, VLA, $3.6\cm$ observations by \citet{Zapata2001} detected a cometary UC\textsc{Hii} region (AFGL 4029-IRS2) to the south of G138.2957+01.5552, two compact sources (S and N) separated by $0.6\arcsec$ in the north-south direction (at the MSX source position) and elongated/extended emission along a position angle of $80\degr$ relative to S, which they concluded to be an ionized jet. The radio source N was determined to be a time-variable, low-mass T-Tauri star close to S. A bipolar $2.122\um$ outflow has previously been detected, comprised of arc-like eastern ($\theta_{PA}=90\degr$) and more compact western emission ($\theta_{PA}=270\degr$), both of which are separated by $\sim 12 \arcsec$ ($0.24\pc$) from the MSX source \citep{Navarete2015}.

\begin{figure*}
\centering
\includegraphics[width=5.79in]{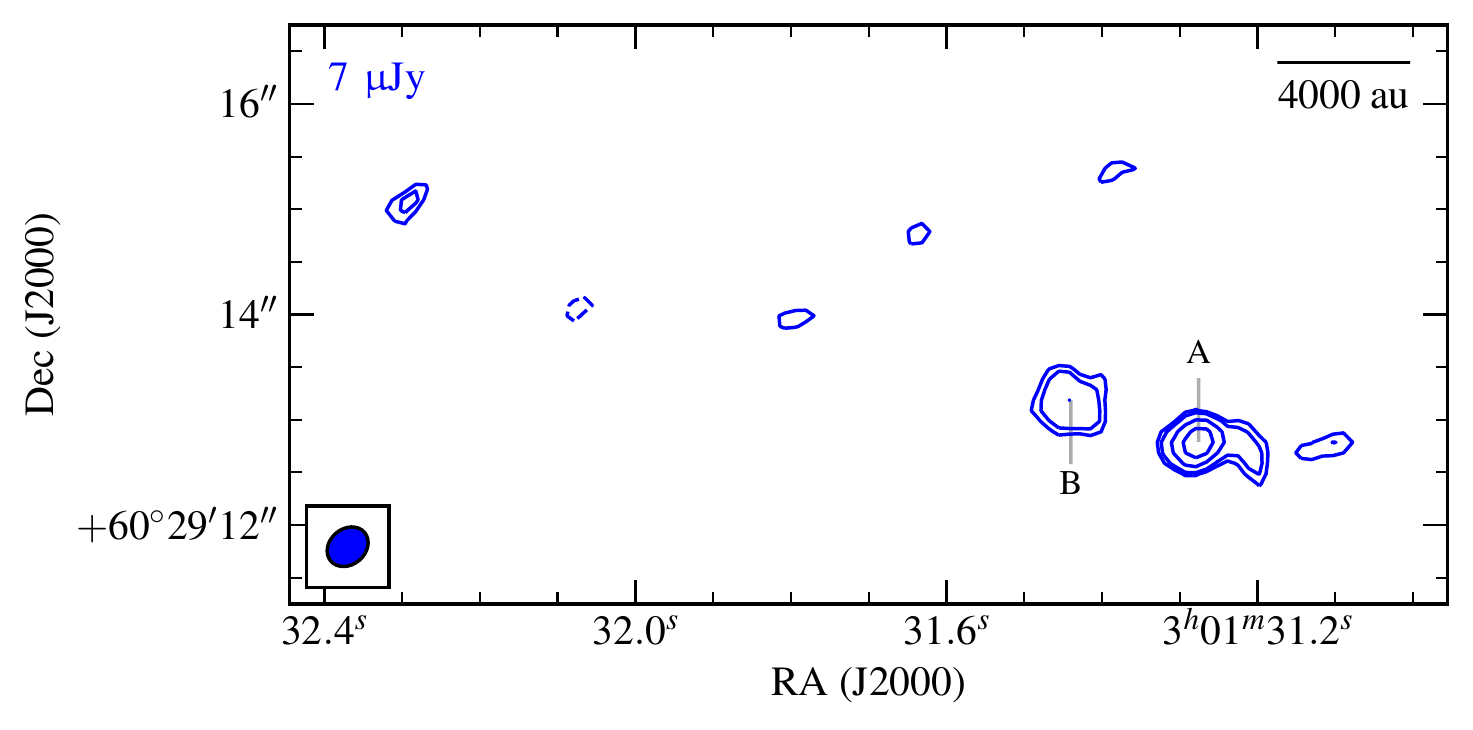}
\caption[Radio image of G138.2957+01.5552 employing a fully natural robustness]{C-band image of G138.2957+01.5552, utilising a robustness of 2 and a restoring beam of $0.421\arcsec\times0.340\arcsec$ at $-50\degr$. Contour levels are set to $(-3, 3, 4, 7, 12, 18) \times \sigma$.}
\label{cplot:G138.2957Robust2}
\end{figure*}

Our C-band radio images shown in \autoref{cplot:G138.2957} detect 6 objects within a $1\arcmin$ field of view. Near-infrared, 2MASS, reddened point sources are associated to C, D and E, meaning they are likely YSOs/T-tauri stars which are members of the cluster established by \citet{Deharveng1997}. A resolved out \textsc{Hii} region is detected, associated to a bright 2MASS source, which is the previously detected cometary UC\textsc{Hii} region. Source A is centred at the pointing centre and is the only radio object also detected in our Q-band images. It is separated from B, which lies at a position angle of $72\degr$ relative to A, by $1.3\arcsec$. Considering the alignment with the established outflow axis we determine B (which appears partially resolved out) to be a lobe of shock emission, especially given its lack of an infrared counterpart. Being located at the MYSO's position and with its spectral index typical of an ionized jet of $0.69\pm0.11$, A is classified as a thermal jet with lobes. In our images however, we detect no emission from the position of source N from 
\citet{Zapata2001}, which is not due to resolving out effects (our beam is practically identical to theirs) or sensitivity issues (image RMS noise is $\sim2$ times lower). We believe that the previously established time-variability is the reason for our non-detection of N. 

Considering the absence of previously detected emission (N) and apparent resolving out effects towards A2, the C-band data was re-imaged with a robustness of 2, the results of which are shown in \autoref{cplot:G138.2957Robust2}. The new, more sensitive robustness shows how B was indeed resolved out in the images of \autoref{cplot:G138.2957}, but also that significant ($>5\sigma$), arc-like emission is also detected at a position angle of $-100\degr$ from A with an \textsc{imfit} derived flux of $128\pm21\uJy$. With this measurement, the flux of A decreases to $71\pm10\uJy$, giving a spectral index of $\alpha=1.0\pm0.1$. For consistency however, we use the value for flux and spectral index from the clean maps with a robustness of $0.5$ in the analysis within the main body of this work. A $4\sigma$ component is also detected $4\arcsec$ to the ENE of A, in line with the jet's outflow axis which, if real, is likely to be another shock surface interacting with the jet.

\subsubsection{G139.9091+00.1969A}
\label{sec:G139.9091}
Associated to the cluster AFGL437 of at least 60 members, this is a well studied object. At the centre of the cluster, three near-infrared sources dominate in terms of flux with both diffuse, knotted and bipolar 2.122$\um$, H$_2$ emission detected at a variety of position angles \citep{Navarete2015} with respect to the MYSO. A relatively uncollimated, CO, molecular outflow was detected oriented roughly on a north (red) - south (blue) axis \citep{Gomez1992}, whose material was hypothesised to be sourced from laminar flow of dragged gas from outflow cavity walls. \citet{WeintraubKastner1996} resolved the northern-most of the three NIR sources into two, dubbed WK34 and WK35, of which WK34 was determined to be a low-luminosity protostar driving the north-south molecular outflow and located at the centre of a similarly oriented reflection nebula \citep{Meakin2005}. From those works, the other NIR sources were dubbed S (our target) and W for the southern and western bright NIR sources respectively. NIR speckle imagery by \citet{Alvarez2004} showed a monopolar reflection nebula centred on S and extended towards the SE ($\theta_{\rm PA}=135\pm10\degr$), which highlights the position of an outflow cavity.

We detect three C-band sources, all of which are associated with near-infrared objects in 2MASS imagery and only one of which (the \textsc{Hii} region) is not detected at Q-band. The radio object A \citep[the southern-most NIR source, S, from][]{WeintraubKastner1996} is extended at a position angle of $62\degr$ in the direction of Ab, which is located $0.3\arcsec$ from A at a position angle of $58\degr$. To the south-west of A is the low-SNR source, Ac, separated by $0.7\arcsec$ (at $\theta_{PA}=241\degr$). Calculated values of $\alpha$ and $\gamma$, for A, are $0.42\pm0.08$ and $-0.81\pm0.19$ respectively, which agree well with the models of \citet{Reynolds1986}. However, due to the extended emission being perpendicular to the defined outflow cavities, we believe this source is a photo-evaporative disc wind. The discrepancy from the expected value of $\alpha\sim0.6$ is likely due to resolving out effects moving from C to Q-bands. At Q-band the source seems roughly quadrupolar, and in light of the disc wind classification, is likely tracing the NE portion of the wind.

Approximately $9\arcsec$ north-west of A is a spherical \textsc{Hii} region (the NIR source, W), with a C-band flux of $20.5\pm1.2\mJy$ which is therefore (assuming it is optically thin at C-band) powered by a B1 type star with a bolometric luminosity of $11000\Lsol$ \citep{Davies2011}. This agrees well with the $10000\Lsol$ \citep{Lumsden2013} calculated from the infrared SED of its alias in the RMS survey.

Located $10\arcsec$ north of A is the weakest C/Q-band source, B (also known as WK34), which is unresolved at both frequencies. Derived to have a spectral index of $\alpha=1.13\pm0.16$, it is classified as a nearby YSO, which may be powering a small HC\textsc{Hii} region, or an ionized jet. 

\subsubsection{G141.9996+01.8202}
\label{sec:G141.9996}
Near and mid-infrared surveys show a bright source which saturates in 2MASS, UKIDSS and GLIMPSE images.  \citet{Mitchell1992} observed a CO $J=2\rightarrow1$ outflow, with G141.9996+01.8202 (alias AFGL490) at the centre of the red and blue lobes. The molecular outflow's red lobe peaked $\sim20\arcsec$ to the north-west but extended in an arc round to the north east of AFGL490, with the blue lobe situated at a position angle of $225\degr$ (i.e. anti-parallel to the arc of red emission). A cold envelope elongated at a position angle of $\sim-45\degr$ was observed in the sub-mm continuum, while simultaneous observations of CS $J=2\rightarrow1$ showed evidence for multiple outflows driven by low-mass sources in the envelope \citep{Schreyer2002}. Further PdBI observations of C$^{17}$O $(2-1)$ revealed a rotating, clumpy, molecular disc with an inclination of $35\degr$ and major axis oriented at a position angle of $105\degr$ \citep{Schreyer2006}.

In our C-band images, we detect 3 compact sources labelled A (the MYSO), B and C, all of which are coincident with a NIR source. Only A was detected at Q-band, the images of which show an interesting morphology with a compact radio `core' embedded in (partially) resolved-out emission. Initially it was thought that the extended component was the dominant emission detected at C-band, however A's C-band dimensions ($0.21\arcsec\times0.06\arcsec$ at $58\degr$) do not coincide with the patches of extended emission at Q-band. Values derived for $\alpha$ and $\gamma$ closely agree to those of a spherical, stellar wind \citep{Panagia1975}, thermal jet or disc wind, all of which are supported by the presence of the shock tracer [FeII] at $1.64\um$ in NIR spectra \citep{Cooper2013}. Since the derived position angle of the C-band component is as equally aligned with the inferred disc's major axis, as it is with the molecular outflow, we classify this as a jet candidate. As a further note, considering its relatively evolved status in the literature (in comparison to this sample), it is likely that the stellar wind \citep[especially given IR recombination line profiles seen by][]{Bunn1995}, or photo-evaporative disc wind picture is correct, however more observations are required to distinguish between all three possibilities. As for the nature of the extended emission, under the assumption that the central object has evolved past the MYSO stage and is beginning to produce appreciable Lyman fluxes, it could be sources from ionized cavity walls, disc surfaces or an optically thick \textsc{Hii} region. Both B and C are almost certainly members of the same cluster, however their classification, although almost certainly not extragalactic, is unknown.

\subsubsection{G143.8118$-$01.5699}
\label{sec:G143.8118}
Near-infrared images show a reddened source at the MYSO position which is embedded in a cluster $1.8\arcmin\times1.4\arcmin$ in size \citep{Bica2003}. \citet{Navarete2015} detect no emission in $2.122\um$ H$_2$ towards this object.

Although we do not detect anything at $44\GHz$, we see three, low-flux, compact sources in C-band images, A (G143.8118$-$01.5699), B and C, of which all have near-infrared counterparts. Although A and B are resolved according to \textsc{imfit} routines, their signal to noise is low and consequently the errors on deconvolved position angles/dimensions are large. Considering the relatively unconstrained nature of the emission detected towards all C-band sources, we therefore classify A to be a jet candidate, while B and C, although certainly no extragalactic, are of unknown classification.

\subsubsection{G148.1201+00.2928}
\label{sec:G148.1201}
Another object with a small observation history, the most relevant of which were the H$_2$, $2.122\um$ observations by \citet{Navarete2015} which detected diffuse, knotted and bipolar, shock emission at position angles of $160-215\degr$ and separations between $0.27$ and $1.08\pc$. In 2MASS images, the MYSO is situated in a cluster, with a diffuse reflection nebula emanating from it towards the north-west ($\theta_\mathrm{PA}\sim110\degr$).

At $5.8\GHz$, we detect three sources within the inner $60\arcsec$ of the field of view. One of these sources (A) is coincident with the MYSO from the RMS survey, while the other two are coincident with other NIR objects in the field. The C-band radio source A breaks up into two Q-band sources whose peaks are separated by $0.1\arcsec$ ($320\au$), with the weaker lobe (A2) located at a position angle of $295\degr$ from the other (A), roughly aligned with the NIR reflection nebula. Because both Q-band sources (A and A2) are coincident with the C-band image of A, we combine the Q-band fluxes of A and A2 and find a spectral index of $\alpha=1.13\pm0.14$ for the MYSO. From this information, A and A2 may be a jet/counter-jet system, or a close binary system. If a jet/counterjet, from the models of \citet{Reynolds1986} the derived spectral index would suggest some degree of recombination or acceleration in the flow which, if the separation represents the collimation radius (i.e.\ where the toroidal become dominant over the poloidal component in the launching magnetic field), would be more likely from acceleration. A piece of evidence against the biconical jet scenario would be the position angles of shocked H$_2$ emission observed by \citet{Navarete2015}, which are misaligned with the apparent jet's axis. Due to these considerations, we define A/A2 to be a jet candidate. As for B and C, upper limits on the spectral indices of $<0.63\pm0.12$ and $<-0.01\pm0.05$ do not constrain their natures much and therefore, although not extragalactic, are assigned the unknown classification.

\subsubsection{G160.1452+03.1559}
\label{sec:G160.1452}
Near-infrared UKIDSS images (colourscale in \autoref{cplot:G160.1452} respectively) show that the reddened MYSO is extended in the NIR (especially UFTI imagery), highlighting the possibility of cavities at a position angle of $\sim100\degr$, an interpretation backed up by the detection of a parallel $^{12}$CO $(2-1)$ outflow by \citet{Xu2012}. Radio observations by \citet{SanchezMonge2008} detect a compact radio source centred on the MYSO at $3.6\cm$, $1.3\cm$, $7\mm$ and $1.2\mm$ (IRAM) with a spectral index between $3.6\cm$ and $1.3\cm$ of $\alpha=1.1\pm0.4$. Targeted, near-infrared observations showed a H$_2$, $2.122\um$, collimated outflow parallel to both the molecular outflow's, and radio source's major, axes at a position angle of $126\degr$ which extended over a length of $>0.35\pc$ \citep{Varricatt2010}.

\begin{figure}
\centering
\includegraphics[width=84mm]{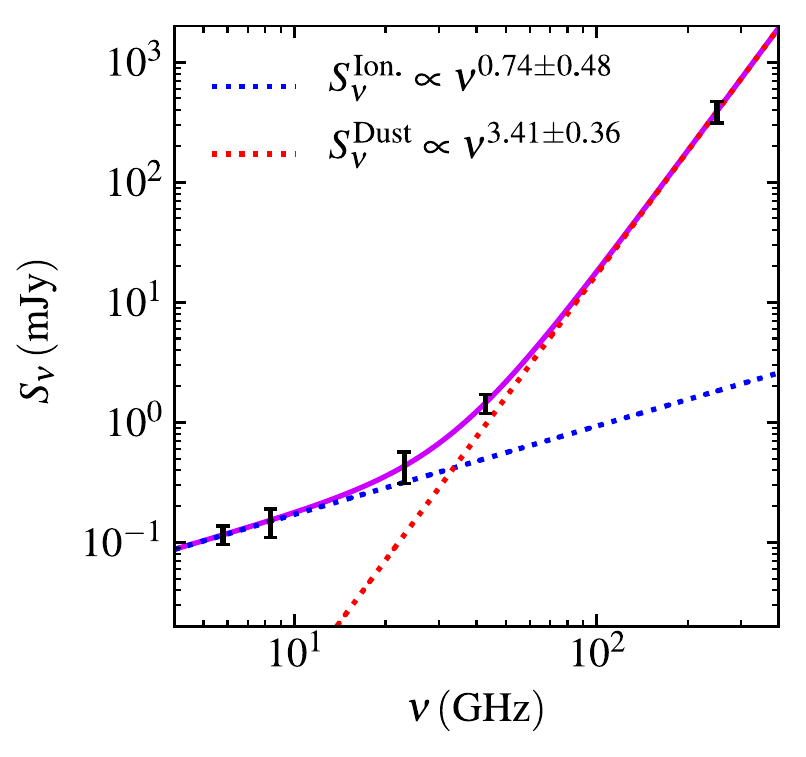}
\caption[Radio SED for source A of G160.1452+03.1559]{Plot of flux against frequency for source `A' of G160.1452+03.1559, utilising data at $8\GHz$, $23\GHz$, $43\GHz$ and $250\GHz$ from \citet{SanchezMonge2008}. The least squares fit is plotted as the purple line for a double power law, with each power law component plotted as a dotted red (dust) and blue (ionized) line. Derived power laws for the dust and ionized contributions are indicated in the top left corner.}
\label{fig:G160.1452SED}
\end{figure}

This source was only observed at C-band, the images of which show two compact sources in the field of view, one situated on the MYSO (A) and the other (B) offset $\sim3\arcsec$ to the SSE. Although B was too low in terms of signal to noise for its dimensions to be derived by \textsc{imfit}, A has a major axis aligned along a position angle of $\sim110\degr$, parallel with the apparent outflow cavities seen in NIR images. In order to try to discern A's nature, the data from \citet{SanchezMonge2008} was combined with our C-band result and fitted with a power law consisting of a dust and an ionized component (assuming no flux variability between the two datasets). The resulting fit has been plotted in \autoref{fig:G160.1452SED}, deriving a spectral index for the ionized component of $\alpha=0.74\pm0.48$, expected of a typical ionized jet. Using the derived power law we calculate a flux contribution from the ionized component, at $43\GHz$, of $500\uJy$ which is equivalent to an optically thin \textsc{Hii} region powered by a ZAMS type B3 star with a luminosity of $\sim2500\pm2000\Lsol$. Comparing this to the IR derived flux of $2100\Lsol$, it appears that the radio flux is also compatible with a HC\textsc{Hii} region with an average electron density of $8\pm7\times10^4\cm^{-3}$. However, because of its extensive observational history, with collimated, H$_2$, shock features, molecular outflows and a full sampling of its radio spectrum, the ionized jet scenario is extremely likely. We therefore have classified A as an ionized jet. Component B, being associated with a reddened near-infrared source is likely a coeval low-mass YSO, however more observations are needed to clarify this. 

\subsubsection{G173.4839+02.4317}
\label{sec:G173.4839}
Our field of view contains two MYSOs from the RMS database, G173.4839+02.4317 and G173.4815+02.4459. In 2MASS images (middle panel of \autoref{cplot:G173.4839}), G173.4839+02.4317 displays as a relatively unreddened, bright source surrounded by a small, young cluster \citep{Ginsburg2009}, while the reddened, near-infrared source, G173.4815+02.4459 shows significant, EGO emission at $4.5\um$ (see GLIMPSE inset of \autoref{cplot:G173.4839}) whose diffusivity extends along a rough north-west to south-east axis. The EGO emission is embedded in a protocluster according to \citet{Ginsburg2009} who also inferred that G173.4815+02.4459 was, in fact, a massive binary system with a separation of $400\au$. H$_2$, $2.122\um$ observations show an outflow at a position angle of $\sim150\degr$ centred on G173.4839+02.4317 \citep[components 7/8 and 1a/b/c from][respectively]{Varricatt2010,Navarete2015}. Most other observations concentrate on G173.4815+02.4459, for example \citet{Beuther2002b} detect 3 condensations in H$^{13}$CO$^+$ $(1-0)$, condensation `1' being centred on G173.4815+02.4459, and condensations `2' and `3' located $16\arcsec$ and $25\arcsec$ to the west of condensation `1' respectively. A highly collimated outflow in CO $(1-0)$, at a position angle of $\sim165\degr$, which terminates in H$_2$ bow-shocks, is detected centred on condensation `1' which is also at the centre of a high-velocity CO $(1-0)$ outflow at a position angle of $\sim130\degr$, parallel to the EGO's diffuse emission. Core `2' powers an SiO $(2-1)$ outflow at a position angle of $\sim15\degr$. Simultaneous $2.6\mm$ observations showed condensation `1' to break into 3 mm-cores, with the brightest core, mm1, being centred on the MYSO and the other two mm-cores, mm2 and mm3, positioned $4\arcsec$ to the east and $8\arcsec$ to the north-west of mm1 respectively. Further sub-mm observations \citep{Beuther2007} resolved mm1 into two further sub-cores (one of which was detected by the VLA at X-band), and mm2 into 4 further sub-cores, creating a complicated picture of this particular massive star-forming site. From the spectral energy distribution across cm, mm and sub-mm wavelengths, mm2 was also hypothesised to be emission from an outflow/jet. At other wavelengths, both the collimated, and high-velocity, molecular outflows are also seen in line emission in the near-infrared \citep{Varricatt2010,Navarete2015}. Perpendicular to the established H$_2$, and high-velocity molecular, outflows, a line of CH$_3$OH masers ($\theta_\mathrm{PA}\sim35\degr$), along a velocity gradient suggestive of Keplerian rotation in an accretion disc, were detected by VLBI observations \citep[taken in 1997,][]{Minier2000}.

\begin{figure}
\centering
\includegraphics[width=84mm]{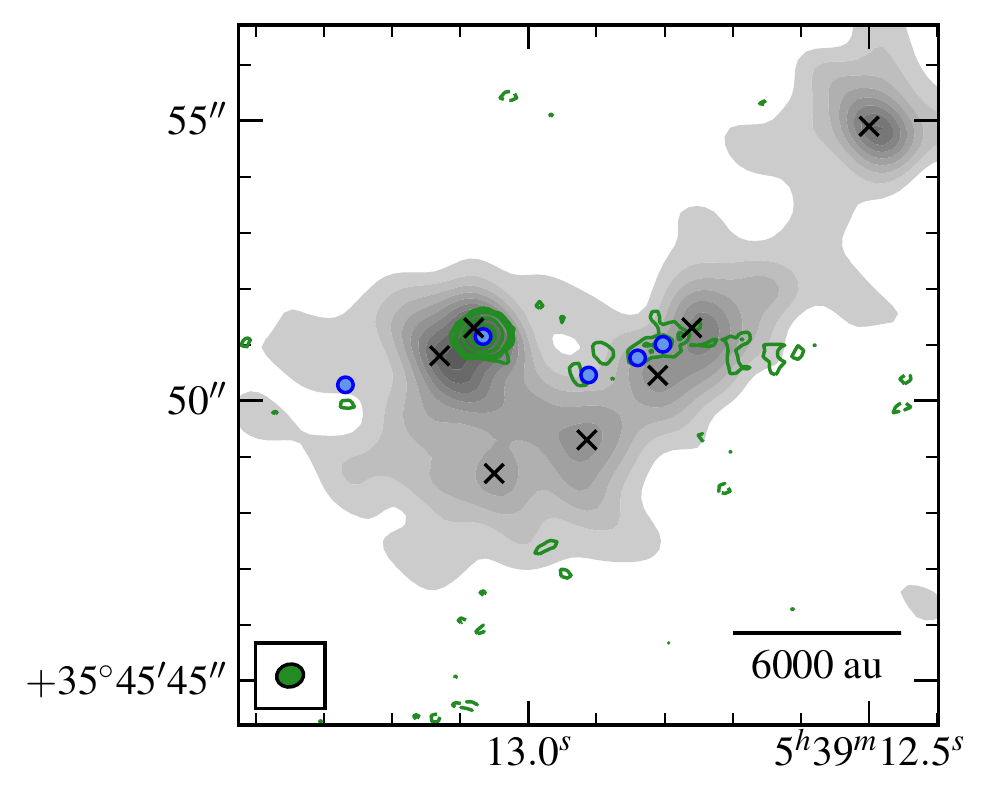}
\caption[Radio image of G173.4815+02.4459 overlayed on top of $1.2\mm$ PdBI observations]{Sub-mm continuum image of G173.4815+02.4459 at $1.2\mm$ (PdBI, greyscale) with $8.4\GHz$ VLA data overlayed (green contours). The `x' markers show the position of the 5 cores from \citet{Beuther2007} and the blue `$\circ$' markers show the positions of the 5 sources, E1, E2, E3, E4 and E5, that we detected at C-band. Grey-scale increases by $10\%$ from $10\%$ to $90\%$ of the peak flux ($58.9\mJy\,\mathrm{beam}^{-1}$), while green contours are set at $-3, 3, 4, 5, 10$ and $20\sigma$ ($\sigma=16.4\uJy\,\mathrm{beam}^{-1}$). Both the $8.4\GHz$ VLA and $1.2\mm$ PdBI data are from \citet{Beuther2007}.}
\label{fig:G173.4815_beuther}
\end{figure}

Within $1\arcmin$ of the two MYSOs in our field of view, we detect a total of 12 different C-band radio sources. Component E1$\rightarrow$E5 are associated with G173.4815+02.4459, while A1 and A2 are associated to G173.4839+02.4317. Components B, C, D, E and F are not associated to any infrared source, while G is coincident with a bright, unreddened 2MASS source. We detect a methanol maser located at E1's position which is coincident with the EGO from GLIMPSE images (top left panel in \autoref{cplot:G173.4839}). High-resolution NIR images show that E2, E3, E4 and E5 are not associated with any near-infrared point sources, only diffuse emission \citep[Figure 1c of][]{Yan2010}. Those same NIR images were however saturated around our component A1 and therefore we are unable to discern if A2 has its own infrared source. In order to better understand the natures of E1$\rightarrow$E5, we plotted their positions on top of previous mm and radio observations in \autoref{fig:G173.4815_beuther} from \citet{Beuther2007}. E1 is located at the position of core mm1a from \citet{Beuther2007} and is also detected at X-band (we calculate $\alpha=1.5\pm0.4$ between C and X-band), while E2 and E5 are not co-located with any $1.2\mm$ emission. There is also diffuse, low-SNR, X-band emission coincident with E2, E3 and E4 with a rough spectral index of $\sim0$. E3 and E4 are located towards the mm-core, mm2b, which was the core with an inferred emission contribution from a jet/outflow. Considering the wealth of previous observations, it is clear that E1 is driving a collimated, outflow and therefore harbours an ionized jet with associated lobes. We believe that E5 is a lobe as a result of jet-shocks on surrounding material, considering its position angle from the MYSO (along which there is another $3\sigma$ lobe $\sim1.1\arcsec$ ESE of E5). As for E2, E3 and E4, there is a variety in position angle offsets from E1. If the massive binary scenario from \citet{Ginsburg2009} is accurate, large precessional shifts may be affecting the north-western side of the jet, providing a mechanism to alter the jet's outflow axis and we tentatively classify E2, E3 and E4 as shock-excited lobes on this basis. Interestingly our source F is also located along a similar axis from E1 as E5 and therefore may be another lobe, however this is uncertain. Sources A2, B, C and D are not easy to classify considering their absence of previous detections and are therefore of unknown classification. From its near-infrared profile and evolved status in the literature, G173.4839+02.4317 is almost certainly a \textsc{Hii} region, especially considering both its strong Br$\gamma$ emission in the near-infrared \citep{Cooper2013} and C-band, radio-inferred bolometric luminosity of $2450\Lsol$ which agrees with that derived from IR studies ($2900\Lsol$). As for G, due to its bright, main sequence near-infrared profile, it is likely to be a main-sequence star. 

\subsubsection{G174.1974$-$00.0763}
\label{sec:G174.1974}
Located in an active star formation region with a rich observational history, the MSX source G174.1974$-$00.0763 is a relatively unreddened, bright near-infrared 2MASS source, with a saturated mid-infrared profile surrounded by diffuse $8\um$ PAH emission, typical of \textsc{Hii} regions. \citet{Carral1999} detect a faint radio source at $8.4\GHz$ with a flux of $0.30\pm0.06\mJy$. Within $60\arcsec$ of the pointing centre is another MYSO, at an earlier stage in evolution, referred to in the literature as AFGL 5142 ($\alpha\,(\mathrm{J2000}) = 05\rahr30\ramin48.02\rasec, \delta\,(\mathrm{J2000}) = 33\degr47\arcmin54.5\arcsec$). In GLIMPSE images, it is at the heart of a dense cluster, while at near-infrared, dust extinction gives it a faint, reddened profile in 2MASS images. AFGL 5142 is listed as possessing a bolometric luminosity of $2300\Lsol$ \citep{Palau2011}, which is revised to $3300\Lsol$ in light of the more accurate distance estimate of $2.14^{+0.051}_{-0.049}\kpc$ by \citet{Burns2017}. AFGL 5142 is positioned at the centre of collimated HCO$^+$ $(1-0)$ and SiO$(\mathrm{v}=0,2-1)$ outflows, along a position angle of $\sim5\degr$ \citep{Hunter1999}, as well as co-located with a compact radio source \citep[$0.83\pm0.15\mJy$ at $8.4\GHz$ using the VLA in A-configuration,][]{Hunter1995}. However, it appears this source may be prone to resolving out effects and/or variability at radio wavelengths (see Figure 2 of \cite{Goddi2006} for a summary), for example a higher flux of $1\mJy$ was reported by \citet{Carpenter1990} for D-configuration $4.86\GHz$ observations (performed 1989), and $1.5\pm0.3\mJy$ reported by \citet{Hunter1995} at $8.6\GHz$ which used B-configuration observations (taken 1998). Later studies at $8.4\GHz$ show that the continuum source breaks down in to 2 separate sources named, CM-1A and CM-1B, and also appears to power a further two molecular outflows at position angles of $35\degr$ and $-60\degr$ \citep{Zhang2007}. PdBI, $1.3\mm$ observations by \citet{Palau2011} showed that two mm-cores were located with the MYSO. Of these, MM1 is attributed to the previous radio detections of the literature, and is elongated at a position angle of $94\degr$ (perpendicular to the north-south outflow), whereas MM2 is situated $\sim1\arcsec$ to its south and may have been previously detected at $8.4\GHz$ \citep[with a peak flux of $0.35\pm0.09\mJy,$][]{Zhang2007}.

\begin{figure}
\centering
\includegraphics[width=84mm]{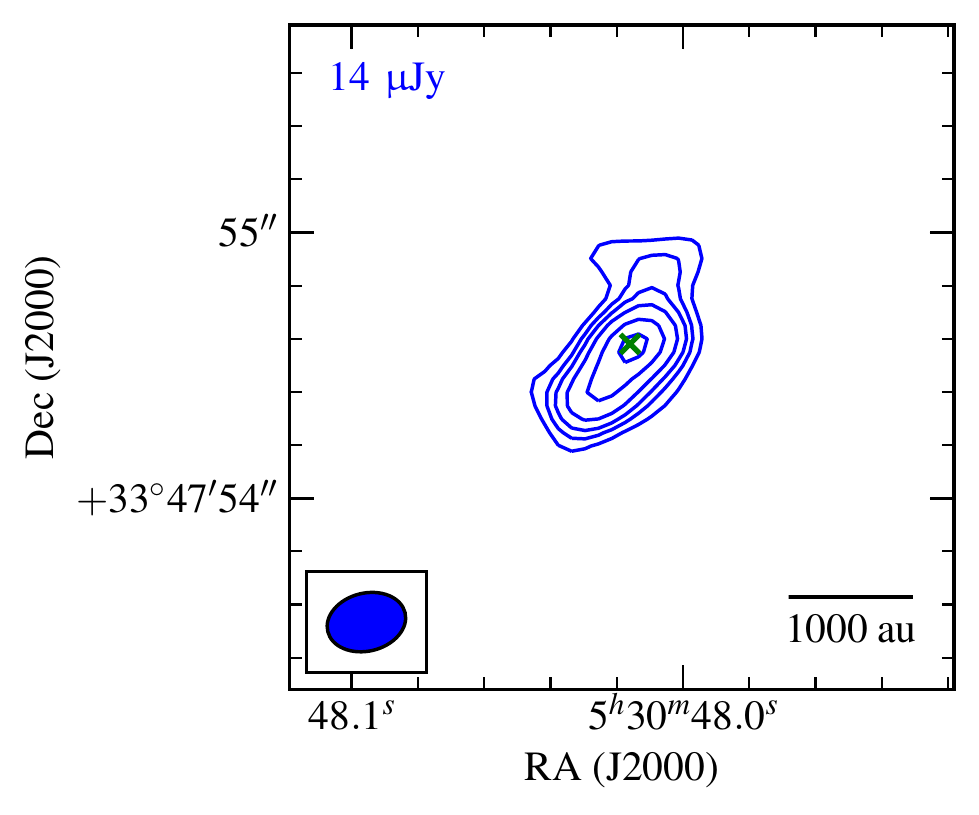}
\caption[Uniform ($R=-1$) radio image of source B (AFGL 5172) from the G174.1974$-$00.0763 field]{Re-imaged C-band data of source B (AFGL 5172) from the G174.1974$-$00.0763 field, utilising a robustness of -1. Contours are set at $-3, 3, 5, 7, 10, 15$ and $20\sigma$ where $\sigma = 14.1\uJy\,\mathrm{beam}^{-1}$. The restoring beam's dimensions were $0.299\arcsec\times0.218\arcsec$ at $\theta_\mathrm{PA} =104\degr$. }
\label{fig:G174.9714_B_Robust-1}
\end{figure}

From the C-band data, we report the detection of 3 radio sources, A, B and C, of which A is located at the same position as G174.1974$-$00.0763, and both B and C are colocated with near-infared/mid-infrared sources, $24\arcsec$ to the east and $51\arcsec$ to the WNW respectively. Using the radio flux at $8.4\GHz$ of $0.30\pm0.06\mJy$ from \citet{Carral1999}, we calculate a spectral index for A of $\alpha=2.3\pm0.7$, typical of optically thick emission from a \textsc{Hii} region. It must be conceded that the flux from the literature used observations from 1994 with the VLA in its most compact configuration, therefore the resolving out of flux and/or variability may affect this calculated value. However, considering its near and mid-infrared appearance, likely optically thick radio spectrum and diffuse $8\um$ emission, it is determined that A is a HC\textsc{Hii} region. As for B, in combination with our C-band flux of $726\pm22\uJy$, we calculate a spectral index for B of $0.4\pm0.5$. However, as previously discussed, this derived spectral index is not very useful in trying to establish whether or not the emission stems from an ionized jet on account of resolving out/variability issues apparent in the literature. Since \citet{Zhang2007} resolve B into two sources, we reimaged the field using a robustness of -1, in order to increase the effective resolution. The results are plotted in \autoref{fig:G174.9714_B_Robust-1} which more reliably shows two components, one elongated at a position angle of $\sim0\degr$ from the peak emission, and the other elongated at a position angle of $\sim-50\degr$. \citet{Zhang2007} recorded almost equal fluxes at $8.4\GHz$ for both of these sources, whereas our $5.8\GHz$ data shows the most southern to be the strongest. This may be due to spectral index effects, or variability, however further observations are required to clarify this. It is believed however that the northern elongation is an ionized jet driving the well establish north-south outflows and we therefore classify B to be an ionized jet. Our detected source C has no previous detections present in the literature, however considering its coincidence with a bright main-sequence star in the near-infrared, it is likely a stellar wind, however more radio observations will be required to determine this.

\subsubsection{G177.7291$-$00.3358}
\label{sec:G177.7291}
K-band, near-infrared, continuum images show a bright point source at the centre of a quadrupolar patch of extended emission with a UIB feature indicative of UV excitation \citep[see Figure 1y of][who associated the morphology to tracing cavity walls]{Ishii2002}. Low-resolution, near-infrared spectra taken with a slit alignment of $\theta_\mathrm{PA}=-70\degr$ (i.e. along the NW-SE, K-band diffusivity) shows shocked, [$\mathrm{FeII}$] emission, a weak Br$\gamma$ line and CO ($v=2-0$) bandhead emission \citep{Cooper2013}, presumably from an accretion disc \citep{Ilee2013}. Non-detection of both methanol (where $3\sigma=150\mJy$), and water, masers towards this source has been recorded by \citet{Fontani2010}.

Two C-band sources are detected in the inner $60\arcsec$ of our field of view, labelled A (the MYSO) and B (located $25\arcsec$ west of A). Component A is completely unresolved at C-band, whereas B has dimensions of $0.16\arcsec\times0.13\arcsec$ with large errors on account of low signal to noise and barely extended structure. Assuming an optically thin \textsc{Hii} region, from the C-band flux for A we infer a bolometric luminosity of $1300\Lsol$ \citep{Davies2011}, close to that derived from the infrared ($2300\Lsol$), especially considering the radio emission is likely still partially optically thick. Therefore neither jet, nor compact \textsc{Hii}, gains precedence and therefore we classify A as a jet candidate. As for B, it is aligned with the outflow cavities' central axis seen at near-infrared wavelengths yet lacks an (mid or near) infrared counterpart, and therefore could be a shocked, Herbig-Haro type object (especially given its resolved dimensions) from a jet. However without more spectral information we must determine it to be of an unknown nature due to its similarity to extragalactic phenomena.

\subsubsection{G183.3485$-$00.5751}
\label{sec:G183.3485}
Near-infrared, K-band images show diffuse emission elongated along a NW-SE axis, with the MYSO's reddened source at its centre. From C$^{18}$O observations, a clump $0.76\pc$ ($75\arcsec$) in diameter was detected with a mass of $250\Msol$ \citep{YuanWeiWu2011}, whose peak is offset to the MYSO's position by $14\arcsec$ to the south-west. No other relevant observations are present in the literature.

Our observations detect two, point-like, C-band sources, one (A) is roughly positioned at the MYSO's location, while the second (B) is co-located with a main sequence (from its near-infrared colours) star. We also detect two methanol maser spots separated from A's location by $0.22\arcsec$ and $0.32\arcsec$, at position angles of $189\degr$ and $341\degr$ respectively. The positioning of the methanol masers, along the edges of the diffuse emission, would suggest them to reside in the cavity walls, rather than in the disc (which would need to be at least $1000\au$ in diameter). From the sparsity of information on this source, it is impossible to definitively classify A and we assign it candidacy as a jet. Source B's positioning with a main sequence star means it is likely a stellar wind, though further radio data, at a variety of frequencies, are needed to clarify this.

\subsubsection{G188.9479+00.8871}
\label{sec:G188.9479}
Located in the star formation region AFGL 5180, G188.9479+00.8871 is commonly referred to as NIRS1 \citep[following the work of][who identified 11 K-band point sources in its proximity]{Tamura1991} and has an established, bipolar, CO outflow at a P.A. of $\sim130\degr$ detected by \citet{Snell1988}. The RMS source is at the centre of a bipolar, $2.122\um$, H$_2$ outflow at a position angle of $110\degr$ \citep[1a/1b from][]{Navarete2015} as well as two diffuse, bipolar, K-band, reflection nebula at position angles of $90\degr$ and $110\degr$ \citep{Tamura1991}. \citet{Saito2006} detect 11 cores in C$^18$O, with core F associated to NIRS1, and two continuum sources at $98\GHz$, of which core F is coincident with one (MCS B) and core E with the other (MCS A, the brightest), which in turn is located with the NIR source, NIRS5. Both of the continuum sources were previously detected as $1.2\mm$ cores by \citet{Minier2005}, with bolometric luminosities of $7.0\times10^3\Lsol$ and $2.4\times10^4\Lsol$ for NIRS1 and NIRS5 respectively \citep[reduced following the more accurate distance estimate of $1.76\pm0.11\kpc$ by][]{Oh2010}. VLBI observations of CH$_3$OH masers show a linear arrangement of maser spots along a position angle of $78\pm7\degr$, with an inferred magnetic field direction parallel to the CO outflow \citep{Surcis2013}.

At C-band we detect 5 distinct radio sources. Sources A and B1/B2 are unresolved and coincident with the C$^{18}$O cores F and E, from \citet{Saito2006}, respectively, while C and D are resolved and have no mm-counterparts in the literature but are associated to near-infrared sources. A methanol maser is also detected to be coincident with continuum source A. Considering the wide variety of outflow phenomena seen, we believe A to be a jet, however without more information at a range of frequencies we can only classify it as a candidate. As for B1 and B2, interestingly the axis running through them in a direction parallel to extended 98$\GHz$ emission from \citet{Saito2006}, which may include contribution from an ionised jet assuming B1/B2 are a jet/lobe pair. As with A, more information is required, and consequently B1 is determined to be a candidate jet with lobes (B2). Component C's coincidence with a near-infrared source classifies it as a cluster member, but the origin of its radio emission is unclear. Due to its extended nature, presence of resolved out emission and mid-infrared appearance in GLIMPSE images, we classify D to be a \textsc{Hii} region.

\subsubsection{G189.0307+00.7821}
\label{sec:G189.0307}
Also known as AFGL 6366S, this source is located with a K-band, reflection nebula, indicative of outflow cavities, along a rough north-west to south-east axis. Approximately $100\arcsec$ to the ENE another YSO, G189.0323+00.8092 (alias AFGL 6366N) from the RMS database ($\Lbol=1.1\times10^4\Lsol$), is within our C-band field of view and is another YSO which has diffuse NIR emission towards its east. \citet{Kurtz1994} detect a $0.6\pm0.06\mJy$, unresolved, $8.4\GHz$ source at $\alpha\,(\mathrm{J2000}) = 06\rahr 08\ramin 40.66\rasec, \delta\,(\mathrm{J2000}) = 21\degr 31\arcmin 07.3\arcsec$ (slightly offset to G189.0307+00.7821's position), with an unlisted $\sim0.25\mJy$ source $\sim2\arcsec$ south west of it. Observations at 98 and 110$\GHz$ show that there are two mm-cores (also detected in C$^{18}$O), one centred on G189.0307+00.7821's coordinates (MCS A) and the other (MCS B) located approximately $8\arcsec$ to it north-east \citep{Saito2008}. Two molecular clumps, one coincident with G189.0307+00.7821 (clump 3a) and the other located $\sim140\arcsec$ to its ENE (clump 3b, positioned with G189.0323+00.8092) were detected in single dish $^{13}$CO ($J=1-0$) observations by \citet{Shimoikura2013}. Along a position angle of $\sim100\degr$, \citet{Wu2010} detected a bipolar, $^{12}$CO ($1-0$) outflow, centred on G189.0307+00.7821, with the blue lobe towards the west, along the same position angle that \citet{Navarete2015} detect a knot of H$_2$ emission, as well as a bipolar H$_2$ outflow at a position angle of $\sim130\degr$.

Within $60\arcsec$ of the pointing centre we detect 7 radio sources in our C-band image. Four of these are located within $8\arcsec$ of the reddened 2MASS source shown in the top left panel of \autoref{cplot:G189.0307}, named A1, A2, B and C, the last of which we also detect a methanol maser towards. Comparing to NIR, UIST images, A1 is associated to the bright K-band source, while A2, B and C do not have a NIR counterpart. In $12\um$ MICHELLE images however, it can be seen that both A1 and C have mid-infrared counterparts, suggesting that C's YSO is more deeply embedded in the natal clump \citep[clump 3a from][]{Shimoikura2013}. It is also relevant to note that component C was previously detected by \citet{Kurtz1994} (see above) and, assuming no variability, therefore has a spectral index of $\alpha=0.4\pm0.4$ between $5.8$ and $8.4\GHz$. As for the natures of these central components, given the CO ($\theta_\mathrm{PA}\sim100\degr$) and H$_2$ outflows centred on A1, a strong case could be made for A2 being a shocked lobe along the jets path. We therefore (through lack of more radio information), classify A1 to be a candidate jet with lobes. Due to C's coincidence with both a methanol maser and mid-infrared source, it is almost certainly another MYSO in the vicinity of G189.0307+00.7821, however the nature of its radio emission is unconstrained. Due to C's deconvolved dimensions, the major axis of which is aligned along a position angle parallel to the axis through B and C, B may be a shocked lobe of emission related to a jet at C. However more information is again needed and we therefore designate C as a candidate jet with lobes. As for the other three sources, both D and F have no near, or mid, infrared counterparts and are located away from the clump's sub-mm emission, but are aligned along the same position angle as A2, relative to A1. Therefore it is quite possible that these are also lobes of shocked emission significantly more separated from the jet's launching site than A2. However, considering the speculative nature of this and that a serendipitous alignment of background sources is a possibility, we therefore classify them as unknown sources. Due to its clear near-infrared profile as a blue, main-sequence star, we classify E as a likely stellar wind.

\begin{figure}
\centering
\includegraphics[width=84mm]{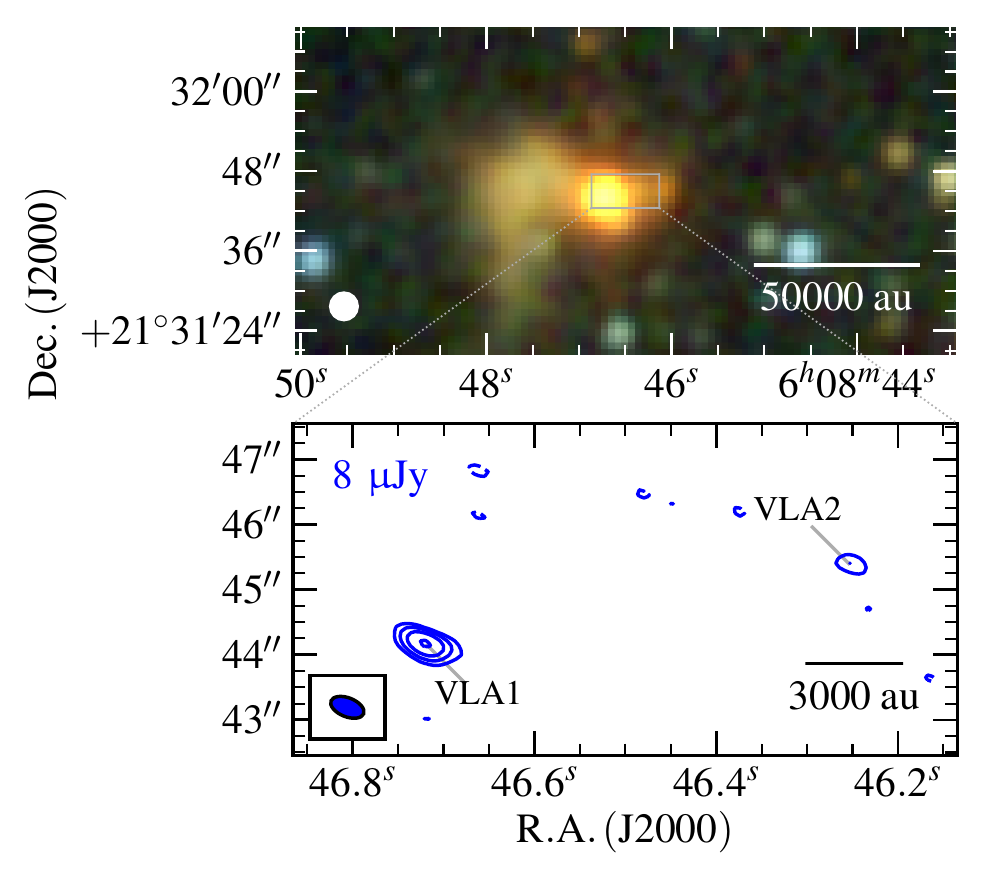}
\caption[Near-infrared and VLA radio images of G189.0323+00.8092]{Near-infrared (top panel; 2MASS, $\mathrm{K,H,J}$ R, G, B colour-scale) and C-band radio map (bottom) of G189.0323+00.8092. Restoring beams are the same as in \autoref{cplot:G189.0307}, while contour levels are set at $(-3, 3, 6, 11, 21)\times\sigma$.}
\label{cplot:G189_0323}
\end{figure}

As previously mentioned, another MYSO (G189.0323+00.8092) is registered in the RMS database $100\arcsec$ to the ENE of G189.0307+00.7821, and coincident with another molecular clump. Inspection of the C-band data outside the central $1\arcmin$, shows a radio source coincident with the second YSO's position (VLA1) and a $6\sigma$ source $\sim7\arcsec$ to its east (VLA2). In \autoref{cplot:G189_0323} these sources are presented in conjunction with a 2MASS, NIR image. Using \textsc{imfit}, we derive VLA1's right ascension to be $06\rahr 08\ramin 46.7192\rasec \pm 0.014\arcsec$, declination to be $+21\degr31\arcmin44.164\arcsec \pm 0.007\arcsec$, with a flux of $212\pm16\uJy$ and dimensions $(287\pm84\mas)\times(76\pm67\mas)$ at $\theta_\mathrm{PA}=69\pm13\degr$. Those same properties derived for the unresolved source VLA2 are $\alpha\,(\mathrm{J2000})=06\rahr 08\ramin 46.2528\rasec \pm 0.054\arcsec$, $\delta\,(\mathrm{J2000})=+21\degr31\arcmin45.383\arcsec \pm 0.015\arcsec$ and a flux of $43\pm13\uJy$. For VLA1, the radio flux is equivalent to an optically thin \textsc{Hii} region powered by a central object with a bolometric luminosity of $1700\Lsol$. Considering the emission is likely still optically thick, this under-luminosity is expected should VLA1 be a \textsc{Hii} region. However it is elongated towards VLA2, and therefore a jet/lobe pair in VLA1/VLA2 is also a plausible scenario. The presence of [FeII] and relatively weak Br$\gamma$ in the NIR spectrum \citep{Cooper2013} would favour this. We therefore classify VLA1 to be a candidate jet with lobes (VLA2), with a better determination of it radio spectral characteristics required for a more definitive conclusion about the radio emission's origin.
 
\subsubsection{G192.6005$-$00.0479}
\label{sec:G192.6005}
Also referred to as S255IR-IRS3, this is another of our sample with a relatively rich observational history and lies in the massive star forming clump, S255IR, itself sandwiched between two, large, classical \textsc{Hii} regions, S255 and S257. Previously, a $1.97\pm0.32\mJy$ source was detected at $15\GHz$ with VLA, B-configuration data taken in 1984 \citep{RengarajanHo1996} on the MYSO's position, though remained unresolved. From near-infrared speckle imaging, two YSOs were apparent, IRS3 and IRS1 ($2.4\arcsec$ to the WNW of IRS1), which powered a bipolar reflection nebulas, IRN 1 (along a NE-SW axis) and IRN 2 (along a north-south axis) respectively. While \citet{Heyer1989} observed a north-south, $^{12}$CO $(J=1-0)$ outflow, \citet{Zinchenko2015} detected an extremely collimated, bipolar, CO $(J=3-2)$ outflow, centred on G192.6005$-$00.0479, at a position angle of $67\degr$. Concurrent continuum observations at $1.3\mm$, showed three cores, SMA1 (G192.6005$-$00.0479), SMA2 and SMA3, the latter two of which are separated from the MYSO by $2\arcsec$ and $5\arcsec$ respectively, to the north-west. Multi-epoch VLBI observations by \citet{Burns2016} both refined distance estimates for this object ($D=1.78^{+0.12}_{-0.11}\kpc$) and observed a jet-driven bow-shock of water maser spots moving in a direction parallel to the overall outflow. Further to this, both episodic accretion and ejection events have been seen in the infrared and radio \citep[][respectively]{Caratti2017,Cesaroni2018}. The NIR, H$_2$, $2.122\um$ survey of \citet{Navarete2015} showed a large amount of diffuse emission associated to the two neighbouring, classical \textsc{Hii} regions, however a collimated, bipolar outflow was seen centred on G192.6005$-$00.0479 at a position angle of $70\degr$ (their lobes 1a, 1b and 1c). A north/south alignment of H$_2$ shock features was also observed, but their direct association to S255IR-SMA1 is less certain.

Our C-band images show the presence of 7 radio sources within $5\arcsec$ of the MYSO, denoted as A$\rightarrow$G, whereby A is positioned at the MYSO's coordinates and was also detected to have a methanol maser. Lobes C, D, F and G are all aligned along an average position angle of $70\degr$ with A, whilst B and E are positioned north of A by $1\arcsec$ and $3\arcsec$ respectively. From the overwhelming evidence for a collimated jet from studies in the literature, A is classified as a jet with lobes. Subsequently, least squares fitting of a simple sinusoid to the separations/position angles of C, D, F and G with respect to A, derives an approximate precession period and angle of $140\yr$ and $21\degr$ respectively \citep[assuming a jet velocity of $660\kmps$,][]{Cesaroni2018}. As for the natures of B and E, it is possible, given the north-south outflow detected by \citet{Heyer1989}, that this may be tracing a second jet, possibly from a close binary near A's location. However, higher resolution observations are required to establish this and B and E are therefore unknown in nature. 

Away from the central object, we detect 2 more compact sources, H and I, the latter of which is separated by $\sim2\arcsec$ from a large extended \textsc{Hii} region $\sim3\arcsec$ in diameter. While I has a near infrared counterpart, H does not, and neither does it have one in mid-infrared images therefore we can not determine its nature. Because component I has its own UKIDSS source ahead of a bright NIR source which is clearly the extended \textsc{Hii} region, it is possible that it is a wind from an main-sequence cluster member, and therefore we classify it as stellar in origin.

\subsubsection{G196.4542$-$01.6777}
\label{sec:G196.4542}
More commonly referred to as S269 IRS2, the reddened, near-infrared emission is comprised of two sources separated by $4.1\arcsec$, designated as IRS 2e and IRS 2w (G196.4542$-$01.6777) by \citet{Eiroa1994}. The same work also discovered a HH object (HH 191 at $\alpha\,\mathrm{(J2000)} = 06\rahr 14\ramin 37.8\rasec, \,\delta\,\mathrm{(J2000)} = +13\degr 49\arcmin 38\arcsec$) separated from IRS 2w, its powering source, by $10\arcsec$ at a position angle of $82\degr$, with an inferred shock velocity of $570\kmps$. \citet{Jiang2003} detect one, possibly two, H$_2$, $2.122\um$ outflows along a south-east/north-west direction from IRS 2e, IRS 2w, or both. Near infrared spectroscopy of IRS 2w reveal H$_2$ and [FeII] shock emission, as well as relatively weak (in comparison to typical UC\textsc{Hii} regions) Br$\gamma$ emission \citep{Cooper2013} from the MYSO.

Our C-band maps of flux show 5 distinct components labelled A1, A2, B$\rightarrow$D, with A1 and A2 centred on the MYSO's (IRS 2w) position. Methanol maser emission is also detected $\sim0\farcs9$ to the south-west of A1. Both A1 and A2 are aligned and elongated at a position angle of $64\degr$, with relatively small errors on deconvolved dimensions. However, considering the fact that the restoring beam itself is oriented at a position angle of $64\degr$, this may be an effect of elongated beams and imperfect cleaning/calibration. That being said, A2 looks unquestionably extended along this position angle and inspection of images show them to be noise-limited rather than dynamic range limited by the presence of residual side lobes. It is worth noting that this alignment is also roughly aligned with the Herbig-Haro object HH 191, but not the H$_2$ outflows. Without significantly more spectral information, we classify A1 to be a candidate jet with lobes (A2). The detected component B is at first glance, coincident with source 47 from \citet{Eiroa1995}, however it is in fact offset by $2\arcsec$ to its south-east and likely associated with a nearby cluster member. By its extended morphology ($\sim0.015\pc\times0.012\pc$) B is likely to be a small \textsc{Hii} region powered by a $4000\Lsol$ B2 type star \citep[assuming optically thin emission,][]{Davies2011}. Source C is separated from A1 by $14\arcsec$ to the east and is coincident with the near infrared source 63 from \citet{Eiroa1995}, which looks like a typical, more-evolved cluster member. As for source D, it is coincident with a bright NIR source and is likely to be a stellar wind, or other relatively evolved phenomena.

%% file: Paper3.bbl
\newcommand{\noop}[1]{}
\begin{thebibliography}{}
\makeatletter
\relax
\def\mn@urlcharsother{\let\do\@makeother \do\$\do\&\do\#\do\^\do\_\do\%\do\~}
\def\mn@doi{\begingroup\mn@urlcharsother \@ifnextchar [ {\mn@doi@}
  {\mn@doi@[]}}
\def\mn@doi@[#1]#2{\def\@tempa{#1}\ifx\@tempa\@empty \href
  {http://dx.doi.org/#2} {doi:#2}\else \href {http://dx.doi.org/#2} {#1}\fi
  \endgroup}
\def\mn@eprint#1#2{\mn@eprint@#1:#2::\@nil}
\def\mn@eprint@arXiv#1{\href {http://arxiv.org/abs/#1} {{\tt arXiv:#1}}}
\def\mn@eprint@dblp#1{\href {http://dblp.uni-trier.de/rec/bibtex/#1.xml}
  {dblp:#1}}
\def\mn@eprint@#1:#2:#3:#4\@nil{\def\@tempa {#1}\def\@tempb {#2}\def\@tempc
  {#3}\ifx \@tempc \@empty \let \@tempc \@tempb \let \@tempb \@tempa \fi \ifx
  \@tempb \@empty \def\@tempb {arXiv}\fi \@ifundefined
  {mn@eprint@\@tempb}{\@tempb:\@tempc}{\expandafter \expandafter \csname
  mn@eprint@\@tempb\endcsname \expandafter{\@tempc}}}

\bibitem[\protect\citeauthoryear{{AMI Consortium} et~al.,}{{AMI Consortium}
  et~al.}{2011}]{AMI2011a}
{AMI Consortium} et~al., 2011, \mn@doi [\mnras]
  {10.1111/j.1365-2966.2011.18755.x}, \href
  {http://adsabs.harvard.edu/abs/2011MNRAS.415..893A} {415, 893}

\bibitem[\protect\citeauthoryear{{Ainsworth}, {Scaife}, {Shimwell},
  {Titterington}  \& {Waldram}}{{Ainsworth} et~al.}{2012}]{AMI2012}
{Ainsworth} R.~E.,  {Scaife} A.~M.~M.,  {Shimwell} T.,  {Titterington} D.,
  {Waldram} E.,  2012, \mn@doi [MNRAS] {10.1111/j.1365-2966.2012.20935.x},
  \href {http://adsabs.harvard.edu/abs/2012MNRAS.423.1089A} {423, 1089}

\bibitem[\protect\citeauthoryear{{Ainsworth}, {Scaife}, {Ray}, {Taylor},
  {Green}  \& {Buckle}}{{Ainsworth} et~al.}{2014}]{Ainsworth2014}
{Ainsworth} R.~E.,  {Scaife} A.~M.~M.,  {Ray} T.~P.,  {Taylor} A.~M.,  {Green}
  D.~A.,   {Buckle} J.~V.,  2014, \mn@doi [\apjl]
  {10.1088/2041-8205/792/1/L18}, \href
  {http://adsabs.harvard.edu/abs/2014ApJ...792L..18A} {792, L18}

\bibitem[\protect\citeauthoryear{Akritas \& Bershady}{Akritas \&
  Bershady}{1996}]{Akritas1996a}
Akritas M.~G.,  Bershady M.~A.,  1996, \mn@doi [ApJ] {10.1086/177901}, 470, 706

\bibitem[\protect\citeauthoryear{Akritas, Murphy  \& LaValley}{Akritas
  et~al.}{1995}]{Akritas1995}
Akritas M.~G.,  Murphy S.~A.,   LaValley M.~P.,  1995, Journal of the American
  Statistical Association, 90, 170

\bibitem[\protect\citeauthoryear{{Alvarez}, {Hoare}, {Glindemann}  \&
  {Richichi}}{{Alvarez} et~al.}{2004}]{Alvarez2004}
{Alvarez} C.,  {Hoare} M.,  {Glindemann} A.,   {Richichi} A.,  2004, \mn@doi
  [\aap] {10.1051/0004-6361:20034438}, \href
  {http://adsabs.harvard.edu/abs/2004A%26A...427..505A} {427, 505}

\bibitem[\protect\citeauthoryear{{Anglada}}{{Anglada}}{1995}]{Anglada1995}
{Anglada} G.,  1995, RMxAC, \href
  {http://adsabs.harvard.edu/abs/1995RMxAC...1...67A} {1, 67}

\bibitem[\protect\citeauthoryear{{Anglada} \& {Rodr{\'{\i}}guez}}{{Anglada} \&
  {Rodr{\'{\i}}guez}}{2002}]{Anglada2002}
{Anglada} G.,  {Rodr{\'{\i}}guez} L.~F.,  2002, RMAA, \href
  {http://adsabs.harvard.edu/abs/2002RMxAA..38...13A} {38, 13}

\bibitem[\protect\citeauthoryear{{Aspin}, {Sandell}  \& {Weintraub}}{{Aspin}
  et~al.}{1994}]{Aspin1994}
{Aspin} C.,  {Sandell} G.,   {Weintraub} D.~A.,  1994, \aap, \href
  {http://adsabs.harvard.edu/abs/1994A\%26A...282L..25A} {282, L25}

\bibitem[\protect\citeauthoryear{{Astropy Collaboration} et~al.,}{{Astropy
  Collaboration} et~al.}{2013}]{Astropy2013}
{Astropy Collaboration} et~al., 2013, \mn@doi [\aap]
  {10.1051/0004-6361/201322068}, \href
  {http://adsabs.harvard.edu/abs/2013A%26A...558A..33A} {558, A33}

\bibitem[\protect\citeauthoryear{{Bartkiewicz}, {Szymczak}, {van Langevelde},
  {Richards}  \& {Pihlstr{\"o}m}}{{Bartkiewicz} et~al.}{2009}]{Bartkiewicz2009}
{Bartkiewicz} A.,  {Szymczak} M.,  {van Langevelde} H.~J.,  {Richards}
  A.~M.~S.,   {Pihlstr{\"o}m} Y.~M.,  2009, \mn@doi [\aap]
  {10.1051/0004-6361/200912250}, \href
  {http://adsabs.harvard.edu/abs/2009A\%26A...502..155B} {502, 155}

\bibitem[\protect\citeauthoryear{{Battersby}, {Bally}, {Jackson}, {Ginsburg},
  {Shirley}, {Schlingman}  \& {Glenn}}{{Battersby}
  et~al.}{2010}]{Battersby2010}
{Battersby} C.,  {Bally} J.,  {Jackson} J.~M.,  {Ginsburg} A.,  {Shirley}
  Y.~L.,  {Schlingman} W.,   {Glenn} J.,  2010, \mn@doi [ApJ]
  {10.1088/0004-637X/721/1/222}, \href
  {http://adsabs.harvard.edu/abs/2010ApJ...721..222B} {721, 222}

\bibitem[\protect\citeauthoryear{{Bell}}{{Bell}}{1978}]{Bell1978}
{Bell} A.~R.,  1978, \mn@doi [MNRAS] {10.1093/mnras/182.2.147}, \href
  {http://adsabs.harvard.edu/abs/1978MNRAS.182..147B} {182, 147}

\bibitem[\protect\citeauthoryear{{Beltr{\'a}n}, {Brand}, {Cesaroni}, {Fontani},
  {Pezzuto}, {Testi}  \& {Molinari}}{{Beltr{\'a}n} et~al.}{2006}]{Beltran2006}
{Beltr{\'a}n} M.~T.,  {Brand} J.,  {Cesaroni} R.,  {Fontani} F.,  {Pezzuto} S.,
   {Testi} L.,   {Molinari} S.,  2006, \mn@doi [A\&A]
  {10.1051/0004-6361:20053999}, \href
  {http://adsabs.harvard.edu/abs/2006A\%26A...447..221B} {447, 221}

\bibitem[\protect\citeauthoryear{{Beltr{\'a}n}, {Cesaroni}, {Moscadelli},
  {S{\'a}nchez-Monge}, {Hirota}  \& {Kumar}}{{Beltr{\'a}n}
  et~al.}{2016}]{Beltran2016}
{Beltr{\'a}n} M.~T.,  {Cesaroni} R.,  {Moscadelli} L.,  {S{\'a}nchez-Monge}
  {\'A}.,  {Hirota} T.,   {Kumar} M.~S.~N.,  2016, \mn@doi [\aap]
  {10.1051/0004-6361/201628588}, \href
  {https://ui.adsabs.harvard.edu/abs/2016A&A...593A..49B} {593, A49}

\bibitem[\protect\citeauthoryear{{Beuther}, {Schilke}, {Sridharan}, {Menten},
  {Walmsley}  \& {Wyrowski}}{{Beuther} et~al.}{2002a}]{Beuther2002}
{Beuther} H.,  {Schilke} P.,  {Sridharan} T.~K.,  {Menten} K.~M.,  {Walmsley}
  C.~M.,   {Wyrowski} F.,  2002a, \mn@doi [A\&A] {10.1051/0004-6361:20011808},
  \href {http://adsabs.harvard.edu/abs/2002A%26A...383..892B} {383, 892}

\bibitem[\protect\citeauthoryear{{Beuther}, {Schilke}, {Gueth}, {McCaughrean},
  {Andersen}, {Sridharan}  \& {Menten}}{{Beuther} et~al.}{2002b}]{Beuther2002b}
{Beuther} H.,  {Schilke} P.,  {Gueth} F.,  {McCaughrean} M.,  {Andersen} M.,
  {Sridharan} T.~K.,   {Menten} K.~M.,  2002b, \mn@doi [\aap]
  {10.1051/0004-6361:20020319}, \href
  {http://adsabs.harvard.edu/abs/2002A\%26A...387..931B} {387, 931}

\bibitem[\protect\citeauthoryear{{Beuther}, {Zhang}, {Hunter}, {Sridharan}  \&
  {Bergin}}{{Beuther} et~al.}{2007}]{Beuther2007}
{Beuther} H.,  {Zhang} Q.,  {Hunter} T.~R.,  {Sridharan} T.~K.,   {Bergin}
  E.~A.,  2007, \mn@doi [\aap] {10.1051/0004-6361:20077992}, \href
  {http://adsabs.harvard.edu/abs/2007A\%26A...473..493B} {473, 493}

\bibitem[\protect\citeauthoryear{{Bica}, {Dutra}, {Soares}  \& {Barbuy}}{{Bica}
  et~al.}{2003}]{Bica2003}
{Bica} E.,  {Dutra} C.~M.,  {Soares} J.,   {Barbuy} B.,  2003, \mn@doi [\aap]
  {10.1051/0004-6361:20030486}, \href
  {http://adsabs.harvard.edu/abs/2003A\%26A...404..223B} {404, 223}

\bibitem[\protect\citeauthoryear{{Blandford} \& {Payne}}{{Blandford} \&
  {Payne}}{1982}]{BlandfordPayne1982}
{Blandford} R.~D.,  {Payne} D.~G.,  1982, MNRAS, \href
  {http://adsabs.harvard.edu/abs/1982MNRAS.199..883B} {199, 883}

\bibitem[\protect\citeauthoryear{{Bonaldi}, {Bonato}, {Galluzzi}, {Harrison},
  {Massardi}, {Kay}, {De Zotti}  \& {Brown}}{{Bonaldi}
  et~al.}{2019}]{Bonaldi2019}
{Bonaldi} A.,  {Bonato} M.,  {Galluzzi} V.,  {Harrison} I.,  {Massardi} M.,
  {Kay} S.,  {De Zotti} G.,   {Brown} M.~L.,  2019, \mn@doi [\mnras]
  {10.1093/mnras/sty2603}, \href
  {https://ui.adsabs.harvard.edu/abs/2019MNRAS.482....2B} {482, 2}

\bibitem[\protect\citeauthoryear{{Bonnell}, {Bate}, {Clarke}  \&
  {Pringle}}{{Bonnell} et~al.}{2001}]{Bonnell2001a}
{Bonnell} I.~A.,  {Bate} M.~R.,  {Clarke} C.~J.,   {Pringle} J.~E.,  2001,
  \mn@doi [MNRAS] {10.1046/j.1365-8711.2001.04270.x}, \href
  {http://adsabs.harvard.edu/abs/2001MNRAS.323..785B} {323, 785}

\bibitem[\protect\citeauthoryear{{Bunn}, {Hoare}  \& {Drew}}{{Bunn}
  et~al.}{1995}]{Bunn1995}
{Bunn} J.~C.,  {Hoare} M.~G.,   {Drew} J.~E.,  1995, \mn@doi [\mnras]
  {10.1093/mnras/272.2.346}, \href
  {http://adsabs.harvard.edu/abs/1995MNRAS.272..346B} {272, 346}

\bibitem[\protect\citeauthoryear{{Burns}, {Handa}, {Nagayama}, {Sunada}  \&
  {Omodaka}}{{Burns} et~al.}{2016}]{Burns2016}
{Burns} R.~A.,  {Handa} T.,  {Nagayama} T.,  {Sunada} K.,   {Omodaka} T.,
  2016, \mn@doi [\mnras] {10.1093/mnras/stw958}, \href
  {http://adsabs.harvard.edu/abs/2016MNRAS.460..283B} {460, 283}

\bibitem[\protect\citeauthoryear{{Burns} et~al.,}{{Burns}
  et~al.}{2017}]{Burns2017}
{Burns} R.~A.,  et~al., 2017, \mn@doi [\mnras] {10.1093/mnras/stx216}, \href
  {http://adsabs.harvard.edu/abs/2017MNRAS.467.2367B} {467, 2367}

\bibitem[\protect\citeauthoryear{{Caratti o Garatti} et~al.,}{{Caratti o
  Garatti} et~al.}{2017}]{Caratti2017}
{Caratti o Garatti} A.,  et~al., 2017, \mn@doi [Nature Physics]
  {10.1038/nphys3942}, \href
  {https://ui.adsabs.harvard.edu/abs/2017NatPh..13..276C} {13, 276}

\bibitem[\protect\citeauthoryear{{Carpenter}, {Snell}  \&
  {Schloerb}}{{Carpenter} et~al.}{1990}]{Carpenter1990}
{Carpenter} J.~M.,  {Snell} R.~L.,   {Schloerb} F.~P.,  1990, \mn@doi [\apj]
  {10.1086/169251}, \href {http://adsabs.harvard.edu/abs/1990ApJ...362..147C}
  {362, 147}

\bibitem[\protect\citeauthoryear{{Carral}, {Kurtz}, {Rodr{\'{\i}}guez},
  {Mart{\'{\i}}}, {Lizano}  \& {Osorio}}{{Carral} et~al.}{1999}]{Carral1999}
{Carral} P.,  {Kurtz} S.,  {Rodr{\'{\i}}guez} L.~F.,  {Mart{\'{\i}}} J.,
  {Lizano} S.,   {Osorio} M.,  1999, \rmxaa, \href
  {http://adsabs.harvard.edu/abs/1999RMxAA..35...97C} {35, 97}

\bibitem[\protect\citeauthoryear{{Carrasco-Gonz{\'a}lez}, {Rodr{\'{\i}}guez},
  {Torrelles}, {Anglada}  \&
  {Gonz{\'a}lez-Mart{\'{\i}}n}}{{Carrasco-Gonz{\'a}lez}
  et~al.}{2010}]{CarrascoGonzalez2010}
{Carrasco-Gonz{\'a}lez} C.,  {Rodr{\'{\i}}guez} L.~F.,  {Torrelles} J.~M.,
  {Anglada} G.,   {Gonz{\'a}lez-Mart{\'{\i}}n} O.,  2010, \mn@doi [ApJ]
  {10.1088/0004-6256/139/6/2433}, \href
  {http://adsabs.harvard.edu/abs/2010AJ....139.2433C} {139, 2433}

\bibitem[\protect\citeauthoryear{{Carrasco-Gonz{\'a}lez}
  et~al.,}{{Carrasco-Gonz{\'a}lez} et~al.}{2015}]{CarrascoGonzalez2015}
{Carrasco-Gonz{\'a}lez} C.,  et~al., 2015, \mn@doi [Science]
  {10.1126/science.aaa7216}, \href
  {http://adsabs.harvard.edu/abs/2015Sci...348..114C} {348, 114}

\bibitem[\protect\citeauthoryear{{Cesaroni} et~al.,}{{Cesaroni}
  et~al.}{2018}]{Cesaroni2018}
{Cesaroni} R.,  et~al., 2018, \mn@doi [\aap] {10.1051/0004-6361/201732238},
  \href {http://adsabs.harvard.edu/abs/2018A%26A...612A.103C} {612, A103}

\bibitem[\protect\citeauthoryear{{Chambers}, {Jackson}, {Rathborne}  \&
  {Simon}}{{Chambers} et~al.}{2009}]{Chambers2009}
{Chambers} E.~T.,  {Jackson} J.~M.,  {Rathborne} J.~M.,   {Simon} R.,  2009,
  \mn@doi [\apjs] {10.1088/0067-0049/181/2/360}, \href
  {http://adsabs.harvard.edu/abs/2009ApJS..181..360C} {181, 360}

\bibitem[\protect\citeauthoryear{{Chen}, {Yao}, {Yang}, {Zeng}  \&
  {Sato}}{{Chen} et~al.}{2009}]{Chen2009}
{Chen} Y.,  {Yao} Y.,  {Yang} J.,  {Zeng} Q.,   {Sato} S.,  2009, \mn@doi
  [\apj] {10.1088/0004-637X/693/1/430}, \href
  {http://adsabs.harvard.edu/abs/2009ApJ...693..430C} {693, 430}

\bibitem[\protect\citeauthoryear{{Chen}, {Keto}, {Zhang}, {Sridharan}, {Liu}
  \& {Su}}{{Chen} et~al.}{2016}]{Chen2016}
{Chen} H.-R.~V.,  {Keto} E.,  {Zhang} Q.,  {Sridharan} T.~K.,  {Liu} S.-Y.,
  {Su} Y.-N.,  2016, \mn@doi [\apj] {10.3847/0004-637X/823/2/125}, \href
  {http://adsabs.harvard.edu/abs/2016ApJ...823..125C} {823, 125}

\bibitem[\protect\citeauthoryear{Choi, Hachisuka, Reid, Xu, Brunthaler, Menten
  \& Dame}{Choi et~al.}{2014}]{Choi2014}
Choi Y.~K.,  Hachisuka K.,  Reid M.~J.,  Xu Y.,  Brunthaler A.,  Menten K.~M.,
   Dame T.~M.,  2014, \mn@doi [ApJ] {10.1088/0004-637x/790/2/99}, 790, 99

\bibitem[\protect\citeauthoryear{{Cooper} et~al.,}{{Cooper}
  et~al.}{2013}]{Cooper2013}
{Cooper} H.~D.~B.,  et~al., 2013, \mn@doi [MNRAS] {10.1093/mnras/sts681}, \href
  {http://adsabs.harvard.edu/abs/2013MNRAS.430.1125C} {430, 1125}

\bibitem[\protect\citeauthoryear{{Cyganowski}, {Brogan}, {Hunter}  \&
  {Churchwell}}{{Cyganowski} et~al.}{2011}]{Cyganowski2011}
{Cyganowski} C.~J.,  {Brogan} C.~L.,  {Hunter} T.~R.,   {Churchwell} E.,  2011,
  \mn@doi [\apj] {10.1088/0004-637X/743/1/56}, \href
  {http://adsabs.harvard.edu/abs/2011ApJ...743...56C} {743, 56}

\bibitem[\protect\citeauthoryear{{Davies}, {Hoare}, {Lumsden}, {Hosokawa},
  {Oudmaijer}, {Urquhart}, {Mottram}  \& {Stead}}{{Davies}
  et~al.}{2011}]{Davies2011}
{Davies} B.,  {Hoare} M.~G.,  {Lumsden} S.~L.,  {Hosokawa} T.,  {Oudmaijer}
  R.~D.,  {Urquhart} J.~S.,  {Mottram} J.~C.,   {Stead} J.,  2011, \mn@doi
  [MNRAS] {10.1111/j.1365-2966.2011.19095.x}, \href
  {http://adsabs.harvard.edu/abs/2011MNRAS.416..972D} {416, 972}

\bibitem[\protect\citeauthoryear{{Deharveng}, {Zavagno}, {Cruz-Gonzalez},
  {Salas}, {Caplan}  \& {Carrasco}}{{Deharveng} et~al.}{1997}]{Deharveng1997}
{Deharveng} L.,  {Zavagno} A.,  {Cruz-Gonzalez} I.,  {Salas} L.,  {Caplan} J.,
   {Carrasco} L.,  1997, \aap, \href
  {http://adsabs.harvard.edu/abs/1997A\%26A...317..459D} {317, 459}

\bibitem[\protect\citeauthoryear{{Draine}}{{Draine}}{2006}]{Draine2006}
{Draine} B.~T.,  2006, \mn@doi [\apj] {10.1086/498130}, \href
  {http://adsabs.harvard.edu/abs/2006ApJ...636.1114D} {636, 1114}

\bibitem[\protect\citeauthoryear{{Eiroa} \& {Casali}}{{Eiroa} \&
  {Casali}}{1995}]{Eiroa1995}
{Eiroa} C.,  {Casali} M.~M.,  1995, \aap, \href
  {http://adsabs.harvard.edu/abs/1995A\%26A...303...87E} {303, 87}

\bibitem[\protect\citeauthoryear{{Eiroa}, {Casali}, {Miranda}  \&
  {Ortiz}}{{Eiroa} et~al.}{1994}]{Eiroa1994}
{Eiroa} C.,  {Casali} M.~M.,  {Miranda} L.~F.,   {Ortiz} E.,  1994, \aap, \href
  {http://adsabs.harvard.edu/abs/1994A\%26A...290..599E} {290, 599}

\bibitem[\protect\citeauthoryear{Fedriani et~al.,}{Fedriani
  et~al.}{2019}]{Fedriani2019}
Fedriani R.,  et~al., 2019, \mn@doi [Nature Communications]
  {10.1038/s41467-019-11595-x}, 10

\bibitem[\protect\citeauthoryear{{Fontani}, {Cesaroni}, {Testi}, {Molinari},
  {Zhang}, {Brand}  \& {Walmsley}}{{Fontani} et~al.}{2004}]{Fontani2004}
{Fontani} F.,  {Cesaroni} R.,  {Testi} L.,  {Molinari} S.,  {Zhang} Q.,
  {Brand} J.,   {Walmsley} C.~M.,  2004, \mn@doi [\aap]
  {10.1051/0004-6361:20035848}, \href
  {http://adsabs.harvard.edu/abs/2004A\%26A...424..179F} {424, 179}

\bibitem[\protect\citeauthoryear{{Fontani}, {Cesaroni}  \& {Furuya}}{{Fontani}
  et~al.}{2010}]{Fontani2010}
{Fontani} F.,  {Cesaroni} R.,   {Furuya} R.~S.,  2010, \mn@doi [\aap]
  {10.1051/0004-6361/200913679}, \href
  {http://adsabs.harvard.edu/abs/2010A\%26A...517A..56F} {517, A56}

\bibitem[\protect\citeauthoryear{{Frank}, {Lery}, {Gardiner}, {Jones}  \&
  {Ryu}}{{Frank} et~al.}{2000}]{Frank2000}
{Frank} A.,  {Lery} T.,  {Gardiner} T.~A.,  {Jones} T.~W.,   {Ryu} D.,  2000,
  \mn@doi [\apj] {10.1086/309298}, \href
  {https://ui.adsabs.harvard.edu/abs/2000ApJ...540..342F} {540, 342}

\bibitem[\protect\citeauthoryear{{Frank} et~al.,}{{Frank}
  et~al.}{2014}]{Frank2014PPVI}
{Frank} A.,  et~al., 2014, \mn@doi [Protostars and Planets VI]
  {10.2458/azu_uapress_9780816531240-ch020}, \href
  {http://adsabs.harvard.edu/abs/2014prpl.conf..451F} {pp 451--474}

\bibitem[\protect\citeauthoryear{{Fujisawa} et~al.,}{{Fujisawa}
  et~al.}{2012}]{Fujisawa2012}
{Fujisawa} K.,  et~al., 2012, \mn@doi [PASJ] {10.1093/pasj/64.1.17}, \href
  {http://adsabs.harvard.edu/abs/2012PASJ...64...17F} {64, 17}

\bibitem[\protect\citeauthoryear{{Furuya}, {Kitamura}, {Wootten}, {Claussen}
  \& {Kawabe}}{{Furuya} et~al.}{2003}]{Furuya2003}
{Furuya} R.~S.,  {Kitamura} Y.,  {Wootten} A.,  {Claussen} M.~J.,   {Kawabe}
  R.,  2003, \mn@doi [ApJS] {10.1086/342749}, \href
  {http://adsabs.harvard.edu/abs/2003ApJS..144...71F} {144, 71}

\bibitem[\protect\citeauthoryear{{Galv{\'a}n-Madrid}, {Zhang}, {Keto}, {Ho},
  {Zapata}, {Rodr{\'{\i}}guez}, {Pineda}  \&
  {V{\'a}zquez-Semadeni}}{{Galv{\'a}n-Madrid} et~al.}{2010}]{GalvanMadrid2010}
{Galv{\'a}n-Madrid} R.,  {Zhang} Q.,  {Keto} E.,  {Ho} P.~T.~P.,  {Zapata}
  L.~A.,  {Rodr{\'{\i}}guez} L.~F.,  {Pineda} J.~E.,   {V{\'a}zquez-Semadeni}
  E.,  2010, \mn@doi [ApJ] {10.1088/0004-637X/725/1/17}, \href
  {http://adsabs.harvard.edu/abs/2010ApJ...725...17G} {725, 17}

\bibitem[\protect\citeauthoryear{{Garay}, {Rodr{\'{\i}}guez}  \& {de
  Gregorio-Monsalvo}}{{Garay} et~al.}{2007}]{Garay2007b}
{Garay} G.,  {Rodr{\'{\i}}guez} L.~F.,   {de Gregorio-Monsalvo} I.,  2007,
  \mn@doi [\apj] {10.1086/520334}, \href
  {http://adsabs.harvard.edu/abs/2007AJ....134..906G} {134, 906}

\bibitem[\protect\citeauthoryear{{Gardiner} \& {Frank}}{{Gardiner} \&
  {Frank}}{2000}]{GardinerFrank2000}
{Gardiner} T.~A.,  {Frank} A.,  2000, \mn@doi [\apjl] {10.1086/317875}, \href
  {https://ui.adsabs.harvard.edu/abs/2000ApJ...545L.153G} {545, L153}

\bibitem[\protect\citeauthoryear{{Gibb}, {Hoare}, {Little}  \& {Wright}}{{Gibb}
  et~al.}{2003}]{GibbHoare2003}
{Gibb} A.~G.,  {Hoare} M.~G.,  {Little} L.~T.,   {Wright} M.~C.~H.,  2003,
  \mn@doi [MNRAS] {10.1046/j.1365-8711.2003.06251.x}, \href
  {http://adsabs.harvard.edu/abs/2003MNRAS.339.1011G} {339, 1011}

\bibitem[\protect\citeauthoryear{{Ginsburg}, {Bally}, {Yan}  \&
  {Williams}}{{Ginsburg} et~al.}{2009}]{Ginsburg2009}
{Ginsburg} A.~G.,  {Bally} J.,  {Yan} C.-H.,   {Williams} J.~P.,  2009, \mn@doi
  [\apj] {10.1088/0004-637X/707/1/310}, \href
  {http://adsabs.harvard.edu/abs/2009ApJ...707..310G} {707, 310}

\bibitem[\protect\citeauthoryear{{Goddi} \& {Moscadelli}}{{Goddi} \&
  {Moscadelli}}{2006}]{Goddi2006}
{Goddi} C.,  {Moscadelli} L.,  2006, \mn@doi [\aap]
  {10.1051/0004-6361:20053721}, \href
  {http://adsabs.harvard.edu/abs/2006A\%26A...447..577G} {447, 577}

\bibitem[\protect\citeauthoryear{{Goddi}, {Moscadelli}, {Alef}, {Tarchi},
  {Brand}  \& {Pani}}{{Goddi} et~al.}{2005}]{Goddi2005}
{Goddi} C.,  {Moscadelli} L.,  {Alef} W.,  {Tarchi} A.,  {Brand} J.,   {Pani}
  M.,  2005, \mn@doi [\aap] {10.1051/0004-6361:20042074}, \href
  {http://adsabs.harvard.edu/abs/2005A\%26A...432..161G} {432, 161}

\bibitem[\protect\citeauthoryear{{Gomez}, {Torrelles}, {Estalella}, {Anglada},
  {Verdes-Montenegro}  \& {Ho}}{{Gomez} et~al.}{1992}]{Gomez1992}
{Gomez} J.~F.,  {Torrelles} J.~M.,  {Estalella} R.,  {Anglada} G.,
  {Verdes-Montenegro} L.,   {Ho} P.~T.~P.,  1992, \mn@doi [\apj]
  {10.1086/171806}, \href {http://adsabs.harvard.edu/abs/1992ApJ...397..492G}
  {397, 492}

\bibitem[\protect\citeauthoryear{{Gottschalk}, {Kothes}, {Matthews},
  {Landecker}  \& {Dent}}{{Gottschalk} et~al.}{2012}]{Gottschalk2012}
{Gottschalk} M.,  {Kothes} R.,  {Matthews} H.~E.,  {Landecker} T.~L.,   {Dent}
  W.~R.~F.,  2012, \mn@doi [A] {10.1051/0004-6361/201118600}, \href
  {http://adsabs.harvard.edu/abs/2012A\%26A...541A..79G} {541, A79}

\bibitem[\protect\citeauthoryear{{Guzm{\'a}n}, {Garay}, {Brooks}  \&
  {Voronkov}}{{Guzm{\'a}n} et~al.}{2012}]{Guzman2012}
{Guzm{\'a}n} A.~E.,  {Garay} G.,  {Brooks} K.~J.,   {Voronkov} M.~A.,  2012,
  \mn@doi [ApJ] {10.1088/0004-637X/753/1/51}, \href
  {http://adsabs.harvard.edu/abs/2012ApJ...753...51G} {753, 51}

\bibitem[\protect\citeauthoryear{{Guzm{\'a}n}, {Garay}, {Rodr{\'{\i}}guez},
  {Contreras}, {Dougados}  \& {Cabrit}}{{Guzm{\'a}n} et~al.}{2016}]{Guzman2016}
{Guzm{\'a}n} A.~E.,  {Garay} G.,  {Rodr{\'{\i}}guez} L.~F.,  {Contreras} Y.,
  {Dougados} C.,   {Cabrit} S.,  2016, \mn@doi [\apj]
  {10.3847/0004-637X/826/2/208}, \href
  {http://adsabs.harvard.edu/abs/2016ApJ...826..208G} {826, 208}

\bibitem[\protect\citeauthoryear{{Hachisuka} et~al.,}{{Hachisuka}
  et~al.}{2006}]{Hachisuka2006}
{Hachisuka} K.,  et~al., 2006, \mn@doi [\apj] {10.1086/502962}, \href
  {http://adsabs.harvard.edu/abs/2006ApJ...645..337H} {645, 337}

\bibitem[\protect\citeauthoryear{{Hartigan}, {Morse}  \& {Raymond}}{{Hartigan}
  et~al.}{1994}]{Hartigan1994}
{Hartigan} P.,  {Morse} J.~A.,   {Raymond} J.,  1994, \mn@doi [ApJ]
  {10.1086/174887}, \href {http://adsabs.harvard.edu/abs/1994ApJ...436..125H}
  {436, 125}

\bibitem[\protect\citeauthoryear{{Heyer}, {Snell}, {Morgan}  \&
  {Schloerb}}{{Heyer} et~al.}{1989}]{Heyer1989}
{Heyer} M.~H.,  {Snell} R.~L.,  {Morgan} J.,   {Schloerb} F.~P.,  1989, \mn@doi
  [\apj] {10.1086/168003}, \href
  {http://adsabs.harvard.edu/abs/1989ApJ...346..220H} {346, 220}

\bibitem[\protect\citeauthoryear{{Hildebrand}}{{Hildebrand}}{1983}]{Hildebrand1983}
{Hildebrand} R.~H.,  1983, QJRAS, \href
  {http://adsabs.harvard.edu/abs/1983QJRAS..24..267H} {24, 267}

\bibitem[\protect\citeauthoryear{{Honma} et~al.,}{{Honma}
  et~al.}{2007}]{Honma2007}
{Honma} M.,  et~al., 2007, \mn@doi [\pasj] {10.1093/pasj/59.5.889}, \href
  {http://adsabs.harvard.edu/abs/2007PASJ...59..889H} {59, 889}

\bibitem[\protect\citeauthoryear{{Hosokawa}, {Yorke}  \& {Omukai}}{{Hosokawa}
  et~al.}{2010}]{Hosokawa2010}
{Hosokawa} T.,  {Yorke} H.~W.,   {Omukai} K.,  2010, ApJ, 721

\bibitem[\protect\citeauthoryear{{Hunter}, {Testi}, {Taylor}, {Tofani}, {Felli}
   \& {Phillips}}{{Hunter} et~al.}{1995}]{Hunter1995}
{Hunter} T.~R.,  {Testi} L.,  {Taylor} G.~B.,  {Tofani} G.,  {Felli} M.,
  {Phillips} T.~G.,  1995, \aap, \href
  {http://adsabs.harvard.edu/abs/1995A\%26A...302..249H} {302, 249}

\bibitem[\protect\citeauthoryear{{Hunter}, {Testi}, {Zhang}  \&
  {Sridharan}}{{Hunter} et~al.}{1999}]{Hunter1999}
{Hunter} T.~R.,  {Testi} L.,  {Zhang} Q.,   {Sridharan} T.~K.,  1999, \mn@doi
  [\aj] {10.1086/300936}, \href
  {http://adsabs.harvard.edu/abs/1999AJ....118..477H} {118, 477}

\bibitem[\protect\citeauthoryear{{Ilee} et~al.,}{{Ilee}
  et~al.}{2013}]{Ilee2013}
{Ilee} J.~D.,  et~al., 2013, \mn@doi [MNRAS] {10.1093/mnras/sts537}, \href
  {http://adsabs.harvard.edu/abs/2013MNRAS.429.2960I} {429, 2960}

\bibitem[\protect\citeauthoryear{{Imai}, {Kameya}, {Sasao}, {Miyoshi},
  {Deguchi}, {Horiuchi}  \& {Asaki}}{{Imai} et~al.}{2000}]{Imai2000}
{Imai} H.,  {Kameya} O.,  {Sasao} T.,  {Miyoshi} M.,  {Deguchi} S.,  {Horiuchi}
  S.,   {Asaki} Y.,  2000, \mn@doi [ApJ] {10.1086/309165}, \href
  {http://adsabs.harvard.edu/abs/2000ApJ...538..751I} {538, 751}

\bibitem[\protect\citeauthoryear{{Ishii}, {Hirao}, {Nagashima}, {Nagata},
  {Sato}  \& {Yao}}{{Ishii} et~al.}{2002}]{Ishii2002}
{Ishii} M.,  {Hirao} T.,  {Nagashima} C.,  {Nagata} T.,  {Sato} S.,   {Yao} Y.,
   2002, \mn@doi [\aj] {10.1086/340961}, \href
  {http://adsabs.harvard.edu/abs/2002AJ....124..430I} {124, 430}

\bibitem[\protect\citeauthoryear{{Jiang} et~al.,}{{Jiang}
  et~al.}{2003}]{Jiang2003}
{Jiang} Z.,  et~al., 2003, \mn@doi [\apj] {10.1086/378150}, \href
  {http://adsabs.harvard.edu/abs/2003ApJ...596.1064J} {596, 1064}

\bibitem[\protect\citeauthoryear{{Kawamura}, {Onishi}, {Yonekura}, {Dobashi},
  {Mizuno}, {Ogawa}  \& {Fukui}}{{Kawamura} et~al.}{1998}]{Kawamura1998}
{Kawamura} A.,  {Onishi} T.,  {Yonekura} Y.,  {Dobashi} K.,  {Mizuno} A.,
  {Ogawa} H.,   {Fukui} Y.,  1998, \mn@doi [\apjs] {10.1086/313119}, \href
  {http://adsabs.harvard.edu/abs/1998ApJS..117..387K} {117, 387}

\bibitem[\protect\citeauthoryear{{Keto}}{{Keto}}{2002}]{Keto2002}
{Keto} E.,  2002, \mn@doi [\apj] {10.1086/343794}, \href
  {http://adsabs.harvard.edu/abs/2002ApJ...580..980K} {580, 980}

\bibitem[\protect\citeauthoryear{{Kroupa}}{{Kroupa}}{2002}]{KroupaIMF}
{Kroupa} P.,  2002, \mn@doi [Science] {10.1126/science.1067524}, \href
  {http://adsabs.harvard.edu/abs/2002Sci...295...82K} {295, 82}

\bibitem[\protect\citeauthoryear{{Kumar}, {Keto}  \& {Clerkin}}{{Kumar}
  et~al.}{2006}]{Kumar2006}
{Kumar} M.~S.~N.,  {Keto} E.,   {Clerkin} E.,  2006, \mn@doi [\aap]
  {10.1051/0004-6361:20053104}, \href
  {http://adsabs.harvard.edu/abs/2006A\%26A...449.1033K} {449, 1033}

\bibitem[\protect\citeauthoryear{{Kurtz}, {Churchwell}  \& {Wood}}{{Kurtz}
  et~al.}{1994}]{Kurtz1994}
{Kurtz} S.,  {Churchwell} E.,   {Wood} D.~O.~S.,  1994, \mn@doi [\apjs]
  {10.1086/191952}, \href {http://adsabs.harvard.edu/abs/1994ApJS...91..659K}
  {91, 659}

\bibitem[\protect\citeauthoryear{{Lee} et~al.,}{{Lee} et~al.}{2013}]{Lee2013}
{Lee} H.-T.,  et~al., 2013, \mn@doi [ApJS] {10.1088/0067-0049/208/2/23}, \href
  {http://adsabs.harvard.edu/abs/2013ApJS..208...23L} {208, 23}

\bibitem[\protect\citeauthoryear{{Lefloch}, {Lazareff}  \& {Castets}}{{Lefloch}
  et~al.}{1997}]{Lefloch1997}
{Lefloch} B.,  {Lazareff} B.,   {Castets} A.,  1997, \aap, \href
  {http://adsabs.harvard.edu/abs/1997A\%26A...324..249L} {324, 249}

\bibitem[\protect\citeauthoryear{{Lodders}}{{Lodders}}{2003}]{Lodders2003}
{Lodders} K.,  2003, \mn@doi [\apj] {10.1086/375492}, \href
  {http://adsabs.harvard.edu/abs/2003ApJ...591.1220L} {591, 1220}

\bibitem[\protect\citeauthoryear{{L{\'o}pez-Sepulcre}, {Cesaroni}  \&
  {Walmsley}}{{L{\'o}pez-Sepulcre} et~al.}{2010}]{LopezSepulcre2010}
{L{\'o}pez-Sepulcre} A.,  {Cesaroni} R.,   {Walmsley} C.~M.,  2010, \mn@doi
  [A\&A] {10.1051/0004-6361/201014252}, \href
  {http://adsabs.harvard.edu/abs/2010A\%26A...517A..66L} {517, A66}

\bibitem[\protect\citeauthoryear{{Lumsden}, {Wheelwright}, {Hoare}, {Oudmaijer}
   \& {Drew}}{{Lumsden} et~al.}{2012}]{Lumsden2012}
{Lumsden} S.~L.,  {Wheelwright} H.~E.,  {Hoare} M.~G.,  {Oudmaijer} R.~D.,
  {Drew} J.~E.,  2012, \mn@doi [\mnras] {10.1111/j.1365-2966.2012.21280.x},
  \href {http://adsabs.harvard.edu/abs/2012MNRAS.424.1088L} {424, 1088}

\bibitem[\protect\citeauthoryear{{Lumsden}, {Hoare}, {Urquhart}, {Oudmaijer},
  {Davies}, {Mottram}, {Cooper}  \& {Moore}}{{Lumsden}
  et~al.}{2013}]{Lumsden2013}
{Lumsden} S.~L.,  {Hoare} M.~G.,  {Urquhart} J.~S.,  {Oudmaijer} R.~D.,
  {Davies} B.,  {Mottram} J.~C.,  {Cooper} H.~D.~B.,   {Moore} T.~J.~T.,  2013,
  \mn@doi [ApJS] {10.1088/0067-0049/208/1/11}, \href
  {http://adsabs.harvard.edu/abs/2013ApJS..208...11L} {208, 11}

\bibitem[\protect\citeauthoryear{{Mallick} et~al.,}{{Mallick}
  et~al.}{2014}]{Mallick2014}
{Mallick} K.~K.,  et~al., 2014, \mn@doi [\mnras] {10.1093/mnras/stu1396}, \href
  {http://adsabs.harvard.edu/abs/2014MNRAS.443.3218M} {443, 3218}

\bibitem[\protect\citeauthoryear{{Marti}, {Rodriguez}  \& {Reipurth}}{{Marti}
  et~al.}{1993}]{Marti1993}
{Marti} J.,  {Rodriguez} L.~F.,   {Reipurth} B.,  1993, \mn@doi [ApJ]
  {10.1086/173227}, \href {http://adsabs.harvard.edu/abs/1993ApJ...416..208M}
  {416, 208}

\bibitem[\protect\citeauthoryear{{Masqu{\'e}}, {Rodr{\'{\i}}guez}, {Trinidad},
  {Kurtz}, {Dzib}, {Rodr{\'{\i}}guez-Rico}  \& {Loinard}}{{Masqu{\'e}}
  et~al.}{2017}]{Masque2017}
{Masqu{\'e}} J.~M.,  {Rodr{\'{\i}}guez} L.~F.,  {Trinidad} M.~A.,  {Kurtz} S.,
  {Dzib} S.~A.,  {Rodr{\'{\i}}guez-Rico} C.~A.,   {Loinard} L.,  2017, \mn@doi
  [\apj] {10.3847/1538-4357/836/1/96}, \href
  {http://adsabs.harvard.edu/abs/2017ApJ...836...96M} {836, 96}

\bibitem[\protect\citeauthoryear{{Maud}, {Moore}, {Lumsden}, {Mottram},
  {Urquhart}  \& {Hoare}}{{Maud} et~al.}{2015}]{Maud2015CO}
{Maud} L.~T.,  {Moore} T.~J.~T.,  {Lumsden} S.~L.,  {Mottram} J.~C.,
  {Urquhart} J.~S.,   {Hoare} M.~G.,  2015, \mn@doi [MNRAS]
  {10.1093/mnras/stv1635}, \href
  {http://adsabs.harvard.edu/abs/2015MNRAS.453..645M} {453, 645}

\bibitem[\protect\citeauthoryear{{McKee} \& {Tan}}{{McKee} \&
  {Tan}}{2003}]{McKeeTan2003}
{McKee} C.~F.,  {Tan} J.~C.,  2003, \mn@doi [ApJ] {10.1086/346149}, \href
  {http://adsabs.harvard.edu/abs/2003ApJ...585..850M} {585, 850}

\bibitem[\protect\citeauthoryear{{McMullin}, {Waters}, {Schiebel}, {Young}  \&
  {Golap}}{{McMullin} et~al.}{2007}]{CASARef}
{McMullin} J.~P.,  {Waters} B.,  {Schiebel} D.,  {Young} W.,   {Golap} K.,
  2007, in {Shaw} R.~A.,  {Hill} F.,   {Bell} D.~J.,  eds,  Astronomical
  Society of the Pacific Conference Series Vol. 376, Astronomical Data Analysis
  Software and Systems XVI. p.~127

\bibitem[\protect\citeauthoryear{{Meakin}, {Hines}  \& {Thompson}}{{Meakin}
  et~al.}{2005}]{Meakin2005}
{Meakin} C.~A.,  {Hines} D.~C.,   {Thompson} R.~I.,  2005, \mn@doi [\apj]
  {10.1086/496969}, \href {http://adsabs.harvard.edu/abs/2005ApJ...634.1146M}
  {634, 1146}

\bibitem[\protect\citeauthoryear{Meyer, Vorobyov, Elbakyan, Stecklum,
  Eislöffel  \& Sobolev}{Meyer et~al.}{2018}]{Meyer2019}
Meyer D. M.-A.,  Vorobyov E.~I.,  Elbakyan V.~G.,  Stecklum B.,  Eislöffel J.,
    Sobolev A.~M.,  2018, \mn@doi [MNRAS] {10.1093/mnras/sty2980}, 482, 5459

\bibitem[\protect\citeauthoryear{{Mezger} \& {Henderson}}{{Mezger} \&
  {Henderson}}{1967}]{MezgerHenderson1967}
{Mezger} P.~G.,  {Henderson} A.~P.,  1967, \mn@doi [ApJ] {10.1086/149030}, 147,
  471

\bibitem[\protect\citeauthoryear{{Minier}, {Booth}  \& {Conway}}{{Minier}
  et~al.}{2000}]{Minier2000}
{Minier} V.,  {Booth} R.~S.,   {Conway} J.~E.,  2000, \aap, \href
  {http://adsabs.harvard.edu/abs/2000A\%26A...362.1093M} {362, 1093}

\bibitem[\protect\citeauthoryear{{Minier}, {Burton}, {Hill}, {Pestalozzi},
  {Purcell}, {Garay}, {Walsh}  \& {Longmore}}{{Minier}
  et~al.}{2005}]{Minier2005}
{Minier} V.,  {Burton} M.~G.,  {Hill} T.,  {Pestalozzi} M.~R.,  {Purcell}
  C.~R.,  {Garay} G.,  {Walsh} A.~J.,   {Longmore} S.,  2005, \mn@doi [\aap]
  {10.1051/0004-6361:20041137}, \href
  {http://adsabs.harvard.edu/abs/2005A\%26A...429..945M} {429, 945}

\bibitem[\protect\citeauthoryear{{Mitchell}, {Hasegawa}  \&
  {Schella}}{{Mitchell} et~al.}{1992}]{Mitchell1992}
{Mitchell} G.~F.,  {Hasegawa} T.~I.,   {Schella} J.,  1992, \mn@doi [\apj]
  {10.1086/171042}, \href {http://adsabs.harvard.edu/abs/1992ApJ...386..604M}
  {386, 604}

\bibitem[\protect\citeauthoryear{{Molinari}, {Testi}, {Rodr{\'{\i}}guez}  \&
  {Zhang}}{{Molinari} et~al.}{2002}]{Molinari2002}
{Molinari} S.,  {Testi} L.,  {Rodr{\'{\i}}guez} L.~F.,   {Zhang} Q.,  2002,
  \mn@doi [\apj] {10.1086/339630}, \href
  {http://adsabs.harvard.edu/abs/2002ApJ...570..758M} {570, 758}

\bibitem[\protect\citeauthoryear{{Moscadelli}, {Reid}, {Menten}, {Brunthaler},
  {Zheng}  \& {Xu}}{{Moscadelli} et~al.}{2009}]{Moscadelli2009}
{Moscadelli} L.,  {Reid} M.~J.,  {Menten} K.~M.,  {Brunthaler} A.,  {Zheng}
  X.~W.,   {Xu} Y.,  2009, \mn@doi [\apj] {10.1088/0004-637X/693/1/406}, \href
  {http://adsabs.harvard.edu/abs/2009ApJ...693..406M} {693, 406}

\bibitem[\protect\citeauthoryear{{Moscadelli} et~al.,}{{Moscadelli}
  et~al.}{2016}]{Moscadelli2016}
{Moscadelli} L.,  et~al., 2016, \mn@doi [A\&A] {10.1051/0004-6361/201526238},
  \href {http://adsabs.harvard.edu/abs/2016A%26A...585A..71M} {585, A71}

\bibitem[\protect\citeauthoryear{{Mottram} et~al.,}{{Mottram}
  et~al.}{2011a}]{Mottram2011a}
{Mottram} J.~C.,  et~al., 2011a, \mn@doi [A\&A] {10.1051/0004-6361/201014479},
  \href {http://adsabs.harvard.edu/abs/2011A%26A...525A.149M} {525, A149}

\bibitem[\protect\citeauthoryear{{Mottram} et~al.,}{{Mottram}
  et~al.}{2011b}]{Mottram2011b}
{Mottram} J.~C.,  et~al., 2011b, \mn@doi [ApJ] {10.1088/2041-8205/730/2/L33},
  \href {http://adsabs.harvard.edu/abs/2011ApJ...730L..33M} {730, L33}

\bibitem[\protect\citeauthoryear{{Murakawa}, {Lumsden}, {Oudmaijer}, {Davies},
  {Wheelwright}, {Hoare}  \& {Ilee}}{{Murakawa} et~al.}{2013}]{Murakawa2013}
{Murakawa} K.,  {Lumsden} S.~L.,  {Oudmaijer} R.~D.,  {Davies} B.,
  {Wheelwright} H.~E.,  {Hoare} M.~G.,   {Ilee} J.~D.,  2013, \mn@doi [\mnras]
  {10.1093/mnras/stt1592}, \href
  {http://adsabs.harvard.edu/abs/2013MNRAS.436..511M} {436, 511}

\bibitem[\protect\citeauthoryear{{Navarete}, {Damineli}, {Barbosa}  \&
  {Blum}}{{Navarete} et~al.}{2015}]{Navarete2015}
{Navarete} F.,  {Damineli} A.,  {Barbosa} C.~L.,   {Blum} R.~D.,  2015, \mn@doi
  [MNRAS] {10.1093/mnras/stv914}, \href
  {http://adsabs.harvard.edu/abs/2015MNRAS.450.4364N} {450, 4364}

\bibitem[\protect\citeauthoryear{Obonyo, Lumsden, Hoare, Purser, Kurtz  \&
  Johnston}{Obonyo et~al.}{2019}]{Obonyo2019}
Obonyo W.~O.,  Lumsden S.~L.,  Hoare M.~G.,  Purser S. J.~D.,  Kurtz S.~E.,
  Johnston K.~G.,  2019, \mn@doi [MNRAS] {10.1093/mnras/stz1091}, 486, 3664

\bibitem[\protect\citeauthoryear{{Ogura}, {Sugitani}  \& {Pickles}}{{Ogura}
  et~al.}{2002}]{Ogura2002}
{Ogura} K.,  {Sugitani} K.,   {Pickles} A.,  2002, \mn@doi [\apj]
  {10.1086/339976}, \href {http://adsabs.harvard.edu/abs/2002AJ....123.2597O}
  {123, 2597}

\bibitem[\protect\citeauthoryear{{Oh}, {Kobayashi}, {Honma}, {Hirota}, {Sato}
  \& {Ueno}}{{Oh} et~al.}{2010}]{Oh2010}
{Oh} C.~S.,  {Kobayashi} H.,  {Honma} M.,  {Hirota} T.,  {Sato} K.,   {Ueno}
  Y.,  2010, \mn@doi [\pasj] {10.1093/pasj/62.1.101}, \href
  {http://adsabs.harvard.edu/abs/2010PASJ...62..101O} {62, 101}

\bibitem[\protect\citeauthoryear{Osorio et~al.,}{Osorio
  et~al.}{2017}]{Osorio2017}
Osorio M.,  et~al., 2017, \mn@doi [ApJ] {10.3847/1538-4357/aa6975}, 840, 36

\bibitem[\protect\citeauthoryear{{Palau} et~al.,}{{Palau}
  et~al.}{2011}]{Palau2011}
{Palau} A.,  et~al., 2011, \mn@doi [\apjl] {10.1088/2041-8205/743/2/L32}, \href
  {http://adsabs.harvard.edu/abs/2011ApJ...743L..32P} {743, L32}

\bibitem[\protect\citeauthoryear{{Palau} et~al.,}{{Palau}
  et~al.}{2013}]{Palau2013}
{Palau} A.,  et~al., 2013, \mn@doi [\apj] {10.1088/0004-637X/762/2/120}, \href
  {http://adsabs.harvard.edu/abs/2013ApJ...762..120P} {762, 120}

\bibitem[\protect\citeauthoryear{{Panagia} \& {Felli}}{{Panagia} \&
  {Felli}}{1975}]{Panagia1975}
{Panagia} N.,  {Felli} M.,  1975, \aap, \href
  {http://adsabs.harvard.edu/abs/1975A\%26A....39....1P} {39, 1}

\bibitem[\protect\citeauthoryear{{Pelletier} \& {Pudritz}}{{Pelletier} \&
  {Pudritz}}{1992}]{Pelletier1992}
{Pelletier} G.,  {Pudritz} R.~E.,  1992, \mn@doi [\apj] {10.1086/171565}, \href
  {http://adsabs.harvard.edu/abs/1992ApJ...394..117P} {394, 117}

\bibitem[\protect\citeauthoryear{Proga, Stone  \& Drew}{Proga
  et~al.}{1998}]{Proga1998}
Proga D.,  Stone J.~M.,   Drew J.~E.,  1998, \mn@doi [MNRAS]
  {10.1046/j.1365-8711.1998.01337.x}, 295, 595

\bibitem[\protect\citeauthoryear{{Purcell} et~al.,}{{Purcell}
  et~al.}{2013}]{Purcell2013}
{Purcell} C.~R.,  et~al., 2013, \mn@doi [\apjs] {10.1088/0067-0049/205/1/1},
  \href {http://adsabs.harvard.edu/abs/2013ApJS..205....1P} {205, 1}

\bibitem[\protect\citeauthoryear{{Purser} et~al.,}{{Purser}
  et~al.}{2016}]{Purser2016}
{Purser} S. J.~D.,  et~al., 2016, \mn@doi [MNRAS] {10.1093/mnras/stw1027}, 460,
  1039

\bibitem[\protect\citeauthoryear{Purser, Ainsworth, Ray, Green, Taylor  \&
  Scaife}{Purser et~al.}{2018}]{Purser2018b}
Purser S. J.~D.,  Ainsworth R.~E.,  Ray T.~P.,  Green D.~A.,  Taylor A.~M.,
  Scaife A. M.~M.,  2018, \mn@doi [MNRAS] {10.1093/mnras/sty2649}, 481, 5532

\bibitem[\protect\citeauthoryear{{Rathborne}, {Jackson}  \&
  {Simon}}{{Rathborne} et~al.}{2006}]{Rathborne2006}
{Rathborne} J.~M.,  {Jackson} J.~M.,   {Simon} R.,  2006, \mn@doi [\apj]
  {10.1086/500423}, \href
  {http://ukads.nottingham.ac.uk/abs/2006ApJ...641..389R} {641, 389}

\bibitem[\protect\citeauthoryear{{Rathborne}, {Simon}  \&
  {Jackson}}{{Rathborne} et~al.}{2007}]{Rathborne2007}
{Rathborne} J.~M.,  {Simon} R.,   {Jackson} J.~M.,  2007, \mn@doi [ApJ]
  {10.1086/513178}, \href {http://adsabs.harvard.edu/abs/2007ApJ...662.1082R}
  {662, 1082}

\bibitem[\protect\citeauthoryear{{Rathborne}, {Jackson}, {Chambers},
  {Stojimirovic}, {Simon}, {Shipman}  \& {Frieswijk}}{{Rathborne}
  et~al.}{2010}]{Rathborne2010}
{Rathborne} J.~M.,  {Jackson} J.~M.,  {Chambers} E.~T.,  {Stojimirovic} I.,
  {Simon} R.,  {Shipman} R.,   {Frieswijk} W.,  2010, \mn@doi [\apj]
  {10.1088/0004-637X/715/1/310}, \href
  {http://adsabs.harvard.edu/abs/2010ApJ...715..310R} {715, 310}

\bibitem[\protect\citeauthoryear{{Ray}, {Poetzel}, {Solf}  \& {Mundt}}{{Ray}
  et~al.}{1990}]{Ray1990}
{Ray} T.~P.,  {Poetzel} R.,  {Solf} J.,   {Mundt} R.,  1990, \mn@doi [\apjl]
  {10.1086/185762}, \href {http://adsabs.harvard.edu/abs/1990ApJ...357L..45R}
  {357, L45}

\bibitem[\protect\citeauthoryear{{Reid} et~al.,}{{Reid}
  et~al.}{2019}]{Reid2019}
{Reid} M.~J.,  et~al., 2019, \mn@doi [\apj] {10.3847/1538-4357/ab4a11}, \href
  {https://ui.adsabs.harvard.edu/abs/2019ApJ...885..131R} {885, 131}

\bibitem[\protect\citeauthoryear{{Rengarajan} \& {Ho}}{{Rengarajan} \&
  {Ho}}{1996}]{RengarajanHo1996}
{Rengarajan} T.~N.,  {Ho} P.~T.~P.,  1996, \mn@doi [\apj] {10.1086/177425},
  \href {http://adsabs.harvard.edu/abs/1996ApJ...465..363R} {465, 363}

\bibitem[\protect\citeauthoryear{{Reynolds}}{{Reynolds}}{1986}]{Reynolds1986}
{Reynolds} S.~P.,  1986, \mn@doi [ApJ] {10.1086/164209}, \href
  {http://adsabs.harvard.edu/abs/1986ApJ...304..713R} {304, 713}

\bibitem[\protect\citeauthoryear{{Rod{\'o}n}, {Beuther}, {Megeath}  \& {van der
  Tak}}{{Rod{\'o}n} et~al.}{2008}]{Rodon2008}
{Rod{\'o}n} J.~A.,  {Beuther} H.,  {Megeath} S.~T.,   {van der Tak} F.~F.~S.,
  2008, \mn@doi [\aap] {10.1051/0004-6361:200810158}, \href
  {http://adsabs.harvard.edu/abs/2008A\%26A...490..213R} {490, 213}

\bibitem[\protect\citeauthoryear{{Rodr\'{i}guez-Esnard}, {Mignes}  \&
  {Trinidad}}{{Rodr\'{i}guez-Esnard} et~al.}{2014}]{RodriguezEsnard2014}
{Rodr\'{i}guez-Esnard} T.,  {Mignes} V.,   {Trinidad} M.~A.,  2014, \apj, 788,
  176

\bibitem[\protect\citeauthoryear{{Rodr{\'{\i}}guez-Kamenetzky},
  {Carrasco-Gonz{\'a}lez}, {Araudo}, {Torrelles}, {Anglada}, {Mart{\'{\i}}},
  {Rodr{\'{\i}}guez}  \& {Valotto}}{{Rodr{\'{\i}}guez-Kamenetzky}
  et~al.}{2016}]{RodriguezKamenetzky2016}
{Rodr{\'{\i}}guez-Kamenetzky} A.,  {Carrasco-Gonz{\'a}lez} C.,  {Araudo} A.,
  {Torrelles} J.~M.,  {Anglada} G.,  {Mart{\'{\i}}} J.,  {Rodr{\'{\i}}guez}
  L.~F.,   {Valotto} C.,  2016, \mn@doi [\apj] {10.3847/0004-637X/818/1/27},
  \href {http://adsabs.harvard.edu/abs/2016ApJ...818...27R} {818, 27}

\bibitem[\protect\citeauthoryear{{Rodr{\'\i}guez}, {Trinidad}  \&
  {Migenes}}{{Rodr{\'\i}guez} et~al.}{2012}]{Rodriguez2012b}
{Rodr{\'\i}guez} T.,  {Trinidad} M.~A.,   {Migenes} V.,  2012, \mn@doi [\apj]
  {10.1088/0004-637X/755/2/100}, \href
  {https://ui.adsabs.harvard.edu/abs/2012ApJ...755..100R} {755, 100}

\bibitem[\protect\citeauthoryear{{Rosero} et~al.,}{{Rosero}
  et~al.}{2016}]{Rosero2016}
{Rosero} V.,  et~al., 2016, \mn@doi [\apjs] {10.3847/1538-4365/227/2/25}, \href
  {https://ui.adsabs.harvard.edu/abs/2016ApJS..227...25R} {227, 25}

\bibitem[\protect\citeauthoryear{Rosero et~al.,}{Rosero
  et~al.}{2019}]{Rosero2019}
Rosero V.,  et~al., 2019, \mn@doi [ApJ] {10.3847/1538-4357/ab2595}, 880, 99

\bibitem[\protect\citeauthoryear{{Rygl}, {Brunthaler}, {Reid}, {Menten}, {van
  Langevelde}  \& {Xu}}{{Rygl} et~al.}{2010}]{Rygl2010}
{Rygl} K.~L.~J.,  {Brunthaler} A.,  {Reid} M.~J.,  {Menten} K.~M.,  {van
  Langevelde} H.~J.,   {Xu} Y.,  2010, \mn@doi [A\&A]
  {10.1051/0004-6361/200913135}, \href
  {http://adsabs.harvard.edu/abs/2010A\%26A...511A...2R} {511, A2}

\bibitem[\protect\citeauthoryear{{Rygl} et~al.,}{{Rygl}
  et~al.}{2012}]{Rygl2012}
{Rygl} K.~L.~J.,  et~al., 2012, \mn@doi [\aap] {10.1051/0004-6361/201118211},
  \href {http://adsabs.harvard.edu/abs/2012A\%26A...539A..79R} {539, A79}

\bibitem[\protect\citeauthoryear{{Rygl} et~al.,}{{Rygl}
  et~al.}{2014}]{Rygl2014}
{Rygl} K.~L.~J.,  et~al., 2014, \mn@doi [\mnras] {10.1093/mnras/stu300}, \href
  {http://adsabs.harvard.edu/abs/2014MNRAS.440..427R} {440, 427}

\bibitem[\protect\citeauthoryear{{Saito}, {Saito}, {Moriguchi}  \&
  {Fukui}}{{Saito} et~al.}{2006}]{Saito2006}
{Saito} H.,  {Saito} M.,  {Moriguchi} Y.,   {Fukui} Y.,  2006, \mn@doi [\pasj]
  {10.1093/pasj/58.2.343}, \href
  {http://adsabs.harvard.edu/abs/2006PASJ...58..343S} {58, 343}

\bibitem[\protect\citeauthoryear{{Saito}, {Saito}, {Yonekura}  \&
  {Nakamura}}{{Saito} et~al.}{2008}]{Saito2008}
{Saito} H.,  {Saito} M.,  {Yonekura} Y.,   {Nakamura} F.,  2008, \mn@doi
  [\apjs] {10.1086/590146}, \href
  {http://adsabs.harvard.edu/abs/2008ApJS..178..302S} {178, 302}

\bibitem[\protect\citeauthoryear{{S{\'a}nchez-Monge}, {Palau}, {Estalella},
  {Beltr{\'a}n}  \& {Girart}}{{S{\'a}nchez-Monge}
  et~al.}{2008}]{SanchezMonge2008}
{S{\'a}nchez-Monge} {\'A}.,  {Palau} A.,  {Estalella} R.,  {Beltr{\'a}n} M.~T.,
    {Girart} J.~M.,  2008, \mn@doi [\aap] {10.1051/0004-6361:20078406}, \href
  {http://adsabs.harvard.edu/abs/2008A\%26A...485..497S} {485, 497}

\bibitem[\protect\citeauthoryear{{S{\'a}nchez-Monge}
  et~al.,}{{S{\'a}nchez-Monge} et~al.}{2014}]{SanchezMonge2014}
{S{\'a}nchez-Monge} {\'A}.,  et~al., 2014, \mn@doi [A\&A]
  {10.1051/0004-6361/201424032}, \href
  {http://adsabs.harvard.edu/abs/2014A\%26A...569A..11S} {569, A11}

\bibitem[\protect\citeauthoryear{{Sandell}, {Weintraub}  \&
  {Hamidouche}}{{Sandell} et~al.}{2011}]{Sandell2011}
{Sandell} G.,  {Weintraub} D.~A.,   {Hamidouche} M.,  2011, \mn@doi [\apj]
  {10.1088/0004-637X/727/1/26}, \href
  {http://adsabs.harvard.edu/abs/2011ApJ...727...26S} {727, 26}

\bibitem[\protect\citeauthoryear{{Sanna, A.}, {Moscadelli, L.}, {Goddi, C.},
  {Krishnan, V.}  \& {Massi, F.}}{{Sanna, A.} et~al.}{2018}]{Sanna2018}
{Sanna, A.} {Moscadelli, L.} {Goddi, C.} {Krishnan, V.}  {Massi, F.} 2018,
  \mn@doi [A\&A] {10.1051/0004-6361/201833573}, 619, A107

\bibitem[\protect\citeauthoryear{{Sanna, A.} et~al.,}{{Sanna, A.}
  et~al.}{2019}]{Sanna2019}
{Sanna, A.} et~al., 2019, \mn@doi [A\&A] {10.1051/0004-6361/201834551}, 623, L3

\bibitem[\protect\citeauthoryear{{Sanna}, {Moscadelli}, {Cesaroni}, {Caratti o
  Garatti}, {Goddi}  \& {Carrasco-Gonz{\'a}lez}}{{Sanna}
  et~al.}{2016}]{Sanna2016}
{Sanna} A.,  {Moscadelli} L.,  {Cesaroni} R.,  {Caratti o Garatti} A.,  {Goddi}
  C.,   {Carrasco-Gonz{\'a}lez} C.,  2016, \mn@doi [A\&A]
  {10.1051/0004-6361/201629544}, \href
  {http://adsabs.harvard.edu/abs/2016A%26A...596L...2S} {596, L2}

\bibitem[\protect\citeauthoryear{{Schreyer}, {Henning}, {van der Tak},
  {Boonman}  \& {van Dishoeck}}{{Schreyer} et~al.}{2002}]{Schreyer2002}
{Schreyer} K.,  {Henning} T.,  {van der Tak} F.~F.~S.,  {Boonman} A.~M.~S.,
  {van Dishoeck} E.~F.,  2002, \mn@doi [\aap] {10.1051/0004-6361:20021160},
  \href {http://adsabs.harvard.edu/abs/2002A\%26A...394..561S} {394, 561}

\bibitem[\protect\citeauthoryear{{Schreyer}, {Semenov}, {Henning}  \&
  {Forbrich}}{{Schreyer} et~al.}{2006}]{Schreyer2006}
{Schreyer} K.,  {Semenov} D.,  {Henning} T.,   {Forbrich} J.,  2006, \mn@doi
  [\apjl] {10.1086/500732}, \href
  {http://adsabs.harvard.edu/abs/2006ApJ...637L.129S} {637, L129}

\bibitem[\protect\citeauthoryear{{Schulz}, {Black}, {Lada}, {Ulich}, {Martin},
  {Snell}  \& {Erickson}}{{Schulz} et~al.}{1989}]{Schulz1989}
{Schulz} A.,  {Black} J.~H.,  {Lada} C.~J.,  {Ulich} B.~L.,  {Martin} R.~N.,
  {Snell} R.~L.,   {Erickson} N.~J.,  1989, \mn@doi [\apj] {10.1086/167492},
  \href {http://adsabs.harvard.edu/abs/1989ApJ...341..288S} {341, 288}

\bibitem[\protect\citeauthoryear{{Shepherd}, {Testi}  \& {Stark}}{{Shepherd}
  et~al.}{2003}]{Shepherd2003}
{Shepherd} D.~S.,  {Testi} L.,   {Stark} D.~P.,  2003, \mn@doi [\apj]
  {10.1086/345743}, \href {http://adsabs.harvard.edu/abs/2003ApJ...584..882S}
  {584, 882}

\bibitem[\protect\citeauthoryear{{Shimoikura} et~al.,}{{Shimoikura}
  et~al.}{2013}]{Shimoikura2013}
{Shimoikura} T.,  et~al., 2013, \mn@doi [\apj] {10.1088/0004-637X/768/1/72},
  \href {http://adsabs.harvard.edu/abs/2013ApJ...768...72S} {768, 72}

\bibitem[\protect\citeauthoryear{Shu \& Shang}{Shu \& Shang}{1997}]{Shu1997}
Shu F.~H.,  Shang H.,  1997, \mn@doi [Symposium - International Astronomical
  Union] {10.1017/S0074180900061672}, 182, 225–239

\bibitem[\protect\citeauthoryear{{Shu}, {Najita}, {Ostriker}, {Wilkin}, {Ruden}
   \& {Lizano}}{{Shu} et~al.}{1994}]{Shu1994}
{Shu} F.,  {Najita} J.,  {Ostriker} E.,  {Wilkin} F.,  {Ruden} S.,   {Lizano}
  S.,  1994, \mn@doi [\apj] {10.1086/174363}, \href
  {http://adsabs.harvard.edu/abs/1994ApJ...429..781S} {429, 781}

\bibitem[\protect\citeauthoryear{{Simon}, {Rathborne}, {Shah}, {Jackson}  \&
  {Chambers}}{{Simon} et~al.}{2006}]{Simon2006}
{Simon} R.,  {Rathborne} J.~M.,  {Shah} R.~Y.,  {Jackson} J.~M.,   {Chambers}
  E.~T.,  2006, \mn@doi [ApJ] {10.1086/508915}, \href
  {http://adsabs.harvard.edu/abs/2006ApJ...653.1325S} {653, 1325}

\bibitem[\protect\citeauthoryear{{Smith} \& {Fischer}}{{Smith} \&
  {Fischer}}{1992}]{Smith1992}
{Smith} H.~A.,  {Fischer} J.,  1992, \mn@doi [\apjl] {10.1086/186586}, \href
  {http://adsabs.harvard.edu/abs/1992ApJ...398L..99S} {398, L99}

\bibitem[\protect\citeauthoryear{{Snell}, {Huang}, {Dickman}  \&
  {Claussen}}{{Snell} et~al.}{1988}]{Snell1988}
{Snell} R.~L.,  {Huang} Y.-L.,  {Dickman} R.~L.,   {Claussen} M.~J.,  1988,
  \mn@doi [\apj] {10.1086/166056}, \href
  {http://adsabs.harvard.edu/abs/1988ApJ...325..853S} {325, 853}

\bibitem[\protect\citeauthoryear{{Sridharan}, {Beuther}, {Schilke}, {Menten}
  \& {Wyrowski}}{{Sridharan} et~al.}{2002}]{Sridharan2002}
{Sridharan} T.~K.,  {Beuther} H.,  {Schilke} P.,  {Menten} K.~M.,   {Wyrowski}
  F.,  2002, \mn@doi [\apj] {10.1086/338332}, \href
  {http://adsabs.harvard.edu/abs/2002ApJ...566..931S} {566, 931}

\bibitem[\protect\citeauthoryear{{Su}, {Zhang}  \& {Lim}}{{Su}
  et~al.}{2004}]{Su2004}
{Su} Y.-N.,  {Zhang} Q.,   {Lim} J.,  2004, \mn@doi [\apj] {10.1086/381880},
  \href {http://adsabs.harvard.edu/abs/2004ApJ...604..258S} {604, 258}

\bibitem[\protect\citeauthoryear{{Sugitani}, {Fukui}, {Mizuni}  \&
  {Ohashi}}{{Sugitani} et~al.}{1989}]{Sugitani1989}
{Sugitani} K.,  {Fukui} Y.,  {Mizuni} A.,   {Ohashi} N.,  1989, \mn@doi [\apjl]
  {10.1086/185491}, \href {http://adsabs.harvard.edu/abs/1989ApJ...342L..87S}
  {342, L87}

\bibitem[\protect\citeauthoryear{{Surcis}, {Vlemmings}, {van Langevelde},
  {Hutawarakorn Kramer}  \& {Quiroga-Nu{\~n}ez}}{{Surcis}
  et~al.}{2013}]{Surcis2013}
{Surcis} G.,  {Vlemmings} W.~H.~T.,  {van Langevelde} H.~J.,  {Hutawarakorn
  Kramer} B.,   {Quiroga-Nu{\~n}ez} L.~H.,  2013, \mn@doi [A\&A]
  {10.1051/0004-6361/201321501}, \href
  {http://adsabs.harvard.edu/abs/2013A\%26A...556A..73S} {556, A73}

\bibitem[\protect\citeauthoryear{{Surcis}, {Vlemmings}, {van Langevelde},
  {Hutawarakorn Kramer}, {Bartkiewicz}  \& {Blasi}}{{Surcis}
  et~al.}{2015}]{Surcis2015}
{Surcis} G.,  {Vlemmings} W.~H.~T.,  {van Langevelde} H.~J.,  {Hutawarakorn
  Kramer} B.,  {Bartkiewicz} A.,   {Blasi} M.~G.,  2015, \mn@doi [\aap]
  {10.1051/0004-6361/201425420}, \href
  {http://adsabs.harvard.edu/abs/2015A\%26A...578A.102S} {578, A102}

\bibitem[\protect\citeauthoryear{{Tamura}, {Gatley}, {Joyce}, {Ueno}, {Suto}
  \& {Sekiguchi}}{{Tamura} et~al.}{1991}]{Tamura1991}
{Tamura} M.,  {Gatley} I.,  {Joyce} R.~R.,  {Ueno} M.,  {Suto} H.,
  {Sekiguchi} M.,  1991, \mn@doi [\apj] {10.1086/170462}, \href
  {http://adsabs.harvard.edu/abs/1991ApJ...378..611T} {378, 611}

\bibitem[\protect\citeauthoryear{{Thompson}}{{Thompson}}{1984}]{Thompson1984}
{Thompson} R.~I.,  1984, \mn@doi [ApJ] {10.1086/162287}, \href
  {http://adsabs.harvard.edu/abs/1984ApJ...283..165T} {283, 165}

\bibitem[\protect\citeauthoryear{{Tieftrunk}, {Gaume}, {Claussen}, {Wilson}  \&
  {Johnston}}{{Tieftrunk} et~al.}{1997}]{Tieftrunk1997}
{Tieftrunk} A.~R.,  {Gaume} R.~A.,  {Claussen} M.~J.,  {Wilson} T.~L.,
  {Johnston} K.~J.,  1997, \aap, \href
  {http://adsabs.harvard.edu/abs/1997A\%26A...318..931T} {318, 931}

\bibitem[\protect\citeauthoryear{{Tofani}, {Felli}, {Taylor}  \&
  {Hunter}}{{Tofani} et~al.}{1995}]{Tofani1995}
{Tofani} G.,  {Felli} M.,  {Taylor} G.~B.,   {Hunter} T.~R.,  1995, \aaps,
  \href {http://adsabs.harvard.edu/abs/1995A\%26AS..112..299T} {112, 299}

\bibitem[\protect\citeauthoryear{{Trinidad}, {Curiel}, {Torrelles},
  {Rodr{\'{\i}}guez}, {Migenes}  \& {Patel}}{{Trinidad}
  et~al.}{2006}]{Trinidad2006}
{Trinidad} M.~A.,  {Curiel} S.,  {Torrelles} J.~M.,  {Rodr{\'{\i}}guez} L.~F.,
  {Migenes} V.,   {Patel} N.,  2006, \mn@doi [\apj] {10.1086/507127}, \href
  {http://adsabs.harvard.edu/abs/2006AJ....132.1918T} {132, 1918}

\bibitem[\protect\citeauthoryear{{Urquhart, J. S.} et~al.,}{{Urquhart, J. S.}
  et~al.}{2008}]{Urquhart2008}
{Urquhart, J. S.} et~al., 2008, \mn@doi [A\&A] {10.1051/0004-6361:200809415},
  487, 253

\bibitem[\protect\citeauthoryear{{Urquhart} et~al.,}{{Urquhart}
  et~al.}{2009}]{Urquhart2009VLA}
{Urquhart} J.~S.,  et~al., 2009, \mn@doi [A\&A] {10.1051/0004-6361/200912108},
  \href {http://adsabs.harvard.edu/abs/2009A%26A...501..539U} {501, 539}

\bibitem[\protect\citeauthoryear{{Urquhart} et~al.,}{{Urquhart}
  et~al.}{2011}]{Urquhart2011}
{Urquhart} J.~S.,  et~al., 2011, \mn@doi [\mnras]
  {10.1111/j.1365-2966.2011.19594.x}, \href
  {http://adsabs.harvard.edu/abs/2011MNRAS.418.1689U} {418, 1689}

\bibitem[\protect\citeauthoryear{{Urquhart} et~al.,}{{Urquhart}
  et~al.}{2014}]{Urquhart2014}
{Urquhart} J.~S.,  et~al., 2014, \mn@doi [MNRAS] {10.1093/mnras/stu1207}, \href
  {http://adsabs.harvard.edu/abs/2014MNRAS.443.1555U} {443, 1555}

\bibitem[\protect\citeauthoryear{{Varricatt}, {Davis}, {Ramsay}  \&
  {Todd}}{{Varricatt} et~al.}{2010}]{Varricatt2010}
{Varricatt} W.~P.,  {Davis} C.~J.,  {Ramsay} S.,   {Todd} S.~P.,  2010, \mn@doi
  [\mnras] {10.1111/j.1365-2966.2010.16356.x}, \href
  {http://adsabs.harvard.edu/abs/2010MNRAS.404..661V} {404, 661}

\bibitem[\protect\citeauthoryear{{Vink}, {de Koter}  \& {Lamers}}{{Vink}
  et~al.}{2001}]{Vink2001}
{Vink} J.~S.,  {de Koter} A.,   {Lamers} H.~J.~G.~L.~M.,  2001, \mn@doi [\aap]
  {10.1051/0004-6361:20010127}, \href
  {http://adsabs.harvard.edu/abs/2001A\%26A...369..574V} {369, 574}

\bibitem[\protect\citeauthoryear{{Viti}, {Collings}, {Dever}, {McCoustra}  \&
  {Williams}}{{Viti} et~al.}{2004}]{Viti2004}
{Viti} S.,  {Collings} M.~P.,  {Dever} J.~W.,  {McCoustra} M. R.~S.,
  {Williams} D.~A.,  2004, \mn@doi [MNRAS] {10.1111/j.1365-2966.2004.08273.x},
  354, 1141

\bibitem[\protect\citeauthoryear{{Wang}, {Zhang}, {Pillai}, {Wyrowski}  \&
  {Wu}}{{Wang} et~al.}{2008}]{Wang2008}
{Wang} Y.,  {Zhang} Q.,  {Pillai} T.,  {Wyrowski} F.,   {Wu} Y.,  2008, \mn@doi
  [\apjl] {10.1086/524949}, \href
  {http://adsabs.harvard.edu/abs/2008ApJ...672L..33W} {672, L33}

\bibitem[\protect\citeauthoryear{{Wang} et~al.,}{{Wang}
  et~al.}{2016}]{Wang2016}
{Wang} Y.,  et~al., 2016, \mn@doi [\aap] {10.1051/0004-6361/201526637}, \href
  {http://adsabs.harvard.edu/abs/2016A\%26A...587A..69W} {587, A69}

\bibitem[\protect\citeauthoryear{{Weintraub} \& {Kastner}}{{Weintraub} \&
  {Kastner}}{1996}]{WeintraubKastner1996}
{Weintraub} D.~A.,  {Kastner} J.~H.,  1996, in {Roberge} W.~G.,  {Whittet}
  D.~C.~B.,  eds,  Astronomical Society of the Pacific Conference Series Vol.
  97, Polarimetry of the Interstellar Medium. p.~345

\bibitem[\protect\citeauthoryear{Wenger, Balser, Anderson  \& Bania}{Wenger
  et~al.}{2018}]{Wenger2018}
Wenger T.~V.,  Balser D.~S.,  Anderson L.~D.,   Bania T.~M.,  2018, \mn@doi
  [ApJ] {10.3847/1538-4357/aaaec8}, 856, 52

\bibitem[\protect\citeauthoryear{{Wilson}, {Boboltz}, {Gaume}  \&
  {Megeath}}{{Wilson} et~al.}{2003}]{Wilson2003}
{Wilson} T.~L.,  {Boboltz} D.~A.,  {Gaume} R.~A.,   {Megeath} S.~T.,  2003,
  \mn@doi [\apj] {10.1086/378233}, \href
  {http://adsabs.harvard.edu/abs/2003ApJ...597..434W} {597, 434}

\bibitem[\protect\citeauthoryear{{Wright} \& {Barlow}}{{Wright} \&
  {Barlow}}{1975}]{WrightBarlow1975}
{Wright} A.~E.,  {Barlow} M.~J.,  1975, \mn@doi [\mnras]
  {10.1093/mnras/170.1.41}, \href
  {http://adsabs.harvard.edu/abs/1975MNRAS.170...41W} {170, 41}

\bibitem[\protect\citeauthoryear{{Wu, Y. W.} et~al.,}{{Wu, Y. W.}
  et~al.}{2014}]{Wu2014}
{Wu, Y. W.} et~al., 2014, \mn@doi [A\&A] {10.1051/0004-6361/201322765}, 566,
  A17

\bibitem[\protect\citeauthoryear{{Wu}, {Xu}, {Pandian}, {Yang}, {Henkel},
  {Menten}  \& {Zhang}}{{Wu} et~al.}{2010}]{Wu2010}
{Wu} Y.~W.,  {Xu} Y.,  {Pandian} J.~D.,  {Yang} J.,  {Henkel} C.,  {Menten}
  K.~M.,   {Zhang} S.~B.,  2010, \mn@doi [\apj] {10.1088/0004-637X/720/1/392},
  \href {http://adsabs.harvard.edu/abs/2010ApJ...720..392W} {720, 392}

\bibitem[\protect\citeauthoryear{{Wu}, {Xu}  \& {Yang}}{{Wu}
  et~al.}{2011}]{YuanWeiWu2011}
{Wu} Y.-W.,  {Xu} Y.,   {Yang} J.,  2011, \mn@doi [Research in Astronomy and
  Astrophysics] {10.1088/1674-4527/11/2/002}, \href
  {http://adsabs.harvard.edu/abs/2011RAA....11..137W} {11, 137}

\bibitem[\protect\citeauthoryear{{Xu}, {Wang}  \& {Qin}}{{Xu}
  et~al.}{2012}]{Xu2012}
{Xu} J.-L.,  {Wang} J.-J.,   {Qin} S.-L.,  2012, \mn@doi [\aap]
  {10.1051/0004-6361/201118657}, \href
  {http://adsabs.harvard.edu/abs/2012A\%26A...540L..13X} {540, L13}

\bibitem[\protect\citeauthoryear{{Yan}, {Minh}, {Wang}, {Su}  \&
  {Ginsburg}}{{Yan} et~al.}{2010}]{Yan2010}
{Yan} C.-H.,  {Minh} Y.~C.,  {Wang} S.-Y.,  {Su} Y.-N.,   {Ginsburg} A.,  2010,
  \mn@doi [\apj] {10.1088/0004-637X/720/1/1}, \href
  {http://adsabs.harvard.edu/abs/2010ApJ...720....1Y} {720, 1}

\bibitem[\protect\citeauthoryear{{Zapata}, {Rodr{\'{\i}}guez}  \&
  {Kurtz}}{{Zapata} et~al.}{2001}]{Zapata2001}
{Zapata} L.~A.,  {Rodr{\'{\i}}guez} L.~F.,   {Kurtz} S.~E.,  2001, \rmxaa,
  \href {http://adsabs.harvard.edu/abs/2001RMxAA..37...83Z} {37, 83}

\bibitem[\protect\citeauthoryear{{Zhang}, {Hunter}, {Brand}, {Sridharan},
  {Cesaroni}, {Molinari}, {Wang}  \& {Kramer}}{{Zhang}
  et~al.}{2005}]{Zhang2005}
{Zhang} Q.,  {Hunter} T.~R.,  {Brand} J.,  {Sridharan} T.~K.,  {Cesaroni} R.,
  {Molinari} S.,  {Wang} J.,   {Kramer} M.,  2005, \mn@doi [\apj]
  {10.1086/429660}, \href {http://adsabs.harvard.edu/abs/2005ApJ...625..864Z}
  {625, 864}

\bibitem[\protect\citeauthoryear{{Zhang}, {Hunter}, {Beuther}, {Sridharan},
  {Liu}, {Su}, {Chen}  \& {Chen}}{{Zhang} et~al.}{2007}]{Zhang2007}
{Zhang} Q.,  {Hunter} T.~R.,  {Beuther} H.,  {Sridharan} T.~K.,  {Liu} S.-Y.,
  {Su} Y.-N.,  {Chen} H.-R.,   {Chen} Y.,  2007, \mn@doi [\apj]
  {10.1086/511381}, \href {http://adsabs.harvard.edu/abs/2007ApJ...658.1152Z}
  {658, 1152}

\bibitem[\protect\citeauthoryear{{Zhang}, {Zheng}, {Reid}, {Menten}, {Xu},
  {Moscadelli}  \& {Brunthaler}}{{Zhang} et~al.}{2009}]{Zhang2009}
{Zhang} B.,  {Zheng} X.~W.,  {Reid} M.~J.,  {Menten} K.~M.,  {Xu} Y.,
  {Moscadelli} L.,   {Brunthaler} A.,  2009, \mn@doi [ApJ]
  {10.1088/0004-637X/693/1/419}, \href
  {http://adsabs.harvard.edu/abs/2009ApJ...693..419Z} {693, 419}

\bibitem[\protect\citeauthoryear{{Zhang} et~al.,}{{Zhang}
  et~al.}{2013}]{Zhang2013}
{Zhang} Y.,  et~al., 2013, \mn@doi [ApJ] {10.1088/0004-637X/767/1/58}, \href
  {http://adsabs.harvard.edu/abs/2013ApJ...767...58Z} {767, 58}

\bibitem[\protect\citeauthoryear{{Zinchenko} et~al.,}{{Zinchenko}
  et~al.}{2015}]{Zinchenko2015}
{Zinchenko} I.,  et~al., 2015, \mn@doi [\apj] {10.1088/0004-637X/810/1/10},
  \href {http://adsabs.harvard.edu/abs/2015ApJ...810...10Z} {810, 10}

\bibitem[\protect\citeauthoryear{{van der Tak}, {Tuthill}  \& {Danchi}}{{van
  der Tak} et~al.}{2005}]{VanDerTak2005b}
{van der Tak} F.~F.~S.,  {Tuthill} P.~G.,   {Danchi} W.~C.,  2005, \mn@doi
  [A\&A] {10.1051/0004-6361:20041595}, \href
  {http://adsabs.harvard.edu/abs/2005A\%26A...431..993V} {431, 993}

\makeatother
\end{thebibliography}
